\documentclass[a4paper,12pt,numbered, print,index]{Classes/PhDThesisPSnPDF}

\input{Preamble/preamble}

\title{Gravitational Effective Field Theories and Black Hole Mechanics}

\author{Iain Davies}

\dept{Department of Applied Mathematics and Theoretical Physics}

\university{University of Cambridge}
\crest{\includegraphics[width=0.2\textwidth]{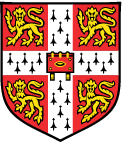}}

\supervisor{Prof. Harvey S. Reall}

\degreetitle{Doctor of Philosophy}

\college{Trinity College}

\degreedate{August 2024} 

\subject{Gravity} \keywords{{General Relativity} {Black Holes} {Black Hole Entropy} {Effective Field Theory}}

\ifdefineAbstract
\fi

\ifdefineChapter
\fi

\begin{document}

\frontmatter

\maketitle


\begin{declaration}

This thesis is the result of my own work and includes nothing which is the outcome of work done in collaboration except as declared in the preface and specified in the text. It is not substantially the same as any work that has already been submitted, or is being concurrently submitted, for any degree, diploma or other qualification at the University of Cambridge or any other University or similar institution except as declared in the preface and specified in the text. It does not exceed the prescribed word limit for the relevant Degree Committee.


\end{declaration}
\begin{abstract}
General Relativity (GR) has been immensely successful at predicting gravitational phenomena at the energy scales we have observed thus far. However, we know GR will lose its predictive power at sufficiently high energy scales due to its nonrenormalizability. Therefore, we should think of GR as the low energy limit of some unknown UV-complete theory of gravity which we have been unable to probe so far, and thus we should develop the mathematical theory around possible corrections to GR. One approach is the framework of effective field theory (EFT), where the Lagrangian is treated as a series of higher derivative terms scaling with some UV parameter beyond which we have "integrated out" the unknown physics. 

This thesis is concerned with studying two issues in gravitational EFTs. First, we would like these theories to be mathematically healthy and thus should ask whether their equations are well-posed. Previous work has answered this in the affirmative for certain EFTs of gravity and the simplest form of matter, a scalar field. Here, we will demonstrate this remains true with the inclusion of a more complicated matter field, the electromagnetic field. Specifically, we show that a "modified harmonic" gauge produces a strongly hyperbolic formulation of the leading order Einstein-Maxwell EFT so long as the higher derivative terms are small. 

Second, the thermodynamic nature of black holes is independent of the theory of gravity, therefore we should expect the laws of black hole mechanics to remain valid in EFT. Here we will demonstrate this is indeed the case for EFTs of gravity and broad classes of matter fields. We show that the zeroth law can be proved for EFTs of gravity, electromagnetism and a charged or uncharged scalar field. We find we must modify the statements of the first and second laws by generalizing the formula of black hole entropy from the Bekenstein-Hawking entropy used in GR. For stationary black holes, the Wald entropy is a sufficient definition to satisfy the first law in EFT. For dynamical black holes we show that with further corrections we can prove a non-perturbative second law. We discuss the gauge dependence of this definition of dynamical black hole entropy and provide explicit constructions for specific EFTs.
\end{abstract}


\begin{acknowledgements}      

My first thanks must go to my supervisor, Prof Harvey Reall. The most productive and enlightening hours of my last four years were, without exception, those spent in discussion with Harvey. His expert guidance, generosity of time and gentle-but-firm prompting were precisely the invaluable resources I needed to produce this thesis. 

Next, to my family. My parents have been my most enduring supporters, in any walk of life I have taken. To my father, I owe a strong faith in my own problem solving abilities and a belief in what is right in the world. My mother has shown me how to treat others with empathy and dignity, and I aspire to her emotional and academic intelligence. My grandfather Keri instilled in me a love of mathematics and physics from as far back as I can remember. My granny Jean taught me the art of public speaking and made sure I was in tune with my Scottish roots. I am proud to call my sister one of my closest friends, and I am excited to see what she will set her sights on to make the world a better place. I thank all of them for making me who I am today.

Trinity College has been a place of welcome, beauty and awe to me since I first set foot there as an undergraduate in 2016. My various Directors of Studies, Tutors, and most of all, my Tutorial Administrator Lynn Clift, have provided support and opportunity to me whenever I needed it. I will sincerely miss having Great Court to wander through, and the feeling of security that came along with it. 

I would also like to thank STFC for providing the financial support for my studies.

Finally, I have been extremely lucky to have been surrounded by so many remarkable people during my time in Cambridge. My fellow mathematicians, Will, Toma and Isti, have tirelessly helped with every academic question I've had, no matter how silly or basic, and made my time in the CMS joyful. The Hare and Hounds, whilst supposedly a running club, has been for me an exceptional group of friends, housemates and my most trusted confidantes. To name but a few, Su-Min, Izzi, Lawrence, Joe, Pete, Niamh, Phil and Jeremy have made my time incredibly special and given me a deep sense of belonging. I am certain those relationships will endure, be that through runs in the countryside, pub quizzing successes or Burns Night ceilidhs, long after we ever leave this wonderful place.

\end{acknowledgements}


\tableofcontents





\mainmatter


\chapter{Introduction: Effective Field Theories of Gravity}  

\ifpdf
    \graphicspath{{Chapter1/Figs/Raster/}{Chapter1/Figs/PDF/}{Chapter1/Figs/}}
\else
    \graphicspath{{Chapter1/Figs/Vector/}{Chapter1/Figs/}}
\fi

In this Chapter we review effective field theories of gravity and summarise the main results in this thesis.

\section{General Relativity}

General Relativity (GR) has had exceptional success in describing the gravitational behaviour of our universe. In 1915, Einstein first demonstrated its predictive power when he proved it resolved longstanding contradictions between the Newtonian theory of gravity and the astronomical observations of the orbit of Mercury. In the more than one hundred years since, it has become established as one of our fundamental theories of physics, predicting phenomena such as black holes, gravitational lensing, gravitational redshift and gravitational waves that are all confirmed by observation today. Furthermore, it provides the basis of our theory of cosmology which charts our universe's history from the Big Bang to the accelerated expansion we see today.

Aside from its physical importance, it also has a remarkable mathematical beauty. Its core concept is that space and time can be described by a manifold, $\MM$, in the language of differential geometry. A particular coordinate system, $x^\mu$, may locally describe a particular observer's experience of the manifold, but this will be different for a different observer. The geometrical properties of the manifold are determined by the metric, $g_{\mu \nu}$, which encodes how observers measure space and time. Thus, our universe is given by a "spacetime", $(\MM, g)$. The equations governing the spacetime in General Relativity are the "Einstein equations"
\be
    R_{\mu \nu} - \frac{1}{2} R g_{\mu \nu} + \Lambda g_{\mu\nu}= \frac{8\pi G}{c^4} T_{\mu \nu}
\ee
$R_{\mu \nu}$ and $R$ are the Ricci tensor and scalar - geometrical quantities constructed from $g_{\mu \nu}$ - whilst $\Lambda$ is the "cosmological constant" and $T_{\mu \nu}$ is the energy-momentum tensor of the matter in the universe.  Therefore the Einstein equations state that matter bends the geometry of spacetime around it. These equations also carry an important and elegant mathematical symmetry: they are "covariant", meaning they transform in a fixed tensorial way under a change of coordinates. Thus, even though spacetime may look different to different observers, covariance allows us to solve for the metric in any coordinate system and straightforwardly transform to any other if necessary.

Solving the Einstein equations in general is no mean feat, because hidden underneath the deceptively pretty mathematical notation are 10 non-linear second order partial differential equations. Under certain approximations such as spherical symmetry, time translation invariance or the exclusion of matter, they can be solved exactly. Examples include the Schwarzschild, the Reissner-Nordstrom and the Kerr-Newman metrics which describe isolated, stationary gravitational objects, and the Friedmann–Lemaître–Robertson–Walker metric which describes a homogeneous and isotropic cosmology. Such exact solutions allow us to make some predictions which can be measured against observation - for example the correct form of Mercury's orbit that so mystified 19th Century astronomers can be found by treating Mercury as a point particle free-falling in the spherically symmetric Schwarzschild spacetime with the central body having the sun's mass. Under more general conditions, the field of Numerical Relativity is dedicated to numerically solving the Einstein equations to as fine a precision as possible. Such simulations have provided us with templates of gravitational waves which we have successfully compared to the data from gravitational wave detectors.

Before we go about actually solving equations however, we can ask more general questions about the equations themselves:
\begin{enumerate}
    \item Are the equations "well-posed"? In other words, do they have a unique solution that depends continuously on initial data? 
    \item Does the form of the equations allow us to prove general properties about solutions or classes of solutions?
\end{enumerate}

The first question is crucial to ask of any physical theory if one hopes to use it to predict reality. Loosely speaking, well-posedness says that if I change my starting conditions slightly then the final solution also only changes by a small amount. If this did not hold, then any difference in our starting conditions from reality (and there always will be some) may lead to a solution that bears no resemblance to that of reality. We would be unable to put any faith in our exact solutions or our numerical simulations. 

Fortunately for us, the Einstein equations are well-posed, so long as suitable coordinates are chosen to work in. Choquet-Bruhat \cite{Bruhat:1952} was the first to prove this by using "harmonic" coordinates and a mathematically satisfying method: certain quantities are added to the Einstein equations to bring them into a manifestly well-posed form, and then those quantities are shown to vanish in harmonic coordinates. There are also formulations more suited to numerical simulation such as the BSSN formulation \cite{Baumgarte:1998te,Shibata:1995we}, that have been proved to be well-posed \cite{Sarbach:2002bt,Nagy:2004td}. 

The second question is a common approach to simplifying or classifying physical systems. Simple examples are conservation laws in classical mechanics. Without explicitly solving the equations of motion, we can often show that solutions must have constant energy, momentum, angular momentum etc. An example in General Relativity is the Positive Mass Theorem \cite{Schoen:1981}\cite{Witten:1981}, which states that the mass of any asymptotically flat end of any solution to the Einstein equations is non-negative (for a suitable definition of "mass" and under certain energy conditions and technical assumptions).

A set of properties we will be particularly concerned with in this thesis is the "laws of black hole mechanics". First proved by Bardeen, Carter and Hawking \cite{Bardeen:1973} in 1973, these are statements regarding properties of "black hole" solutions to the Einstein equations. The proofs again use the form of the Einstein equations, rather than explicit solutions, to prove general properties that all black holes must have. The physical importance of these laws was hinted at in \cite{Bardeen:1973} when they noted their striking resemblance to the laws of thermodynamics. This naturally leads to the question of whether this is more than coincidence, which was confirmed by Hawking \cite{Hawking:1975} in his discovery of Hawking radiation. By treating matter quantum mechanically on a classical black hole spacetime, he showed that a black hole radiates exactly like a thermodynamic object with a temperature consistent with the analogous quantity in the laws of black hole mechanics. Thus, as predicted by General Relativity, black holes are thermodynamic objects with temperature, energy and entropy.

\subsection{Beyond General Relativity}

In short then, General Relativity is a mature, successful physical theory of gravity, standing on strong mathematical foundations. However, there are longstanding mysteries in gravitational physics which GR does not provide any insight into. The so-called "horizon problem" and "flatness problem" of cosmology can be resolved via a rapid period of acceleration in the early universe known as inflation, but exactly how this is driven in GR remains unanswered. "Dark matter" plays a crucial role in the formation of galaxies and the expansion of the universe, but its origin and description remain unknown. The value of the cosmological constant as measured by astronomical observations is much, much smaller (at least 60 order of magnitude \cite{Bousso:2008}) than expected from calculations of the vacuum energy of the standard model. 

To investigate these cosmological mysteries, many "modified" theories of gravity have been proposed that aim to produce the observed gravitational phenomena at cosmological scales. See \cite{Clifton:2011jh} for a thorough review of the area. Most proposals make a modification of the Einstein-Hilbert action of GR to some more complicated action involving the metric and other gravitational fields and matter fields, $\psi_I$:
\begin{equation}
    S_{E-H}= \frac{c^3}{16\pi G}\int \diff^d x \sqrt{-g}(R-2\Lambda) \quad \rightarrow \quad S= \int \diff^d x \sqrt{-g} \LL(g_{\mu\nu},\psi_I) 
\end{equation}
Most proposals stick to the core symmetry that underpins GR: diffeomorphism invariance. In these cases $\LL(g_{\mu\nu},\psi_I)$ must be a scalar under coordinate changes. Examples include $f(R)$ gravity \cite{Nojiri:2006ri}, scalar-tensor theories such as Brans-Dicke theory \cite{Nariai:1969} and Horndeski theory \cite{Charmousis:2011bf}, and higher dimensional theories such as Kaluza-Klein theory \cite{Freund:1982pg}. Constraining these theories using astronomical and gravitational wave data is a highly active area of research, but we do not yet have a candidate that has emerged as an obvious improvement on GR. Additionally, many of these theories suffer from instabilities or so-called "ghost" degrees of freedom, and proofs of the well-posedness of their equations of motion are lacking, thus they often stand on shaky mathematical ground.  

Perhaps more fundamentally than the phenomenological concerns of the above, GR predicts its own failure through the formation of singularities - points where curvatures and forces become infinite - and thus some deeper theory must take its place in these scenarios. GR is a classical field theory, and is not formulated in the mathematical language of operators that is needed to describe quantum mechanical effects that would presumably become important in the high energy regime near a singularity. One can hope that - like the other famous classical field theory, electromagnetism - GR can be formulated as a quantum field theory, however this presents many difficulties. For example, all other quantum field theories rely on a fixed background spacetime, whereas in GR the spacetime is the fundamental dynamical object. In quantum theories, states evolve through some fixed notion of time according to the Hamiltonian operator, whereas in GR the very notion of time is dependent on the state of the matter via the Einstein equations. Another major stumbling block is the fact that GR is "non-renormalizable" as a quantum field theory, meaning that it loses all predictive power at high energies.

With such manifest difficulties, we must change perspective. Rather than treating GR (or one of the modified theories above) as the fundamental theory of gravity to be quantized, we should instead view GR merely as the classical low-energy limit of some deeper underlying quantum gravity theory that is valid at arbitrarily high energies (so-called UV-complete). Proposals for such quantum gravity theories include string theory and loop quantum gravity. Experimentally probing any of these theories has proven extremely difficult, however there is hope for the future as our gravitational wave detectors improve and we are able to detect higher-energy gravitational events. In this case, we should study the corrections to GR these UV-complete theories will produce, and whether objects like black holes behave as expected. Developing the mathematical theory around these corrections will also help when trying to constrain these theories through observation. In this thesis, we shall use the framework of Effective Field Theory (EFT) to perform these studies. 

\subsection{Effective Field Theories}

A feature of the physics of the universe is that it can be split into distinct energy scales. We need not know the detailed mechanics of quark-to-quark interactions to describe the classical dynamics of billiard balls, which in turn are governed by different equations from the formation of galaxies. Qualitatively different structures emerge at varying energy scales, with each scale not interacting with the others. The underlying theory of physics describing the universe is assumed to be the same everywhere, so why can we neglect high energy physics when we are probing lower energy phenomena? This feature is called \textit{decoupling} and can be described mathematically by Effective Field Theories.

Suppose we have some grand, UV-complete theory, $\LL(\phi,\psi)$, which describes the physics of two fields $\phi$ and $\psi$. Suppose the fields have mass parameters $m_{\phi}$ and $m_{\psi}$, which roughly correspond to the lowest possible energy states of the fields. Generally, these scales will be very different, say $m_{\phi}\ll m_{\psi}$, in which case $\psi$ is some "heavy" high energy field, and $\phi$ is some "light" low energy field. If we can only probe the physics observationally at low energies, then by conservation of energy, we will only be able to study the behaviour of $\phi$, for example $\phi-\phi$ scattering amplitudes or its classical behaviour. The effect of $\psi$ on the physics of $\phi$ will still be there (because, for example, $\psi$ can be virtually created then destroyed in a scattering process), but we are unable to measure $\psi$ directly. This can be represented by "integrating out" the $\psi$ dependence in the theory's path integral, resulting in an "effective" Lagrangian $\LL_{\text{eff}}(\phi)$:
\begin{equation}
    \int \diff \phi \diff \psi \exp\left[i\int \diff^d x \LL(\phi,\psi)\right] = \int \diff \phi \exp\left[i\int \diff^d x \LL_{\text{eff}}(\phi)\right]
\end{equation}
Why should we expect $\LL_{\text{eff}}(\phi)$ to still be local? The answer is due to the uncertainty principle: virtual production of particles with energy $\Delta E \sim m_{\psi}$ can only result in the particles existing for $\Delta t \sim 1/m_{\psi}$, which is effectively instantaneous to our low energy probes. Similarly, the effects must be local in space by considering the momentum of $\psi$ particles. Therefore, $\LL_{\text{eff}}(\phi)$ will be a sum of local operators made from $\phi$. These operators can be of any dimension, but since $1/m_{\psi}$ sets the scale of the locality, the coefficients of the operators will have appropriate factors of $1/m_{\psi}$ in order to make them the same dimension overall:
\begin{equation}\label{Leff}
    \LL_{\text{eff}}(\phi) = A \sum_{D} \frac{c_D}{m^D_{\psi}}O_D(\phi)
\end{equation}
where $O_D(\phi)$ is some local operator of dimension $D$, $c_{D}$ are dimensionless constants, and $A$ is some overall dimensionful factor. If we know that $\phi$ obeys particular symmetries, then the operators present in (\ref{Leff}) must also obey those symmetries. Physical observables such as scattering amplitudes of $\phi$ at energy $E$ will then be a power series in $E/m_{\psi}$, and we can truncate $\LL_{\text{eff}}(\phi)$ to any order in $1/m_{\psi}$ to get predictions accurate up to that order in $E/m_{\psi}$. 

The crucial fact about this procedure is that the form of (\ref{Leff}) is completely agnostic about the details of the UV-complete theory $\LL(\phi,\psi)$.  Only the precise values of the coefficients $c_D$ depend on the details of $\LL(\phi,\psi)$, but in theory we can measure the value of the $c_D$ up to some order by performing sufficiently accurate observations of the low energy behaviour of $\phi$. In particular, even though (\ref{Leff}) may have infinitely many terms, to obtain predictions up to a particular order in $E/m_{\psi}$, we only need finitely many terms. Thus we can see how high energy physics naturally decouples from low energy physics, so long as we only probe at a particular scale.

\subsection{Example: Integrating out the Electron from QED}\label{QEDIntegration}

Let us now discuss a specific example of this procedure. We will do this in a setting where we already have the UV-complete theory: the interactions between photons and electrons in flat space, which are determined by quantum electrodynamics (QED). In this theory, photons are described by a massless spin-one gauge field $A_\mu$, and electrons are described by a Dirac spinor field $\psi$. The QED Lagrangian is
\begin{equation}
    \LL(A_\mu,\psi) = -\frac{1}{4} F_{\mu\nu}F^{\mu\nu} + \bar{\psi}\left[i \gamma^\mu (\partial_\mu+ie A_\mu)-m_e\right]\psi
\end{equation}
where $F_{\mu\nu} = \partial_\mu A_\nu-\partial_\nu A_\mu$, and $-e$ and $m_e$ are respectively the charge and mass of the electron. The two particles in this theory have very different mass parameters: the photon is massless whereas $m_e$ is non-zero. Therefore, we can treat the photon as a light field and the electron as a heavy field. If we only want to study the physics of the photon at energies $E\ll m_e$ then we should be able to integrate out the electron from $\LL(A_\mu,\psi)$ to produce an effective action $\LL_{eff}(A_\mu)$. This effective action should predict exactly the same photon physics as the QED action, order-by-order in $E/m_e$.

What will $\LL_{eff}(A_\mu)$ look like? The QED action has various symmetries: Lorentz invariance, invariance under a gauge transformation $A_\mu \rightarrow A_\mu + \partial_\mu \chi$, charge conjugation invariance and parity invariance. The effective action will inherit these symmetries. Moreover, it will be a sum of terms of increasing operator dimension with associated factors of $1/m_e$. Therefore it will be of the form
\begin{equation}
    \LL_{eff}(A_\mu) = \LL_4 + \frac{1}{m_e^2}\LL_6 + \frac{1}{m_e^4}\LL_8+...
\end{equation}
where each $\LL_n$ is a sum of terms with dimension $n$ that obey the above symmetries\footnote{Terms with odd dimension are forbidden by these symmetries.}. The only possible dimension 4 term is $F_{\mu\nu}F^{\mu\nu}$, and we can always rescale $\LL_{eff}$ to take
\begin{equation}
    \LL_4 = -\frac{1}{4} F_{\mu\nu}F^{\mu\nu}
\end{equation}
There is also only one possible term in $\LL_6$ (up to total derivatives) which is of the form
\begin{equation}
    \LL_6 = a_1 \partial^\mu F_{\mu\nu} \partial_\rho F^{\rho \nu}
\end{equation}
where $a_1$ is a dimensionless constant. Terms such as $F^{\mu\nu}\partial^\rho \partial_\rho  F_{\mu\nu}$ can be rewritten in terms of the above using the identity $\partial_\rho F_{\mu\nu} + \partial_\mu F_{\nu\rho}+\partial_\nu F_{\rho\mu}=0$ and integrating by parts. Similarly the set of terms in $\LL_8$ can be expressed as 
\begin{equation}
    \LL_8 = b_1 (F_{\mu\nu}F^{\mu\nu})^2 + b_2 (F_{\mu\nu}\Tilde{F}^{\mu\nu})^2 + (\partial^4 F^2 \text{ terms})
\end{equation}
where $\Tilde{F}_{\mu\nu} = \frac{1}{2}\epsilon_{\mu\nu\rho\sigma}F^{\rho\sigma}$ and $b_1$, $b_2$ are dimensionless constants. Note that terms with odd numbers of $F$'s are forbidden by charge-conjugation invariance, which sends $F_{\mu\nu} \rightarrow -F_{\mu\nu}$. 

One can continue to write down higher and higher order terms in $\LL_{eff}$. However, it turns out that we can simplify $\LL_6$ and $\LL_8$ further by making a field redefinition. If we define $A'_\mu$ by
\begin{equation}
    A_\mu = A'_\mu + \frac{1}{m^2_e} T_\mu
\end{equation}
for some tensor $T_\mu$, then $\LL_4$ can be rewritten as
\begin{equation}
    \LL_4 = -\frac{1}{4} F'_{\mu\nu}F'^{\mu\nu} -\frac{1}{m_e^2} F'^{\mu\nu} \partial_\mu T_\nu + O\left(\frac{1}{m_e^4}\right)
\end{equation}
Integrating by parts, the first order change in $\LL_4$ is $\frac{1}{m_e^2} (\partial_\mu F'^{\mu\nu})  T_\nu$. Therefore we can pick $T_\mu$ to eliminate any term proportional to $\partial_\mu F'^{\mu\nu}$ in the effective action, order-by-order in $1/m_e$. For example, choosing $T_\mu = -a_1 \partial^\rho F'_{\rho\nu}$ will eliminate the remaining term in $\LL_6$. Similarly, all $\partial_4 F^2$ terms in $\LL_8$ can be eliminated in this way \cite{Burgess:2020}. Therefore, dropping the 's, we can write our effective action as
\begin{equation}\label{photonEFT}
    \LL_{eff} = -\frac{1}{4} F_{\mu\nu}F^{\mu\nu} + \frac{b_1}{m_e^4}(F_{\mu\nu}F^{\mu\nu})^2+ \frac{b_2}{m_e^4} (F_{\mu\nu}\Tilde{F}^{\mu\nu})^2 + ...
\end{equation}
with the ellipse denoting terms with dimension 10 or above. 

Note that up until this point, we have not used the specifics of the QED Lagrangian apart from its symmetries. \textit{Any} UV-complete theory of the photon and electron which shares those symmetries will produce an effective action of the form (\ref{photonEFT}). It is only the specific values of the constants $b_1, b_2,...$ that will depend on the specifics of QED. In order to calculate what these constants are in the case of QED, there are two routes forward: i) actually perform the "integrating out" of the electron in the path integral, or ii) compare predictions of QED to that of the effective action. The first is exceedingly difficult; the second is only tediously difficult.

\begin{figure}
    \centering
    \includegraphics[width=0.35\linewidth]{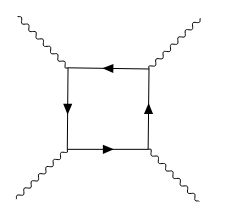}
    \caption{Box Graphs of QED: The Feynman graphs contributing the leading order contributions to $2\rightarrow 2$ photon scattering in QED.}
    \label{fig:boxgraph}
\end{figure}

\begin{figure}
    \centering
    \includegraphics[width=0.35\linewidth]{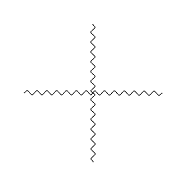}
    \caption{Four Photon Vertex in EFT: The Feynman graphs contributing the leading order contributions to $2\rightarrow 2$ photon scattering in $\LL_{eff}$. The vertex can represent either of the interactions arising from $\frac{b_1}{m_e^4}(F_{\mu\nu}F^{\mu\nu})^2$ or $\frac{b_2}{m_e^4} (F_{\mu\nu}\Tilde{F}^{\mu\nu})^2$.}
    \label{fig:photonphoton}
\end{figure}

We can sketch how one would calculate $b_1$ and $b_2$ through the second route. This is done by comparing the predictions of the $2\rightarrow 2$ photon scattering amplitudes in the two theories, to leading order in $E/m_e$, where $E$ is the centre-of-mass energy of the photons. In QED, the leading order contributions to this scattering process come from box graphs of the form in Figure \ref{fig:boxgraph} while in the effective action they come from the vertex interactions represented by the two terms in $\LL_8$ (see Figure \ref{fig:photonphoton}).  In both cases, one finds the amplitudes go as $(E/m_e)^4$. Agreement of the two calculations implies \cite{Burgess:2020}
\begin{equation}
    b_1 = \frac{\alpha^2}{90}, \quad \quad b_2 = \frac{7\alpha^2}{360}
\end{equation}
where $\alpha$ is the fine-structure constant. With these values of the $b_1$, $b_2$, the effective action (\ref{photonEFT}) truncated at the ellipse is known as the Euler-Heisenberg EFT; it describes the physics of the photon to order $(E/m_e)^4$ in the regime $E\ll m_e$.

EFT techniques are used in many other areas of QED \cite{Caswell:1985ui}, as well as QCD \cite{Hong:1998tn}, Landau theory \cite{Polchinski:1992ed}, condensed matter physics \cite{Shankar:1996} and in gravity, as we shall discuss next. A thorough review of the EFT procedure can be found in \cite{Burgess:2020} or \cite{Donoghue:1992dd}.

\subsection{Effective Field Theory Corrections to GR}\label{EFTCorstoGR}

In the above example, we constructed a low energy EFT from a known UV-complete theory. However, the general form of that EFT was independent of the specifics of the UV-complete theory. This is the great usefulness of EFTs: we can parameterize and quantify corrections to a known low-energy theory in the absence of any knowledge about its UV completion. Importantly for our purposes, this is exactly the framework we need for describing corrections to GR! We now discuss how this is achieved; see \cite{Burgess:2004} or \cite{Donoghue:2012zc} for thorough reviews of gravitational EFTs.

We know that GR matches observation to excellent precision on energy scales such as those of planetary orbits in our solar system and the deflection of light around the sun. Therefore, the lowest order terms in any effective field theory of gravity (at least in regimes where other light matter fields can be neglected) must be those from the Einstein-Hilbert action
\begin{equation}
    \frac{\LL_{\text{eff}}(g_{\mu\nu})}{\sqrt{-g}} = \frac{c^3}{16\pi G} (-2\Lambda + R+...) 
\end{equation}
The metric has vanishing mass term, hence its mass parameter is $m_{g} = 0$. It is indeed a "light" field. What should we take as the "heavy" mass scale $m$ (our equivalent of $m_{\psi}$ from above)? This can be whatever is the smallest scale we have integrated out, for example the electron mass if we are studying applications at energies less than the mass of all elementary particles. Regardless, for convenience we shall define $l=1/m$ and call $l$ our "UV-scale".

$R$ has mass dimension 2 as it involves two derivatives of the metric, which is itself dimensionless. The EFT corrections in the ellipsis can include all higher mass dimension quantities made from $g_{\mu\nu}$, i.e., terms with more derivatives. However, we know that GR has a symmetry which we expect to hold in the UV-complete theory: diffeomorphism invariance. Thus we restrict ourselves to including any possible scalar made from contractions or covariant derivatives of the Riemann tensor\footnote{In a $d$-dimensional spacetime, we should also include the volume form $\epsilon_{\mu_1... \mu_d}$. In even dimensions, only terms with even numbers of derivatives can appear in a pure gravity EFT, but in odd dimensions we can use $\epsilon_{\mu_1... \mu_d}$ to construct terms with an odd number of derivatives.}:
\begin{multline} \label{vacuumgravEFT}
    \frac{\LL_{\text{eff}}(g_{\mu\nu})}{\sqrt{-g}} = \frac{c^3}{16\pi G} \big[-2\Lambda + R + l^2( a_1R_{\mu\nu}R^{\mu\nu} + a_2R^2 + a_3R_{\mu\nu\rho\sigma}R^{\mu\nu\rho\sigma} + a_4\nabla_\mu \nabla^\mu R) +\\
    l^4(a_5R^3 + ...) + ... \big]
\end{multline}
Note that total derivative terms such as $\nabla_\mu \nabla^\mu R$ are redundant and can be ignored so long as we are only concerned with observables like the classical equations of motion or perturbative scattering amplitudes that do not depend on the topology of the spacetime. Furthermore, the higher derivative terms are not unique under a field redefinition of the metric, a fact which we can use to eliminate more coefficients in a similar fashion to the low-energy photon EFT example of the previous section. A field redefinition of the form $g_{\mu \nu} \rightarrow g_{\mu \nu}  + l^2( a \Lambda g_{\mu \nu}+ b R g_{\mu \nu} + c R_{\mu \nu})$ with appropriate constants $a, b,c$ can be used to push anything proportional to $R_{\mu\nu}$ up to arbitrarily high order in $l$ \cite{Burgess:2004}. This can be used to bring the 4-derivative terms into the form of a single term given by the Einstein-Gauss-Bonnet (EGB) Lagrangian \footnote{This process may also renormalize the values of $\Lambda$ and $G$.}: 
\begin{equation}
    \frac{\LL_{\text{eff}}(g_{\mu\nu})}{\sqrt{-g}} = \frac{c^3}{16\pi G} \big[ -2 \Lambda + R + \frac{1}{16} k l^2 \delta^{\rho_1 \rho_2 \rho_3 \rho_4}_{\sigma_1 \sigma_2 \sigma_3 \sigma_4} R_{\rho_1 \rho_2}\,^{\sigma_1 \sigma_2} R_{\rho_3 \rho_4}\, ^{\sigma_3 \sigma_4} + ... \big]
\end{equation}
where the generalized Kronecker delta is
\begin{equation}
    \delta^{\rho_1 ... \rho_n}_{\sigma_1 ... \sigma_n}= n! \delta^{\rho_1}_{[\sigma_1} ... \delta^{\rho_n}_{\sigma_n]}
\end{equation}
For general spacetime dimension $d>4$, this is as far as we can go, in which case EGB gravity is the leading order EFT correction to GR. However, in $d=4$ dimensions the 4-derivative EGB term is topological and can be neglected\footnote{In fact, for $d=4$ there is another topological term of the form $R_{\mu\nu\rho\sigma} R_{\alpha \beta}{}^{\rho\sigma} \epsilon^{\mu\nu\alpha \beta}$ which we ignore for the same reasons.}. Therefore we can eliminate all order $l^2$ contributions to $\mathcal{L}_{eff}$. We can perform a similar procedure \cite{Endlich:2017, Cano:2019} to reduce the set of 6-derivative terms to the following:
\begin{equation}
    \frac{\LL_{\text{eff}}(g_{\mu\nu})}{\sqrt{-g}} = \frac{c^3}{16\pi G} \big[ -2 \Lambda + R + l^4( k_{1} R_{\mu \nu \kappa \lambda} R^{\kappa \lambda \chi \eta} R_{\chi \eta}\,^{\mu \nu} + k_{2} R_{\mu \nu \kappa \lambda} R^{\kappa \lambda \chi \eta} R_{\chi \eta \rho \sigma} \epsilon^{\mu \nu \rho \sigma} ) + ... \big]
\end{equation}

Note the above reduction only holds if we are in vacuum, i.e., in a regime where other light matter fields can be neglected. We could instead look at situations where such matter fields are important. For example around a highly charged black hole, we should include the electromagnetic field $F_{\mu\nu}$. In this case, $\LL_{\text{eff}}(g_{\mu\nu}, F_{\rho\sigma})$ will be the Einstein-Maxwell Lagrangian supplemented by all higher mass dimension contractions and covariant derivatives of $R_{\mu\nu\rho\sigma}$ and $F_{\mu\nu}$. 

Finally, we note that we have included the cosmological constant $\Lambda$ here without any factors of $l$. Naively in EFT, we should expect 0-derivative terms to come with a factor $1/l^2$ relative to the 2-derivative $R$. However, for somewhat mysterious reasons (see the cosmological constant problem), observations find that $\Lambda$ is extremely small and so we will assume it instead scales as $l^0$.

\subsection{Validity of Classical Effective Field Theory}\label{EFTValidityDef}

Above, we heuristically formulated effective field theory in terms of quantum field theory. However there may be a regime where both the classical approximation is valid and the effective field theory corrections make an observable effect. Such a regime is often assumed when making practical applications to gravity, for example in cosmology and in numerical simulations of black hole mergers in modified theories of gravity \cite{East:2020hgw}. We shall assume the existence of this regime
in this thesis. 

From a classical perspective, "integrating out" $\psi$ from our UV-complete theory $\LL(\phi,\psi)$ is equivalent to eliminating $\psi$ by formally solving the classical equation of motion for $\psi$ order-by-order in $l$ in terms of $\phi$. Under what conditions is this expansion in $l$ valid? Furthermore, the EFT equation of motion for $\phi$ arising from $\LL_{\text{eff}}$ will generally involve higher than second order derivatives - can we avoid needing additional initial data and can a well-posed initial value problem be formulated? In general, the EFT equations will result in "runaway" solutions that blow up in time - can we identify and eliminate such unphysical solutions? Moreover, are the classical solutions of the EFT theory even close to some classical solution of the underlying UV theory? Broadly, these are unanswered questions in the full generality of gravitational EFTs, although there has been much discussion in the area \cite{Flanagan:1996gw} \cite{Burgess:2014lwa} \cite{Solomon:2017nlh} \cite{Allwright:2018rut} \cite{Cayuso:2023xbc}  \cite{Reall:2021ebq}. 

There is, however, a well-motivated proposal for the "regime of validity" of a classical EFT in which these questions are all expected to be answered in the affirmative. We saw that we should expect our effective Lagrangian $\LL_{\text{eff}}(\phi)$ to be local if and only if we probe the light field $\phi$ at energy scales much less than the mass scale we have integrated out. From a classical perspective, this corresponds to a classical solution $\phi(x^\mu)$ varying on length/time scales that are much longer than the UV-scale $l$. We can formulate this mathematically as follows. Assume that we have a 1-parameter family of solutions $\phi(x^\mu,L)$ to the EFT equations of motion on some region of spacetime $\mathcal{R}$, such that any $n$-derivative quantity made from $\phi$ is uniformly bounded above by $c_n/L^n$, i.e.,
\begin{equation}
    \sup_{x^\mu \in \mathcal{R}} |\partial^n \phi(x^\mu,L)| \leq c_n/L^n
\end{equation}
for some dimensionless ($L$-independent) constant $c_n$. $L$ can represent the smallest length/time scale associated to the solution, e.g. some timescale in the dynamics or some typical length scale. Then $\phi(x^\mu,L)$ is said to lie in the regime of validity of the EFT if $L\gg l$.

Any solution that lies outside this regime must have some scale that is comparable to the UV-scale. In that scenario however, we should not expect our effective action to be valid because the energies of the heavy and light fields cannot be separated! Thus such solutions should be immediately disregarded because they are solutions to a theory of physics inapplicable to their situation. If we are in a scenario where we trust our EFT, then solutions that lie in the regime of validity are the only sensible ones to consider. 

Note that when the light fields are gauge fields, we must add to the above definition that it holds in some choice of gauge. For example in the case of the metric, we say it lies in the regime of validity if there is some particular choice of coordinates in which the above conditions hold. It does not need to hold in all coordinate systems - indeed one can always make a diffeomorphism that will take the solution out of the regime of validity, for example by adding highly oscillatory behaviour or a coordinate singularity. In this sense the definition is not covariant, but this seems unavoidable in any notion of regime of validity of EFT due to its very purpose being to describe a separation of scales.

Mathematically, this definition of regime of validity is useful because it provides us with a dimensionless small parameter, $l/L$, which we can use to bound the size of terms in the EFT equations of motion or solution. For example, the EFT equation of motion for the metric in the general vacuum gravity EFT (\ref{vacuumgravEFT}) is of the form
\begin{equation}
    \Lambda g_{\mu\nu} + R_{\mu\nu} -\frac{1}{2}R g_{\mu\nu} + \sum_{n=1}^\infty l^n E_{\mu\nu}^{(n)} = 0
\end{equation}
where $E_{\mu\nu}^{(n)}$ comes from the $(n+2)$-derivative part of $\LL_{\text{eff}}(g_{\mu\nu})$ and hence is a sum of $(n+2)$-derivative terms itself. Therefore, if $g_{\mu\nu}$ lies in the regime of validity of the EFT in some coordinate system then in that coordinate system $l^n E_{\mu\nu}^{(n)} = O(l^n/L^{n+2})$ and so the higher derivative terms are less and less important. Furthermore, in practice we will only ever know finitely many of the coefficients in $\LL_{\text{eff}}$, so there will be some maximal $N$ for which we know all the terms with $N+1$ or fewer derivatives. In this case we only fully know the $\Lambda g_{\mu\nu} + R_{\mu\nu} -\frac{1}{2}R g_{\mu\nu} + \sum_{n=1}^{N-1} l^n E_{\mu\nu}^{(n)}$ part of the equations of motion, but in the regime of validity we can still state
\be 
E_{\mu\nu} \equiv \Lambda g_{\mu\nu} + R_{\mu\nu} -\frac{1}{2}R g_{\mu\nu} + \sum_{n=1}^{N-1} l^n E_{\mu\nu}^{(n)} = O\left(\frac{l^N}{L^{N+2}}\right)
\ee
We can then make progress on proving properties of these equations, even though we do not know the precise value of the right hand side. In general, we will suppress the explicit $L$-dependence and just write the RHS as $O(l^N)$. The factors of $L$ can be reinstated by dimensional analysis.

For a particular toy EFT model of a scalar field in a "Mexican hat" potential, \cite{Reall:2021ebq} has rigorously demonstrated that in the regime of validity all of the above questions regarding the mathematical soundness of the EFT equations can be resolved. The definition has also been used in applications such as black hole entropy \cite{Hollands:2022} \cite{Deo:2023} and in proving well-posedness \cite{Figueras:2024bba} of gravitational EFTs. In this thesis, we shall not consider general solutions to gravitational EFT equations of motion, but only ones that lie within the regime of validity of the EFT.

\section{Summary of Main Results}

From the above discussion we can see that gravitational EFTs are the natural framework in which to study corrections to GR on energy scales we might hope to observe in the foreseeable future. Therefore, we should ask of such theories the same two questions we asked of GR at the start of this Chapter: 1) are the equations well-posed, and 2) can we prove general properties about classes of solutions?

In Chapter \ref{ChapterWellposedness}, we shall study the first question. Previous work \cite{Kovacs:2020pns,Kovacs:2020ywu} has shown that certain EFTs of gravity and the simplest form of matter, a scalar field, are well-posed in the regime of validity of the EFT. Specifically, they have shown that a "modified harmonic" gauge casts the equations of motion into a strongly hyperbolic form so long as the higher derivative terms are small (as will be the case in the regime of validity). Here, we shall consider the case of a more complicated matter field: the electromagnetic field. We shall demonstrate that a generalization of modified harmonic gauge can be applied to the leading order Einstein-Maxwell EFT to produce strongly hyperbolic equations so long as the higher derivative terms are small, and thus they are well-posed. 

In the remainder of this thesis we shall study the second question. Specifically, we will ask whether the laws of black hole mechanics still hold for black hole solutions of gravitational EFTs. We shall find the answer is yes in the regime of validity of the EFT. The investigation of this is structured as follows. In Chapter \ref{ChapterLaws} we introduce the laws of black hole mechanics in detail, explain the motivation for their validity in beyond-GR theories, and sketch previous attempts to prove them in specific cases. In Chapter \ref{ChapterZerothLaw}, we provide a proof of the zeroth law in EFTs of gravity coupled to electromagnetism and a charged or uncharged scalar field. In Chapter \ref{ChapterSecondLaw}, we demonstrate that a new proposal for dynamical black hole entropy satisfies the second law in these EFTs up to the same accuracy to which the EFT is known. In Chapter \ref{ChapterBHEntropy}, we investigate the gauge dependence of this definition of dynamical black hole entropy, and explicitly compute its form in specific EFTs. Finally, in Chapter \ref{ChapterConclusion} we provide some concluding remarks.


\chapter{Well-Posed Formulation of Einstein-Maxwell Effective Field Theory}\label{ChapterWellposedness}

\ifpdf
    \graphicspath{{Chapter2/Figs/Raster/}{Chapter2/Figs/PDF/}{Chapter2/Figs/}}
\else
    \graphicspath{{Chapter2/Figs/Vector/}{Chapter2/Figs/}}
\fi

The contents of this chapter are the results of original research conducted by the author of this thesis in collaboration with Harvey Reall. It is based on work published in \cite{Davies:2022}.

\section{Introduction}

The only fundamental fields for which the classical approximation has been observed to be useful are the gravitational and electromagnetic fields. To an excellent approximation these are described by conventional Einstein-Maxwell theory but as discussed in the previous Chapter, this theory will be modified by higher order effective field theory (EFT) corrections. For example, such corrections arise from Quantum Electrodynamics (QED). In this Chapter we will consider the initial value problem for Einstein-Maxwell theory with the leading order EFT corrections. 

In EFT we write the effective action as an expansion involving terms with increasing numbers of derivatives (in units with $c=1$):\footnote{
Dimensional analysis suggests that this should be viewed as a double expansion ordered by increasing number of $F_{\mu\nu}$ factors, and increasing number of derivatives.}
\be
 S = \int \diff^4 x \sqrt{-g} \left[\frac{1}{16\pi G}\left(  -2\Lambda + R \right) - \frac{1}{4} F_{\mu\nu} F^{\mu\nu} + \LL_4 + \LL_6 + \ldots \right]
\ee
where (at least locally) $F=dA$ with $A_\mu$ the vector potential and the scalar $\LL_n$ is a polynomial in derivatives of the fields, where each term contains a total of $n$ derivatives of the fields $(g_{\mu\nu},A_\rho)$. For example $\LL_4$ contains terms such as $R^2$, $R_{\mu\nu} R^{\mu\nu}$ and $(F_{\mu\nu}F^{\mu\nu})^2$.

If we truncate the above theory by discarding terms with $6$ or more derivatives then the resulting 4-derivative theory describes the leading EFT corrections to conventional Einstein-Maxwell theory. However, $\LL_4$ gives terms in the equations of motion containing third or fourth derivatives of the fields $(g_{\mu\nu},A_\rho)$. This is problematic for two reasons. First, the mathematical properties of the equations of motion are very sensitive to the terms with the most derivatives. If these terms do not have a nice algebraic structure then the initial value problem will not be well-posed. Second, even if these equations admit a well-posed formulation, as is the case for 4-derivative corrections to pure gravity in 4d \cite{Noakes:1983xd}, the higher order nature of the equations of motion means that additional initial data are required, which means that the equations describe spurious (massive) degrees in addition to the two fields present in the EFT. 

One way around these problems is to treat the higher derivative terms perturbatively, i.e., construct solutions as expansions in the coefficients of the higher derivative terms. However, there are situations where such expansions exhibit secular growth, leading perturbation theory to break down in a situation when EFT should remain valid \cite{Flanagan:1996gw}. If a formulation of the equations could be found that admits a well-posed initial value problem then we would not be restricted to constructing solutions perturbatively. 

To make progress, we exploit the fact that in EFT, the higher derivative terms in the Lagrangian are not unique, but can be adjusted order by order in the UV-scale by using field redefinitions. This enables one to freely adjust the coefficients of terms in the Lagrangian that are proportional to equations of motion. For example, as we saw in the previous Chapter, in the case of pure gravity the coefficients of the terms $R^2$ and $R_{\mu\nu}R^{\mu\nu}$ can be adjusted by a field redefinition of the form $g_{\mu \nu} \rightarrow g_{\mu \nu}  + l^2( a \Lambda g_{\mu \nu}+ b R g_{\mu \nu} + c R_{\mu \nu})$ with suitable choices of $a,b,c$. This can be used to make $L_4$ proportional to the Euler density (of the Gauss-Bonnet invariant). In 4d this term is topological, i.e., it does not affect the equations of motion, and so this shows that one can eliminate 4-derivative corrections in 4d pure gravity.

Similarly, in Einstein-Maxwell theory, field redefinitions can be used to bring $\LL_4$ to the form (neglecting topological terms)
\be
\label{L4}
 \LL_4 = c_1 X^2 + c_2 Y^2 + c_3 R_{\mu\nu\rho\sigma} \tilde{F}^{\mu\nu} \tilde{F}^{\rho\sigma} + c_4 XY + c_5 R_{\mu\nu\rho\sigma} F^{\mu\nu} \tilde{F}^{\rho\sigma}
\ee
where
\be
 \tilde{F}_{\mu\nu} = \frac{1}{2} \epsilon_{\mu\nu\rho\sigma} F^{\rho\sigma}
\ee
and
\be
X = F_{\mu\nu} F^{\mu\nu} \qquad \qquad Y = F_{\mu\nu} \tilde{F}^{\mu\nu}
\ee
The terms with coefficients $c_1$, $c_2$ and $c_3$ are symmetric under space-time orientation reversal (i.e., under parity or time-reversal) whereas the terms with coefficients $c_4$ and $c_5$ break this symmetry. In the parity-symmetric case, the above form of the Lagrangian can be determined from results in \cite{Deser:1974cz}. Ref. \cite{Jones:2019nev} discusses the parity violating terms (for a more general class of theories).

It is well-known that $c_1,c_2,c_3$ receive contributions from QED effects. In flat spacetime, at energies well below the electron mass $m$, QED predicts corrections to Maxwell theory described by the Euler-Heisenberg EFT which has specific values for $c_1$ and $c_2$ proportional to $\alpha^2/m^4$ where $\alpha$ is the fine-structure constant, as we saw in \ref{QEDIntegration}. In curved spacetime, the term with coefficient $c_3$ also arises from ``integrating out" the electron in this way, with $c_3 \propto \alpha/m^2$ \cite{Drummond:1979pp}. 

The nice thing about using field redefinitions to write $L_4$ as above is that all of the terms except for the last one give rise to second order equations of motion (for the $c_3$ term this follows from \cite{Horndeski:1980}). In particular, if we restrict to a theory with $c_5=0$ (e.g. a parity symmetric theory) then the equations of motion are second order and we can hope that the theory admits a well-posed initial value problem. 

If we ignore gravity and just consider the 4-derivative corrections to Maxwell theory (the terms quadratic in $X,Y$) then it can be shown that the equations of motion can be written as a first order symmetric hyperbolic system for $F_{\mu\nu}$, which ensures a well-posed initial value problem \cite{Abalos:2015gha}. This result holds only when the 4-derivative corrections are small, i.e., as required if the solution is within the regime of validity of EFT. 

With dynamical gravity, well-posedness is much more complicated. There are no gauge-invariant observables for gravity so any formulation of the equations requires a choice of gauge. By ``formulation" we mean a choice of gauge plus a way of  gauge-fixing the equations. Even for the 2-derivative vacuum Einstein equation it is well-known that many formulations do not give a well-posed initial value problem (the same is true for the Maxwell equations viewed as equations for $A_\mu$). For example, the ADM formulation of the Einstein equation is only weakly hyperbolic \cite{Kidder:2001tz}, which is not enough to ensure a well-posed initial value problem. Choquet-Bruhat \cite{Bruhat:1952} was the first to show that a well-posed formulation existed by proving the harmonic gauge formulation met this criteria. There are also modifications of the ADM formulation, such as the BSSN formulation \cite{Baumgarte:1998te,Shibata:1995we}, that are strongly hyperbolic \cite{Sarbach:2002bt,Nagy:2004td}, which is sufficient to admit a well-posed initial value problem. 

Even if one considers a formulation of the equations of motion that gives a well-posed initial value problem for 2-derivative Einstein-Maxwell theory, there is no guarantee that well-posedness will persist when one deforms the theory to include 4-derivative corrections, no matter how small. This has been seen in recent work on the EFT of gravity coupled to a scalar field. 
In this EFT, one considers gravity minimally coupled to the scalar field and then extends this 2-derivative theory by including 4-derivative corrections. Field redefinitions can be used to write (parity symmetric) 4-derivative terms in a form that gives second order equations of motion. The simplest strongly hyperbolic formulation of the 2-derivative theory is based on harmonic gauge. However, if one includes 4-derivative corrections then this formulation is only weakly hyperbolic, even for arbitrarily small 4-derivative terms, so the initial value problem is not well-posed \cite{Papallo:2017qvl,Papallo:2017ddx}.\footnote{This EFT is a Horndeski theory, i.e., a diffeomorphism invariant scalar-tensor theory with second order equations of motion. Strongly hyperbolic BSSN-like formulations have been found for a certain subset of Horndeski theories \cite{Kovacs:2019jqj} but this subset does not include the EFT we are discussing.} Note that this problem is not apparent when equations are linearized around a Minkowski background, or even around some non-trivial backgrounds (e.g. a static spherically symmetric black hole). But it is apparent when the equations are linearized around a {\it generic} weakly curved background \cite{Papallo:2017qvl,Papallo:2017ddx}.

Fortunately it has been shown that there exists a deformation of the harmonic gauge formulation that {\it does} give strongly hyperbolic equations when the 4-derivative terms are small \cite{Kovacs:2020pns,Kovacs:2020ywu}. This smallness requirement is not a concern because it is also required for the validity of EFT\footnote{
Requiring that the EFT arises from a consistent UV theory may impose restrictions on the coupling constants of the theory. However, the result of \cite{Kovacs:2020pns,Kovacs:2020ywu} demonstrates that no such restrictions arise from the requirement that the theory admits a well-posed initial value problem within the regime of validity of EFT. We will see that the same is true for the class of theories considered in this Chapter.
}. This ``modified harmonic gauge" formulation gives a well-posed initial value problem for the gravity-scalar EFT in 4d, as well as for the EFT of pure gravity in higher dimensions (where the Euler density is not topological). 

In this Chapter we will consider the theory \eqref{L4} with $c_5=0$, which describes Einstein-Maxwell theory with the leading (parity-symmetric) higher derivative EFT corrections. The similarity with the Einstein-scalar case strongly suggests that a conventional harmonic/Lorenz gauge formulation of this theory will be only weakly hyperbolic even when the $4$-derivative terms are small.\footnote{
As the Einstein-scalar case, we expect that one can see this by studying the theory linearized around a {\it generic} background solution.} However, we can adapt the modified harmonic gauge formulation to this EFT. We will show that the resulting equations are strongly hyperbolic when the 4-derivative terms are small, and therefore this formulation admits a well-posed initial value problem when it is within the regime of validity of EFT. Our results apply also to the larger class of ($c_5=0$) theories obtained by replacing the terms quadratic in $X,Y$ with an arbitrary smooth function $f(X,Y)$ satisfying $f(0,0)=f_X(0,0)=f_Y(0,0)=0$. This includes, for example, the Born-Infeld Lagrangian for nonlinear electrodynamics. 

This Chapter is organised as follows. In Section \ref{Initial Value Problem} we set up our initial value problem and explain what we mean by well-posedness in this context. In Section \ref{ModifHarmGau} we
describe the modified harmonic gauge formulation. In Section \ref{Strong Hyperbolicity} we briefly review the notion of strong hyperbolicity. In Section \ref{Principal Symbol} we determine the principal symbol of our equations of motion and describe their symmetries. In Section \ref{SecMotivation} we discuss the motivation behind the modified harmonic gauge, and sketch the steps in our proof of strong hyperbolicity. In Section \ref{Proof} we detail this proof following arguments very close to those of \cite{Kovacs:2020ywu}. In Section \ref{Conclusion} we make a few concluding remarks.

\section{The Initial Value Problem and Well-Posedness}\label{Initial Value Problem}

In this Section we will review the notions of an initial value problem and its well-posedness, and discuss the additional complications of posing an initial value problem for our Einstein-Maxwell EFT that arise from it being a relativistic gauge theory. 

\subsection{The Initial Value Problem for Non-Gauge Fields}\label{nongauge}

Suppose we would like to study the evolution of a collection of classical fields $u_I$, with $I=1, \ldots, N$. For simplicity, let us suppose there are no gauges involved and that we have a fixed notion of time and space, $x^\mu = (x^0,x^i)$. After an initial instant of time, $x^0=0$, suppose the evolution of $u_I$ is governed by a set of equations of motion involving $u_I$ and its derivatives. In this Chapter, we shall be concerned with second order equations of motion, i.e. those of the form
\begin{equation}
    E^I(x, u, \partial_\mu u, \partial_{\mu} \partial_\nu u) = 0
\end{equation}
Since we have $N$ unknown fields, we expect $N$ such equations for the system to be fully determined. For second order equations, we expect to have to prescribe "initial data" consisting of $(u, \partial_0 u)$ on the "initial data slice" $x^0=0$. The "initial value problem" for $u_I$ is then to solve the equations of motion given such initial data.

For a sensible physical theory, we should expect two criteria to hold for this problem: i) there exists a unique solution, i.e. the evolution is completely determined by the initial data, and ii) the solution depends continuously\footnote{By "continuously" we mean in some normed function space, however this is a technical detail we shall not delve into.} on the initial data, i.e. a "small change" in the initial data produces a corresponding "small change" in the final solution. The initial value problem is said to be "well-posed" if it meets both of these criteria. In this Chapter, we shall only concern ourselves with proving "local" well-posedness, which means the solution is only proved to exist for some amount of time, $0\leq x^0 < T$, no matter how small.

Famous second order, well-posed initial value problems include the wave equation and the Klein-Gordon equation.

A common method of proving local well-posedness of initial value problems of the above form is to show that the equations of motion satisfy a condition called "strong hyperbolicity". This will be discussed in Section \ref{Strong Hyperbolicity}.

\subsection{The Initial Value Problem for Einstein-Maxwell EFT}\label{IVPforEMEFT}

Let us now turn to the scenario we actually desire to study: a theory of a metric $g_{\mu\nu}$ and a Maxwell potential $A_\mu$, whose action is given by
\begin{equation} \label{actionwellposedness}
    S = \frac{1}{16\pi G} \int \diff^4 x \sqrt{-g} \left[ -2\Lambda + R - \frac{1}{4} F_{\mu\nu} F^{\mu\nu} + f(X,Y) + c_3 R_{\mu\nu\rho\sigma} \tilde{F}^{\mu\nu} \tilde{F}^{\rho\sigma} \right]
\end{equation}
where $f(0,0)=f_X(0,0)=f_Y(0,0)=0$. This includes our ($c_5=0$) Einstein-Maxwell EFT, where we have rescaled the Maxwell field to scale out an overall factor of $16 \pi G$, so $A_\mu$ is now dimensionless.

\subsubsection{Equations of Motion}

The equations of motion are derived from varying $S$ with respect to $A_\mu$ and $g_{\mu\nu}$. We define variations
\begin{equation}
    E^{\mu} \equiv -\frac{16 \pi G}{\sqrt{-g}} \frac{\delta S}{\delta A_\mu}, \,\,\,\,\,\,
    E^{\mu\nu} \equiv -\frac{16 \pi G}{\sqrt{-g}} \frac{\delta S}{\delta g_{\mu\nu}}
\end{equation}
The equations of motion are these variations set to 0. Explicitly these are given by
\begin{multline}\label{Chapt2Eom1}
    E^{\mu\nu}\equiv \Lambda g^{\mu\nu} + G^{\mu\nu} +\frac{1}{2} \left( \frac{1}{4} g^{\mu\nu}F^{\rho\sigma}F_{\rho\sigma}-F^{\mu\rho}F^{\nu}_{\,\,\,\rho} \right)\\
        - \frac{1}{2}g^{\mu\nu}\left(f-Y\partial_Y f\right) + 2F^{\mu\alpha} F^{\nu}_{\,\,\,\alpha}\partial_X f\\
    - \frac{1}{2}c_3\, g^{\nu\alpha}\,\delta^{\mu\lambda\rho\sigma}_{\alpha\beta\gamma\delta}\,\left(F_{\lambda\tau}F^{\beta\tau}R_{\rho\sigma}^{\,\,\,\,\,\,\gamma\delta}+\nabla^{\delta}F_{\lambda\rho}\nabla_{\sigma}F^{\beta\gamma}\right) = 0
\end{multline}
\begin{multline}\label{Chapt2Eom2}
    E^\mu \equiv \nabla_{\nu}F^{\mu\nu}( 1 - 4\partial_X f)\\ - 2 \nabla_\nu F_{\alpha\beta} \left( 4F^{\mu\nu}F^{\alpha\beta}\partial_X^2 f + 2\left( F^{\mu\nu}F_{\gamma\delta}\epsilon^{\alpha\beta\gamma\delta} + \epsilon^{\mu\nu\rho\sigma}F_{\rho\sigma}F^{\alpha\beta} \right)\partial_X\partial_Y f + \epsilon^{\mu\nu\rho\sigma}\epsilon^{\alpha\beta\gamma\delta}F_{\rho\sigma}F_{\gamma\delta}\partial^2_Y f\right)\\ + c_3\,\delta^{\mu\nu\rho\sigma}_{\alpha\beta\gamma\delta}\,\nabla_{\nu}F^{\alpha\beta}R_{\rho\sigma}^{\,\,\,\,\,\,\gamma\delta}=0
\end{multline}

As noted in the Introduction, these equations contain at most second derivatives of $(g_{\mu\nu}, A_{\rho})$, and so are second order. However, there are two additional complications in defining a well-posed initial value problem for this theory. 

The first is the question of how to cast the equations of motion as genuine evolution equations when the very notion of time and space is one of the things we are trying to solve. How does one construct an initial data slice and what initial data should be prescribed on it? The second complication is that there is gauge redundancy in how $(g_{\mu\nu}, A_{\rho})$ physically describe the world: two sets of solutions that are related by a diffeomorphism $x^\mu \rightarrow \tilde{x}^\mu$ and/or an electromagnetic gauge transformation $A_\mu \rightarrow A_\mu + \partial_\mu \chi$ describe the same physical universe. Therefore we should not expect there to be a unique solution to this initial value problem; the solution should only be unique up to a gauge transformation.

\subsubsection{Defining the Initial Value Problem and Prescribing Initial Data}

The standard way to resolve these interconnected issues for the metric is to assume our spacetime $(\MM,g_{\mu\nu})$ is globally hyperbolic (since these are the only spacetimes we expect to be predictive), meaning it contains a Cauchy surface $\Sigma$. In this case, it can be proved \cite{Wald:1984rg} that we can foliate $(\MM,g_{\mu\nu})$ by Cauchy surfaces $\Sigma_t$ parameterized by a global time function $t$ with $\Sigma = \Sigma_0$. The metric can then be decomposed in the standard $3+1$ split with time coordinate $x^0=t$:
\begin{equation}
    g = -N^2 \diff t^2 + h_{ij}\left( \diff x^i + N^i \diff t \right)\left( \diff x^j + N^j \diff t\right)
\end{equation} 
$N>0$ is called the "lapse function", $N^i$ is called the "shift vector" and $h_{ij}$ is the induced 3-dimensional Riemannian metric on $\Sigma_t$. The timelike future-directed unit normal to $\Sigma_t$ is $n_{\mu} = - N (\diff t)_{\mu}$. $h_{ij}$ can be identified with the spacetime tensor
\begin{equation}\label{inducedmetric}
    h_{\mu\nu} \equiv g_{\mu\nu} + n_{\mu}n_\nu
\end{equation}
which defines a projection operator $h_{\mu}^{\nu}$ onto $\Sigma_t$. Via a diffeomorphism, $N$ and $N^i$ can be taken to be whatever we like, hence they can absorb the gauge redundancy inherent in the metric. The true dynamical variable should be regarded as $h_{ij}$. By identifying the surfaces $\Sigma_t$ with $\Sigma$, we can view solving for $(\MM,g_{\mu\nu})$ as the time evolution of the Riemannian metric $h_{ij}(t)$ on a fixed three-dimensional manifold $\Sigma$. $h_{ij}(0)$ is one piece of the initial data for this problem. Since the equations of motion are second order, we should also have to prescribe its first time derivative $\partial_0 h_{ij}(0)$. We want to represent this by a spacetime tensor in a similar fashion to how $h_{ij}$ can be identified with (\ref{inducedmetric}). This is achieved through the "extrinsic curvature tensor" defined by
\begin{equation}\label{extrcurv}
    K_{\mu\nu} \equiv h_{\mu}^{\,\,\rho} \nabla_\rho n_\nu
\end{equation}
It is invariant under projection by $h_{\mu}^{\nu}$ and so can be identified with $K_{ij}$, which can be shown to be
\begin{equation}
    K_{ij} = \frac{1}{2N}\left(\partial_0 h_{ij} - D_{i}N_j -D_{j} N_i\right) 
\end{equation}
Hence $K_{ij}(0)$ can be taken as the second part of the initial data for the evolution of $h_{ij}(t)$. 

We can also fit prescribing initial data for the Maxwell field into this framework. As mentioned above, $A_{\mu}$ is only physical up to an electromagnetic gauge transformation. The difference with the metric, however, is that there is an electromagnetic gauge-independent quantity which contains all of the classical physical information: $F_{\mu\nu} = \partial_\mu A_\nu - \partial_\nu A_\mu$. We can decompose this into the electric and magnetic fields as measured by the "Eulerian" observer (an observer travelling with 4-velocity $n^\mu$):
\begin{equation}\label{electromagnetfield}
    E^\mu \equiv F^{\mu\nu}n_{\nu} \quad \quad B^{\mu} \equiv -\frac{1}{2} \epsilon^{\mu\nu\rho\sigma}n_\nu F_{\rho \sigma}
\end{equation}
which can be inverted to give
\begin{equation}
    F_{\mu\nu} = n_\mu E_\nu - n_\nu E_\mu - \epsilon_{\mu\nu\rho\sigma} B^{\rho} n^\sigma
\end{equation}
and hence $E^\mu$ and $B^\mu$ completely determine $F_{\mu\nu}$. Furthermore, note that $E^\mu$ and $B^\mu$ are invariant under projection by $h_\mu^\nu$ and hence they can be identified with the 3-vectors $E^i$ and $B^i$ on $\Sigma_t$. Thus, similarly to the metric case, we can view the equations of motion as evolution equations for $E^i(t)$ and $B^i(t)$. The equations viewed in this way are first order so should only need initial data $E^i(0)$ and $B^i(0)$.

The above suggests that initial data for the initial value problem for $(g_{\mu\nu}, A_{\rho})$ should consist of the quintuple $(\Sigma, h_{ij}, K_{ij}, E^i, B^i)$, where ($\Sigma$, $h_{ij})$ is a Riemannian 3-manifold, $K_{ij}$ is a symmetric 3-tensor on $\Sigma$, and $E^i$ and $B^i$ are 3-vectors on $\Sigma$. The initial value problem is then to show that the equations of motion (\ref{Chapt2Eom1}-\ref{Chapt2Eom2}) have a solution $(g_{\mu\nu}, A_{\rho})$ that contains a Cauchy surface which can be identified with $\Sigma$ and which has induced quantities $(h_{\mu\nu}, K_{\mu\nu}, E^\mu, B^\mu)$ (defined by (\ref{inducedmetric},\ref{extrcurv}, \ref{electromagnetfield})) such that their identifications on $\Sigma$ are $(h_{ij}, K_{ij}, E^i, B^i)$. Furthermore, we should expect this solution to be unique up to a gauge transformation.

\subsubsection{Initial Data Constraints}

We can ask, should we expect solutions to exist given any choice of initial data $(\Sigma, h_{ij}, K_{ij}, E^i, B^i)$? For example, the Klein-Gordon equation has a well-posed initial value problem for any initial data (up to regularity assumptions we shall not delve into). However, in the case of a metric and Maxwell field this is not the case - the initial data must satisfy a set of constraint equations.

The first constraint is otherwise known as Gauss's law for magnetism and is a condition on $B^i$. It arises from the fact that $B^i$ needs to be identified with the 4-tensor $B^\mu = -\frac{1}{2} \epsilon^{\mu\nu\rho\sigma} n_\nu F_{\rho\sigma}$. Since $F = \diff A$, we have that $\nabla_{[\mu} F_{\nu \rho]}=0$. After a bit of algebra, one can show these imply we must have
\begin{equation}\label{Gauss}
    D_i B^i = 0
\end{equation}
where $D_i$ is the covariant derivative associated with $h_{ij}$. This is therefore a constraint our initial data for $B^i$ must satisfy on $\Sigma$ if we have any hope of producing a solution.

The remaining constraints come directly from the equations of motion. A solution must satisfy
\begin{equation}\label{constraints}
    n_\mu E^\mu=0, \quad \quad n_\mu E^{\mu\nu}=0
\end{equation}
on $\Sigma$. However, in (\ref{Chapt2Eom1}) and (\ref{Chapt2Eom2}) the anti-symmetries of the generalized Kronecker delta imply that $n_\mu E^\mu$ and $n_\mu E^{\mu\nu}$ do not contain any second $t$ derivatives of $g_{\mu\nu}$ or first $t$ derivatives of $F_{\mu\nu}$. Therefore, on $\Sigma$, (\ref{constraints}) are equations that only involve the prescribed initial data (and $N$ and $N^i$ but we are free to pick these), and so are constraints our initial data must satisfy. In conventional Einstein-Maxwell theory, $n_\mu E^{\mu\nu}=0$ are the familiar constraint equations of GR and $n_\mu E^\mu=0$ is Gauss's law. 

\subsubsection{Well-Posedness}

To summarise, we would like to study the initial value problem for $(g_{\mu\nu}, A_\rho)$ with evolution equations (\ref{Chapt2Eom1}) and (\ref{Chapt2Eom2}), given initial data $(\Sigma, h_{ij}, K_{ij}, E^i, B^i)$ that satisfies the constraint equations (\ref{Gauss}) and (\ref{constraints}). We shall hereafter refer to this as the Einstein-Maxwell EFT initial value problem. We would like to prove this initial value problem is locally well-posed in the sense that a solution exists for some time $0\leq t<T$ in some $3+1$ coordinates, depends continuously on the initial data, and is unique up to a gauge transformation. This is indeed what we shall prove in this Chapter, with the additional assumption that the initial data is sufficiently {\it weakly coupled} (defined precisely in section \ref{Weak}), i.e. the higher derivative terms are small compared to the $2$-derivative terms. This will be the case if the solution lies in the regime of validity of the EFT so is a physically reasonable assumption to make.

\section{Modified Harmonic Gauge}\label{ModifHarmGau}

\subsection{Picking a Formulation to Prove Well-Posedness}

In order to prove local well-posedness of the Einstein-Maxwell EFT initial value problem described above, we will actually study a different, "gauge-fixed" set of equations of motion of the form
\begin{equation}\label{gaugefixed}
    E^{\mu\nu} + E^{\mu\nu}_{GF} = 0, \quad\quad\quad E^{\mu} + E^{\mu}_{GF} = 0
\end{equation}
In this Chapter, we will demonstrate that for a particular, carefully chosen choice of $E^{\mu\nu}_{GF}$ and $E^{\mu}_{GF}$, we can do three things that together provide a proof of local well-posedness of the Einstein-Maxwell EFT initial value problem:
\begin{enumerate}
    \item We will prove the gauge-fixed equations are strongly hyperbolic in $3+1$ coordinates (strong hyperbolicity will be defined in \ref{Strong Hyperbolicity}). This means they have a locally well-posed initial value problem for $(g_{\mu\nu},A_\rho)$ in the sense of non-gauge fields discussed in \ref{nongauge}, i.e. they have a unique solution given initial data $(g_{\mu\nu}, \partial_0 g_{\mu\nu}, A_\rho, \partial_0 A_\rho)$ on the initial data slice $x^0=0$, the solution exists for some time $0\leq x^0<T$, and this solution depends continuously on the initial data.\\ \label{WPListItem1}
    \item We relate this to the Einstein-Maxwell EFT initial value problem by the following argument. Given $(\Sigma, h_{ij}, K_{ij}, E^i, B^i)$ satisfying the constraint equations, we will show that we can prescribe initial data $(g_{\mu\nu}, \partial_0 g_{\mu\nu}, A_\rho, \partial_0 A_\rho)$ for (\ref{gaugefixed}) on $x^0=0$ such that i) the slice $x^0=0$ has induced metric, extrinsic curvature, electric field and magnetic field $(h_{ij}, K_{ij}, E^i, B^i)$, and ii) the solution to (\ref{gaugefixed}) satisfies $E^{\mu\nu}_{GF} = E^{\mu}_{GF} =0$ everywhere, hence is also a solution to $E^{\mu\nu} = E^{\mu} =0$. Therefore Step \ref{WPListItem1} implies the Einstein-Maxwell initial value problem has a solution that exists for some time $0\leq x^0<T$ and depends continuously on the initial data.\\ \label{WPListItem2}
    \item Finally, to show uniqueness up to a gauge transformation of the solution to the Einstein-Maxwell initial value problem, we shall prove that given two solutions with the same initial data $(\Sigma, h_{ij}, K_{ij}, E^i, B^i)$, we can bring them via gauge transformations into solutions of the gauge-fixed equations (\ref{gaugefixed}) with the same initial data $(g_{\mu\nu}, \partial_0 g_{\mu\nu}, A_\rho, \partial_0 A_\rho)$ on $x^0=0$. Uniqueness of solutions to the gauge-fixed equations implies these solutions are the same in these gauges. \label{WPListItem3}
\end{enumerate}

The procedure outlined above explains how to construct a "formulation" of the Einstein-Maxwell EFT equations: first gauge-fixing the equations by adding particular terms and then "picking a gauge" by taking a particular choice of initial data $(g_{\mu\nu}, \partial_0 g_{\mu\nu}, A_\rho, \partial_0 A_\rho)$ given $(\Sigma, h_{ij}, K_{ij}, E^i, B^i)$. The formulation we will work in to prove all of the above is "modified harmonic gauge" formulation, which we shall define below. We will demonstrate Steps \ref{WPListItem2} and \ref{WPListItem3} in conjunction with this. Demonstrating Step \ref{WPListItem1} is by far the lengthiest step, and so we postpone this to the remaining sections of this Chapter.

\subsection{Modified Harmonic Gauge}

\subsubsection{Auxiliary Metrics}

\begin{figure}[H]
\centering
\begin{subfigure}{0.5\linewidth}
\centering
\begin{tikzpicture}[tdplot_main_coords,scale=1.0]
  \coordinate (O) at (0,0,0);
  \coneback[surface3]{-3}{3.5}{-15}
  \coneback[surface2]{-3}{3}{-10}
  \coneback[surface1]{-3}{2.5}{-5}
  \conefront[surface3]{-3}{3.5}{-15}
  \conefront[surface2]{-3}{3}{-10}
  \conefront[surface1]{-3}{2.5}{-5}
  \coneback[surface3]{3}{3.5}{15}
  \coneback[surface2]{3}{3}{10}
  \coneback[surface1]{3}{2.5}{5}
  \conefront[surface3]{3}{3.5}{15}
  \conefront[surface2]{3}{3}{10}
  \conefront[surface1]{3}{2.5}{5}
  \node[above,color=blue!70!black] (1) at (0.5,2,3.3){$g^{\mu\nu}$};
  \node[above,color=red!70!black] (2) at (0.5,3,3.2){${\tilde g}^{\mu\nu}$};
  \node[above,color=green!70!black] (3) at (0.5,4,3.1){${\hat g}^{\mu\nu}$};
\end{tikzpicture}
\caption{}
\label{fig:cones1}
\end{subfigure}%
\begin{subfigure}{0.5\linewidth}
\centering
\begin{tikzpicture}[tdplot_main_coords,scale=1.0]
  \coordinate (O) at (0,0,0);
  \coneback[surface1]{-3}{2.5}{-15}
  \coneback[surface2]{-3}{2}{-10}
  \coneback[surface3]{-3}{1.5}{-5}
  \conefront[surface1]{-3}{2.5}{-15}
  \conefront[surface2]{-3}{2}{-10}
  \conefront[surface3]{-3}{1.5}{-5}
  \coneback[surface1]{3}{2.5}{15}
  \coneback[surface2]{3}{2}{10}
  \coneback[surface3]{3}{1.5}{5}
  \conefront[surface1]{3}{2.5}{15}
  \conefront[surface2]{3}{2}{10}
  \conefront[surface3]{3}{1.5}{5}
  \node[above,color=blue!70!black] (1) at (0.5,3,3.3){$g_{\mu\nu}$};
  \node[above,color=red!70!black] (2) at (0.5,1.5,3.4){$({\tilde g}^{-1})_{\mu\nu}$};
  \node[above,color=green!70!black] (3) at (0.5,-0.5,3.6){$({\hat g}^{-1})_{\mu\nu}$};
\end{tikzpicture}
\caption{}
\label{fig:cones2}
\end{subfigure}
 Reproduced from \cite{Kovacs:2020ywu} with permission of authors. \caption{(a) Cotangent space at a point, showing the
null cones of $g^{\mu\nu}$, $\tilde{g}^{\mu\nu}$ and $\hat{g}^{\mu\nu}$. 
(b) Tangent space at a point, showing the null cones of $g_{\mu\nu}$, $(\tilde{g}^{-1})_{\mu\nu}$ and $(\hat{g}^{-1})_{\mu\nu}$.}
\label{fig:cones}
\end{figure}
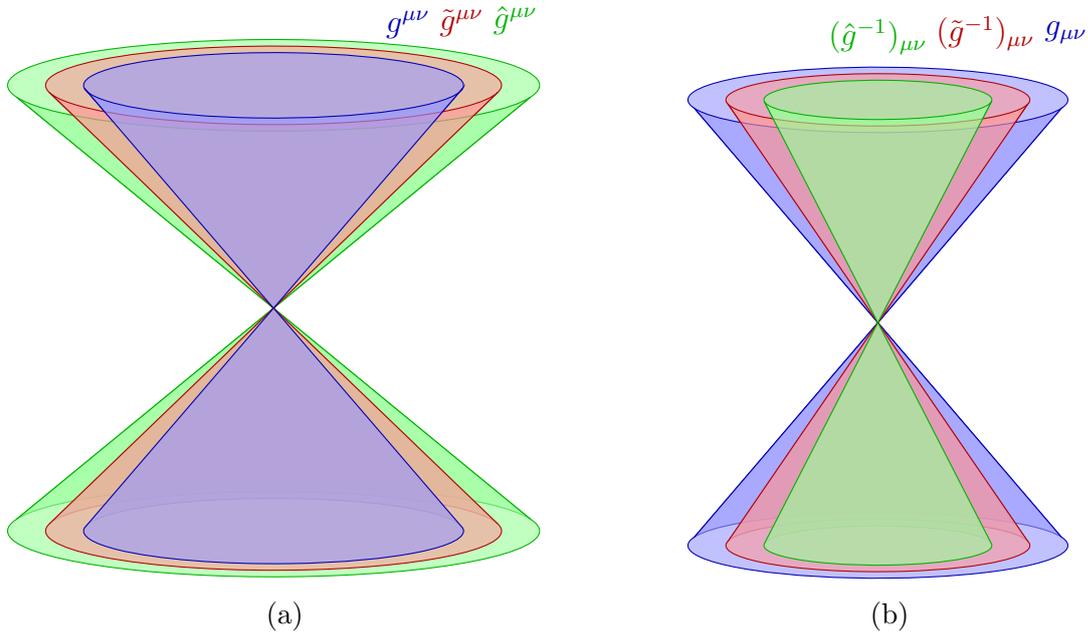

Modified harmonic gauge formulation\footnote{
It would be more accurate to refer to our formulation as ``modified harmonic/Lorenz" gauge, since it modifies the harmonic gauge condition on the metric and the Lorenz gauge condition on the Maxwell field. However, we will stick with the shorter name which was introduced in \cite{Kovacs:2020ywu} for pure gravity and gravity-scalar EFTs.} requires the introduction of two auxiliary metrics, $\tilde{g}^{\mu\nu}$ and $\hat{g}^{\mu\nu}$. These are completely unphysical, introduced only for purposes of fixing the gauge. Any index raising and lowering is done using the physical metric as usual. The motivation behind introducing these auxiliary metrics and some explanation of why they make the modified harmonic gauge strongly hyperbolic is given in Section \ref{SecMotivation} after we have introduced the relevant ideas.

The only properties we require of the auxiliary metrics are that the (cotangent space) null cones of $\tilde{g}^{\mu\nu}$,  $\hat{g}^{\mu\nu}$ and $g^{\mu\nu}$ are nested as shown in Fig. \ref{fig:cones1}, with the the null cone of $g^{\mu\nu}$ lying inside the null cone of $\tilde{g}^{\mu\nu}$, which lies inside the null cone of $\hat{g}^{\mu\nu}$.\footnote{
Ref. \cite{Kovacs:2020ywu} discusses alternative choices for the ordering of the nested cones.
} This nested structure ensures that any covector that is causal with respect to $g^{\mu\nu}$ is timelike with respect to $\tilde{g}^{\mu\nu}$ or $\hat{g}^{\mu\nu}$. This implies that if $\Sigma$ is spacelike with respect to $g^{\mu\nu}$, then it is also spacelike with respect to $\tilde{g}^{\mu\nu}$ and $\hat{g}^{\mu\nu}$. Ref. \cite{Kovacs:2020ywu} provides examples for how to construct such auxiliary metrics, such as $\tilde{g}^{\mu\nu} = g^{\mu\nu} - a n^\mu n^\nu$, $\hat{g}^{\mu\nu} = g^{\mu\nu} - b n^\mu n^\nu$ where $n^\mu$ is the unit normal to surfaces of constant $x^0$, and $a,b$ are functions chosen to take values in a certain range. In the tangent space, the null cones are also nested, with the ordering reversed, as shown in Fig. 1(b).

\subsubsection{The Gauge-Fixed Equations}

Let us define two important quantities
\begin{align}
    H^\mu &\equiv \tilde{g}^{\rho\sigma}\nabla_\rho \nabla_\sigma x^\mu\\
    H &\equiv -\tilde{g}^{\nu\sigma}\nabla_\sigma A_\nu
\end{align}
Quantities without tildes are calculated using the physical metric $g_{\mu\nu}$, and in $H^\mu$ the wave operator $\tilde{g}^{\rho\sigma}\nabla_\rho \nabla_\sigma$ is acting on $x^\mu$ as it does on scalars.

The gauge-fixed equations of motion for our formulation are taken to be
\begin{align}
    E^{\mu\nu}_{mhg} &\equiv E^{\mu\nu}+\hat{P}_{\alpha}^{\,\,\,\beta\mu\nu}\partial_\beta H^\alpha=0 \label{eq:gfg}\\
    E^\mu_{mhg} &\equiv E^\mu +\hat{g}^{\mu\nu}\nabla_\nu H=0 \label{eq:gfm}
\end{align}
where
\begin{equation}
    \hat{P}_{\alpha}^{\,\,\,\beta\mu\nu} = \delta_\alpha^{(\mu}\hat{g}^{\nu)\beta}-\frac{1}{2}\delta_\alpha^\beta \hat{g}^{\mu\nu}
\end{equation}
We shall prove Step \ref{WPListItem1} (strong hyperbolicity of these equations) in Section \ref{Proof}. The equations of standard harmonic/Lorenz gauge formulation would result from choosing $\hat{g}^{\mu\nu} = \tilde{g}^{\mu\nu}=g^{\mu\nu}$. However, the similarity with the Einstein-scalar case \cite{Papallo:2017qvl,Papallo:2017ddx} strongly suggests that this would result in equations that are only weakly hyperbolic, so would not be suitable for proving well-posedness.

\subsubsection{Picking a Gauge}

We now turn to Step \ref{WPListItem2}, which is to pick a particular gauge that will relate certain solutions of the gauge-fixed equations to solutions of the Einstein-Maxwell initial value problem. In Section \ref{IVPforEMEFT}, we discussed how there is gauge redundancy in how $(g_{\mu\nu},A_{\rho})$ physically describe the world. Thus, as should be expected, a particular prescription of constraint-equation-satisfying $(\Sigma, h_{ij}, K_{ij}, E^i, B^i)$ does not uniquely prescribe what every component of $(g_{\mu\nu}, \partial_0 g_{\mu\nu},A_{\rho}, \partial_0 A_\rho)$ need to be on $\Sigma$ in any particular $3+1$ coordinate system or electromagnetic gauge. 

For the metric, $g_{ij}(0)=h_{ij}(0)$ must match the prescribed $h_{ij}$, however we are free to pick the other components $g_{00} = -N^2 + N_i N^i$ and $g_{0i} = N_i$ on $x^0=0$ via an arbitrary choice of the lapse and shift. Once these are chosen, $\partial_0 h_{ij}(0)$ must be such that $K_{ij}(0)$ matches the prescribed $K_{ij}$ through the formula (\ref{extrcurv}). 

For the Maxwell potential, using the formulas (\ref{electromagnetfield}), on $\Sigma$, $A_\mu$ must be such that 
\begin{equation}
    E^i = - N g^{i\mu} g^{0\nu}\left( \partial_\mu A_{\nu} - \partial_\nu A_\mu\right), \quad\quad \quad B^i = -\frac{N}{2}\epsilon^{0ijk}\left( \partial_i A_{j} - \partial_i A_j\right)
\end{equation}
There are many ways to choose $A_\mu$ on $\Sigma$ to satisfy these. Of particular importance to us is that $\partial_0 A_0$ is completely unconstrained so can be freely chosen on $\Sigma$.

Taking a particular choice for these free parameters prescribes a particular choice of $(g_{\mu\nu}, \partial_0 g_{\mu\nu},A_{\rho}, \partial_0 A_\rho)$ on $x^0=0$. This is what we mean by picking a gauge. Modified harmonic gauge is defined by
\be
 H^\mu = H = 0
\ee
Let us demonstrate that we can do this on $x^0=0$. Expanding $H$, \begin{equation}
    H = -\tilde{g}^{0 0} \partial_0 A_0 + ...
\end{equation}
where the ellipsis denotes terms that do not depend on $\partial_0 A_0$. The surface $x^0=0$ is spacelike w.r.t $g^{\mu\nu}$ by the definition of $3+1$ coordinates, hence it is spacelike w.r.t. $\tilde{g}^{\mu\nu}$, hence $\tilde{g}^{0 0}\neq 0$. Thus we can always pick $\partial_0 A_0(0)$ such that $H=0$ on $x^0 = 0$. Similarly, expanding $H_\mu$,
\begin{equation}
    H_\mu = - \tilde{g}^{\rho \sigma}(2 \partial_\rho g_{\mu \sigma} - \partial_\mu g_{\rho \sigma})
\end{equation}
so $H_i = -2 \tilde{g}^{00}\partial_0 g_{0i} + ...$ where the ellipsis denotes terms that do not depend on $\partial_0 g_{0\mu}$. Thus we can choose $\partial_0 N_i$ on $x^0=0$ to set $H_i=0$. Finally, $H_0 = - \tilde{g}^{00}\partial_0 g_{0} + ...$ where the ellipsis denotes terms that do not depend on $\partial_0 g_{00}$. Hence we can choose $\partial_0 N$ on $x^0=0$ to set $H_0=0$. Thus we can set $H^\mu=H=0$ on $x^0=0$ with our choice of gauge.

Note that in picking this gauge we have never made a choice of the initial lapse and shift, only their first time derivatives. For technical reasons, in the later proof of strong hyperbolicity we shall take a choice of $N(0)$ and $N^i(0)$ such that $\partial/\partial x^0$ is timelike w.r.t. all three metrics on $x^0=0$. If this is satisfied initially then, by continuity, it will hold in a neighbourhood of the initial data slice. 

\subsubsection{Propagation of Gauge Condition}

Let us now demonstrate the second part of Step \ref{WPListItem2}, which is that with this prescription of initial data $(g_{\mu\nu}, \partial_0 g_{\mu\nu},A_{\rho}, \partial_0 A_\rho)$, the solution to the gauge-fixed equations of motion has $H^\mu=H=0$ everywhere. Due to gauge invariance of the action, the expressions for $E^\mu$ and $E^{\mu\nu}$ satisfy the Bianchi identities 
\begin{align}
    \nabla_\mu E^\mu &= 0 \label{bianchi1}\\
    \nabla_\nu E^{\mu\nu} -\frac{1}{2}F^{\mu\nu} E_\nu &=  0\label{bianchi2}
\end{align}
for any configuration of $(g_{\mu\nu},A_{\rho})$ (see Appendix \ref{AppendixBianchi} for derivations of these). Therefore, let us take the divergence of (\ref{eq:gfm}) and use the Bianchi identity (\ref{bianchi1}) to get
\begin{equation} \label{eq:divbianchi}
    0 = \nabla_{\mu}E^{\mu}_{mhg} = \hat{g}^{\mu\nu}\nabla_\mu \nabla_\nu H + (\nabla_\mu \hat{g}^{\mu\nu})\nabla_\nu H 
\end{equation}
which is a linear wave equation for $H$ with wave operator  $\hat{g}^{\mu\nu}\partial_\mu \partial_\nu$. Equations of this type have a unique solution in $\hat{D}(\Sigma)$ (the domain of dependence of $\Sigma$ with respect to $\hat{g}^{\mu\nu}$) for given initial data $H$ and $n_\mu\hat{g}^{\mu\nu}\partial_\nu H$ on $\Sigma$, so long as $\Sigma$ is spacelike w.r.t. $\hat{g}^{\mu\nu}$, which it is by the definition of $\hat{g}^{\mu\nu}$. With initial data $(g_{\mu\nu}, \partial_0 g_{\mu\nu},A_{\rho}, \partial_0 A_\rho)$ given by our modified harmonic gauge choice, we know that $H=0$ on $\Sigma$. Furthermore, the prescribed $(\Sigma, h_{ij}, K_{ij}, E^i, B^i)$ which we built our initial data around must satisfy the constraint equation $n_\mu E^\mu=0$ on $\Sigma$. Hence we have 
\begin{equation}
    0 = n_\mu E^{\mu}_{mhg} = n_\mu \hat{g}^{\mu\nu} \partial_\nu H
\end{equation}
on $\Sigma$, and hence the initial data for $H$ is trivial and the unique solution to (\ref{eq:divbianchi}) throughout $\hat{D}(\Sigma)$ is $H=0$ . Similarly we can show $H^\mu = 0$ throughout $\hat{D}(\Sigma)$ using the Bianchi identity (\ref{bianchi2}) to recast $\nabla_\nu E^{\mu\nu}_{mhg}=0$ as a linear wave equation for $H^\mu$, which has trivial initial data on $\Sigma$ by the constraint equation $n_\mu E^{\mu\nu}=0$. Finally, since the causal cone of $g^{\mu\nu}$ lies inside that of $\hat{g}^{\mu\nu}$, we have $D(\Sigma)\subset \hat{D}(\Sigma)$, where $D(\Sigma)$ is the domain of dependence w.r.t. $g^{\mu\nu}$. Since $\Sigma$ is a Cauchy surface for our solution, $D(\Sigma)$ covers the entire region on which the solution exists. Thus this solution to the gauge-fixed equations is also a solution to the Einstein-Maxwell initial value problem. This completes Step \ref{WPListItem2}.

\subsection{Uniqueness of Solution up to Gauge}

Let us now turn to Step \ref{WPListItem3}, which is to use modified harmonic gauge to demonstrate that solutions of the Einstein-Maxwell initial value problem are unique up to a gauge transformation, given initial data $(\Sigma, h_{ij}, K_{ij}, E^i, B^i)$ that satisfies the constraint equations.

Let $(g^H_{\mu\nu}, A^{H}_\rho)$ be the corresponding solution to $E^\mu = E^{\mu\nu}=0$ that we generate by solving the gauge-fixed equations $E_{mhg}^\mu = E^{\mu\nu}_{mhg}=0$ in modified harmonic gauge with initial data $(g^H_{\mu\nu}, \partial_0 g^H_{\mu\nu},A^H_{\rho}, \partial_0 A^H_\rho)$. Let $(g_{\mu\nu}, A_\rho)$ be any other solution to $E^\mu = E^{\mu\nu}=0$ which induces the same $(\Sigma, h_{ij}, K_{ij}, E^i, B^i)$. For the latter solution, via a choice of $3+1$ coordinates we can take $N$ and $N^i$ to be whatever we like, hence we can choose coordinates $x^\mu$ such that $(g_{0\mu}, \partial_0 g_{0\mu})$ agrees with $(g^H_{0\mu}, \partial_0 g^H_{0\mu})$ on $\Sigma$. Since $g^H_{\mu\nu}$ and $g_{\mu\nu}$ induce the same $h_{ij}$ and $K_{ij}$ on $\Sigma$, in these coordinates $(g_{ij}, \partial_0 g_{ij})$ must also agree with $(g^H_{ij}, \partial_0 g^H_{ij})$ on $\Sigma$. Therefore, all components of $(g_{\mu\nu}, \partial_0 g_{\mu\nu})$ and $(g^H_{\mu\nu}, \partial_0 g^H_{\mu\nu})$ agree on $\Sigma$. To $(g_{\mu\nu}, A_\rho)$, we now make a further coordinate transformation $x^\mu \rightarrow y^\mu$, where $y^\mu$ everywhere satisfies
\begin{equation}\label{harmoniccoords}
    \tilde{g}^{\rho\sigma}\nabla_\rho \nabla_\sigma y^\mu = 0
\end{equation}
and, on $\Sigma$, $y^\mu = x^\mu$ and $\partial y^\mu / \partial x^0 = \delta^\mu_0$. (\ref{harmoniccoords}) are linear wave equations for the $y^\mu$ with prescribed initial data on $\Sigma$, hence they have a unique solution for the same reasons as in the propagation of the gauge condition above. Therefore we have found modified harmonic coordinates for $(g_{\mu\nu}, A_\rho)$, i.e. a coordinate system in which $H^\mu=0$ everywhere. Furthermore, the choice of initial data for $y^\mu$ ensures that $(g_{\mu\nu}, \partial_0 g_{\mu\nu})$ still agrees with $(g^H_{\mu\nu}, \partial_0 g^H_{\mu\nu})$ on $\Sigma$.

We now perform a similar procedure to the Maxwell potential $A^\mu$. Since $A_\mu$ and $A^H_\mu$ induce the same $(E^i,B^i)$ on $\Sigma$, we must have that $F_{\mu\nu}$ agrees with $F^H_{\mu\nu}$ on $\Sigma$. This implies that we can pick some electromagnetic gauge for $A_\mu$ such that $(A_\mu, \partial_0 A_\mu)$ agrees with $(A^H_\mu, \partial_0 A^H_\mu)$ on $\Sigma$\footnote{To prove this, pick any gauge for $A_\mu$ and let $V_i = A_i^H-A_i$ on $\Sigma$. Since $F^H_{ij}=F_{ij}$ on $\Sigma$, we have that $\diff_\Sigma V=0$, where $\diff_\Sigma$ is the exterior derivative on $\Sigma$. Therefore, by the Poincar\'e Lemma on $\Sigma$, (at least locally) there exists a function $\phi(y^i)$ such that $V_i = \partial_i \phi$. Now define
\begin{equation}
    \chi(y^0,y^i) = \phi(y^i)+ y^0 \left(A_0^H-A_0\right)(0,y^i) + \frac{1}{2}(y^0)^2\left(\partial_0 A^H_0 - \partial_0 A_0\right)(0,y^i)
\end{equation}
and make the gauge transformation $A'_\mu = A_\mu + \partial_\mu \chi$. By construction, on $\Sigma$ (which is $y^0=0$), $A'_i = A^H_i$, $A_0'=A^H_0$ and $\partial_0 A'_0 = \partial_0 A^H_0$. Finally, on $\Sigma$, $\partial_0 A'_i = \partial_0 A_i + \partial_i A^H_0-\partial_i A_0 = \partial_i A^H_0 + F_{0i} = \partial_i A^H_0 + F^H_{0i} = \partial_0 A^H_i$, so $(A'_\mu, \partial_0 A'_\mu)$ agrees with $(A^H_\mu, \partial_0 A^H_\mu)$ on $\Sigma$.}. We now make the further electromagnetic gauge transformation $A'_\mu = A_\mu + \partial_\mu \chi$, where $\chi$ satisfies
\begin{equation}
    \tilde{g}^{\nu \sigma}\nabla_\sigma \nabla_\nu \chi = - \tilde{g}^{\nu \sigma}\nabla_\sigma A_\nu 
\end{equation}
with $\chi=0$ and $\partial_0 \chi=0$ on $\Sigma$. Again this is a linear wave equation for $\chi$ with prescribed initial data and so has a unique solution. This particular choice of $\chi$ means that
\begin{equation}
    - \tilde{g}^{\nu \sigma}\nabla_\sigma A'_\nu = 0 
\end{equation}
and so $A'_\mu$ is in modified harmonic gauge everywhere, i.e. $H=0$ everywhere. Furthermore the initial data is chosen so that $(A'_\mu, \partial_0 A'_\mu)$ still agrees with $(A^H_\mu, \partial_0 A^H_\mu)$ on $\Sigma$. 

To summarise, dropping the $'$s, we have been able to put $(g_{\mu\nu}, A_\rho)$ into modified harmonic gauge such that $(g_{\mu\nu}, \partial_0 g_{\mu\nu},A_\mu, \partial_0 A_\mu)$ agrees with $(g^H_{\mu\nu}, \partial_0 g^H_{\mu\nu},A^H_\mu, \partial_0 A^H_\mu)$ on $\Sigma$. Therefore, in this gauge $(g_{\mu\nu}, A_\rho)$ satisfies the gauge-fixed equations $E^{\mu\nu}_{mhg}=E^\mu_{mhg}=0$ with initial data $(g^H_{\mu\nu}, \partial_0 g^H_{\mu\nu},A^H_\mu, \partial_0 A^H_\mu)$ on $\Sigma$. The gauge-fixed equations are strongly hyperbolic (Step \ref{WPListItem1}, which we are yet to prove) which means they have a unique solution given such initial data. $(g^H_{\mu\nu}, A^H_\rho)$ also satisfies the gauge-fixed equations with the same initial data and hence $(g_{\mu\nu}, A_\rho)$ and $(g^H_{\mu\nu}, A^H_\rho)$ must be the same! Therefore, solutions to the Einstein-Maxwell EFT initial value problem are unique up to a gauge transformation.

The remainder of this Chapter is dedicated to proving Step 1, which is that the gauge-fixed equations of motion are strongly hyperbolic.

\section{Strong Hyperbolicity}\label{Strong Hyperbolicity}

In this Section, we review the definition of strongly hyperbolic second order equations of motion. Suppose we have an initial value problem for a collection of fields $u_I$, $I=1,...,N$, given initial data $(u, \partial_0 u)$. In the case of our gauge fixed equations for $u_I = (g_{\mu\nu},A_\rho)$, $N=10 + 4 = 14$ (10 for the independent components of a 4 $\times$ 4 symmetric matrix and 4 for the components of a 4-vector).

We restrict our attention to equations which are linear in $\partial_0\partial_0 u$, in which case $E^I$ can be written in the form
    \begin{equation} \label{invert1}
        E^I = A^{IJ}(x,u,\partial_\mu u,\partial_0 \partial_i u, \partial_i \partial_j u)\partial_0^2 u_J + F^I(x,u,\partial_\mu u,\partial_0 \partial_i u, \partial_i \partial_j u)
    \end{equation}
Our gauge-fixed equations of motion can be shown to be of this form. Furthermore, we will restrict ourselves to choosing initial data $(u, \partial_0 u)$ such that the slice $x_0 = 0$ is non-characteristic, meaning that $A^{IJ}$ is invertible there. For standard 2-derivative Einstein-Maxwell theory in modified harmonic gauge, it can be shown that the matrix $A^{IJ}$ is invertible on surfaces of constant $x^0$ provided the surfaces are spacelike. By continuity of $\det A^{IJ}$, when we include the higher derivative EFT terms this will still be the case for a spacelike initial surface if the initial data is sufficiently weakly coupled, i.e., the higher derivative terms are small compared to the $2$-derivative terms. Thus these restrictions do not exclude us from using this definition on our gauge-fixed equations in weak coupling. The notion of weak coupling will be defined more precisely in section \ref{Weak}.  By continuity, invertibility of $A^{IJ}$ will continue to hold in a neighbourhood of $x^0=0$. 

For an arbitrary covector $\xi_\mu$, the "principal symbol" $\mathcal{P}(\xi)^{IJ}$ is a $N \times N$ matrix defined by
\begin{equation}
    \mathcal{P}(\xi)^{IJ} \equiv \frac{\partial E^I}{\partial (\partial_\mu \partial_\nu u_J)} \xi_\mu \xi_\nu
\end{equation}
It encodes the coefficients of the second-derivative terms when the equations of motion are linearized around a background. $\mathcal{P}(\xi)^{IJ}$ is quadratic in $\xi_\mu$ and so we can write it in the following form
\begin{equation} \label{eq:ABCdef}
    \mathcal{P}(\xi)^{IJ}=\xi_0^2 A^{IJ} +\xi_0 B(\xi_i)^{IJ}+ C(\xi_i)^{IJ}  
\end{equation}
Here $A^{IJ}$ is the same as in ($\ref{invert1}$) by the definition of $\mathcal{P}(\xi)^{IJ}$. The $N \times N$ matrices $A^{IJ}$, $B^{IJ}$ and $C^{IJ}$ have additional suppressed arguments $(x,u,\partial_\mu u,\partial_0 \partial_i u, \partial_i \partial_j u)$ but not $\partial_0\partial_0 u$ by the linearity condition (\ref{invert1}).

We now define the matrix
\begin{equation} \label{Mmatrix}
    M(\xi_i) = \begin{pmatrix}
        0 & I\\
        -A^{-1}C & -A^{-1}B
    \end{pmatrix}
\end{equation}
Let $\xi_i$ be unit with respect to some smooth Riemannian metric $G^{ij}$ on surfaces of constant $x^0$. Then the system of equations ($\ref{invert1}$) is {\it strongly hyperbolic} if for any such $\xi_i$, there exists a positive definite matrix $K(\xi_i)$ that depends smoothly on $\xi_i$ and its other suppressed arguments $(x,u,\partial_\mu u,\partial_0 \partial_i u, \partial_i \partial_j u)$ such that
\begin{equation} \label{eq:symmet}
    KM = M^\dagger K
\end{equation}
and there exists a positive constant $\lambda$ such that
\begin{equation}
    \lambda^{-1}I<K<\lambda I
\end{equation}
$K(\xi)$ is called the "symmetrizer".

It can be shown \cite{Kovacs:2020ywu} that in order to prove the initial value problem for $u_I$ is locally well-posed, it is sufficient to show two things :
\begin{itemize}
    \item The equations of motion for $u_I$ are strongly hyperbolic,\\
    \item AND $M$ is invertible.
\end{itemize}
Thus strong hyperbolicity is useful because it recasts an analysis problem (well-posedness) into a linear algebra problem!

\section{The Principal Symbol} \label{Principal Symbol}

\subsection{Principal Symbol}

The principal symbol is a crucial part of the definition of strong hyperbolicity. Let us calculate it for our modified harmonic gauge-fixed equations of motion. The principal symbol for our equations acts on a vector of the form\footnote{Here the superscript ``T'' denotes a transpose, i.e., this is a column vector.}
\begin{equation}
    T_I=(t_{\mu\nu}, s_\rho)^T
\end{equation}
where $t_{\mu\nu}$ is symmetric. Indices $I, J, ...$ take values from 1 to 14. In geometric optics, $T_I$ describes the polarisation of high frequency gravitoelectromagnetic waves. 

We label the blocks of the principal symbol\footnote{Here we are following the notation of \cite{Reall:2021voz}. ``g" stands for gravitational and ``m" stands for matter. In our case the matter is a Maxwell field.} as
\begin{equation}
    \mathcal{P}(\xi)^{IJ} = \mathcal{P}^{IJ\gamma\delta}\xi_\gamma \xi_\delta = \begin{pmatrix}
        \mathcal{P}_{gg}(\xi)^{\mu\nu\rho\sigma} & \mathcal{P}_{gm}(\xi)^{\mu\nu\rho} \\
        \mathcal{P}_{mg}(\xi)^{\mu\rho\sigma} & \mathcal{P}_{mm}(\xi)^{\mu\rho}
    \end{pmatrix}
\end{equation}
where
\begin{align}
    \mathcal{P}_{gg}(\xi)^{\mu\nu\rho\sigma} = \frac{\partial E_{mhg}^{\mu\nu}}{\partial (\partial_\alpha\partial_\beta g_{\rho\sigma})}\xi_\alpha\xi_\beta \quad\quad&\quad\quad \mathcal{P}_{gm}(\xi)^{\mu\nu\rho} = \frac{\partial E_{mhg}^{\mu\nu}}{\partial (\partial_\alpha\partial_\beta A_{\rho})}\xi_\alpha\xi_\beta\\
    \mathcal{P}_{mg}(\xi)^{\mu\rho\sigma} = \frac{\partial E_{mhg}^{\mu}}{\partial (\partial_\alpha\partial_\beta g_{\rho\sigma})}\xi_\alpha\xi_\beta \quad\quad&\quad\quad \mathcal{P}_{mm}(\xi)^{\mu\rho} = \frac{\partial E_{mhg}^{\mu}}{\partial (\partial_\alpha\partial_\beta A_{\rho})}\xi_\alpha\xi_\beta
\end{align}
and where we have suppressed the dependence on $(x,u,\partial_\mu u,\partial_0 \partial_i u, \partial_i \partial_j u)$. We decompose $\mathcal{P}(\xi)$ into a part $\mathcal{P}_{\star}(\xi)$ coming from $E^\mu$ and $E^{\mu\nu}$, and a gauge-fixing part $\mathcal{P}_{GF}(\xi)$ with
\begin{equation}\label{PrincSym1}
    \mathcal{P}(\xi)^{IJ} = \mathcal{P}_{\star}(\xi)^{IJ} + \mathcal{P}_{GF}(\xi)^{IJ} 
\end{equation}
\begin{equation}
    \mathcal{P}_{\star}(\xi)^{IJ} = \begin{pmatrix}
        \mathcal{P}_{gg\star}(\xi)^{\mu\nu\rho\sigma} & \mathcal{P}_{gm\star}(\xi)^{\mu\nu\rho} \\
        \mathcal{P}_{mg\star}(\xi)^{\mu\rho\sigma} & \mathcal{P}_{mm\star}(\xi)^{\mu\rho}
    \end{pmatrix}
\end{equation}
\begin{equation}
    \mathcal{P}_{GF}(\xi)^{IJ} = \begin{pmatrix}
        -\hat{P}_{\alpha}^{\,\,\,\gamma\mu\nu} \tilde{P}^{\alpha\delta\rho\sigma} \xi_\gamma \xi_\delta & 0\\
        0 & -\hat{g}^{\mu\gamma}\tilde{g}^{\rho\delta}\xi_\gamma \xi_\delta
    \end{pmatrix}
\end{equation}
where
\begin{equation}
    \tilde{P}_{\alpha}^{\,\,\,\beta\mu\nu} = \delta_\alpha^{(\mu}\tilde{g}^{\nu)\beta}-\frac{1}{2}\delta_\alpha^\beta \tilde{g}^{\mu\nu}
\end{equation}
Furthermore, we split $\mathcal{P}_\star(\xi)$ into the standard Einstein-Maxwell terms (i.e., those coming from the first three terms in (\ref{actionwellposedness})) and the higher-derivative terms:
\begin{equation}
    \mathcal{P}_\star(\xi)^{IJ} = \mathcal{P}_{\star}^{EM}(\xi)^{IJ} + \delta\mathcal{P}_{\star}(\xi)^{IJ} 
\end{equation}
where
\begin{align}\label{PrincSym2}
    \mathcal{P}_{\star}^{EM}(\xi)^{IJ} &= \begin{pmatrix}
        (-\frac{1}{2}g^{\gamma\delta}P^{\mu\nu\rho\sigma}+P_{\alpha}^{\,\,\,\gamma\mu\nu}P^{\alpha\delta\rho\sigma})\xi_\gamma \xi_\delta & 0 \\
        0 & (-g^{\mu\rho}g^{\gamma\delta}+g^{\mu\gamma}g^{\rho\delta})\xi_\gamma \xi_\delta
    \end{pmatrix} \\
    &\equiv \begin{pmatrix}
        \mathcal{P}_{\star}^{E}(\xi)^{\mu\nu\rho\sigma} & 0 \\
        0 & \mathcal{P}_{\star}^{M}(\xi)^{\mu\rho}
    \end{pmatrix}
\end{align}
\begin{align} \label{deltaP}
    \delta\mathcal{P}_{\star}(\xi)^{IJ} &= \begin{pmatrix}
        -c_3 T^{\mu\rho\lambda\alpha\nu\sigma\eta\beta}F_{\lambda\tau}F_\eta^{\,\,\,\tau}\xi_\alpha\xi_\beta & -2c_3 T^{\mu\gamma\lambda\alpha\nu\rho\eta\beta}\nabla_\eta F_{\lambda\gamma} \xi_\alpha\xi_\beta \\
        -2c_3 T^{\rho\gamma\lambda\alpha\sigma\mu\eta\beta}\nabla_\eta F_{\lambda\gamma} \xi_\alpha\xi_\beta & \left( 2c_3T^{\mu\alpha\lambda\eta\rho\beta\gamma\delta}R_{\lambda\eta\gamma\delta} + M^{\mu\rho\alpha\beta} \right)\xi_\alpha\xi_\beta
    \end{pmatrix}
\end{align}
where
\begin{align}
    P_{\alpha}^{\,\,\,\beta\mu\nu} &= \delta_\alpha^{(\mu}g^{\nu)\beta}-\frac{1}{2}\delta_\alpha^\beta g^{\mu\nu}\\
    T^{\mu\rho\lambda\alpha\nu\sigma\eta\beta} &= \frac{1}{2}\left(\epsilon^{\mu\rho\lambda\alpha}\epsilon^{\nu\sigma\eta\beta} + \epsilon^{\nu\rho\lambda\alpha}\epsilon^{\mu\sigma\eta\beta}\right)
\end{align}
and
\begin{multline}
    M^{\mu\rho\alpha\beta} = 4\left(g^{\mu\rho}g^{\alpha\beta}-g^{\mu\alpha}g^{\rho\beta}\right)\partial_X f + 16F^{\mu\alpha}F^{\rho\beta}\partial_X^2 f\\ + 8\left( F^{\mu\alpha}\epsilon^{\rho\beta\gamma\delta}F_{\gamma\delta} + F^{\rho\beta}\epsilon^{\mu\alpha\gamma\delta}F_{\gamma\delta} \right)\partial_X\partial_Y f + 4\epsilon^{\mu\alpha\lambda\eta}\epsilon^{\rho\beta\gamma\delta}F_{\lambda\eta}F_{\gamma\delta}\partial^2_Y f
\end{multline}

\subsection{Symmetries of the Principal Symbol}\label{subsecsymms}

The precise expression for $\delta\mathcal{P}_{\star}(\xi)^{IJ}$ will be unimportant for the proof of strong hyperbolicity. However, we will make heavy use of the symmetries of the $\mathcal{P}_{\star}(\xi)^{IJ}$, which we detail here. The following symmetries are immediate from its definition:
\begin{gather}
    \mathcal{P}_{gg\star}(\xi)^{\mu\nu\rho\sigma} = \mathcal{P}_{gg\star}(\xi)^{(\mu\nu)\rho\sigma} = \mathcal{P}_{gg\star}(\xi)^{\mu\nu(\rho\sigma)} \label{eq:symgg}\\  \mathcal{P}_{gm\star}(\xi)^{\mu\nu\rho} = \mathcal{P}_{gm\star}(\xi)^{(\mu\nu)\rho} \label{eq:symgm}\\ \mathcal{P}_{mg\star}(\xi)^{\rho\mu\nu} = \mathcal{P}_{mg\star}(\xi)^{\rho(\mu\nu)} \label{eq:symmg}
\end{gather}
In \cite{Reall:2021voz}, it is shown that the fact $E^\mu$ and $E^{\mu\nu}$ are derived from an action principle leads to the following symmetries:
\begin{align}
    \mathcal{P}_{gg\star}(\xi)^{\mu\nu\rho\sigma} &= \mathcal{P}_{gg\star}(\xi)^{\rho\sigma\mu\nu}\\  \mathcal{P}_{gm\star}(\xi)^{\mu\nu\rho} &= \mathcal{P}_{mg\star}(\xi)^{\rho\mu\nu}\\  
    \mathcal{P}_{mm\star}(\xi)^{\mu\rho} &= \mathcal{P}_{mm\star}(\xi)^{\rho\mu}
\end{align}
In particular this means that $\mathcal{P}_{\star}(\xi)^{IJ}$ is symmetric.

Finally, the Bianchi identities (\ref{eq:big}) and (\ref{eq:bim}) together with the symmetries above can also be shown to put conditions on the principal symbol (given in \cite{Reall:2021voz} in an equivalent form), namely
\begin{align}
    \mathcal{P}_{gg\star}(\xi)^{\mu\nu\rho\sigma}\xi_\nu =\,&0 \label{eq:sgg}\\ 
    \mathcal{P}_{gm\star}(\xi)^{\mu\nu\rho}\xi_\nu =\,&0 \label{eq:sgm}\\  
    \mathcal{P}_{mg\star}(\xi)^{\mu\rho\sigma}\xi_\mu =\,&0 \label{eq:smg}\\  
    \mathcal{P}_{mm\star}(\xi)^{\mu\rho}\xi_\rho =\,&0  \label{eq:smm}
\end{align}
\subsection{Weak Coupling} \label{Weak}

The principal symbol is quadratic in $\xi_\mu$ so we can write e.g. ${\cal P}_\star(\xi)^{IJ} = P^{IJ\mu\nu}_\star \xi_\mu \xi_\nu$. We say that the theory is {\it weakly coupled} in some region of spacetime if a basis can be chosen in that region such that the components of  $\delta\mathcal{P}_{\star}^{IJ\mu\nu}$ are small compared to the components of $(\mathcal{P}_{\star}^{EM})^{IJ\mu\nu}$. This is the condition that the contribution of the higher derivative terms to the principal symbol is small compared to the contribution from the 2-derivative terms. Note that this is a necessary condition for the solution to be in the regime of validity of EFT, as defined in \ref{EFTValidityDef}. 

We assume that the initial data is chosen so that the theory is weakly coupled on $\Sigma$.  By continuity, any solution arising from such data will remain weakly coupled at least for a small time. However, there is no guarantee that the solution will remain weakly coupled for all time e.g. weak coupling would fail if a curvature singularity forms. Under such circumstances, well-posedness may fail in the strongly coupled region but EFT would not be valid in this region anyway. 

\section{Motivation Behind Modified Harmonic Gauge}\label{SecMotivation}

Before diving into the proof, we try to explain the motivation behind introducing these strange, unphysical, auxiliary metrics $\tilde{g}^{\mu\nu}$ and $\hat{g}^{\mu\nu}$, and sketch the steps we will perform in the next section. 

Proving strong hyperbolicity means finding a symmetrizer $K(\xi_i)$ for the matrix $M(\xi_i)$ defined by (\ref{Mmatrix}). The standard way of proving this is to show that $M(\xi_i)$ is diagonalizable with real eigenvalues and eigenvectors that depend smoothly on $\xi_i$. This is because, in this case, a suitable symmetrizer $K(\xi_i)$ can be defined by $K=(S^{-1})^\dagger S^{-1}$, where $S$ is the matrix whose columns are the eigenvectors. Conversely, strong hyperbolicity implies that $M(\xi_i)$ is diagonalizable with real eigenvalues. Therefore, the eigenvalue problem for $M(\xi_i)$ is the sensible place to start when proving strong hyperbolicity. 

In our case $M(\xi_i)$ is a $28 \times 28$ matrix, therefore we are looking for up to $28$ linearly independent eigenvectors. $M$ acts on vectors of the form $v=(T_I, T'_I)^T$ where $T_I=(t_{\mu\nu}, s_\rho)^T$ and $T'_I=(t'_{\mu\nu}, s'_\rho)^T$. It is straightforward to show that any eigenvector of $M(\xi_i)$ with eigenvalue $\xi_0$ is of the form $(T_I, \xi_0 T_I)^T$ where $T_I$ satisfies
\begin{equation}
    \mathcal{P}(\xi)^{IJ} T_J = 0
\end{equation}
where $\xi_\mu = (\xi_0, \xi_i)$.

Let us first consider standard 2-derivative Einstein-Maxwell theory (i.e. with no EFT corrections) in standard harmonic gauge (i.e. with $\hat{g}^{\mu\nu} = \tilde{g}^{\mu\nu}=g^{\mu\nu}$). The principal symbol of this formulation can be read off from (\ref{PrincSym1}-\ref{PrincSym2}), and is simply
\begin{equation}
    \mathcal{P}(\xi)^{IJ} = \begin{pmatrix}
        -\frac{1}{2}P^{\mu\nu\rho\sigma}g^{\gamma\delta} \xi_\gamma \xi_\delta & 0 \\
        0 & -g^{\mu\rho}g^{\gamma\delta}\xi_\gamma \xi_\delta
    \end{pmatrix} \\
\end{equation}
One can see that if $\xi_0$ satisfies
\begin{equation}\label{xinull}
    g^{\mu\nu} \xi_\mu \xi_\nu = 0
\end{equation}
then $\mathcal{P}(\xi)^{IJ}$ = 0, in which case $\mathcal{P}(\xi)^{IJ} T_J = 0$ for any $T_I$. Given $\xi_i$, there are two real solutions to (\ref{xinull})\footnote{This follows because (\ref{xinull}) is a quadratic with determinant $4((g^{0i}\xi_i)^2 - g^{00} g^{i j}\xi_i\xi_j)$. $g^{00}<0$ because surfaces of constant $x^0$ are spacelike, and $g^{i j}$ is positive definite by our gauge choice that $\partial/\partial x^0$ is timelike, therefore the determinant is positive.}, which we shall label $\xi_0^+$ and $\xi_0^-$. These depend smoothly on $\xi_i$. We can trivially construct a basis of eigenvectors $(T_I, \xi_0^{\pm} T_I)$ that depend smoothly on $\xi_i$ by picking any fixed (14-dimensional) basis for the $T_I$. Hence $M(\xi_i)$ is diagonalizable with real eigenvalues and eigenvectors that depend smoothly on $\xi_i$, and so standard Einstein-Maxwell theory in standard harmonic gauge is strongly hyperbolic. 

An important point to note in the above is that each of the eigenvalues $\xi_0^{\pm}$ is highly degenerate: each has a 14-dimensional space of eigenvectors. Degeneracy is generically a sign that a matrix is not diagonalizable, however because the principal symbol is so simple (essentially because standard Einstein-Maxwell theory in standard harmonic gauge is simply a collection of non-linear wave equations), it is trivial to construct a basis of eigenvectors. 

Let us now turn on the higher derivative EFT terms, but stay in standard harmonic gauge to see what goes wrong. In weak coupling, the principal symbol will be a small deformation of that of standard Einstein-Maxwell, however because $\mathcal{P}_{\star}(\xi)^{IJ}$ is a much more complicated quantity, we no longer have such control over the eigenvalues and eigenvectors of $M(\xi_i)$. $\mathcal{P}_{\star}(\xi)^{IJ}$ is so complicated in fact, that in practice all one has to go off are the symmetries and properties given in Section \ref{subsecsymms}. The equations (\ref{eq:sgg}-\ref{eq:smm}) suggest contractions with $\xi_\mu$ might be promising, and indeed a guess of the form $T_I = (\xi_{(\mu} X_{\nu)}, \lambda \xi_\rho)$ satisfies 
\begin{equation}
    \mathcal{P}_{\star}(\xi)^{IJ} T_J = 0
\end{equation}
for all $X_\mu$, $\lambda$ and $\xi_\mu$. For this choice of $T_I$ to give an eigenvector by satisfying $\mathcal{P}(\xi)^{IJ} T_J = 0$, we would also need $\mathcal{P}_{GF}(\xi)^{IJ} T_J = 0$, which one can show in standard harmonic gauge happens if and only if
\begin{equation}
    g^{\mu\nu} \xi_\mu \xi_\nu=0
\end{equation}
Thus we have found that $\xi_0^{\pm}$ are still eigenvalues and that we can construct 5 linearly independent eigenvectors for each of them via the 5 independent choices of $(X_\mu,\lambda)$. We call these "pure gauge" eigenvectors because they arise from residual gauge freedoms in our gauge-fixed equations of motion. The total of 10 is well short, however, of the 28 linearly independent eigenvectors we would need to show diagonalizability. Moreover, we do not know whether the eigenvectors we have found even span the generalized eigenspace\footnote{The generalized eigenspace corresponding to a matrix $A$ and eigenvalue $\lambda$ is the space of vectors $x$ such that there exists a positive integer $m$ with $(A-\lambda I)^m x = 0$. In the Jordan decomposition of $A$, each Jordan block is associated with a generalized eigenspace.} associated with $\xi_0^{\pm}$. If the generalized eigenspace is larger than the space of true eigenvectors then $M(\xi_i)$ is not diagonalizable and strong hyperbolicity fails. 

In standard harmonic gauge GR with higher derivative Lovelock or Horndeski terms, \cite{Papallo:2017qvl,Papallo:2017ddx} have shown that the pure gauge eigenvectors are the only eigenvectors with eigenvalues $\xi_0^{\pm}$, but that the generalized eigenspaces associated with $\xi_0^{\pm}$ have double the dimension of the span of the pure gauge eigenvectors. The similarity of those equations to our scenario of Einstein-Maxwell EFT suggests the same will occur, and thus a standard harmonic gauge formulation will fail to be strongly hyperbolic. The critical stumbling block is the high degeneracy of the $\xi_0^{\pm}$ eigenvalues and the inability to construct enough eigenvectors.

Reducing this degeneracy is exactly the motivation behind introducing the auxiliary metrics $\tilde{g}^{\mu\nu}$ and $\hat{g}^{\mu\nu}$ in modified harmonic gauge. Let us now sketch out how it achieves this by first considering standard Einstein-Maxwell theory in modified harmonic gauge. In the next section we will show that instead of $\xi_0^{\pm}$ being the only eigenvalues of $M(\xi_i)$ (as they were in standard harmonic gauge), there are now 6 eigenvalues:
\begin{enumerate}
    \item $\tilde{\xi}_0^{\pm}$, the two real solutions to $\tilde{g}^{\mu\nu} \xi_\mu \xi_\nu=0$. \\
    \item $\hat{\xi}_0^{\pm}$, the two real solutions to $\hat{g}^{\mu\nu} \xi_\mu \xi_\nu=0$.\\
    \item $\xi_0^{\pm}$, the two real solutions to $g^{\mu\nu} \xi_\mu \xi_\nu=0$.
\end{enumerate}
Furthermore, we will explicitly construct a complete basis of eigenvectors of these eigenvalues, with $\tilde{\xi}_0^{\pm}$ each having degeneracy 5, $\hat{\xi}_0^{\pm}$ each having degeneracy 5, $\xi_0^{\pm}$ each having degeneracy 4, and with all eigenvalues and eigenvectors depending smoothly on $\xi_i$. Thus standard Einstein-Maxwell theory in modified harmonic gauge is strongly hyperbolic. Moreover, the degeneracies of the $\xi_0^{\pm}$ eigenvalues have been greatly reduced.

When we turn on the higher derivative EFT terms at weak coupling, $M(\xi_i)$ will be a small deformation of the corresponding matrix of the 2-derivative theory. In particular, the eigenvalues of $M(\xi_i)$ will be close to the ones above, and can be sorted into 6 groups corresponding to which eigenvalue they approach in the standard 2-derivative (i.e., $c_3\rightarrow0$, $f\rightarrow0$) limit. We call these groups the $\tilde{\xi}_0^{+}$-group, the $\tilde{\xi}_0^{-}$-group etc. The space, $V$, of vectors on which $M(\xi_i)$ acts can be decomposed into
\begin{equation}
    V = \tilde{V}^{+} \oplus \hat{V}^{+} \oplus V^{+} \oplus \tilde{V}^{-} \oplus \hat{V}^{-} \oplus V^{-}
\end{equation}
where $V^{+}$, $V^{-}$, etc., denote the direct sums of the generalized (complex) eigenspaces of all the eigenvalues in the $\xi_0^+$, $\xi_0^{-}$ group etc. For sufficiently weak coupling, the spaces $\tilde{V}^{+}$, $\tilde{V}^{-}$ etc. must have the same dimensions as the corresponding eigenspaces in standard 2-derivative E-M theory. Therefore, $\tilde{V}^{\pm}$ and $\hat{V}^{\pm}$ are all five dimensional and $V^{\pm}$ are four dimensional. 

To see why this is helpful, let us look at what has happened to the pure gauge eigenvectors from earlier, i.e. ones of the form $T_I = (\xi_{(\mu} X_{\nu)}, \lambda \xi_\rho)$. These automatically set $\mathcal{P}_{\star}(\xi)^{IJ} T_J = 0$ as discussed above. In order to be an eigenvector, we also need $\mathcal{P}_{GF}(\xi)^{IJ} T_J = 0$, which, due to the positioning of $\tilde{g}^{\mu\nu}$ in modified harmonic gauge, now happens iff
\begin{equation}
    \tilde{g}^{\mu\nu} \xi_\mu \xi_\nu=0
\end{equation}
i.e. $\tilde{\xi}_0^{\pm}$ are still eigenvalues, and hence must be in the $\tilde{\xi}_0^{\pm}$ groups. Moreover, we have explicitly constructed 5 eigenvectors for each, which must span $\tilde{V}^{\pm}$ because we know $\tilde{V}^{\pm}$ are 5-dimensional. Therefore $\tilde{V}^{\pm}$ are true eigenspaces (with eigenvectors that depend smoothly on $\xi_i$) with only one eigenvalue each, and the non-diagonalizability associated with the pure gauge eigenvectors has disappeared! 

Where have the extra dimensions that were in the non-diagonalizable generalized eigenspaces associated with $\xi_0^{\pm}$ in standard harmonic gauge gone? It turns out they now make up $\hat{V}^{\pm}$, which we shall show still have only eigenvalues $\hat{\xi}_0^{\pm}$ and are true eigenspaces with eigenvectors that depend smoothly on $\xi_i$. In standard harmonic gauge $\tilde{V}^{\pm}$ and $\hat{V}^{\pm}$ have the same eigenvalues; it is this extra degeneracy which breaks the diagonalizability. In modified harmonic gauge, these spaces are split up by having distinct eigenvalues and then become true eigenspaces. 

Finally, we will prove that $V^{\pm}$ are also true eigenspaces and thus $M(\xi_i)$ is diagonalizable. However, we are not able to construct the eigenvalues associated with $V^{\pm}$, and it may be the case that the eigenvalues and eigenvectors do not depend smoothly on $\xi_i$. Nevertheless, we will demonstrate how to construct a symmetrizer for this space, which will complete the proof of strong hyperbolicity.

\section{Proof of Strong Hyperbolicity} \label{Proof}

\subsection{Characteristic equation}

\label{char_eq}

If $M(\xi_i)$ (defined by (\ref{Mmatrix})) is diagonalizable with real eigenvalues and eigenvectors that depend smoothly on $\xi_i$ then a symmetrizer $K(\xi_i)$ can be defined by $K=(S^{-1})^\dagger S^{-1}$ where $S$ is the matrix whose columns are the eigenvectors. We will therefore start by considering the eigenvalue problem for $M(\xi_i)$, following the approach of \cite{Kovacs:2020ywu}. We will first prove diagonalizability and smoothness for the modified harmonic gauge formulation of standard 2-derivative Einstein-Maxwell, thus establishing the strong hyperbolicity of this formulation. We then consider the weakly coupled EFT. In this case we will not demonstrate smoothness of all eigenvectors but nevertheless we will explain how a symmetrizer can still be constructed. 

$M$ acts on vectors of the form $v=(T_I, T'_I)^T$ where $T_I=(t_{\mu\nu}, s_\rho)^T$ and $T'_I=(t'_{\mu\nu}, s'_\rho)^T$. As mentioned above, it is straightforward to show that any eigenvector of $M(\xi_i)$ with eigenvalue $\xi_0$ is of the form $(T_I, \xi_0 T_I)^T$ where $T_I$ satisfies
\begin{equation} \label{chareq}
    \mathcal{P}(\xi)^{IJ} T_J = 0
\end{equation}
where $\xi_\mu = (\xi_0, \xi_i)$. This is called the {\it characteristic equation}. If this equation admits a non-zero solution $T_I$ then $\xi_\mu$ is called a {\it characteristic covector} or simply characteristic. In geometric optics, characteristics arise as wavevectors of high frequency waves, with $T_I$ describing the polarisation.

The characteristic equation can be rewritten as 
\begin{equation} \label{eq:eve}
    \begin{pmatrix}
        \mathcal{P}_{gg}(\xi)^{\mu\nu\rho\sigma}t_{\rho\sigma}+ \mathcal{P}_{gm}(\xi)^{\mu\nu\rho}s_\rho \\
        \mathcal{P}_{mg}(\xi)^{\mu\rho\sigma}t_{\rho\sigma} + \mathcal{P}_{mm}(\xi)^{\mu\rho}s_\rho
    \end{pmatrix} = \begin{pmatrix}
        0 \\
        0
    \end{pmatrix}
\end{equation}
We contract the first row of this equation with $\xi_\nu$ and the second row with $\xi_\mu$, and use (\ref{eq:sgg}-\ref{eq:smm}) to see that the non-gauge-fixing parts now vanish. After expanding the gauge-fixing parts we get two equations:
\begin{align}
    -\frac{1}{2}\left(\hat{g}^{\nu\gamma}\xi_\nu\xi_\gamma\right)&\left(g^{\mu\beta} \tilde{P}_{\beta}^{\,\,\,\delta\rho\sigma} \xi_\delta t_{\rho\sigma} \right) = 0\\
    -\left(\hat{g}^{\mu\rho}\xi_\mu\xi_\rho\right)&\left(\tilde{g}^{\sigma\nu} \xi_\sigma s_\nu \right) = 0
\end{align}
Therefore we can split the analysis into two cases:
\renewcommand{\labelenumi}{\Roman{enumi}} 
\begin{enumerate}
    \item $\hat{g}^{\nu\gamma}\xi_\nu\xi_\gamma\neq  0 \Rightarrow g^{\mu\beta} \tilde{P}_{\beta}^{\,\,\,\delta\rho\sigma} \xi_\delta t_{\rho\sigma} = 0$ AND $\tilde{g}^{\sigma\nu} \xi_\sigma s_\nu = 0$
    \item $\hat{g}^{\nu\gamma}\xi_\nu\xi_\gamma = 0$
\end{enumerate}
Note that Case I implies that $\mathcal{P}_{GF}(\xi)^{IJ}T_J = 0$, i.e., $T_I$ ``satisfies the gauge conditions''. 

\subsection{Standard Einstein-Maxwell Theory} \label{Standard E-M}
We start our analysis with standard $c_3 = f = 0$ Einstein-Maxwell theory. Let's consider each case above.

\textbf{Case I}: This is defined by $\hat{g}^{\nu\gamma}\xi_\nu\xi_\gamma\neq  0$ which implies 
\begin{equation} \label{eq:gauge1}
    g^{\mu\beta} \tilde{P}_{\beta}^{\,\,\,\delta\rho\sigma} \xi_\delta t_{\rho\sigma} = 0
\end{equation}
and
\begin{equation} \label{eq:gauge2}
    \tilde{g}^{\sigma\nu} \xi_\sigma s_\nu = 0
\end{equation}
As noted above, these imply $\mathcal{P}_{GF}(\xi)^{IJ}T_J = 0$. Substituting this back into (\ref{chareq}) gives
\begin{equation}
    \mathcal{P}_\star(\xi)^{IJ} T_J = 0
\end{equation}
In standard E-M theory, $\mathcal{P}_\star(\xi) = \mathcal{P}_\star^{EM}(\xi)$ which is block diagonal and so this reduces to
\begin{equation} \label{eq:charEins}
    \mathcal{P}_{\star}^{E}(\xi)^{\mu\nu\rho\sigma}t_{\rho\sigma} = 0
\end{equation}
and
\begin{equation} \label{eq:charMax}
    \mathcal{P}_{\star}^{M}(\xi)^{\mu\rho} s_{\rho} = 0
\end{equation}
Hence we can use results derived in \cite{Kovacs:2020ywu} for the Einstein part, supplemented with their Maxwell equivalents. We split into two further cases:

\textbf{Subcase Ia}: This is defined by $g^{\gamma\delta}\xi_\gamma\xi_\delta\neq  0$. Expanding (\ref{eq:charMax}) gives 
\begin{equation}
    s^\mu = \xi^\mu \left(\frac{\xi^\rho s_\rho}{g^{\gamma\delta}\xi_\gamma\xi_\delta} \right) 
\end{equation}
Therefore $s_\rho = \lambda \xi_\rho$ for some $\lambda$. Substituting this into (\ref{eq:gauge2}) gives $\lambda \tilde{g}^{\sigma\nu} \xi_\sigma \xi_\nu = 0$. Similarly for the Einstein parts, \cite{Kovacs:2020ywu} shows that in the case $g^{\gamma\delta}\xi_\gamma\xi_\delta\neq  0$, the equations (\ref{eq:gauge1}) and (\ref{eq:charEins}) imply $t_{\mu\nu} = \xi_{(\mu}X_{\nu)}$ and $X^\mu \tilde{g}^{\sigma\nu} \xi_\sigma \xi_\nu = 0$ for some $X_\mu$. Taken together, for non-zero $T_I$ we have that 
\begin{equation} \label{eq:gaugexi}
    \tilde{g}^{\sigma\nu} \xi_\sigma \xi_\nu = 0
\end{equation}
and $T_I = (\xi_{(\mu}X_{\nu)}, \lambda \xi_\rho)^T$. Note that our assumption that the null cones of $\tilde{g}^{\mu\nu}$ and $g^{\mu\nu}$ do not intersect ensures that (\ref{eq:gaugexi}) is consistent with $g^{\gamma\delta}\xi_\gamma\xi_\delta\neq  0$.

Surfaces of constant $x^0$ are spacelike with respect to $g^{\mu\nu}$ and hence spacelike with respect to $\tilde{g}^{\mu\nu}$. Therefore, equation (\ref{eq:gaugexi}) has two real solutions $\tilde{\xi}^{\pm}_0$ that depend smoothly on $\xi_i$. The associated characteristic covectors are labelled $\tilde{\xi}^{\pm}_\mu = (\tilde{\xi}^{\pm}_0,\xi_i)$. The two solutions can be distinguished by the convention $\mp \tilde{g}^{0\nu}\tilde{\xi}^{\pm}_{\nu}>0$.

There are $4+1 = 5$ linearly independent eigenvectors associated to each eigenvalue $\tilde{\xi}^{\pm}_0$, given by the arbitrary choices of $X_\mu$ and $\lambda$. We call these "pure gauge'' eigenvectors because they arise from a 4-dimensional residual gauge freedom in $g_{\mu\nu}$ and a 1-dimensional residual gauge freedom in $A_\mu$ in the gauge-fixed equations of motion (\ref{eq:gfg}) and (\ref{eq:gfm}). These eigenvectors form $5$-dimensional eigenspaces, which we denote as $\tilde{V}^\pm$.

\textbf{Subcase Ib}: This is defined by 
\begin{equation} \label{eq:physxi}
    g^{\sigma\nu} \xi_\sigma \xi_\nu = 0
\end{equation}
Again this has two real solutions $\xi^{\pm}_0$ with $\mp g^{0\nu}\xi^{\pm}_{\nu}>0$ and characteristic covector $\xi^{\pm}_\mu = (\xi^{\pm}_0,\xi_i)$. We will find the dimension of the space of eigenvectors. Starting with the Maxwell part, the equation $\mathcal{P}_{\star}^{M}(\xi^{\pm}_\mu)^{\mu\rho} s_{\rho} = 0$ reduces to 
\begin{equation} \label{eq:caseib}
    g^{\sigma\nu} \xi^{\pm}_\sigma s_\nu = 0
\end{equation}
But the Case I condition (\ref{eq:gauge2}) is
\begin{equation} \label{eq:caseib2}
    \tilde{g}^{\sigma\nu} \xi^{\pm}_\sigma s_\nu = 0
\end{equation}
Hence the only requirements on the polarisation $s_\rho$ are that it is orthogonal to $\xi^{\pm}_\mu$ with respect to both $g^{\mu\nu}$ and $\tilde{g}^{\mu\nu}$. These are the "physical" photon polarisations, and for each eigenvalue $\xi^{\pm}_0$, the corresponding $T_I = (0,s_\rho)$ form a 2-dimensional eigenspace that depends smoothly on $\xi_i$.

For the metric part, \cite{Kovacs:2020ywu} proves a similar statement in this case. For each eigenvalue $\xi^{\pm}_0$, there are 2 linearly independent eigenvectors with $T_I = (t_{\mu\nu},0)$ that depend smoothly on $\xi_i$. These polarisations are transverse with respect to $g^{\mu\nu}$ and $\tilde{g}^{\mu\nu}$ in the sense that $P_{\beta}^{\,\,\,\delta\rho\sigma} \xi^{\pm}_\delta t_{\rho\sigma} = 0$ and $\tilde{P}_{\beta}^{\,\,\,\delta\rho\sigma} \xi^{\pm}_\delta t_{\rho\sigma} = 0$, and correspond to physical polarisations of the metric.

Therefore there is a 4-dimensional eigenspace $V^{\pm}$ for each eigenvalue $\xi^{\pm}_0$, with eigenvectors depending smoothly on $\xi_i$.

\textbf{Case II}: This is defined by 
\begin{equation} \label{eq:violxi}
    \hat{g}^{\sigma\nu} \xi_\sigma \xi_\nu = 0
\end{equation}
Once again, this has two real solutions $\hat{\xi}^{\pm}_0$ which we distinguish by $\mp \hat{g}^{0\nu}\hat{\xi}^{\pm}_{\nu}>0$. Since the characteristic covectors $\hat{\xi}^{\pm}_\mu = (\hat{\xi}^{\pm}_0, \xi_i)$ are the same as those for (\ref{eq:divbianchi}), we call the corresponding eigenvectors "gauge condition-violating". (In geometric optics these correspond to high frequency solutions of the gauge-fixed equations that violate the gauge condition.)

We first look at $\hat{\xi}^{+}_0$ and construct its eigenvectors. Since $\hat{\xi}^{+}_\mu$ is null with respect to $\hat{g}^{\mu\nu}$, it is spacelike with respect to $g^{\mu\nu}$. Therefore we can introduce a basis $\{e_0^\mu,e_1^\mu,e_2^\mu,e_3^\mu\}$ which is orthonormal with respect to $g_{\mu\nu}$ and $e_1^\mu \propto \hat{\xi}^{+\mu}$ (recall that indices are raised using $g^{\mu\nu}$). This basis can be chosen to depend smoothly on $\xi_i$ \cite{Kovacs:2020ywu}. 

Define indices $A, B, ...$ to take values 0, 2, 3. In this basis we can write a general symmetric tensor as
\begin{equation} \label{eq:tmu}
    t_{\mu\nu} = \hat{\xi}^{+}_{(\mu} X_{\nu)} + t_{AB} e_\mu^A e_\nu^B
\end{equation}
and a general covector as 
\begin{equation} \label{eq:srho}
    s_\rho = \lambda\hat{\xi}^{+}_\rho + s_C e_\rho^C
\end{equation}
By the conditions (\ref{eq:sgg}-\ref{eq:smm}) and symmetries (\ref{eq:symgg}-\ref{eq:symmg}) the only non-vanishing components of $\mathcal{P}_{\star}(\hat{\xi}^{+})$ are those with $A, B, ...$ indices. To construct the eigenvectors, we start by considering solutions $(t_{AB}, s_C)$ to the following
\begin{equation} \label{eq:mat}
    \begin{pmatrix}
        \mathcal{P}^E_{\star}(\hat{\xi}^{+})^{ABCD} & 0 \\
        0 & \mathcal{P}^M_{\star}(\hat{\xi}^{+})^{AC}
    \end{pmatrix}
    \begin{pmatrix}
        t_{CD} \\
        s_C
    \end{pmatrix} = \begin{pmatrix}
        \hat{P}_{\alpha}^{\,\,\,\beta AB} \hat{\xi}^{+}_{\beta} v^{\alpha} \\
        \hat{g}^{A\alpha}\hat{\xi}^{+}_{\alpha}w
    \end{pmatrix}
\end{equation}
where $(v^\alpha,w)$ is a fixed constant vector. We claim this can be solved uniquely. Consider an element $(r_{AB}, p_C)$ of the kernel of the matrix on the left hand side:
\begin{align}
    \mathcal{P}_{\star}^{E}(\hat{\xi}^{+})^{ABCD} r_{CD} &= 0 \\
    \mathcal{P}_{\star}^{M}(\hat{\xi}^{+})^{A C} p_C = 0 \label{eq:ker}
\end{align}
In \cite{Kovacs:2020ywu} it is shown that $\mathcal{P}_{\star}^{E}(\hat{\xi}^{+})^{ABCD}$ has trivial kernel and so $r_{AB}=0$. We can do the same for $\mathcal{P}_{\star M}(\hat{\xi}^{+})^{A B}$ using a similar argument: (\ref{eq:ker}) implies that 
\begin{equation}
     \mathcal{P}_{\star}^{M}(\hat{\xi}^{+})^{\mu\nu} p_\nu = 0
\end{equation}
for any $p_1$. Expanding implies
\begin{equation}
    p^\mu (g^{\gamma\delta}\hat{\xi}^{+}_\gamma \hat{\xi}^{+}_\delta) = \hat{\xi}^{+\mu} (g^{\gamma\delta}\hat{\xi}^{+}_\gamma p_\delta)
\end{equation}
The null cones of $g^{\gamma\delta}$ and $\hat{g}^{\gamma\delta}$ do not intersect and so $g^{\gamma\delta}\hat{\xi}^{+}_\gamma \hat{\xi}^{+}_\delta \neq 0$. This means the above implies $p^\mu \propto \hat{\xi}^{+\mu}$, and hence in our orthonormal basis, $p_A = 0$, which establishes the result.

Therefore the kernel of the matrix on the left hand side of (\ref{eq:mat}) is trivial and so there is a unique solution $(t_{AB}(v^\alpha,w), s_C(v^\alpha,w))$ to (\ref{eq:mat}). This solution depends smoothly on $(v^\alpha, w)$, $\xi_i$ and $g_{\mu\nu}$ since both sides of (\ref{eq:mat}) depend smoothly on these things.

We use these values for $(t_{AB}(v^\alpha,w), s_C(v^\alpha,w))$ in our definitions of $(t_{\mu\nu},s_\rho)$ defined by (\ref{eq:tmu}) and (\ref{eq:srho}). This implies that 
\begin{equation} \label{eq:fix}
    \mathcal{P}_{\star}(\hat{\xi}^{+})^{IJ}  \begin{pmatrix}
        t_{\rho\sigma}(v^\gamma,w) \\
        s_\rho (v^\gamma,w)
    \end{pmatrix} = \begin{pmatrix}
        \hat{P}_{\alpha}^{\,\,\,\beta\mu\nu} \hat{\xi}^{+}_{\beta} v^{\alpha} \\
        \hat{g}^{\mu\alpha}\hat{\xi}^{+}_{\alpha}w
    \end{pmatrix} 
\end{equation}
since the components with 1-indices vanish on the top line and bottom lines because both sides have vanishing contractions with $\hat{\xi}^{+}_\mu$.

Now, in our definitions (\ref{eq:tmu}) and (\ref{eq:srho}) we judiciously chose
\begin{equation}
    X^\mu(v,t_{AB}) = \frac{2}{\tilde{g}^{\gamma\delta}\hat{\xi}^{+}_\gamma \hat{\xi}^{+}_\delta}\left(v^\mu - \tilde{P}^{\mu\nu AB} \hat{\xi}^{+}_{\nu} t_{AB}\right)
\end{equation}
and
\begin{equation}
    \lambda(w,s_C) = \frac{1}{\tilde{g}^{\gamma\delta}\hat{\xi}^{+}_\gamma \hat{\xi}^{+}_\delta}\left(w-\tilde{g}^{\gamma A}\hat{\xi}^{+}_\gamma s_A\right)
\end{equation}
which also depend smoothly on their arguments (and $\xi_i$). These choices imply
\begin{align}
    \tilde{P}^{\mu\nu\rho\sigma}\hat{\xi}^{+}_{\nu} t_{\rho\sigma} &= \frac{1}{2}\left(\tilde{g}^{\gamma\delta}\hat{\xi}^{+}_\gamma \hat{\xi}^{+}_\delta\right)X^\mu + \tilde{P}^{\mu\nu AB} \hat{\xi}^{+}_{\nu} t_{AB} \\ \nonumber
    &= v^\mu - \tilde{P}^{\mu\nu AB} \hat{\xi}^{+}_{\nu} t_{AB} + \tilde{P}^{\mu\nu AB} \hat{\xi}^{+}_{\nu} t_{AB} \\ \nonumber
    &= v^\mu \label{eq:vmu}
\end{align}
and 
\begin{align}
    \tilde{g}^{\gamma\nu}\hat{\xi}^{+}_\gamma s_\nu &= \lambda \tilde{g}^{\gamma\nu}\hat{\xi}^{+}_\gamma \hat{\xi}^{+}_\nu + \tilde{g}^{\gamma A}\hat{\xi}^{+}_\gamma s_A \\ \nonumber
    &= w \label{eq:w}
\end{align}
which imply that $T_I = (t_{\mu\nu}(v^\alpha, w), s_\rho(v^\alpha, w))$ satisfies
\begin{align}
    \mathcal{P}(\hat{\xi}^{+})^{IJ} T_J &= \mathcal{P}_\star(\hat{\xi}^{+})^{IJ} T_J + \mathcal{P}_{GF}(\hat{\xi}^{+})^{IJ} T_J \\
    &= \begin{pmatrix}
        \hat{P}_{\alpha}^{\,\,\,\beta\mu\nu} \hat{\xi}^{+}_{\beta} v^{\alpha} \\
        \hat{g}^{\mu\alpha}\hat{\xi}^{+}_{\alpha}w
    \end{pmatrix} - \begin{pmatrix}
        \hat{P}_{\alpha}^{\,\,\,\gamma\mu\nu} \hat{\xi}^{+}_\gamma \tilde{P}^{\alpha\delta\rho\sigma} \hat{\xi}^{+}_\delta t_{\rho\sigma} \\
        \hat{g}^{\mu\gamma}\tilde{g}^{\nu\delta}\hat{\xi}^{+}_\gamma \hat{\xi}^{+}_\delta s_\nu
    \end{pmatrix} \nonumber\\
    &= \begin{pmatrix}
        \hat{P}_{\alpha}^{\,\,\,\beta\mu\nu} \hat{\xi}^{+}_{\beta} v^{\alpha} \\
        \hat{g}^{\mu\alpha}\hat{\xi}^{+}_{\alpha}w
    \end{pmatrix} - \begin{pmatrix}
        \hat{P}_{\alpha}^{\,\,\,\beta\mu\nu} \hat{\xi}^{+}_{\beta} v^{\alpha} \\
        \hat{g}^{\mu\alpha}\hat{\xi}^{+}_{\alpha}w
    \end{pmatrix} \nonumber\\
    &= 0 \nonumber
\end{align}
where the second equality comes from (\ref{eq:fix}) and the third equality comes from (\ref{eq:vmu}) and (\ref{eq:w}). 

Hence for every $(v^\alpha,w)$ we have constructed a smoothly varying eigenvector $(T_I(v^\alpha,w), \hat{\xi}^{+}_0 T_I(v^\alpha,w))$ of $M$ with eigenvalue $\hat{\xi}^{+}_0$. If we pick a set of 5 linearly independent choices of $(v^\alpha, w)$ then the corresponding $t_{AB}$ and $s_C$ will be linearly independent by the triviality of the kernel of the LHS of (\ref{eq:mat}), and hence the corresponding eigenvectors will be linearly independent. Label the 5-dimensional span of these eigenvectors by $\hat{V}^{+}$. We can repeat all the above steps with $\hat{\xi}_0^{-}$ to get the same result for $\hat{V}^{-}$. We claim that $\hat{V}^{\pm}$ contain all the eigenvectors with eigenvalue $\hat{\xi}_0^{\pm}$ by counting the dimensions of the eigenspaces we have found so far:
\begin{equation}
    \dim \tilde{V}^+ + \dim \tilde{V}^- + \dim V^+ + \dim V^- + \dim \hat{V}^+ + \dim \hat{V}^- = 5 + 5 + 4 + 4 + 5 + 5 = 28
\end{equation}
$M$ is a 28 by 28 matrix, and hence there are no more eigenvectors to find. Therefore $\hat{V}^{\pm}$ are the total eigenspaces for $\hat{\xi}_0^{\pm}$.

To summarise, we have found that, for standard 2-derivative Einstein-Maxwell, $M(\xi)$ has 6 distinct eigenvalues, $\tilde{\xi}_0^{\pm}$, $\xi_0^{\pm}$ and $\hat{\xi}_0^{\pm}$ which are all real. Furthermore it has a complete set of eigenvectors that depend smoothly on $\xi_i$. Therefore the modified-harmonic-gauge formulation of standard Einstein-Maxwell is strongly hyperbolic.

Now, as mentioned in Section \ref{Strong Hyperbolicity}, the argument that strong hyperbolicity implies well-posedness requires that we also show that $M$ is invertible. This holds if $C^{IJ}$ is invertible, which is equivalent to the condition that $\xi_0 \neq 0$ for any characteristic covector. However, we chose our spacetime foliation such that $\partial/\partial x^0$ is timelike with respect to $g^{\mu\nu}$, $\tilde{g}^{\mu\nu}$ and $\hat{g}^{\mu\nu}$. This means that a covector with $\xi_0 = 0$ is spacelike with respect to all three metrics since it would be orthogonal to $\partial/\partial x^0$. But as we found above, the characteristic covectors are null with respect to one of the three metrics, and hence $M$ is invertible for standard E-M theory. By continuity, $M$ will remain invertible when we include higher derivative terms, assuming weak coupling. 

\subsection{Weakly Coupled Einstein-Maxwell EFT}

Now we consider our theory including the higher derivative terms. 
At weak coupling, $M(\xi_i)$ is a small deformation of the corresponding matrix of the 2-derivative theory. The continuity of this deformation will be used in the following to show that many of the above results still hold. In particular, the eigenvalues of $M(\xi_i)$ will be close to those discussed above, and can be sorted into 6 groups corresponding to which eigenvalue they approach in the standard 2-derivative (i.e., $c_3\rightarrow0$, $f\rightarrow0$) limit. As in \cite{Kovacs:2020ywu}, we call these groups the $\tilde{\xi}_0^{+}$-group, the $\tilde{\xi}_0^{-}$-group etc. We do not know whether these eigenvalues are real so we view $M(\xi_i)$ as acting on the 28 dimensional space $V$ of {\it complex} vectors of the form $v=(T_I, T'_I)^T$. 

Following the argument of \cite{Kovacs:2020ywu} we can decompose $V$ as follows
\begin{equation}
    V = \tilde{V}^{+} \oplus \hat{V}^{+} \oplus V^{+} \oplus \tilde{V}^{-} \oplus \hat{V}^{-} \oplus V^{-}
\end{equation}
where $V^+$ is the sum of all generalized eigenspaces\footnote{The generalized eigenspace corresponding to a matrix $A$ and eigenvalue $\lambda$ is the space of vectors $x$ such that there exists a positive integer $m$ with $(A-\lambda I)^m x = 0$. In the Jordan decomposition of $A$, each Jordan block is associated with a generalized eigenspace.} associated with eigenvalues in the $\xi_0^+$ group and similarly for the other spaces. The spaces $\tilde{V}^{+}$, $\tilde{V}^{-}$ etc. must have the same dimensions as the corresponding eigenspaces in standard 2-derivative E-M theory. Therefore, $\tilde{V}^{\pm}$ and $\hat{V}^{\pm}$ are all five dimensional and $V^{\pm}$ are four dimensional. These vector spaces are complex. 

Recall (section \ref{char_eq}) that the analysis of the characteristic equation splits into two cases. We will see that $\tilde{V}^{\pm}$ correspond to eigenvectors in Case I arising from the same residual gauge invariance as in the standard 2-derivative theory. $V^{\pm}$ correspond to the remaining eigenvectors in Case I, which are the physical eigenvectors. $\hat{V}^{\pm}$ correspond to gauge-condition-violating eigenvectors in Case II. 
\subsection{\texorpdfstring{$\tilde{V}^{\pm}$}{Tilde V}}

These are the spaces associated with the $\tilde{\xi}^{\pm}_0$-groups of eigenvalues, where $\tilde{\xi}^{\pm}_0$ are the two real solutions to $\tilde{g}^{\mu\nu}\tilde{\xi}^{\pm}_\mu\tilde{\xi}^{\pm}_\nu = 0$. However, the weakly coupled theory still has the same residual gauge freedoms in $g_{\mu\nu}$ and $A_\mu$ as the 2-derivative theory. As such it turns out that $\tilde{\xi}^{\pm}_0$ are still eigenvalues of the weakly coupled theory with the same eigenvectors $v=(T_I, \xi_0 T_I)^T$ of the form
\begin{equation}
    T_I = (\tilde{\xi}^{\pm}_{(\mu} X_{\nu)}, \lambda \tilde{\xi}^{\pm}_\rho)
\end{equation}
for arbitrary $X_\nu$ and $\lambda$. To show this, note that $T_I$ satisfies $\mathcal{P}_{GF}(\tilde{\xi}^{\pm})^{IJ}T_J = 0$, and also 
\begin{align}
    \mathcal{P}_{\star}(\tilde{\xi}^{\pm})^{IJ} T_J &= \begin{pmatrix}
        \mathcal{P}_{gg\star}(\tilde{\xi}^{\pm})^{\mu\nu\rho\sigma}\tilde{\xi}^{\pm}_{(\rho} X_{\sigma)}+ \lambda \mathcal{P}_{gm\star}(\tilde{\xi}^{\pm})^{\mu\nu\rho} \tilde{\xi}^{\pm}_\rho \\
        \mathcal{P}_{mg\star}(\tilde{\xi}^{\pm})^{\mu\rho\sigma}\tilde{\xi}^{\pm}_{(\rho} X_{\sigma)} + \lambda\mathcal{P}_{mm\star}(\tilde{\xi}^{\pm})^{\mu\rho}\tilde{\xi}^{\pm}_\rho 
    \end{pmatrix} 
    = \begin{pmatrix}
        0 \\
        0
    \end{pmatrix}
\end{align}
where the second equality follows from conditions (\ref{eq:sgg}-\ref{eq:smm}) and symmetries (\ref{eq:symgg}) and (\ref{eq:symmg}). Therefore $\mathcal{P}(\tilde{\xi}^{\pm})^{IJ}T_J = 0$. Hence $\tilde{V}^{\pm}$ are genuine eigenspaces (rather than generalized eigenspaces) with eigenvalues $\tilde{\xi}^{\pm}_0$, and eigenvectors that depend smoothly on $\xi_i$. 

\subsection{\texorpdfstring{$\hat{V}^{\pm}$}{Hat V}}

These are the spaces associated with the $\hat{\xi}^{\pm}_0$-groups of eigenvalues where $\hat{\xi}^{\pm}_0$ are the two real solutions to $\hat{g}^{\mu\nu}\hat{\xi}^{\pm}_\mu\hat{\xi}^{\pm}_\nu = 0$. We'll now show that our above construction of eigenvectors with eigenvalue $\hat{\xi}^{\pm}_0$ in the 2-derivative theory extends to the weakly coupled higher-derivative theory with only minor modifications.

We introduce the same basis as before, and take indices A,B,... to take values 0,2,3. The only line which needs changing is equation (\ref{eq:mat}), as we now want to find solutions $(t_{AB}, s_C)$ to

\begin{equation} \label{eq:matmod}
    \begin{pmatrix}
        \mathcal{P}_{gg\star}(\hat{\xi}^{+})^{ABCD} & \mathcal{P}_{gm\star}(\hat{\xi}^{+})^{ABC} \\
        \mathcal{P}_{mg\star}(\hat{\xi}^{+})^{ACD} & \mathcal{P}_{mm\star}(\hat{\xi}^{+})^{AC}
    \end{pmatrix}
    \begin{pmatrix}
        t_{CD} \\
        s_C
    \end{pmatrix} = \begin{pmatrix}
        \hat{P}_{\alpha}^{\,\,\,\beta AB} \hat{\xi}^{+}_{\beta} v^{\alpha} \\
        \hat{g}^{A\alpha}\hat{\xi}^{+}_{\alpha}w
    \end{pmatrix}
\end{equation}
However in standard E-M theory, we found that the kernel of the matrix on the left hand side is trivial and so its determinant is non-zero. By continuity, its determinant is also non-zero for sufficiently weak coupling, and so its kernel is still trivial and there is still a unique solution $(t_{AB}(v^\alpha,w), s_C(v^\alpha,w))$ to (\ref{eq:matmod}). Both sides of (\ref{eq:matmod}) still depend smoothly on $(v^\alpha, w)$, $\xi_i$, $g_{\mu\nu}$ and $A_\mu$ and their derivatives, and so the solution also depends smoothly on these things.

The rest of the construction follows the same steps as for the 2-derivative theory, and hence $\hat{V}^{\pm}$ are genuine eigenspaces with eigenvalues $\hat{\xi}^{\pm}_0$ and eigenvectors that depend smoothly on $\xi_i$ and the fields and their derivatives.

\subsection{\texorpdfstring{$V^{\pm}$}{V}}

These are the spaces associated with the $\xi^{\pm}_0$-groups of eigenvalues where $\xi^{\pm}_0$ are the two real solutions to $g^{\mu\nu}\xi^{\pm}_\mu\xi^{\pm}_\nu = 0$. Since we are only considering weak coupling, we can assume that these eigenvalues are sufficiently close to $\xi^\pm_0$ so that 
 $\hat{g}^{\sigma\nu}\xi_\sigma \xi_\nu \neq 0$ and $\tilde{g}^{\sigma\nu}\xi_\sigma \xi_\nu \neq 0$. Therefore the eigenvalues and eigenvectors in the 4-dimensional generalized eigenspaces $V^{\pm}$ are those in Case I that don't also satisfy $\tilde{g}^{\sigma\nu}\xi_\sigma \xi_\nu = 0$. We will show that $V^{\pm}$ are genuine eigenspaces by closely following the argument in \cite{Kovacs:2020ywu} for Horndeski theories.

The first step of the argument is to establish that the deformed eigenvalues are real. We proceed by defining
\begin{equation}
    H^{\pm}_\star = \pm
    \begin{pmatrix}
        B_\star & A_\star \\
        A_\star & 0 
    \end{pmatrix}
\end{equation}
where $A_\star$ and $B_\star$ are defined as in (\ref{eq:ABCdef}) but by only using the non-gauge-fixing parts of the principal symbol. $H^{\pm}_\star$ is Hermitian since $\mathcal{P}_{\star}(\xi)^{IJ}$ is symmetric and real. We then define the Hermitian form $(,)_\pm$ on vectors $v^{(i)} = (T^{(i)}_I, T'^{(i)}_I)$ in $V^\pm$ (viewed as 4-dimensional complex vector spaces) by
\begin{equation} \label{eq:8}
    (v^{(1)},v^{(2)})_\pm = v^{(1)\dagger} H^{\pm}_\star v^{(2)}
\end{equation}
We show this is positive definite for standard E-M theory, and hence by continuity it is positive definite for sufficiently weakly coupled E-M theory. In standard E-M, $V^\pm$ are genuine eigenspaces each with one eigenvalue $\xi^{\pm}_0$ and eigenvectors $v^{(i)} = (t_{\mu\nu}^{(i)},s_\rho^{(i)},\xi^{\pm}_0 t_{\mu\nu}^{(i)},\xi^{\pm}_0 s_\rho^{(i)})$ satisfying the equations which define Subcase Ib. $A_\star$ and $B_\star$ are also block diagonal so the Hermitian form splits into a gravitational part and a Maxwell part:
\begin{equation}
    (v^{(1)},v^{(2)})_\pm = \pm \left[ t^{(1)\ast}_{\mu\nu}\left(2\xi^{\pm}_0 A_\star + B_\star\right)^{\mu\nu\rho\sigma} t^{(2)}_{\rho \sigma} + s^{(1)\ast}_{\mu}\left(2\xi^{\pm}_0 A_\star + B_\star\right)^{\mu\rho} s^{(2)}_{\rho} \right]
\end{equation}
In \cite{Kovacs:2020ywu} it is shown that the gravitational part simplifies to $-g^{0\nu}\xi^{\pm}_\nu t^{(1)\ast}_{\mu\nu}P^{\mu\nu\rho\sigma}t^{(2)}_{\rho \sigma}$ where
\be
P_{\alpha}^{\,\,\,\beta\mu\nu} = \delta_\alpha^{(\mu}g^{\nu)\beta}-\frac{1}{2}\delta_\alpha^\beta g^{\mu\nu}
\ee
Using (\ref{eq:caseib}), we can also reduce the Maxwell part, leading to
\begin{equation}
    (v^{(1)},v^{(2)})_\pm = \mp g^{0\nu}\xi^{\pm}_\nu\left[ t^{(1)\ast}_{\mu\nu}P^{\mu\nu\rho\sigma}t^{(2)}_{\rho \sigma} + 2 s^{(1)\ast}_{\mu}g^{\mu\rho} s^{(2)}_{\rho} \right]
\end{equation}
To simplify further, we pick a tangent space basis $(e_0)^\mu = \xi^{\pm\mu}$, $(e_1)^\mu \propto \xi^{\mp\mu}$, $(e_2)^\mu$ and $(e_3)^\mu$ with
\begin{equation}
    g(e_0,e_1) = 1, \,\, g(e_{\hat{i}},e_{\hat{j}}) = \delta_{\hat{i}\hat{j}}
\end{equation}
where $\hat{i}, \hat{j} = 2,3$ and all other contractions vanish. In this basis, equation (\ref{eq:caseib}) becomes $s_0 = 0$, and so $s^{(1)\ast}_{\mu}g^{\mu\rho} s^{(2)}_{\rho} = s^{(1)\ast}_{\hat{i}}s^{(2)}_{\hat{i}}$. Similarly for the gravitational part, \cite{Kovacs:2020ywu} shows that the conditions defining Subcase Ib imply that all components of $t_{\mu\nu}$ either vanish or depend linearly on the traceless quantity $t_{\hat{i}\hat{j}}$. Furthermore, they show that $t^{(1)\ast}_{\mu\nu}P^{\mu\nu\rho\sigma}t^{(2)}_{\rho \sigma} = t^{(1)\ast}_{\hat{i}\hat{j}}t^{(2)}_{\hat{i}\hat{j}}$. Therefore
\begin{equation}
    (v^{(1)},v^{(2)})_\pm = \mp \xi^{\pm0}\left( t_{\hat{i}\hat{j}}^{(1)\ast}\, t_{\hat{i}\hat{j}}^{(2)}+2s_{\hat{i}}^{(1)\ast}\, s_{\hat{i}}^{(2)}\right)
\end{equation}
Our convention was that $\mp \xi^{\pm0} > 0$ so this is non-negative for $v^{(1)}=v^{(2)}=v$. Suppose that $(v,v)_\pm=0$. Then $s_{\hat{i}}=0$ and so equation (\ref{eq:caseib2}) becomes $\tilde{g}^{11}s_1 = 0$. But $0\neq \tilde{g}^{\sigma\nu}\xi^{\pm}_\sigma \xi^{\pm}_\nu = \tilde{g}^{11}$, and hence this implies $s_1 = 0$. Therefore $s_\rho = 0$. Similarly $t_{\hat{i}\hat{j}}=0$ implies that $t_{\mu\nu} = 0$ by Subcase Ib conditions \cite{Kovacs:2020ywu}. Hence $(v,v)_\pm\ge0$ with equality iff $v=0$. Therefore we have shown that the Hermitian form is positive definite for $2$-derivative E-M theory, and hence, by continuity, also for weakly coupled E-M theory.

In \cite{Kovacs:2020ywu}, it is shown that the existence of this positive definite form on the complex vector space $V^\pm$ implies the eigenvalues in the $\xi_0^\pm$-group are real so long as a) $\mathcal{P}_{\star}(\xi)^{IJ}$ is symmetric (which we have) and b) the corresponding eigenvectors are in the kernel of $\mathcal{P}_{GF}(\xi)^{IJ}$. The second condition follows for $V^{\pm}$ by Case I conditions. Hence the eigenvalues in the $\xi_0^\pm$-groups are real.

We now proceed to show diagonalizability of $M(\xi_i)$. Since we've already shown that $\tilde{V}^\pm$ and $\hat{V}^\pm$ are genuine eigenspaces, we just need to show that $M(\xi_i)$ is diagonalizable within $V^\pm$. To this end, let $\xi_0$ be an eigenvalue in the $\xi_0^\pm$ group and consider a {\it left} eigenvector of $M(\xi_i)$ with this eigenvalue. One can show these are of the form
\begin{equation}
    w = \left(T_I, \xi_0 T_I\right) \begin{pmatrix}
        B & A \\
        A & 0
    \end{pmatrix}
\end{equation}
where
\begin{equation}
    T_I \mathcal{P}(\xi)^{IJ} = 0
\end{equation}
Now, using the symmetries of $\mathcal{P}_{\star}(\xi)^{IJ}$ one can show that a family of left eigenvectors with eigenvalue $\hat{\xi}_0^{\pm}$ is given by 
\begin{equation} \label{eq:fam}
    T_I = (\hat{\xi}_{(\mu}^{\pm}X_{\nu)}, \lambda \hat{\xi}_\rho^{\pm})
\end{equation}
for arbitrary $X_\nu$ and $\lambda$. Then by considering the Jordan normal form of $M$, we have that $w$ must be orthogonal to any vector $v=(u_I,u'_I)^T$ in any of $V^{\pm}$ or $\tilde{V}^{\pm}$, i.e.,
\begin{align}
    0 = wv &= \left(T_I, \hat{\xi}_0^{\pm} T_I\right) \begin{pmatrix}
        B^{IJ} & A^{IJ} \\
        A^{IJ} & 0
    \end{pmatrix} \begin{pmatrix}
        u_J \\
        u'_J
    \end{pmatrix} \\
    &= T_I(B^{IJ}+ \hat{\xi}_0^{\pm} A^{IJ})u_J + T_I A^{IJ} u'_J
\end{align}
By expanding $T_I$ through (\ref{eq:fam}) and using the fact that $X_\nu$ and $\lambda$ are arbitrary we get the two following conditions on $(u_I, u'_I)$:
\begin{gather}
    0 = \hat{\xi}_\nu^{\pm} (B^{\mu\nu J} + \hat{\xi}_0^{\pm} A^{\mu\nu J})u_J + \hat{\xi}_\nu^{\pm} A^{\mu\nu J} u'_J \\
    0 = \hat{\xi}_\mu^{\pm} (B^{\mu J} + \hat{\xi}_0^{\pm} A^{\mu J})u_J + \hat{\xi}_\mu^{\pm} A^{\mu J} u'_J
\end{gather}
We can eliminate $(\hat{\xi}_0^{\pm})^2$ using $\hat{g}^{\sigma\nu}\hat{\xi}_\sigma \hat{\xi}_\nu = 0$ to get
\begin{gather}
    \hat{\xi}_0^{\pm}R^\mu +S^\mu = 0 \label{eq:vec} \\ 
    \hat{\xi}_0^{\pm}R +S = 0 \label{eq:sca}
\end{gather}
where
\begin{gather}
    R^\mu \equiv -2(\hat{g}^{00})^{-1}\hat{g}^{0i}\xi_i A^{\mu 0J}u_J + B^{\mu 0 J}u_J + \xi_i A^{\mu i J}u_J + A^{\mu 0 J}u'_J \label{eq:Rmu} \\
    R \equiv -2(\hat{g}^{00})^{-1}\hat{g}^{0i}\xi_i A^{0J}u_J + B^{0 J}u_J + \xi_i A^{i J}u_J + A^{0 J}u'_J \label{eq:R} \\
    S^\mu \equiv -(\hat{g}^{00})^{-1}\hat{g}^{ij}\xi_i\xi_j A^{\mu 0 J}u_J+\xi_i B^{\mu i J}u_J + \xi_i A^{\mu i J}u'_J \\
    S \equiv -(\hat{g}^{00})^{-1}\hat{g}^{ij}\xi_i\xi_j A^{0 J}u_J+\xi_i B^{i J}u_J + \xi_i A^{i J}u'_J
\end{gather}
Note that none of $R^\mu$, $R$, $S^\mu$ or $S$ depend on $\hat{\xi}^{\pm}_0$. Therefore, since (\ref{eq:vec}) and (\ref{eq:sca}) are true for both signs $\pm$, they imply
\begin{equation}
    R^\mu = 0 = S^\mu,\,\,\,\,\,\, R = 0 = S
\end{equation}
Now, we can match co-efficients of powers of $\xi_0$ in the Bianchi-style identities (\ref{eq:sgg}-\ref{eq:smm}) to get the following identities on $A_\star$ and $B_\star$ (and on $C_\star$ but these aren't relevant to our argument):
\begin{align}
    A_\star^{\mu 0 I} &= 0 \\
    \xi_i A_\star^{\mu i I} + B_\star^{\mu 0 I} &= 0 \\
    A_\star^{0 I} &= 0 \\
    \xi_i A_\star^{i I} + B_\star^{0 I} &= 0
\end{align}
We can plug these into (\ref{eq:Rmu}) and (\ref{eq:R}) and see that all the $\star$ terms vanish in $R^\mu$ and $R$. Hence they only depend on gauge-fixing terms which are block diagonal. Writing $u_I = (t_{\mu\nu}, s_\rho)$ and $u'_I = (t'_{\mu\nu}, s'_\rho)$ and expanding $A_{GF}$ and $B_{GF}$, the equations $R^\mu=0$ and $R=0$ reduce to
\begin{gather}
    \tilde{P}_{\beta}^{\,\,\,i\rho\sigma} \xi_i t_{\rho\sigma} + \tilde{P}_{\beta}^{\,\,\,0\rho\sigma} t'_{\rho\sigma} = 0 \label{eq:trho} \\
    \tilde{g}^{\nu i} \xi_i s_\nu + \tilde{g}^{\nu 0}s'_\nu = 0 \label{eq:snu}
\end{gather}
The first condition is the same as in \cite{Kovacs:2020ywu}, whilst the second condition is its Maxwell equivalent. 

These are the conditions we will need to show diagonalizability of $V^\pm$. Let us start with $V^+$. For contradiction, assume we have a non-trivial Jordan block so there exists $w = (u_I,u'_I)^T \in V^+$ such that 
\begin{equation} \label{eq:v}
    v = (M(\xi_i) - \xi_0)w \neq 0
\end{equation}
where $v = (T_I, \xi_0 T_I) \in V^+$ is an eigenvector of $M(\xi_i)$ with eigenvalue $\xi_0$. Write $T_I = (r_{\mu\nu}, p_\rho)$. Now, (\ref{eq:v}) is equivalent to the two equations
\begin{gather}
    u'_I = \xi_0 u_I + T_I \label{eq:uI} \\
    \mathcal{P}(\xi)^{IJ}u_J = - (2\xi_0 A + B)^{IJ}T_J \label{eq:Pfull}
\end{gather}
Note that we can substitute (\ref{eq:uI}) into the conditions (\ref{eq:trho}) and (\ref{eq:snu}) to get 
\begin{gather}
    \tilde{P}_{\beta}^{\,\,\,\mu\rho\sigma} \xi_\mu t_{\rho\sigma} + \tilde{P}_{\beta}^{\,\,\,0\rho\sigma} r_{\rho\sigma} = 0 \label{eq:trho2} \\
    \tilde{g}^{\nu \mu} \xi_\mu s_\nu + \tilde{g}^{\nu 0}p_\nu = 0 \label{eq:snu2}
\end{gather}
We can also rewrite (\ref{eq:Pfull}) by extracting the gauge-fixing terms:
\begin{equation}
    \mathcal{P}_{\star}(\xi)^{IJ}u_J = - \mathcal{P}_{GF}(\xi)^{IJ}u_J - (2\xi_0 A_{GF} + B_{GF})^{IJ}T_J - (2\xi_0 A_\star + B_\star)^{IJ}T_J \label{eq:gaug}
\end{equation}
The gauge-fixing terms are block-diagonal so they split into a gravitational part and a Maxwell part. In \cite{Kovacs:2020ywu}, it is shown that (\ref{eq:trho2}) implies the gravitational gauge-fixing part vanishes. The Maxwell parts vanish similarly since
\begin{align}
    - (\mathcal{P}_{GF}(\xi)^{\mu\nu}s_\nu &+ (2\xi_0 A_{GF} + B_{GF})^{\mu\nu}p_\nu) \notag\\
    &= \hat{g}^{\mu\gamma}\tilde{g}^{\nu\delta}\xi_\gamma \xi_\delta s_\nu +(2\xi_0\hat{g}^{\mu 0}\tilde{g}^{\nu 0} + \xi_i\hat{g}^{\mu i}\tilde{g}^{\nu 0} + \xi_i\hat{g}^{\mu 0}\tilde{g}^{\nu i})p_\nu \notag\\
    &= \hat{g}^{\mu\gamma}\tilde{g}^{\nu\delta}\xi_\gamma \xi_\delta s_\nu + \hat{g}^{\mu\gamma}\tilde{g}^{\nu 0}\xi_\gamma p_\nu + \hat{g}^{\mu 0}\tilde{g}^{\nu\gamma}\xi_\gamma p_\nu \notag\\
    &= \hat{g}^{\mu 0}\tilde{g}^{\nu\gamma}\xi_\gamma p_\nu \notag\\
    &= 0
\end{align}
where the third equality follows from (\ref{eq:snu2}) and the fourth from using the Case I condition on the eigenvector $v$. Hence all the gauge-fixing terms in (\ref{eq:gaug}) vanish, and contracting the remaining terms with $T_I^{\ast}$ gives
\begin{align}
    T_I^{\ast}\mathcal{P}_{\star}(\xi)^{IJ}u_J &= - T_I^{\ast}(2\xi_0 A_\star + B_\star)^{IJ}T_J \notag\\
    &= (v,v)_+
\end{align}
But by symmetry of $\mathcal{P}_{\star}(\xi)^{IJ}$, the left-hand side can be written as
\begin{equation}
    u_J\mathcal{P}_{\star}(\xi)^{JI}T_I^{\ast} = u_J(\mathcal{P}_{\star}(\xi)^{JI}T_I)^{\ast} = 0
\end{equation}
where the second equality follows because $v = (T_I, \xi_0 T_I)$ is a Case I eigenvector of $M$ and therefore $\mathcal{P}(\xi)^{IJ}T_J = 0$ and $\mathcal{P}_{GF}(\xi)^{IJ}T_J = 0$. But this implies that $(v,v)_+ = 0$, and since $(,)_+$ is positive definite this means $v=0$. This is a contradiction, and hence our assumption of a non-trivial Jordan block must be false, i.e., $M(\xi_i)$ must be diagonalizable in $V^+$. Repeating the arguments for $V^-$ gives us the same result. 

\subsection{Construction of Symmetrizer}

To summarize, we have found that $\tilde{V}^\pm$ and $\hat{V}^\pm$ are spaces of smoothly varying eigenvectors of $M(\xi_i)$ with real eigenvalues. We have also found that $V^\pm$ have bases of eigenvectors with real eigenvalues. However, the eigenvectors may not have smooth dependence on $\xi_i$ at points where eigenvalues cross, and so it is not obvious that the symmetrizer built from the eigenvectors will be smooth. 

Instead, as in \cite{Kovacs:2020ywu}, we will show that $H_\star^{\pm}$ is a symmetrizer for $M(\xi_i)$ within $V^{\pm}$. Let $v^{(1)} = (T_I^{(1)}, \xi^{(1)}_0 T^{(1)}_I)^T$ and $v^{(2)}= (T_I^{(2)}, \xi^{(2)}_0 T^{(2)}_I)^T$ be eigenvectors in $V^{\pm}$ with eigenvalues $\xi^{(1)}_0$ and $\xi^{(2)}_0$. Since all the eigenvalues associated with $V^{\pm}$ are real, we can take $v^{(1)}$ and $v^{(2)}$ to be real. Then
\begin{align}
    v^{(1)T}\left(M^T H_\star^{\pm} - H_\star^{\pm} M \right) v^{(2)} &= \left(\xi^{(1)}_0 - \xi^{(2)}_0\right)v^{(1)T}H_\star^{\pm}v^{(2)} \notag\\
    &= T_I^{(1)}\left( \left(\xi^{(1)2}_0 - \xi^{(2)2}_0\right) A_\star + \left(\xi^{(1)}_0 - \xi^{(2)}_0\right) B_\star\right)^{IJ}T^{(2)}_J \notag\\
    &= T_I^{(1)}\left(\mathcal{P}_{\star}(\xi^{(1)}) - \mathcal{P}_{\star}(\xi^{(2)}) \right)^{IJ}T^{(2)}_J = 0
\end{align}
The final equality follows because $v^{(1)}$ and $v^{(2)}$ are Case I eigenvectors, and so $\mathcal{P}_{\star}(\xi^{(1)})^{IJ}T_J^{(1)} = 0$ and $\mathcal{P}_{\star}(\xi^{(2)})^{IJ}T_J^{(2)} = 0$. Now, since eigenvectors form a basis of $V^{\pm}$, it follows that $H_\star^{\pm}$ is a symmetrizer for $M(\xi_i)$ within $V^{\pm}$. In particular, it depends smoothly on all its arguments and is positive definite.

We can construct the total symmetrizer in $V$ in an identical fashion to \cite{Kovacs:2020ywu}. Let $\{v^{\pm}_1, ..., v^{\pm}_4\}$ be a smooth basis for $V^{\pm}$, and let $\{\tilde{v}^{\pm}_1, ..., \tilde{v}^{\pm}_5\}$ and $\{\hat{v}^{\pm}_1, ..., \hat{v}^{\pm}_5\}$ be the smooth eigenvector bases constructed above for $\tilde{V}^{\pm}$ and $\hat{V}^{\pm}$. Let $S$ be the matrix whose columns are these basis vectors. Then the symmetrizer is given by
\begin{equation}
    K(\xi_i) = (S^{-1})^T \begin{pmatrix}
        \mathcal{H}^+_\star & 0 & 0 & 0 & 0 & 0 \\
        0 & I_5 & 0 & 0 & 0 & 0 \\
        0 & 0 & I_5 & 0 & 0 & 0 \\
        0 & 0 & 0 & \mathcal{H}^-_\star & 0 & 0 \\
        0 & 0 & 0 & 0 & I_5 & 0 \\
        0 & 0 & 0 & 0 & 0 & I_5 
    \end{pmatrix}
    S^{-1}
\end{equation}
where $\mathcal{H}^\pm_\star$ are 4x4 matrices with components 
\begin{equation}
    (\mathcal{H}^\pm_\star)_{AB} = (v_A^\pm)^T H^\pm_\star v_B^\pm
\end{equation}
$K(\xi_i)$ has smooth dependence on $\xi_i$, the fields and all their derivatives, is positive definite and satisfies equation (\ref{eq:symmet}). Therefore, at weak coupling, our Einstein-Maxwell EFT is strongly hyperbolic and hence admits a well-posed initial value problem. 

\section{Conclusion} \label{Conclusion}

We have considered Einstein-Maxwell theory extended by the leading (4-derivative) EFT corrections. We have used the methods of \cite{Kovacs:2020ywu} to prove that the modified harmonic gauge formulation of this theory admits a well-posed initial value problem when the initial data is weakly coupled, i.e., when the 4-derivative terms in the equations of motion are initially small compared to the 2-derivative terms. Note that our result concerns {\it local} well-posedness, i.e., it guarantees existence of a solution only for a small interval of time. Over long time intervals, the fields may become large (e.g. if a singularity forms), in which case the theory would not be weakly coupled and well-posedness is likely to fail. From an EFT perspective this is fine because there is no reason to trust the theory if the fields become large. 

It is interesting to ask how large the higher derivative terms can become before strong hyperbolicity fails. Ref. \cite{Reall:2021voz} discussed this question for the case of the scalar-tensor EFT. In that case, it was shown that one can define a characteristic cone associated purely with the physical degrees of freedom. Using this cone one can define a notion of weak hyperbolicity that is independent of any gauge-fixing procedure. When this condition is satisfied, it was suggested that a sufficient condition for the modified harmonic gauge formulation to be strongly hyperbolic might be that the null cones of $\tilde{g}^{\mu\nu}$ and $\hat{g}^{\mu\nu}$ should lie strictly outside the characteristic cone. The same might be true for the Einstein-Maxwell EFT that we have considered. But determining whether or not this is the case, whether for scalar-tensor or Einstein-Maxwell EFT, will require new ideas.

\section{Appendix} \label{AppendixChap2}

\subsection{Bianchi Identities}\label{AppendixBianchi}

If we make a variation $g_{\mu\nu} \rightarrow g_{\mu\nu}+ \delta g_{\mu\nu}$, $A_{\mu} \rightarrow A_{\mu}+ \delta A_{\mu}$, then the first order variation of our action $S$ is
\begin{equation}\label{actionvariation}
    \delta S = -\frac{1}{16 \pi G} \int \diff^4 x \sqrt{-g} \left( E^{\mu\nu}\delta g_{\mu\nu} + E^\mu \delta A_\mu \right)
\end{equation}

For our Einstein-Maxwell EFT, $S$ is invariant under diffeomorphisms and electromagnetic gauge transformations. An infinitesimal diffeomorphism $x^\mu \rightarrow x^\mu - X^\mu$ leads to a corresponding variation $g_{\mu\nu} \rightarrow g_{\mu\nu}+ \LL_X g_{\mu\nu}$, $A_{\mu} \rightarrow A_{\mu}+ \LL_X A_{\mu}$. Hence, if we plug in $\delta g_{\mu\nu} = \LL_X g_{\mu\nu} = 2\nabla_{(\mu}X_{\nu)}$, $\delta A_{\mu} = \LL_X A_\mu = X^\nu \nabla_\nu A_\mu + A_\nu \nabla_\mu X^\nu$ into (\ref{actionvariation}), then $\delta S$ must vanish for all $X^\mu, g_{\mu\nu}, A_\mu$:
\begin{equation}
    0 = \int \diff^4 x \sqrt{-g}\left[ 2 E^{\mu\nu}\nabla_{\mu}X_{\nu} + E^\mu (X^\nu \nabla_\nu A_\mu + A_\nu \nabla_\mu X^\nu) \right]
\end{equation}
We can integrate this by parts and do some rearranging to get
\begin{equation}
    0 = \int \diff^4 x \sqrt{-g} X_\nu \left( 2 \nabla_\mu E^{\mu\nu} + A^\nu \nabla_\mu E^\mu  + E_\mu F^{\mu\nu} \right)
\end{equation}
Since this must hold for all $X^\mu, g_{\mu\nu}, A_\mu$, we have the first Bianchi identity
\begin{equation} \label{eq:6}
    \nabla_\nu E^{\mu\nu}-\frac{1}{2}F^\mu_{\,\,\,\nu} E^\nu + \frac{1}{2}A^\mu \nabla_\nu E^\nu = 0
\end{equation}
Furthermore, by considering an infinitesimal gauge transformation $A_\mu \rightarrow A_\mu + \nabla_\mu \chi$ we get 
\begin{equation}
    0 = \int \diff^4 x \sqrt{-g} E^\mu \nabla_\mu \chi 
\end{equation}
and so integrating by parts we get the second Bianchi identity
\begin{equation} \label{eq:bim}
    \nabla_\mu E^\mu = 0
\end{equation}
Plugging this into (\ref{eq:6}) we get
\begin{equation} \label{eq:big}
    \nabla_\nu E^{\mu\nu}-\frac{1}{2}F^\mu_{\,\,\,\nu} E^\nu = 0
\end{equation}
These equations hold for any field configuration.

\chapter{The Laws of Black Hole Mechanics}\label{ChapterLaws}

\ifpdf
    \graphicspath{{Chapter3/Figs/Raster/}{Chapter3/Figs/PDF/}{Chapter3/Figs/}}
\else
    \graphicspath{{Chapter3/Figs/Vector/}{Chapter3/Figs/}}
\fi

In this Chapter we review the laws of black hole mechanics.

\section{Black Holes}

The Einstein equations famously admit solutions with "black hole" regions \cite{Hawking:1973}:
\begin{definition}
    A black hole region, $\mathcal{B}$, of an asymptotically flat spacetime is a region from which no future-directed null curve can escape to future null infinity. 
\end{definition}

There are many known exact black hole solutions, such as the highly symmetric Schwarzschild and Kerr-Newman metrics, the dynamical Oppenheimer–Snyder solution, the multi-black hole Majumdar-Papapetrou solution, and the exotic 5D black ring of Emparan and Reall \cite{Emparan:2002}. The formation of black holes from the gravitational collapse of matter in GR is demonstrated by the singularity theorems of Penrose \cite{Penrose:1965} \cite{Penrose:1969}. Furthermore, there is now overwhelming observational evidence for their existence, for example from the Event Horizon Telescope \cite{Akiyama:2019} or from the observation of stellar orbits around our galactic centre \cite{Ghez:1998} \cite{Gillessen:2009}.

An important geometric feature of a black hole, whether in GR or beyond, is the event horizon:
\begin{definition}
    The event horizon, $\cH$, is the boundary of a black hole region. 
\end{definition}

Event horizons are always null hypersurfaces.
 
A black hole solution is "stationary" if it has a Killing vector field $k^a$ that is timelike at null infinity, and it is "static" if $k^a$ is hypersurface-orthogonal. It is assumed that all physical black hole solutions settle down to stationary states after e.g. gravitational collapse or binary merger. Therefore, the properties of stationary black holes are important to classify. In GR with matter satisfying appropriate energy conditions, many results are known. For example, the horizon of a stationary black hole is smooth \cite{Chrusciel:2008} and, in $d=4$ dimensions, has topology $\mathbb{S}^2$ \cite{Hawking:1972}. Another example is the "Rigidity Theorem":

\begin{theorem}
    Rigidity Theorem (Hawking 1972 \cite{Hawking:1972}): The event horizon, $\cH$, of a stationary, analytic black hole solution of the Einstein equations with matter satisfying the Dominant Energy Condition is a Killing Horizon with Killing vector field $\xi^a$. 
\end{theorem}

The original proof by Hawking considers only $d=4$ dimensions, however the $d>4$ case have been proved in \cite{Hollands:2006rj} and \cite{Moncrief:2008mr}. If the spacetime is static then $\xi^a$ can be taken to be $k^a$, otherwise the black hole is said to be rotating and $\xi^a$ is a linear combination of $k^a$ and some other commuting Killing vector fields, $\psi^a_j$, with closed orbits of $2\pi$. By convention we normalise $\xi^a$ so that 
\begin{equation}
    \xi^a = k^a + \sum_j \Omega^j_\cH \psi_j^a
\end{equation}
for some constants $\Omega^j_\cH$ we call the angular velocities of the black hole with respect to $\psi^a_j$. Later work has partially eliminated the analyticity assumption \cite{Friedrich:1998} \cite{Racz:2000}, and, at the expense of adding a technical smallness assumption, eliminated it entirely \cite{Ionescu:2007vk}. Furthermore, the Rigidity Theorem has recently been generalized to vacuum gravity EFTs \cite{Hollands:2022ajj}. Going forward, we will assume it holds in all the EFTs we study, even if this has not been explicitly proven.

The implication of the Rigidity Theorem is that for any stationary black hole, there exists a Killing vector field $\xi^a$ which is a null normal to $\cH$. Therefore $\xi^a \xi_a$ is constant on $\cH$, and hence $\diff (\xi^a \xi_a)$ is normal to $\cH$. Therefore, there exists some function $\kappa$ defined on $\cH$ such that
\begin{align}
    \nabla_a (\xi^b \xi_b) \big|_{\cH}=& -2 \kappa \xi_a\\
    \implies \xi^b \nabla_b \xi_a \big|_{\cH}=& \kappa \xi_a
\end{align}
where Killing's equation was used to get to the second line. This $\kappa$ is known as the "surface gravity" and measures the failure of $\xi^a$ to be an affinely parameterized tangent vector to the horizon generators.

\section{The Laws of Black Hole Mechanics}

An important set of theorems governing the classical properties of black holes in GR is the laws of black hole mechanics. The laws consist of the following three statements, first proved (with matter satisfying appropriate energy conditions) by \cite{Bardeen:1973}:

\textbf{The Zeroth Law:} The surface gravity $\kappa$ of a stationary black hole is constant on $\cH$.

\textbf{The First Law:} Linearized perturbations of a stationary black hole satisfy
\begin{equation}
    \delta M = \frac{\kappa}{8 \pi G} \delta A + \sum_j \Omega^j_\cH \delta J_j + ...
\end{equation}
where $M$ is the ADM mass, $A$ is horizon area, $J_j$ is the ADM angular momentum with respect to $\psi_j^a$, and the ellipsis denotes extra terms that may arise if matter fields are present. For example, if the black hole has electric charge $Q$ then the term $\Phi \delta Q$ appears, where $\Phi$ is the "electric potential", another constant on $\cH$.

\textbf{The Second Law:} The horizon area $A$ of any (stationary or non-stationary) black hole is a non-decreasing function of time.

We give proofs of the zeroth and second laws in GR (at least under certain conditions) at the end of this chapter. In the original formulation of the laws, Bardeen et al. also hypothesised a third law: that a stationary black hole with $\kappa=0$ (a so-called "extremal" black hole) cannot be formed by a finite sequence of operations. However, there has been no mathematical proof of such a statement, and there has even been a counterexample recently proposed by \cite{Kehle:2022uvc}. Thus we will not include it in further discussion.

\section{Hawking Radiation}

There is a very striking resemblance between the laws as written above and the laws of thermodynamics. In particular, if we identify stationarity with thermal equilibrium, and make the identifications 
\begin{equation}\label{identifications}
    T=\lambda \kappa, \quad S= A/(8\pi G \lambda), \quad E=M, \quad \mu = \Omega_\cH 
\end{equation}
for any constant $\lambda$, then the first law of black hole mechanics becomes the first law of thermodynamics (for simplicity ignoring additional charges and taking $d=4$ dimensions so there is only one angular Killing vector $\psi^a$):
\begin{equation}
    \diff E = T \diff S + \mu \diff J
\end{equation}
where $E$ is the energy, $T$ the temperature, $S$ the entropy, $J$ the angular momentum, and $\mu$ is the chemical potential that ensures angular momentum conservation. Similarly, the zeroth law becomes the zeroth law of thermodynamics - that a thermal system in equilibrium has constant temperature - and the second law becomes the second law of thermodynamics - that entropy is always non-decreasing.

If this identification is to be taken more seriously than mere coincidence, then one must expect black holes to radiate at temperature $T$ just like a thermodynamic black box. This is rather surprising since, classically, no particle can be emitted from a black hole! However, Hawking grappled with this dilemma and in \cite{Hawking:1975} proved that quantum mechanical effects cause the black hole to radiate particles at a temperature (in units with $c=\hbar = k =1$)
\begin{equation}
    T_H= \frac{\kappa}{2\pi}
\end{equation}
Hawking considered the quantum field theory of a simple scalar field in curved spacetime (without going as far as quantizing the gravitational field itself) and proved that an observer far away from a Schwarzschild black hole at late times after gravitational collapse sees a blackbody spectrum of particles at temperature $T_H$. Furthermore, he showed this is robust to the black hole rotating or having charge, and holds for other matter models such as the electromagnetic field or fermionic fields. More generally, Kay and Wald \cite{Kay:1988mu} showed that a free scalar field in a stationary Hadamard state
across the horizon must be thermal at the Hawking temperature for any globally hyperbolic spacetime with a bifurcate Killing horizon. Of particular importance, none of these derivations rely on the precise form of the Einstein equations, and thus it is reasonable to expect they are independent of the theory of gravity.

Thus, black holes are indeed thermodynamic objects, and in the stationary case have temperature $T_H$. Therefore $\lambda = \frac{1}{2\pi}$ in (\ref{identifications}), and so black holes in GR should have entropy given by the Bekenstein-Hawking entropy
\be
    S_{BH} = A/4G
\ee
Through the process of Hawking radiation, the black hole will lose energy and shrink, which would seem to violate the second law. However, this neglects the entropy of the matter emitted by the black hole, which the results of \cite{Hawking:1975} and \cite{Bekenstein:1974ax} show compensates for the loss of entropy of the black hole. Therefore, we should think of the second law as the classical limit of a "Generalized Second Law" stating that the sum of black hole entropy and the entropy of any matter surrounding it is non-decreasing.

All of this strongly suggests that the laws of black hole mechanics ought to hold in some shape or form in any physically reasonable beyond-GR theory of gravity. However, the structure of the Einstein equations is a crucial component of the proofs in GR, and thus generalizing them to beyond-GR theories of gravity is highly non-trivial. Nevertheless, the rest of this thesis will show that all three can be proved in the setting of gravitational EFTs with a wide range of matter models.

\section{Gaussian Null Coordinates}

Before we discuss proofs of the laws in detail, we introduce important coordinate systems that can be used to describe the geometry near the horizon of a black hole. Since the laws of black hole mechanics are concerned with quantities defined on $\cH$, we will use these coordinate systems extensively in our proofs. They are two appropriate choices of Gaussian Null Coordinates (GNCs), a class of coordinate systems first systematically analyzed by \cite{Moncrief:1983xua}. The first choice of GNCs applies to both the stationary and dynamical setting, whilst the second choice will only be used in the stationary case to prove the zeroth law.

As mentioned above, $\cH$ is a null hypersurface. We assume $\cH$ is smooth and has generators that extend to infinite affine parameter to the future. In the stationary settings of the zeroth and first laws, this assumption has been proved to be true in GR \cite{Chrusciel:2008} and we assume it holds in gravitational EFTs as well. It is not generally true when considering dynamical black holes as in the second law, for example in a black hole merger. However, as discussed below and in Chapter \ref{ChapterSecondLaw}, we will only concern ourselves with the dynamical situation of a black hole settling down to equilibrium, as envisioned in \cite{Wall:2015}, \cite{Bhattacharyya:2021jhr} and \cite{Hollands:2022}. In this setting the assumption is reasonable. 

\subsection{Affinely Parameterized GNCs} \label{APGNCs}

Take a spacelike cross-section $C$ of $\cH$ which the horizon generators intersect exactly once, and take $x^A$ to be a co-dimension 2 co-ordinate chart on $C$. Let the null geodesic generators have affine parameter $v$ and future directed tangent vector $l^a$ such that $l = \partial_{v}$ and $v=0$ on $C$. We can transport $C$ along the null geodesic generators a parameter distance $v$ to obtain a foliation $C(v)$ of $\cH$. Finally, we uniquely define the null vector field $n^a$ on $\cH$ by $n \cdot (\partial/\partial x^A) =0$ and $n \cdot l = 1$. The co-ordinates $(r, v, x^A)$ are then assigned to the point affine parameter distance $r$ along the null geodesic starting at the point on $\cH$ with coordinates $(v,x^A)$ and with tangent $n^a$ there. The geodesic equations can be used to show that the metric in these GNCs is of the form 
\begin{equation}
    g = 2 \text{d}v \text{d}r - r^2 \alpha(r,v,x^C) \text{d}v^2 -2 r\beta_{A}(r,v,x^C) \text{d}v \text{d}x^A + \mu_{A B}(r,v,x^C) \text{d}x^A \text{d}x^B
\end{equation}
The non-zero inverse metric components are $g^{vr} = 1$, $g^{rr} = r^2 \alpha + r^2 \mu^{A B} \beta_A \beta_B$, $g^{r A} = r \mu^{A B} \beta_B$ and $g^{A B} = \mu^{A B}$, where $\mu^{A B}$ is the inverse of $\mu_{A B}$.

This choice of coordinates will be referred to as "affinely parameterized GNCs". $\cH$ is the surface $r=0$, and $C$ is the surface $r=v=0$. We raise and lower $A, B, C, ...$ indices with $\mu^{A B}$ and $\mu_{A B}$. We denote the induced volume form on $C(v)$ by $\epsilon_{A_1 ... A_{d-2}} = \epsilon_{r v A_1 ... A_{d-2}}$ where $d$ is the dimension of the spacetime. The covariant derivative on $C(v)$ with respect to $\mu_{A B}$ is denoted by $D_{A}$. We also define
\begin{equation}
    K_{A B} \equiv \frac{1}{2} \partial_{v}{\mu_{A B}}, \quad \bar{K}_{A B} \equiv \frac{1}{2} \partial_{r}{ \mu_{A B} }, \quad K \equiv K^{A}\,_{A}, \quad \bar{K} \equiv \bar{K}^{A}\,_{A}
\end{equation}
$K_{AB}$ describes the expansion and shear of the horizon generators. $\bar{K}_{AB}$ describes the expansion and shear of the ingoing null geodesics orthogonal to a horizon cut $C(v)$.

Affinely parameterized GNCs are not unique: we are free to change the affine parameter on each generator of $\cH$ by $v'=v/a(x^A)$ with arbitrary $a(x^A)>0$. This will lead to a change $(v,r, x^A)\rightarrow (v',r', x'^A)$ with $v' = v/a(x^A) + O(r)$, $r'=a(x^A)r+O(r^2)$, $x'^A = x^A + O(r)$ near the horizon. Details of how this transformation changes the quantities above are given in \cite{Hollands:2022}, and will be discussed in Chapter \ref{ChapterBHEntropy}. The remaining freedom in our affinely parameterized GNCs is to change our coordinate chart $x^A$ on $C$, however all calculations in this thesis are manifestly covariant in $A, B,...$ indices and so this freedom will not change any of the expressions.

\subsubsection{Boost Weight}\label{BW}
An important concept in affinely parameterized GNCs is "boost weight". Suppose we take $a$ to be constant and consider the rescaling $v' = v / a$, $r'=a r$, which will produce a new choice of affinely parameterized GNCs above. If a quantity $T$ transforms as $T'(v',r',x^A)=a^b T(v,r,x^A)$, then $T$ is said to have boost weight $b$. See \cite{Hollands:2022} for a full definition. Some important facts are stated here:
\begin{itemize}
    \item A tensor component $T^{\mu_1 ... \mu_n}_{\beta_1 ... \beta_m}$ has boost weight given by the sum of $+1$ for each $v$ subscript and each $r$ superscript, and $-1$ for each $r$ subscript and $v$ superscript. $A, B, ...$ indices contribute 0. E.g. $T^{A}_{v v r B}$ has boost weight $+1$.
    \item $\alpha$, $\beta_{A}$, and $\mu_{A B}$ have boost weight 0. $K_{A B}$ and $\bar{K}_{A B}$ have boost weight $+1$ and $-1$ respectively.
    \item  If $T$ has boost weight $b$, then $D_{A_1}{... D_{A_n}{ \partial_{v}^{p} \partial_{r}^q T } }$ has boost weight $b+p-q$.
    \item If $X_{i}$ has boost weight $b_i$ and $T= \prod_{i} X_{i}$, then $T$ has boost weight $b=\sum_{i} b_{i}$.
\end{itemize}

In Lemma 2.1 of \cite{Hollands:2022}, it is proved that boost weight is independent of the choice of affinely parameterized GNCs on $\cH$. More precisely, a quantity of certain boost weight in $(r, v, x^A)$ GNCs on $\cH$ can be written as the sum of terms of the same boost weight in $(r', v', x'^A)$ GNCs on $\cH$, where $v'=v/a(x^A)$ on $\cH$.

\subsection{Killing Vector GNCs} \label{KVGNCs}

In the stationary setting of the zeroth law we will also use another choice of GNCs, hereafter referred to as \textit{Killing vector GNCs}. 

We assume the Rigidity Theorem holds in our gravitational EFTs, and thus for a stationary black hole there exists a Killing vector field $\xi^a$ that generates $\cH$. Therefore we can take a similar construction to the above, except we take the null geodesic generators to have non-affinely parameterized, future-directed tangent vector $\xi=\partial_{\tau}$, where $\tau$ is the Killing parameter. In the notation of \cite{Bhat:2022}, this leads to coordinates $(\rho, \tau, x^A)$ with metric of the form
\begin{equation}
    g = 2 \text{d}\tau \text{d}\rho - \rho X(\rho,x^C) \text{d}\tau^2 +2 \rho \omega_{A}(\rho,x^C) \text{d}\tau \text{d}x^A + h_{A B}(\rho,x^C) \text{d}x^A \text{d}x^B
\end{equation}
The non-zero inverse metric components are $g^{\tau\rho} = 1$, $g^{\rho\rho} = \rho X + \rho^2 h^{A B} \omega_A \omega_B$, $g^{\rho A} = - \rho h^{A B} \omega_B$ and $g^{A B} = h^{A B}$, where $h^{A B}$ is the inverse of $h_{A B}$.

$\cH$ is the surface $\rho=0$, and $C$ is the surface $\rho=\tau=0$. In these co-ordinates we raise $A, B, C,...$ indices with $h^{A B}$ and $h_{A B}$ and denote the induced volume form on $C(\tau)$ by $\varepsilon_{A_1 ... A_{d-2}} = \epsilon_{\rho \tau A_1 ... A_{d-2}}$. The covariant derivative on $C(\tau)$ with respect to $h_{A B}$ is denoted by $\DD_{A}$. 

The differences between the two GNCs are two-fold. Firstly, since $\xi=\partial_{\tau}$ is a Killing vector, the unknown metric coefficients $X, \omega_A$ and $h_{A B}$ are independent of $\tau$. Secondly, the fact that $\tau$ is not necessarily an affine parameter means the coefficient of $\text{d}\tau^2$ only comes with a factor of $\rho$, whereas $\text{d}v^2$ comes with a factor of $r^2$ in the affinely parameterized GNCs.

A relationship between these two forms of GNCs is crucial to prove the zeroth law in gravitational EFTs, as we shall see in Chapter \ref{ChapterZerothLaw}.

\section{Proofs of the Laws in GR and Beyond}

We now discuss proofs of the laws of black hole mechanics in GR, and attempts to generalize them to various beyond-GR theories. For a more thorough review, see \cite{Sarkar:2019}.

\subsection{The Zeroth Law}\label{ZerothLawBackground}

The zeroth law of black hole mechanics in GR is the statement that the surface gravity, $\kappa$, of a stationary black hole is constant on $\cH$, where $\kappa$ is defined by the equation

\begin{equation}\label{surfacegravdef}
    \xi^\beta \nabla_\beta \xi^\alpha \Big|_\cH = \kappa \xi^\alpha.
\end{equation}

Since Hawking radiation has the same temperature $T_H = \kappa /(2\pi)$ regardless of the theory of gravity, we should expect the statement of the zeroth law to hold unchanged in physically reasonable beyond-GR theories of gravity. To see where the difficulties lie in proving such a statement, let us work in the Killing vector GNCs of Section \ref{KVGNCs}. One can compute the 
$\alpha =\tau$ component of both sides of equation (\ref{surfacegravdef}) and find

\begin{equation}
    \kappa = \frac{1}{2} X(\rho, x^C)\Big|_{\rho=0}
\end{equation}

From this we see that $\kappa$ is clearly independent of $\tau$. Therefore, proving the zeroth law is equivalent to showing that
\begin{equation}
    \partial_{A} X(\rho, x^C)\Big|_{\rho=0} = 0
\end{equation}
One now does some experimenting and evaluates components of the Ricci tensor in Killing vector GNCs on the horizon. In particular, one finds (the latter is Raychaudhuri's equation on the horizon of a stationary black hole)
\be \label{RicciComps}
    R_{\tau A}\Big|_{\rho=0} = -\frac{1}{2} \partial_A X, \quad \quad R_{\tau \tau}\Big|_{\rho=0}=0
\ee
Therefore, to proceed in GR, one uses the Einstein equations to get\footnote{This is essentially equivalent to the steps in the original proof by Bardeen et al., but phrased in the language of Chapter \ref{ChapterZerothLaw} of this thesis.}
\be
    \partial_A X\Big|_{\rho=0} = -16\pi G T_{\tau A}\Big|_{\rho=0}, \quad T_{\tau\tau}\Big|_{\rho=0}=0
\ee
If we are in vacuum GR then we are done. If there is a non-zero $T_{\mu \nu}$, we now assume it satisfies the Dominant Energy Condition (DEC): that $-T^{\mu}_{\,\,\,\,\nu}V^{\nu}$ is a future-directed causal vector (or zero) for all future-directed causal vectors $V^{\mu}$. This means $T^{\mu}_{\,\,\,\,\nu}\xi^{\nu}|_{\cH}$ is causal, but by the above $T^{\mu}_{\,\,\,\,\nu}\xi^{\nu}\xi_{\mu}|_{\cH} = T_{\tau\tau}|_{\rho=0}=0$. One can show this implies $T^{\mu}_{\,\,\,\,\nu}\xi^{\nu}|_{\cH}$ is proportional to $\xi^{\mu}$ which means $T_{\tau A}|_{\rho=0}=0$, and hence the zeroth law holds. (In fact, we only need $T_{\mu\nu}$ \textit{minus any parts proportional to the metric} to satisfy the DEC because $g_{\tau \tau}|_{\rho=0}=0$ and $g_{\tau A}|_{\rho=0}=0$. This allows for e.g. the inclusion of a negative cosmological constant which is excluded by the DEC.) 

If we are considering some beyond-GR theory then there will be additional terms modifying the Einstein equations: $R_{a b} - 1/2 R g_{a b} + H_{a b} = 8\pi G T_{a b}$ for some tensor $H_{a b}$. Plugging this into the first equation of (\ref{RicciComps}) will give $\partial_A X|_{\rho=0} = 2H_{\tau A}-16\pi G T_{\tau A}|_{\rho=0}$, and so we must additionally deal with $H_{\tau A}|_{\rho =0}$ in a proof of the zeroth law. Since this could be some complicated contraction of curvature tensors and matter fields, this is in principle not easy. We will have to hope for some structure in our equations of motion that allows us to do this.

\subsubsection{Ghosh and Sarkar - Lovelock Gravity}

The first work to achieve this in a large class of beyond-GR theories was Ghosh and Sarkar \cite{Ghosh:2020dkk} who studied Lovelock
gravity - the most general diffeomorphism-invariant theory of vacuum gravity with second order equations of motion. The Lovelock Lagrangian is
\begin{equation}
    \mathcal{L} = \sum_{n=1}^{[(d-1)/2]} \alpha_n \frac{1}{2^n} \delta^{a_1 b_1 ... a_n b_n}_{c_1 d_1 ... c_n d_n} R_{a_1 b_1}^{\,\,\,\,\,\,\,\,\,\,\,c_1 d_1} ... R_{a_n b_n}^{\,\,\,\,\,\,\,\,\,\,\,\,\,c_n d_n}
\end{equation}
where $\alpha_1=1$ and $\alpha_n$ are arbitrary constants for $n>1$. The $n=1$ term is the Einstein-Hilbert Lagrangian. In $d=4$ there are no further terms in the Lagrangian, and in $d=5$ the only additional term is the Einstein-Gauss-Bonnet term. For general $d$, Ghosh and Sarkar prove that the problematic $H_{\tau A}$ term in this theory is always a linear combination of $\partial_B X$ on the horizon, meaning the equations of motion can be written as
\begin{equation}\label{LanczosEoM}
    (\delta^{A}_{B} - 2^j \alpha_j L^{A}_{B}) \partial_A X\Big|_{\rho=0}  = 0
\end{equation}
where $\alpha_j$ is the smallest non-zero $\alpha_n$ with $n>1$, and $L^{A}_{B}$ is some contraction of GNC quantities tensorial in $A$,$B$,... indices. To proceed, they assume $L^A_B$ and $X$ are analytic in the $\alpha_n$ with $n>1$. They motivate this assumption with the argument the black hole solution should have a smooth limit to GR, but it is more restrictive than would be desired. Working order-by-order in the $\alpha_n$, they prove the only solution to (\ref{LanczosEoM}) is $\partial_A X\big|_{\rho=0}=0$ and thus the zeroth law holds.

Furthermore, \cite{Reall:2021voz} has shown that all Horndeski theories (diffeomorphism-invariant theories of gravity and a scalar field that have second order equations of motion) have equations of motion with the same structure as (\ref{LanczosEoM}). Therefore the same argument as above can be applied if one assumes analyticity in the coefficients. Alternatively, as argued in \cite{Reall:2021voz}, if the theory is sufficiently "weakly coupled" (meaning the higher derivative terms are small compared to the leading order 2-derivative terms, as will happen if the theory is treated as an EFT and the solution is in the regime of validity), then the matrix multiplying $\partial_A X|_{\rho=0}$ will be invertible and hence the only solution is $\partial_A X|_{\rho=0} = 0$. 

\subsubsection{Bhattacharyya et al. - EFTs of Vacuum Gravity}

Not long after the above result was published, Bhattacharyya et al. \cite{Bhat:2022} were able to design a much more general method that proved the zeroth law for any diffeomorphism-invariant EFT of vacuum gravity (a result that also follows from the analysis of \cite{Hollands:2022ajj}):
\begin{equation}
    \mathcal{L} = R + \sum_{n=1}^{\infty} l^{n} \mathcal{L}_{n+2}
\end{equation}
where $\mathcal{L}_{n+2}$ is a sum of arbitrary $(n+2)$-derivative terms constructed covariantly from the metric. The equations of motion for this theory are
\be \label{vacuumgraveom}
    E_{\mu\nu}[g_{\alpha\beta}] \equiv E^{(0)}_{\mu\nu}[g_{\alpha\beta}]+\sum_{n=1}^{\infty}l^{n} E^{(n)}_{\mu\nu}[g_{\alpha\beta}]=0
\ee
where $E^{(0)}_{\mu\nu} = R_{\mu\nu}-\frac{1}{2}R g_{\mu\nu}$ and $l^{n} E^{(n)}_{\mu\nu}$ is the variation of $l^{n} \mathcal{L}_{2n+2}$. The only undesirable assumption the proof makes is that the stationary black hole solution $g_{\alpha\beta}$ is analytic in the UV scale $l$:
\begin{equation}
    g_{\alpha \beta} = g_{\alpha \beta}^{(0)} + l g_{\alpha \beta}^{(1)} + l^2 g_{\alpha \beta}^{(2)}+ \ldots
\end{equation}
In particular, the Killing vector GNC metric components can be written as series in $l$, e.g.
\be
    X=X^{(0)}+l X^{(1)} + l^2 X^{(2)}+...
\ee
Using the notation
\be
    f^{[n]} = \sum_{m=0}^{n}l^m f^{(n)},
\ee
the proof consists of working through the following four steps:
\begin{enumerate}
    \item At order $l^0$, the $(\tau A)$ component of the equation of motion, $E_{\tau A}[g_{\alpha\beta}]=0$, on the horizon is
    \be \label{Bhatzero}
        E^{(0)}_{\tau A}[g^{(0)}_{\alpha\beta}]\Big|_{\rho=0} = -\frac{1}{2} \partial_A X^{(0)} = 0
    \ee
    and so $\kappa^{(0)}$ is constant.
    \item Proceed inductively in powers of $l$ by assuming $\kappa^{[k-1]}$ is constant. Prove that the coordinate transformation
    \begin{equation}\label{Bhatcoord}
    \rho = r\kappa^{[k-1]} v, \quad \quad \tau = \frac{1}{\kappa^{[k-1]}} \log\left(\kappa^{[k-1]} v\right).
    \end{equation}
    brings $g^{[k-1]}_{\alpha\beta}$ into affinely parameterized GNC form, and that all positive boost weight quantities made from $g^{[k-1]}_{\alpha\beta}$ vanish on the horizon. 
    \item At order $l^k$, the higher derivative parts of $E_{\mu \nu}[g_{\alpha\beta}]$ can only depend on $g^{[k-1]}_{\alpha\beta}$, so we can write
    \be \label{Bhatlk}
    \text{At order} \,\, l^k, \quad E_{\tau A}[g_{\alpha\beta}]\Big|_{\rho=0} = -\frac{1}{2}l^k \partial_A X^{(k)} + \sum_{n=1}^{k}l^{n} E^{(n)}_{\tau A}[g^{[k-1]}_{\alpha\beta}] = 0
    \ee
    \item Under the coordinate transformation (\ref{Bhatcoord}),
    \be
        E^{(n)}_{\tau A}[g^{[k-1]}_{\alpha\beta}]\Big|_{\rho=0} = e^{\kappa^{[k-1]}\tau} E_{v A}^{(n)}[g^{[k-1]}_{\alpha\beta}]\Big|_{r=0}.
    \ee
    $E_{v A}^{(n)}[g^{[k-1]}_{\alpha\beta}]$ is a boost weight $+1$ quantity made from $g^{[k-1]}_{\alpha\beta}$, hence by step 2 it vanishes on the horizon. Therefore (\ref{Bhatlk}) simplifies to $-\frac{1}{2}l^k \partial_A X^{(k)}\big|_{\rho=0} = 0$ and hence the induction proceeds.
\end{enumerate}
This is the inspiration for the proofs in Chapter \ref{ChapterZerothLaw}, where we will show how to replace the analyticity assumption with the condition that the solution lies in the regime of validity of the EFT. The above is a slick argument that exemplifies how powerful the notion of boost weight can be when considering quantities on the horizon of a stationary black hole. The result provides a significant step forward in proving the validity of the zeroth law in vacuum gravity beyond-GR theories. Bhattacharyya et al predicted the method could be generalized to EFTs of gravity and matter fields, so long as the 2-derivative energy-momentum tensor $T_{\mu\nu}$ satisfies the DEC (up to parts proportional to $g_{\mu\nu}$ as discussed above). In Chapter \ref{ChapterZerothLaw}, we will show this is indeed the case for the EFT of gravity, electromagnetism and a scalar field. 

\subsubsection{Wald and R\'{a}cz - Impose Additional Properties or Symmetries}

Finally, we note there are alternative proofs of the zeroth law that avoid use of the equations of motions, however they assume additional properties or symmetries of the spacetime. For example, Wald and R\'{a}cz \cite{Racz:1995nh} prove the surface gravity is a non-zero constant if and only if the horizon can be extended to include a bifurcation surface (a cross-section on which the horizon Killing vector $\xi$ vanishes). They also show the zeroth law holds if the spacetime is (a) static, or (b) stationary, axisymmetric and has a $t-\phi$ reflection symmetry. These proofs still hold in arbitrary beyond-GR theories, but do not cover the full generality of stationary black hole solutions.

\subsection{The First Law}

The first law of black hole mechanics concerns linearized perturbations of stationary black holes. It relates how various fundamental quantities describing the black hole change under the perturbation. In GR it has the form
\begin{equation}\label{firstlawformula}
    \text{d}M = \frac{\kappa}{8 \pi G} \text{d}A + \sum_j \Omega^j_\cH \text{d}J_j + ...
\end{equation}
where the ellipsis denotes extra contributions that can arise when matter fields are present. 

There are two different formulations of the first law. In the "comparison version", we take a stationary black hole spacetime and consider a linearized perturbation of it to some nearby stationary black hole spacetime. Comparing the two spacetimes, (\ref{firstlawformula}) relates the difference in ADM mass $\diff M$, ADM momenta $J_j$ and associated matter charges to the difference in horizon area $\diff A$ of the two black holes. This was proved for stationary, axisymmetric perturbations in GR with fluid matter by \cite{Bardeen:1973}, and later for general perturbations in a wide class of matter fields by \cite{Sudarsky:1992ty}.

In the "physical process version", we instead take a stationary black hole, throw in some small amount of matter and wait until it settles down to equilibrium again. (\ref{firstlawformula}) then relates the mass of the matter, the angular momenta of the matter and the associated charges of the matter to the change in horizon area of the black hole. This was first proved in \cite{Hawking:1972hy} for vacuum GR, and then by \cite{Gao:2001} in Einstein-Maxwell theory. 

We refrain from giving a full proof of the first law in GR, as it will not inform anything else in this thesis. However, we emphasise the point that, much like the zeroth law, the proofs rely at crucial points on the precise form of the Einstein equations. For example, all proofs of the physical process version at some point turn to the Raychaudhuri equation on the horizon, which in affinely parameterized GNCs states
\begin{equation}\label{RaychEq}
    \partial_v K = - K^{A B} K_{A B} -R_{vv}
\end{equation}
We integrate this against $v\sqrt{\mu}$ over the horizon, integrate by parts and use the fact $\partial_{v}{\sqrt{\mu}} = \sqrt{\mu} K$ to relate the left hand side to the difference in horizon area between the initial and final equilibrium states, $\diff A$. The steps thus far apply to any theory of gravity (indeed getting (\ref{RaychEq}) is merely an exercise in expanding $R_{vv}$ in affinely parameterized GNCs on the horizon), however to relate the right hand side to the properties of the infalling matter, one equates $R_{v v}|_{\cH} = 8\pi G T_{vv}$ using the Einstein equations, and then manipulates the integrals into forms that can be interpreted as the mass, angular momentum, etc. of the matter. 

In beyond-GR theories, the additional terms that must be added to the Einstein equations will result in extra terms on the right hand side whose interpretation is unclear. One could hope these extra terms can be shown to vanish, just like has been shown for the zeroth law. However, a look back at the thermodynamic equivalence of the laws suggests it need not be so simple. We expected the zeroth law to hold unmodified in beyond-GR theories because Hawking radiation predicts black holes radiate with a temperature $T=\kappa/(2\pi)$ regardless of the theory of gravity. On the other hand, our interpretation of $A/(4G)$ as the entropy of the black hole followed only from the fact that it satisfies the first and second laws of black hole mechanics in GR. Without a deeper, theory independent calculation of black hole entropy, we should not necessarily expect $A/(4G)$ to still satisfy the first or second laws in beyond-GR theories. Instead, our philosophy should be to find some quantity that does satisfy the laws and interpret it as black hole entropy.

\subsubsection{The Wald Entropy}\label{WaldEntropy}

For stationary black holes in arbitrary diffeomorphism-invariant theories of gravity, the definitive answer to this problem is the "Wald entropy" \cite{Wald:1993nt}. This is constructed using a novel approach designed by Wald known as the covariant phase space formalism. This formalism studies the Lagrangian and its associated symplectic potential under a variation, and constructs a $(d-2)$-form $\mathbf{Q}_{\xi}$ called the Noether charge, which corresponds to the diffeomorphism invariance under an infinitesimal coordinate change in the $\xi$ direction. The Wald entropy of a stationary black hole is then defined by 
\begin{equation}
    S_{\text{Wald}} = \int_C \textbf{Q}_{\xi}
\end{equation}
There are ambiguities in this definition, however these do not affect the value of $S_{\text{Wald}}$ for stationary black holes \cite{Jacobson:1993vj} \cite{Iyer:1994ys}.  In affinely parameterized GNCs, it can be written as \cite{Wall:2015} (in units with $16\pi G = 1$)
\begin{equation} \label{WaldEntropyDef}
    S_{\text{Wald}} = -8\pi \int_C \text{d}^{d-2}x \sqrt{\mu} \frac{\delta \mathcal{L}}{\delta R_{r v r v}}
\end{equation}
where $\frac{\delta \mathcal{L}}{\delta R_{\mu \nu \rho \sigma}}$ denotes the variation of $\LL$ with respect to $R_{\mu\nu\rho\sigma}$ as if it were an independent field.

The original paper by Wald \cite{Wald:1993nt} proved that the Wald entropy satisfies the comparison version of the first law. Furthermore, it has been shown \cite{Mishra:2017sqs} to satisfy the physical process version of the first law. It also reduces to $S_{BH}$ for standard 2-derivative GR. Therefore, the Wald entropy provides a completely satisfactory definition of stationary black hole entropy in diffeomorphism invariant beyond-GR theories (and therefore, importantly for this thesis, in arbitrary gravitational EFTs).


\subsection{The Second Law - Dynamical Black Hole Entropy}

The second law of black hole mechanics in GR is the statement that the horizon area $A$ of any black hole is a non-decreasing function of time. Here we provide a simple proof under a particular set of assumptions, namely that the black hole has smooth horizon with future-complete generators and settles down to equilibrium at late times. These assumptions are not applicable to a highly dynamical situation such as a merger, however they do apply to scenarios such as the settling down period after a merger or gravitational collapse, or interaction with weak gravitational waves. The latter scenarios are the ones envisioned in Chapter \ref{ChapterSecondLaw}, and so this proof provides some motivation for the methods used there. There are proofs of the second law in GR under much more general assumptions, e.g. \cite{Chrusciel:2000cu}.

Under the above assumptions we can use our affinely parameterized GNCs to describe the geometry near the horizon. The area of a horizon cross-section $C(v)$ is 
\begin{equation}
    A(v) = \int_{C(v)} \diff^{d-2} x \, \sqrt{\mu} 
\end{equation}
where we recall $v$ is a future-directed affine parameter along the horizon generators, and thus can be taken as a notion of time. We again use $\partial_{v}{\sqrt{\mu}} = \sqrt{\mu} K$, so
\begin{align}
    \frac{\diff A}{\diff v} & = \int_{C(v)}\diff^{d-2} x \, \sqrt{\mu} K\\
    & = - \int_{C(v)}\diff^{d-2} x \, \sqrt{\mu} \int_{v}^{\infty} \diff v' \partial_v K(v',x^A) 
\end{align}
where in the second line we used our assumption that the black hole settles down to equilibrium, meaning the expansion of the generators $K\rightarrow 0$ as $v\rightarrow \infty$. We now use Raychaudhuri’s equation (\ref{RaychEq}) to get 
\begin{equation}
    \frac{\diff A}{\diff v} = \int_{C(v)}\diff^{d-2} x \, \sqrt{\mu} \int_{v}^{\infty} \diff v' \left( K^{A B}K_{A B} + R_{v v}\right)(v',x^A) 
\end{equation}
The first term, $K^{A B}K_{A B}$, is manifestly non-negative. In GR with matter satisfying the null energy condition, the Einstein equations imply $R_{v v}|_{\cH} = 8\pi G T_{vv} \geq 0$, and hence $\frac{\diff A}{\diff v}\geq 0$ which proves the second law.

Just like in the zeroth and first laws, we can see that the additional terms to be added to the Einstein equations in a beyond-GR theory will invalidate this proof: we no longer have reason to state $R_{vv}$ is non-negative. However, all hope is not lost: as we saw in the first law, the entropy of the black hole is no longer proportional to $A$, so we should not expect $A(v)$ to satisfy the second law anyway. We should instead try to construct a definition of \textit{dynamical} black hole entropy that does satisfy the second law. Moreover, we should aim to define an entropy for a dynamical black hole that (i) depends only on the geometry of a cross-section of the event horizon, (ii) agrees with the Wald entropy in equilibrium, and (iii) increases in any dynamical process.

\subsubsection{The Iyer-Wald Entropy}\label{IyerWaldEntropy}

One could hope that the Wald entropy itself satisfies these properties. Unfortunately, on arbitrary horizon cross-sections of a dynamical black hole, it turns out the Wald entropy definition suffers from ambiguities (named JKM ambiguities after Jacobson, Kang and Myers \cite{Jacobson:1993vj}). Iyer and Wald \cite{Iyer:1994ys} defined a particular prescription, known as the "Iyer-Wald entropy", which they proposed as a definition of dynamical black hole entropy. This prescription takes the formula (\ref{WaldEntropyDef}) on a horizon cross-section $C(v)$ and expands $\frac{\delta \mathcal{L}}{\delta R_{r v r v}}$ in affinely parameterized GNC quantities, but only keeps terms that are zero boost weight:
\begin{equation}
    S_{IW}(v) = 4 \pi \int_{C(v)} \text{d}^{d-2}x \sqrt{\mu} \, s^v_{IW}
\end{equation}
where $s^v_{IW} = -2 \frac{\delta \mathcal{L}}{\delta R_{r v r v}}|_{\text{Zero Boost Weight Terms Only}}$. The zeroth law implies that positive boost weight terms vanish on the horizon of a stationary black hole (as we shall prove in Chapter \ref{ChapterZerothLaw}), and hence $S_{IW}=S_{Wald}$ for stationary black holes. Furthermore, Iyer and Wald proved that it is independent of the choice of affinely parameterized GNCs. Therefore, it satisfies properties (i) and (ii). The motivation behind this particular prescription is that it satisfies the comparison version of the first law even under dynamical (rather than just stationary) perturbations of a stationary black hole, so long as it is evaluated on the bifurcation surface. However, whether it satisfies the second law is an open question.

\subsubsection{The Jacobson-Kang-Myers Entropy}

For the class of gravitational theories with Lagrangians that are polynomials in the Ricci scalar, $\LL = f(R)$, Jacobson, Kang and Myers have defined an entropy which satisfies all three properties (i-iii) \cite{Jacobson:1995uq}:
\begin{equation}
    S_{JKM}(v) = 4\pi \int_{C(v)} \text{d}^{d-2}x \sqrt{\mu} f'(R)
\end{equation}
This agrees with $S_{\text{Wald}}$ for stationary black holes, but differs from the Iyer-Wald entropy in dynamical scenarios, which suggests that $S_{IW}(v)$ will not satisfy (iii) in general. 

\subsubsection{The Iyer-Wald-Wall Entropy}\label{IWWExplanation}

More recent attempts have focused on finding an entropy that satisfies (iii) up to a given order in perturbation theory around a stationary black hole. Affinely parameterized GNCs provide a simple platform to find the order of perturbation of a particular term on the horizon in this scenario because of the fact that positive boost weight terms vanish on the horizon of a stationary black hole. This means the number of factors of positive boost weight quantities in a term equals its order of perturbation. For example, $K_{A B}$ has boost weight $+1$ and so a term such as $K_{A B} K^{A B}$ is quadratic order on the horizon.  

The Iyer-Wald-Wall (IWW) entropy was devised by Wall \cite{Wall:2015} (and formalised by Bhattacharyya et al. \cite{Bhattacharyya:2021jhr}) as a correction to the Iyer-Wald entropy, and satisfies the second law to \textit{linear order}\footnote{
See \cite{Chatterjee:2011wj, Sarkar:2013swa, Bhattacharjee:2015yaa} for earlier work establishing a linearized second law in particular theories.
} in perturbations around a stationary black hole for any diffeomorphism invariant theory of gravity and a real scalar field. We now sketch this construction and the proof of the second law. Wall's approach is to study the $E_{v v}$ equation of motion in affinely parameterized GNCs and prove it can always be manipulated into the following form on the horizon:
\begin{equation} \label{IWWdef}
    -E_{v v}\Big|_{\cH} = \partial_{v}\left[\frac{1}{\sqrt{\mu}} \partial_{v}\left(\sqrt{\mu} s^{v}_{IWW}\right) + D_{A}{ s^A }\right] + ...
\end{equation}
where the ellipsis denotes terms that are quadratic or higher order in positive boost weight quantities (and hence quadratic order in perturbations around a stationary black hole). $(s^{v}_{IWW}, s^A)$ is denoted the IWW entropy current. The IWW entropy of the horizon cross-section $C(v)$ is then defined as
\begin{equation}
    S_{IWW}(v) = 4 \pi \int_{C(v)} \text{d}^{d-2}x \sqrt{\mu} \,s^{v}_{IWW} 
\end{equation}
Schematically, $s^{v}_{IWW} = s^v_{IW} + $ terms that are linear or higher in positive boost weight quantities. $s^{v}_{IWW}$ and $s^A$ are only defined uniquely up to linear order in positive boost weight quantities because any higher order terms can be absorbed into the ellipsis. The higher order terms can be fixed so that $s^v_{IWW}$ is invariant under a change of affinely parameterized GNCs\footnote{The quantity $s^A$ can not be made gauge invariant in general \cite{Bhattacharyya:2022njk,Hollands:2022}.}, as proved in \cite{Hollands:2022}. Therefore, $S_{IWW}(v)$ satisfies (i) and (ii).

Taking the $v$-derivative of $S_{IWW}(v)$ gives
\begin{equation} \label{vderiv}
\begin{split}
    \dot{S}_{IWW} =& 4 \pi\int_{C(v)} \text{d}^{d-2}x \sqrt{\mu}\left[\frac{1}{\sqrt{\mu}} \partial_{v}(\sqrt{\mu} s^{v}_{IWW}) + D_{A}{s^A}\right] \\
    =& - 4 \pi \int_{C(v)} \text{d}^{d-2}x \sqrt{\mu} \int_{v}^{\infty} \text{d}v' \, \partial_{v} \left[ \frac{1}{\sqrt{\mu}} \partial_{v}(\sqrt{\mu} s^{v}_{IWW}) + D_{A}{s^A} \right](v',x)
\end{split}
 \end{equation}
where in the first line we trivially added the total derivative $\sqrt{\mu} D_{A}{s^A}$ to the integrand, and in the second line we assumed the black hole settles down to the stationary black hole solution at late times, so positive boost weight quantities vanish on the horizon as $v\rightarrow \infty$. The integrand can then be swapped for terms that are quadratic or higher in positive boost weight quantities using (\ref{IWWdef}) and the equation of motion $E_{v v}=0$. Thus $\dot{S}_{IWW}$ is quadratic order in perturbations around a stationary black hole and therefore $S_{IWW}$ satisfies the second Law to linear order. Even stronger than that, its change in time vanishes to linear order rather than just being non-negative. 

The original construction by Wall was performed in arbitrary diffeomorphism invariant theories of gravity and a real scalar field, but recent work has shown it can also be constructed in theories with more complicated matter fields or gravitational parts. Of particular relevance to this thesis, the work of Biswas, Dhivakara and Kundu \cite{Biswas:2022} has proved (\ref{IWWdef}) holds in any diffeomorphism-invariant and electromagnetic gauge-independent theory of gravity, electromagnetism and a real (uncharged) scalar field, with $s^v_{IWW}$ also electromagnetic gauge-independent. In addition, \cite{Deo:2023} has shown (\ref{IWWdef}) holds in Chern-Simons theories of gravity. 

\subsubsection{The Hollands-Kov\'acs-Reall Entropy}\label{HKRExplanation}

We saw that $S_{IWW}$ was constant to linear order in perturbations around a stationary black hole. To see a possible increase in the entropy, we must go to quadratic order. This has not been achieved for arbitrary diffeomorphism-invariant theories of gravity, like in the above. However, for arbitrary EFTs of gravity and a scalar field, this is exactly what the extension by Hollands, Kov\'acs and Reall (HKR) achieves in \cite{Hollands:2022}. As we discussed in \ref{EFTValidityDef}, in EFT we will only ever know our effective Lagrangian up to some finite order $l^{N-1}$ in the UV-scale $l$. So long as the solution lies in the regime of validity of EFT, we can bound the terms which we do not know by $O(l^N)$ (suppressing factors of $L$). In this case, we should only expect to be able to prove a second law up to $O(l^N)$ terms, i.e., we should only expect to prove $\dot{S}\geq -O(l^N)$. 

HKR prove that in the regime of validity of the EFT, the ellipsis in (\ref{IWWdef}) can be manipulated into the following form:
\begin{multline}\label{HKRForm}
    -E_{v v}\Big|_{\mathcal{N}} = \partial_{v}\left[\frac{1}{\sqrt{\mu}} \partial_{v}\left(\sqrt{\mu} s^{v}_{IWW}\right) + D_{A}{ s^A }\right] +\\
    \partial_{v}\left[\frac{1}{\sqrt{\mu}} \partial_{v}\left(\sqrt{\mu} \varsigma^{v}\right)\right] + \left(K_{A B}+X_{A B}\right) \left(K^{A B}+X^{A B}\right)+ \frac{1}{2} \left( \partial_v \phi + X\right)^2 + D_{A}{Y^{A}} + O(l^{N})
\end{multline}
$X^{A B}$ and $X$ are at least linear in positive boost weight quantities, and $Y^A$ and $\varsigma^{v}$ are at least quadratic. We shall provide more details of this construction in Chapter \ref{ChapterSecondLaw}. The entropy density is then defined by $s^v_{HKR} = s^v_{IWW} + \zeta^v$, and the HKR entropy is given by
\begin{equation} 
    S_{HKR}(v) = 4 \pi \int_{C(v)} \text{d}^{d-2}x \sqrt{\mu} \,s^{v}_{HKR} 
\end{equation}
Just as in (\ref{vderiv}), we can take the $v$-derivative of this and substitute in (\ref{HKRForm}) to get
\begin{equation} \label{dotSHKR}
        \dot{S}_{HKR}(v)= 4 \pi \int_{C(v)} \text{d}^{d-2} x \sqrt{\mu} \int_{v}^{\infty} \text{d}v' \left[W^2 + D_{A}{Y^{A}} + O(l^{N}) \right](v',x)
\end{equation}
where $W^2= \left(K_{A B} + X_{A B}\right) \left(K^{A B} + X^{A B}\right) + \frac{1}{2} \left( \partial_v \phi + X\right)^2$. Since $W^2$ and $Y^A$ are at least quadratic in positive boost weight, we see that $\dot{S}_{HKR}$ is $O(l^N)$ up to and including linear order in perturbations around a stationary black hole.

To study $\dot{S}_{HKR}$ at quadratic order, let us take the the second variation. $\delta^2 W^2 = (\delta W)^2$ is a positive definite form so must be non-negative. The second term is 
\begin{equation}
    \int_{C(v)} \text{d}^{d-2} x \sqrt{\mu} \int_{v}^{\infty} \text{d}v' D_{A} \delta^2 Y^A
\end{equation}
where $\sqrt{\mu}$ and $D_A$ are evaluated on the stationary solution we are perturbing around. The induced metric $\mu_{A B}$ is independent of $v$ on the horizon for the stationary solution (see (\ref{APGNCsSymmetry}) with $r=0$). Therefore, we can exchange the order of integrations and see the integrand is a total derivative on $C(v)$. Hence this integral vanishes and so $\dot{S}_{HKR}$ is non-negative to quadratic order, modulo $O(l^{N})$ terms. Thus it satisfies the second Law to quadratic order in the sense of $\delta^2 \dot{S}_{HKR} \geq -O(l^N)$.

The HKR procedure adds terms that are at least quadratic in positive boost weight to $S_{IWW}$. Therefore, for stationary black holes $S_{HKR}$ still agrees with the Wald entropy and hence satisfies property (ii). As we shall discuss in Chapter \ref{ChapterBHEntropy}, it is invariant under a change of affinely parameterized GNCs up to and including order $l^4$ terms, so satisfies property (i) to that order. Beyond that order, however, we will find it is not invariant.

In Chapter \ref{ChapterSecondLaw}, we shall show that further additions can be made to $S_{HKR}$ so that a new definition of dynamical black hole entropy satisfies the second law \textit{non-perturbatively} up to $O(l^N)$ terms. Furthermore, we shall show this can be generalized to EFTs of gravity, electromagnetism and a scalar field.

\subsubsection{Hollands, Wald and Zhang}

In very recent work, \cite{Hollands:2024vbe} has proposed another definition of black hole entropy, which we shall call $S_{HWZ}(v)$, for any diffeomorphism-invariant theory of gravity.  The proposal comes at the problem from a different angle, by aiming to construct an entropy that satisfies a "local" physical process version of the first law, meaning the first law is satisfied between any two horizon cross-sections $C(v_1)$, $C(v_2)$. It also satisfies the comparison version of the first law for dynamical perturbations without the need to evaluate it on the bifurcation surface. Furthermore, it agrees with the Wald entropy for stationary black holes and is independent of the choice of coordinates by construction, meaning it satisfies properties (i) and (ii). 

When it comes to property (iii), it satisfies the second law to linear order in perturbations around a stationary black hole, just like $S_{IWW}$. Indeed, whilst it does differ from $S_{IWW}$, they are related by $S_{HWZ} = (1-v\partial_v)S_{IWW}$ to leading order in perturbation. Its drawback as a possible definition of dynamical black hole entropy, however, is that for standard GR it does not agree with $S_{BH}$ and does not satisfy the second law non-perturbatively (unlike $S_{IWW}$ and $S_{HKR}$).

This proposal, therefore, provides an interesting new avenue of research on black hole entropy. It can be seen as a version of the IWW entropy for the apparent horizon (as opposed to the event horizon), at least to linear order in perturbation theory around a stationary black hole and in GR. Working with the apparent horizon comes with the benefit that it only increases after some matter has actually crossed it (as opposed to the event horizon which expands to meet the matter), which matches our physical intuition of how the entropy of a black hole should increase. More work will need to be done to resolve these different interpretations and to reveal the correct physics underneath.

\chapter{The Zeroth Law of Black Hole Mechanics in Effective Field Theory}\label{ChapterZerothLaw}

The contents of this chapter are the results of original research conducted by the author of this thesis. It is based on work published in \cite{Davies:2024fut} and unpublished work.

\section{Introduction}

The zeroth law of black hole mechanics is the statement that the surface gravity, $\kappa$, of a stationary black hole is constant. Equivalently,
\begin{equation}
    \partial_A \kappa = 0
\end{equation}
in Killing vector GNCs $(\tau,\rho,x^A)$. In Chapter \ref{ChapterLaws} we laid out the proof of this statement in GR, and discussed why we should expect it to remain true in physically reasonable beyond-GR theories. We also sketched proofs in specific beyond-GR theories, in particular the proof of Bhattacharyya et al. \cite{Bhat:2022} in arbitrary diffeomorphism-invariant EFTs of vacuum gravity under the assumption of analyticity in the UV scale $l$. In this Chapter, we will generalize this proof in two ways: 1) replacing the analyticity assumption with the condition that the solution lies in the regime of validity of the EFT, and 2) applying it to a much broader class of EFTs, namely the EFT of gravity, electromagnetism and a charged/uncharged scalar field.

\subsection{The Zeroth Law in EFT}

An EFT is, by definition, an approximation to some unknown UV-complete theory in some low energy limit. We assume it can be written as a series in $l$, the UV-scale to which we have integrated out all the unknown physics:
\begin{equation}
    \mathcal{L} = \mathcal{L}_2 + l \LL_3 + l^2 \LL_4 + ... 
\end{equation}
where $\LL_2$ is the classical 2-derivative theory of low energy physics we know well, and $l^{n-2}\LL_n$ with $n>2$ are the higher $n$-derivative corrections with unknown coefficients. We can, in theory, do observations of the physical world to determine what these coefficients are. However, in practice we can only ever determine finitely many of these coefficients up to some finite order $l^{N-1}$. Additionally, we should only consider our EFT to be valid if our light fields lie in the regime of validity, meaning they vary on scales $L\gg l$. In this case, we can bound the terms we do not know in terms of the small parameter $l/L$ 
\begin{equation}
    \mathcal{L} = \mathcal{L}_2 + l \LL_3 + ... + l^{N-1}\LL_{N+1} + O(l^N/L^{N+2})
\end{equation}
with equations of motion
\begin{equation}\label{EFTEOM}
    E_I \equiv E_{I}^{(0)}+ \sum_{n=1}^{N-1} l^n E_{I}^{(n)} = O(l^N/L^{N+2})
\end{equation}
where $I$ is an index over the field components and $l^n E_{I}^{(n)}$ is the variation of $l^n \LL_{n+2}$. In this context, we should not expect to be able to prove anything about the theory exactly, but merely up to $O(l^N/L^{N+2})$ terms. In particular, our goal for the zeroth law is to prove
\begin{equation}
    \partial_A \kappa = O(l^N/L^{N+2})
\end{equation}
rather than $\partial_A \kappa = 0$ exactly. However, the better we know our EFT (or the higher the order we assume it is valid for some yet-to-be-determined coefficients), the better we are able to prove the zeroth law.

\subsection{Regime of Validity of the EFT}\label{assumptionslbehaviour}

Let the metric $g_{\mu\nu}$ and matter fields $\phi_i$ be a stationary black hole solution to the EFT equations of motion. In the proof of Bhattacharyya et al. in vacuum gravity EFTs, it is assumed that the metric is analytic in the UV-scale $l$. However, in the EFT perspective, analyticity in $l$ is rather unnatural because we do not know the theory at order $l^N$, thus should not expect the solution to be determined at that order. We could instead assume that our solution lies in the regime of validity of the EFT and that, like our Lagrangian, the fields and their derivatives have an asymptotic expansion in $l/L$ up to $O((l/L)^N)$, e.g.
\begin{equation}
    g_{\mu\nu} = g_{\mu\nu}^{(0)} + (l/L)g_{\mu\nu}^{(1)} ... + (l/L)^{N-1}g_{\mu\nu}^{(N-1)} + O((l/L)^N)
\end{equation}
Under this assumption, the proof of Bhattacharyya et al. goes through with trivial modification to prove $\partial_A \kappa = O(l^N/L^{N+2})$ for vacuum gravity EFTs. An asymptotic expansion assumption may be reasonable in a stationary setting, and is the assumption taken by \cite{Hollands:2022ajj} to prove the Rigidity Theorem for vacuum gravity EFTs. However, an expansion in $l$ can lead to issues in a time-dependent situation, where secular terms (growing in time) typically arise in such an expansion. A simple example is a quasinormal mode of a linearly perturbed black hole. Such a mode is proportional to $e^{-i\omega t}$ with ${\rm Im}(\omega)<0$. Higher derivative terms will give a perturbative shift in the quasinormal frequency, i.e., $\omega = \omega_0 + l\omega_1/L + ... $. Low lying quasinormal modes certainly lie within the regime of validity of EFT, and decay exponentially in time. However if we expand in $l/L$ we obtain secular growth in the correction term relative to the zeroth order term: $e^{-i\omega t} = e^{-i \omega_0 t}(1- i \omega_1 l t/L + \ldots)$. Therefore, truncating at any order in $l/L$ becomes a poor approximation at late times, even though we have remained in the regime of validity of the EFT. Thus, ideally we would like to avoid taking an expansion in $l/L$ altogether.

In this Chapter, we will instead merely assume that the metric and matter fields lie in the regime of validity of EFT in Killing vector GNCs $x^{\mu}=(\tau,\rho,x^A)$ on the horizon cross-section $C$. Killing vector GNCs are only defined up to a change in $x^A$ co-ordinates, thus we would like to precisely formulate this in a way that is covariant in $A,B,...$ indices. We assume that we have a 1-parameter family of stationary black hole solutions $g_{\mu\nu}(x^\mu,L)$, $\phi_i(x^\mu,L)$ that are smooth in $x^\mu$ on $C$, and whose $n$-derivative quantities are uniformly bounded by $c_n/L^n$ in the following sense. For a covariant quantity $\psi^{B_1...B_q}_{A_1 ... A_p}$ on $C$, define the norm w.r.t. $h_{AB}$:
\begin{equation} \label{hABnorm}
    |\psi|_{h}^2 \equiv h^{A_1 C_1}...h^{A_p C_p}h_{B_1 D_1}... h_{B_q D_q} \psi^{B_1...B_q}_{A_1 ... A_p} \psi^{D_1...D_q}_{C_1 ... C_p}
\end{equation}
Then we assume that any $n$-derivative covariant quantity $\psi^{B_1...B_q}_{A_1 ... A_p}(x^A,L)$ on $C$ made from the Killing vector GNC metric quantities $X$, $\omega_A$, $h_{A B}$, $h^{A B}$ or the matter fields $\phi_i$ satisfies
\begin{equation}
    \sup_C |\psi|_{h} \leq \frac{c_n}{L^n}
\end{equation}
for some dimensionless, $L$-independent constant $c_n$. Note that $X$ and $\omega_A$ count as $1$-derivative quantities since $X = -\partial_\rho g_{\tau\tau}|_C$ and $\omega_A = \partial_\rho g_{\tau A}|_C$, whilst $h_{AB}$ and $h^{AB}$ are 0-derivative quantities. Similarly, the Maxwell field $F_{\mu\nu}$ counts as a 1-derivative quantity since it is the derivative of the Maxwell potential. If the matter fields are gauged, we assume this condition holds in the appropriate gauge used in the proof. These conditions can be interpreted as saying that $L$ is the smallest length/timescale associated to $g_{\mu\nu}$ and $\phi_i$ (for example, the size of the black hole). Then the solution is said to lie in the regime of validity of the EFT if $L\gg l$.   

Fixing $L$, we can WLOG take units with $L=1$, in which case $l$ is a small parameter. We will henceforth suppress any explicit dependence on $L$, and just write a quantity's order with respect to $l/L$ purely in terms of $l$. Factors of $L$ can be reinstated through dimensional analysis. 

We now define some notation that will be useful below. We say that a covariant quantity $\psi^{B_1...B_q}_{A_1 ... A_p}$ on $C$ is $O_{\infty}(l^k)$ if for each $m\geq 0$, there exists a dimensionless, ($l,L$-independent) constant $c_{m}$ such that 
\begin{equation}
    \sup_{C} |\DD^m \psi|_{h} \leq c_{m} l^k
\end{equation}
where $\DD^m \psi$ denotes the $m$-th $\DD_A$ derivative of $\psi^{B_1...B_q}_{A_1 ... A_p}$. Our regime of validity of the EFT can then be expressed as saying that $X$, $\omega_A$, $h_{A B}$, $h^{A B}$, $\phi_i$ and all their $\rho$-derivatives\footnote{We do not need to mention $\tau$-derivatives because $\xi = \partial/\partial \tau$ is the horizon Killing vector, so any $\tau$-derivative will vanish.} are $O_{\infty}(1)$, that they satisfy the EFT equation of motion (\ref{EFTEOM}) such that $E_I = O_{\infty}(l^N)$ on $C$, and that $l$ is a small parameter. Our goal for the zeroth law is to prove
\begin{equation}
    \partial_A \kappa = O_{\infty}(l^N)
\end{equation}

\section{The Zeroth Law in Vacuum Gravity EFTs}

We first detail the proof of the zeroth law in vacuum gravity EFTs, i.e., for Lagrangians of the form\footnote{We have included the cosmological constant $\Lambda$ here without any factors of $l$. Naively in EFT, we should expect 0-derivative terms to come with a factor $1/l^2$ relative to the 2-derivative $R$. This would not have any explicit bearing on the steps in the proof, but would raise questions about whether the solution could lie in the regime of validity of the EFT. As mentioned previously, however, observations find that $\Lambda$ is extremely small and so we will assume that $|\Lambda|\leq 1/L^2 $, i.e., it is of no larger scale than the $2$-derivative terms and is treated as a $2$-derivative quantity.}
\begin{equation}
    \mathcal{L} = -2\Lambda + R + \sum_{n=1}^{N-1} l^{n} \mathcal{L}_{n+2} + O(l^N)
\end{equation}
where $\mathcal{L}_{n+2} = \mathcal{L}_{n+2}(g_{\alpha \beta}, R_{\alpha\beta\gamma\delta}, \nabla_{\mu}R_{\alpha \beta \gamma \delta}, ...)$ is a sum of arbitrary covariant $(n+2)$-derivative terms\footnote{We could also include Chern-Simons terms which are not themselves covariant but produce covariant equations of motion, without changing any step of the proof.}. Many of these steps will carry over when we consider EFTs of gravity and matter fields.

We can write the equations of motion for this theory as
\be
    E_{\mu\nu}[g_{\alpha\beta}] \equiv R_{\mu\nu}-\frac{1}{2}R g_{\mu\nu} + \Lambda g_{\mu\nu} + l H_{\mu\nu}[g_{\alpha\beta}] = O(l^N)
\ee
where $l H_{\mu\nu}$ is the variation of $\sum_{n=1}^{N-1} l^{n} \mathcal{L}_{n+2}$ with respect to $g_{\mu\nu}$, and thus is a sum of contractions of $g_{\mu\nu}$, $g^{\mu\nu}$ and their derivatives.

As we noted in \ref{ZerothLawBackground}, in Killing vector GNCs one can calculate $R_{\tau A}$ on $C$ (equivalent to evaluating on the horizon due to $\tau$-invariance) and find
\begin{equation}
    R_{\tau A}\big|_{C} = -\partial_A \kappa
\end{equation}
Therefore, noting $g_{\tau A}|_{C} = 0$, evaluating the $(\tau A)$ equation of motion on $C$ gives
\begin{equation}
    \partial_A \kappa = l H_{\tau A}[g_{\alpha\beta}]\big|_{C} + O_{\infty}(l^N)
\end{equation}
$H_{\tau A}[g_{\alpha\beta}]\big|_{C}$ is a sum of contractions of $x^A$-covariant quantities made from the metric and its derivatives. Hence, using our regime of validity of the EFT assumptions, we immediately have $\partial_A \kappa = O_{\infty}(l)$. We now proceed by induction and assume $\partial_A \kappa = O_{\infty}(l^k)$ for $k\leq N-1$. Using a coordinate change that brings the metric into affinely parameterized GNC form up to $O_{\infty}(l^k)$ terms, we will show this implies $H_{\tau A}[g_{\alpha\beta}]=O_{\infty}(l^k)$ on $C$ and hence the induction proceeds, terminating at our goal of $\partial_A \kappa = O_{\infty}(l^N)$. 

\subsection{The Coordinate Change}

Recall that the Killing vector GNC quantity $X(\rho,x^A)$ is related to $\kappa(x^A)$ by $X|_{\rho=0} = 2\kappa$. Therefore, by smoothness,
\begin{equation}
    X(\rho,x^A) = 2\kappa(x^A) + \rho f(\rho,x^A)
\end{equation}
where $f(\rho,x^A)$ is regular on the horizon. The metric is thus (with $x^A$ dependence of all quantities suppressed)
\begin{equation}
    g = 2 \text{d}\tau \text{d}\rho - \left[2 \kappa \rho + \rho^2 f(\rho)\right] \text{d}\tau^2 +2 \rho \omega_{A}(\rho) \text{d}x^A \text{d}\tau + h_{A B}(\rho) \text{d}x^A \text{d}x^B
\end{equation}

We now consider the coordinate transformation\footnote{Note this is slightly different from the coordinate transformation used in \cite{Bhat:2022} in that we have $(\kappa v+1)$ where they have $\kappa v$. We have added the $1$ so that $v=0$ corresponds to $\tau=0$, and also to put it in such a form that if the black hole is extremal, i.e., $\kappa=0$, then the transformation is the identity $\rho=r, \tau=v$ rather than undefined.} to $\Tilde{x}^\mu = (r,v,x^C)$, where
\begin{equation}\label{transform}
    \rho = r( \kappa v+1), \quad \tau = \frac{1}{\kappa} \log\left(\kappa v+1\right) \iff r = \rho e^{-\kappa\tau}, \quad v= \frac{1}{\kappa} (e^{\kappa \tau}-1)  
\end{equation}
with the $x^C$ co-ordinates unchanged. In these new co-ordinates, the horizon is $r=0$ and $C$ is $v=r=0$. One calculates the components of the metric in these new coordinates using 
\begin{equation}
    \Tilde{g}_{\mu\nu} = \frac{\partial x^{\alpha}}{\partial \Tilde{x}^{\mu}}\frac{\partial x^{\beta}}{\partial \Tilde{x}^{\nu}} g_{\alpha \beta}
\end{equation}
and finds it can be written in the form
\begin{equation}
    \Tilde{g}_{\mu\nu}=g^{aff}_{\mu\nu} + \partial_{A} \kappa \sigma^{A}_{\mu\nu}
\end{equation}
where $\sigma^{A}_{\mu\nu}$ and all its $v$ and $r$-derivatives are $O_{\infty}(1)$ on $C$ (which follows from our regime of validity of EFT assumptions), and $g^{aff}_{\mu\nu}$ is \begin{equation} 
    g^{aff} = 2 \text{d}v \text{d}r - r^2 f\big(r( \kappa v+1)\big) \text{d}v^2 + 2 r \omega_{A}\big(r( \kappa v+1)\big) \text{d}v \text{d}x^A + h_{A B}\big(r( \kappa v+1)\big) \text{d}x^A \text{d}x^B
\end{equation}
which is in affinely parameterized GNC form (see \ref{APGNCs}) with 
\begin{equation} \label{APGNCsSymmetry}
        \alpha(r,v)= f\big(r(\kappa v+1)\big), \quad \beta_A(r, v)= -\omega_{A}\big(r(\kappa v+1)\big), \quad 
        \mu_{A B}(r,v)=h_{A B}\big(r(\kappa v+1)\big)
\end{equation}
The inverse metric can be decomposed in the same manner. Furthermore, $H_{\mu\nu}$ is a tensor, and so we can calculate $H_{\tau A}[g_{\alpha\beta}]|_{C}$ in these new coordinates:
\begin{align}
    H_{\tau A}[g_{\alpha\beta}]|_{C} &= \frac{\partial \tilde{x}^{\mu}}{\partial \tau}\frac{\partial \tilde{x}^{\nu}}{\partial x^{A}} H_{\mu\nu}[\Tilde{g}_{\alpha\beta}]|_{C}\\ \nonumber
    &= H_{v A}[g^{aff}_{\alpha\beta} + \partial_{B} \kappa \sigma^{B}_{\alpha \beta}]|_{C}\\ \nonumber
    &= H_{v A}[g^{aff}_{\alpha\beta}]|_{C} + O_{\infty}(l^k)
\end{align}
where in the last line we used that $\partial_B\kappa \sigma^{B}_{\mu\nu}$ and all its $v$ and $r$-derivatives are $O_{\infty}(l^k)$ on $C$.

Therefore it just remains to deal with $H_{v A}[g^{aff}_{\alpha\beta}]|_{C}$. But $H_{v A}[g^{aff}_{\alpha\beta}]$ is a tensorial component of contractions of a metric in affinely parameterized GNCs, and thus we can talk about its boost weight. In particular, it has boost weight $+1$ since it has one down $v$-index. We shall now show that every positive boost weight quantity made from $g^{aff}_{\mu\nu}$ vanishes on the horizon and hence we are done.

\subsection{Positive Boost Weight Quantities on the Horizon}\label{PBWQH}

The most basic quantities we can make from a metric $g^{aff}_{\mu\nu}$ in affinely parameterized GNCs are of the form $\partial_{A_1}{... \partial_{A_n}{ \partial_{r}^{p} \partial_{v}^q \varphi } }$ with $\varphi \in \{\alpha, \beta_{A}, \mu_{A B}, \mu^{A B} \}$. Call such terms \textit{building blocks}. On the horizon $r=0$, all contractions of $g^{aff}_{\mu\nu}$, its inverse and their derivatives can be expanded out as sums of products of building blocks (there is no explicit appearance of the coordinates $(v,x^A)$ because they do not appear explicitly in the metric).

The boost weight of a quantity is defined in Section \ref{BW}. Boost weight is additive across products, and hence if we were to expand a positive boost weight quantity in terms of building blocks on the horizon, it would be a sum of products, each of which would have at least one factor of a positive boost weight building block.

The boost weights of these building blocks are
\begin{itemize}
    \item $\alpha$, $\beta_{A}$, $\mu_{A B}$, $\mu^{A B}$ have boost weight 0.
    \item $\partial_v$ derivatives each add $+1$ to the boost weight, $\partial_r$ derivatives each add $-1$, and $\partial_A$ derivatives add $0$.
\end{itemize}

Therefore positive boost weight building blocks are of the form 
\begin{equation} \label{PBB}
    \partial_{A_1}{... \partial_{A_n}{ \partial_{v}^{q} (\partial_{r}\partial_v)^{p}  \varphi } }\,\, \text{with}\,\, \varphi \in \{\partial_{v}\alpha, \partial_{v}\beta_{A}, \partial_{v}\mu_{A B}, \partial_{v}\mu^{A B}\}, \,\, p\geq 0, q\geq0
\end{equation}

Now, for our $g^{aff}_{\mu\nu}$ with GNC quantities (\ref{APGNCsSymmetry}), the $v$-dependence of $\alpha$, $\beta_A$, $\mu_{A B}$ and $\mu^{A B}$ is severely restricted because of the $\tau$-independence of the original Killing vector GNC quantities: $v$ only appears in the combination $rv$. Therefore, taking a $v$-derivative will produce an overall factor of $r$ and so we can write all of the quantities $\varphi \in \{\partial_{v}\alpha, \partial_{v}\beta_{A}, \partial_{v}\mu_{A B}, \partial_{v}\mu^{A B}\}$ in the form
\begin{equation}\label{basicpbw}
    \varphi = r f_{\varphi}\big(r(\kappa v+1)\big)
\end{equation}
Taking a $(\partial_{r}\partial_{v})$ derivative preserves this general form. Any additional $\partial_A$ or $\partial_v$ derivatives do not affect the overall factor of $r$ (and may even bring forward more powers of $r$). Thus, every positive boost weight building block satisfies 
\begin{equation}
        \partial_{A_1}{... \partial_{A_n}{ \partial_{v}^{q} (\partial_{r}\partial_v)^{p}  \varphi } } = r \partial_{A_1}... \partial_{A_n} \partial^{q}_v \left[f_{(\partial_{r}\partial_{v})^{p} \varphi } \big(r(\kappa v+1)\big) \right]
\end{equation}
and hence vanishes on the horizon $r=0$. Thus $H_{v A}[g^{aff}_{\alpha\beta}]|_{C} = 0$ and hence the induction proceeds. This concludes the proof of the zeroth law for vacuum gravity EFTs.

It is worth noting here that if the zeroth law holds exactly, i.e., $\partial_A \kappa \equiv 0$, then $\tilde{g}_{\mu\nu} = g^{aff}_{\mu\nu}$ and hence the above steps would show that all positive boost weight quantities made from the metric in the affinely parameterized GNCs determined by the transformation (\ref{transform}) vanish on the horizon. There are infinitely many choices of affinely parameterized GNCs, all related by $v'=v/a(x^A)$ on $\cH$ for some arbitrary function $a(x^A)>0$, as discussed in Section \ref{APGNCs}. However, we note Lemma 2.1 of \cite{Hollands:2022}, which states that on the horizon a quantity of certain boost weight made out of the metric in $(r, v, x^A)$ GNCs can be written as the sum of terms of the same boost weight in $(r', v', x^A)$ GNCs. This means that all positive boost weight quantities made from the metric in \textit{any} affinely parameterized GNCs would vanish on the horizon if the zeroth law holds exactly. It seems probable that one could show that our result $\partial_A \kappa = O_{\infty}(l^N)$ implies that all positive boost weight quantities made from the metric are $O(l^N)$ on the horizon, but to do so would involve more technical analysis that would stray from the main argument here.

\subsection{Including Matter}\label{IncMatter}

The glaring omission in the above proof is the inclusion of matter. There are many physical scenarios where matter fields are on similar energy scales to the gravitational field, and thus from an EFT point of view we should not ignore these matter fields in the Lagrangian. Is this method of proof robust to the inclusion of physically reasonable matter fields $\phi_j$ in the theory?

Let us see where the difficulties lie by first considering minimally coupled matter. In this case we can represent the matter contribution by some energy-momentum tensor $T_{\mu\nu}[\psi_i, g_{\alpha\beta}]$ (containing no derivatives of the metric) on the right hand side of the equations of motion $R_{\mu\nu}-\frac{1}{2}R g_{\mu\nu} + lH_{\mu\nu}[g_{\alpha\beta}] = T_{\mu\nu} + O(l^N)$. It is reasonable to impose that $T_{\mu\nu}$ satisfies the DEC up to parts proportional to the metric, just as in the proof of the zeroth law in GR. Then $T^{\mu}_{\,\,\,\,\nu} \xi^\nu$ (with possibly some term proportional to $\xi^\mu$ added on, but this will not matter) is causal on the horizon. We can decompose $T^{\mu}_{\,\,\,\,\nu} \xi^\nu$ into the Killing vector GNC basis vectors
\begin{equation}
    T^{\mu}_{\,\,\,\,\nu} \xi^\nu = A \xi^\mu + B \left(\frac{\partial}{\partial \rho}\right)^{\mu} + C^A \left(\frac{\partial}{\partial x^A}\right)^{\mu}
\end{equation}
and calculate this implies
\begin{equation}\label{DEC}
    0 \geq T^{\mu}_{\,\,\,\,\nu} \xi^\nu T_{\mu \sigma} \xi^{\sigma}\big|_{C} = 2AB + h^{A B} C_{A} C_{B} 
\end{equation}
We also have 
\begin{equation}
    T_{\tau\tau}\big|_{C} = B, \quad \quad T_{\tau A}\big|_{C} = C_{A}
\end{equation}
Now, the $(\tau A)$ and $(\tau\tau)$ equations of motion on $C$ give respectively (remembering that $R_{\tau\tau}|_{\cH}=0$)
\begin{align}
    \partial_A \kappa  &= -C_A + l H_{\tau A}[g_{\alpha\beta}]\big|_{C} + O_{\infty}(l^N) \label{TtauA}\\ 
    B & = l H_{\tau\tau}[g_{\alpha\beta}]\big|_{C} + O_{\infty}(l^N) \label{Ttautau}
\end{align}
the second of which tells us $B=O_{\infty}(l)$ on $C$. Plugging this into (\ref{DEC}) and assuming $A=O_{\infty}(1)$ on $C$ gives
\begin{equation}\label{Csquared}
    h^{A B} C_{A} C_{B} \leq -2AB = O_{\infty}(l)
\end{equation}
on $C$. We can plug this into (\ref{TtauA}) to get $|\partial \kappa|_h = O(l^{1/2})$ on $C$, however we are now rather stuck without more information about the matter theory. One can hope $|\partial \kappa|_h = O(l^{1/2})$ somehow implies the stronger result $\partial_A \kappa = O_{\infty}(l^{1/2})$. One can then try to proceed by induction, assuming $\partial_A \kappa = O_{\infty}(l^{k/2})$ and using the same coordinate change as above to show this implies $H_{\tau \tau}[g_{\alpha\beta}]|_{C} = O_{\infty}(l^{k/2})$ and hence $B=O_{\infty}(l^{1+k/2})$. However, even this is not quite enough to complete the induction, because (\ref{Csquared}) would then imply $|C|_h = O(l^{1+k/2})$ which gets us to $|\partial \kappa|_h = O(l^{1/2+k/4})$ by (\ref{TtauA}), which has not raised the order by $l^{1/2}$. 

We can complete the inductive loop however, if we fall back to assuming an asymptotic expansion in $l^{1/2}$ of the metric and $T_{\mu\nu}$,
\begin{equation}
    T_{\mu\nu} = \sum_{n=0}^{2N-1} l^{n/2} T_{\mu\nu}^{(n)} + O(l^N)    
\end{equation}
and assume that each partial sum $\sum_{n=0}^{m} l^{n/2} T_{\mu\nu}^{(n)}$ satisfies the DEC exactly, rather than up to $O(l^{(m+1)/2})$. In this case, comparing powers of $l^{1/2}$ in (\ref{Ttautau}) and using the notation $f^{[m]} = \sum_{n=0}^{m} l^{n/2} f^{(n)}$ gives us
\begin{equation}
    B^{[k+1]} = 0
\end{equation}
on $C$ and so the DEC on $T^{[k+1]}_{\mu\nu}$ gives us
\begin{equation}
    h^{AB} C_A^{[k+1]}C_B^{[k+1]} \leq -2 A^{[k+1]} B^{[k+1]}=0
\end{equation}
on $C$, which implies $C_A^{[k+1]} = 0$. This is precisely what we need to get $\partial_A \kappa = O_{\infty}(l^{(k+1)/2})$ from (\ref{TtauA}) and thus complete the induction.

We can see then, that even with minimal couplings, generalizing the proof of the zeroth law comes with difficulties if we want to do it under the same assumptions on the fields as before. Furthermore, in a scenario where the matter fields are comparable to the gravitational field, it does not make sense to do an EFT expansion in higher derivatives of only the metric. Instead, we should include all non-minimal couplings of matter and the metric in the Lagrangian, ordered by derivatives, with the 2-derivative minimal couplings as the leading order terms. In this case, we should not expect $T_{\mu\nu}$ to satisfy any standard energy condition, and indeed we should not even try to isolate the matter contribution to an energy-momentum tensor since it will contain derivatives of the metric. Instead we should write all the equations of motion, including those of the matter fields, as 
\begin{equation}
    E_{I}[\Phi_J] \equiv E^{(0)}_{I}[\Phi_J] + l H_{I}[\Phi_J]= O(l^N)
\end{equation}
where $\Phi_J = (g_{\mu\nu}, \phi_{j})$ and $E^{(0)}_I$ is the part of the equation of motion coming from the 2-derivative minimal couplings. 

As well as having the same potential difficulties with the minimal couplings as discussed above, there is now an additional issue in proceeding with the induction. Specifically, $H_I[\Phi_J]$ now depends on the matter fields as well as the metric, and thus there will be more positive boost weight quantities that $H_{v A}[g^{aff}_{\alpha\beta}, \phi_j]|_{\cH}$ can depend on. Therefore we cannot state that it vanishes unless we can add to the results in Section \ref{PBWQH} and prove that all positive boost weight quantities made from $\phi_j$ vanish on the horizon as well as those made from $g_{\alpha\beta}$. 

For a scalar field $\phi(x)$, this is trivial if we assume it is Lie derived by the horizon Killing vector. This is because $\phi = \phi(\rho,x^A)$, and so in $(v,r,x^A)$ coordinates $\phi = \phi\left(r(\kappa v+1),x^A \right)$, and hence
\begin{equation}
    \partial_v \phi = r \kappa \partial_\rho\phi\left(r(\kappa v+1),x^A \right)
\end{equation}
which is of the same form as the basic positive boost weight quantities (\ref{basicpbw}) built out of the metric. A general positive boost weight quantity made from $\phi$ will be of the form $\partial_{A_1}... \partial_{A_n} \partial_{v}^{q} (\partial_{r}\partial_v)^{p}  \partial_v \phi$, which again will always come with an overall factor of $r$, and hence vanish on the horizon. 

With a more complicated matter field however, there is no equivalent simple argument. For example, consider a Maxwell field $F_{\mu\nu}$. Again we assume in $(\tau,\rho,x^A)$ coordinates it has no $\tau$ dependence. However, this does not tell us anything about the positive boost weight component $F_{vA}$. We can calculate on the horizon
\begin{equation}
    F_{vA}\big|_{\cH} = \frac{1}{\kappa v+1} F_{\tau A}(\rho, x^A)\big|_{\rho=0}
\end{equation}
which vanishes if and only if $F_{\tau A}|_{\cH}$ vanishes. We have no reason to say this is the case without analyzing the matter theory more carefully.

If the zeroth law is satisfied, there are general arguments given in e.g. \cite{Wall:2024lbd} to show that all positive boost weight components of tensors $T$ that are Lie derived by the horizon Killing vector vanish on the horizon. However, they explicitly use the fact the zeroth law implies the horizon can be extended to include a bifurcation surface in the infinite past, and assume that $T$ is regular on the bifurcation surface. We cannot hope to use a modification of this argument here to show positive boost weight components vanish up to some order in $l$ because if the zeroth law does not hold then the horizon generators terminate in a parallelly propagated singularity in the past \cite{Racz:1992bp}, so we cannot use the existence or regularity of a bifurcation surface at any order.

Instead, we must hope that in specific matter models, the equations of motion for the matter fields allow us to deduce that positive boost weight quantities vanish on the horizon, or at least are sufficiently high order in $l$ that we can advance the induction. In the next sections, we will prove this is indeed the case in the EFT of gravity and some specific matter models. The only fundamental field other than the metric for which we have observed the classical approximation is useful is the Maxwell field, and so it is natural we should include it in our analysis. Scalar fields are also present in many hypothesised gravitational theories, such as Brans-Dicke theory, string theory and inflation, and are used to model matter species in the Standard Model and axionic dark matter, and are thus another important field to include. In the rest of this Chapter we shall prove the zeroth law for the EFT of gravity, electromagnetism and a scalar field. The proof has important differences depending on whether the scalar field is uncharged or charged, so we split the analysis of these two cases into separate sections.

\section{The Zeroth Law in the EFT of Gravity, Electromagnetism and a Real Uncharged Scalar Field}

We first consider the EFT of gravity, electromagnetism and a real uncharged scalar field $\phi$, which we shall refer to as Einstein-Maxwell-Scalar EFT. We assume diffeomorphism invariance and electromagnetic gauge invariance of the Lagrangian, so that it consists only of contractions of $R_{\alpha \beta \gamma \delta}$, $F_{\alpha \beta}$, $\phi$ and their covariant derivatives,
\begin{equation}
    \mathcal{L} = \mathcal{L}(g_{\alpha \beta}, R_{\alpha \beta \gamma \delta}, \nabla_{\mu}R_{\alpha \beta \gamma \delta}, ... , F_{\alpha \beta}, \nabla_{\gamma}F_{\alpha \beta}, ... , \phi, \nabla_{\alpha}\phi, ...  )
\end{equation}
The EFT Lagrangian is again of the form
\begin{equation} \label{actionzerothlaw}
    \LL = \LL_{2} + \sum_{n=1}^{N-1} l^{n} \mathcal{L}_{n+2} + O(l^N)
\end{equation}
where $\mathcal{L}_{n}$ contains all terms with $n$ derivatives. $\phi$ is counted as a 0-derivative term. $F_{\mu\nu}$ is counted as a 1-derivative term because it is the derivative of the Maxwell potential $A_{\mu}$, however we shall not need to involve $A_\mu$ explicitly anywhere in this section.

The most general diffeomorphism-invariant and electromagnetic gauge-invariant\footnote{If we were to merely insist on gauge invariance of the equations of motion rather than the Lagrangian, then the Chern-Simons term $A_\alpha F_{\beta \gamma} F_{\mu\nu} \epsilon^{\alpha\beta\gamma\mu\nu}$ could appear here in $d=5$ dimensions. This can be included in the following analysis without too much difficulty, but to avoid dealing with multiple dimensional cases we shall stick with gauge invariance of the Lagrangian.} 2-derivative Lagrangian, $\LL_2$, can be written as\footnote{We have included the 0-derivative term $V(\phi)$ in the 2-derivative Lagrangian $\LL_2$. As mentioned previously, in EFT we should expect 0-derivative terms to come with a factor $1/l^2$. This would not affect any step of the proof below, but would raise questions about our regime of validity assumptions. Since, like the cosmological constant, we do not observe a large scalar potential, we assume $|V(\phi)|\leq 1/L^2$ on $C$, and similarly for all its $\phi$-derivatives, so $V(\phi)$ is of no larger scale than the 2-derivative terms and is treated as a $2$-derivative quantity.}

\begin{equation} \label{0and2-deriv}
    \mathcal{L}_{2} = R - V(\phi) -\frac{1}{2} \nabla_{\alpha}{\phi} \nabla^{\alpha}{\phi} - \frac{1}{4} c_1(\phi) F_{\alpha\beta} F^{\alpha\beta} + c_2(\phi) F_{\alpha \beta} F_{\gamma \delta} \epsilon^{\alpha \beta \gamma \delta}
\end{equation}
An arbitrary function of $\phi$ multiplying $R$ can be eliminated by redefining the metric, whilst an arbitrary function of $\phi$ multiplying $\nabla_{\alpha}{\phi} \nabla^{\alpha}{\phi}$ can be eliminated by redefining $\phi$ \cite{Weinberg:2008}. Here we have taken units with $16\pi G = 1$ and rescaled $F_{\alpha \beta}$ appropriately. The final term is an axion-like coupling between $\phi$ and $F_{\mu\nu}$ which only appears in $d=4$ where the volume form $\epsilon_{\alpha \beta \gamma \delta}$ has 4 indices; in higher dimensions, it is taken that this term is not present. 

The only sign condition we put on the arbitrary functions $V(\phi), c_1(\phi)$ and $c_2(\phi)$ is that $c_1(\phi)\geq \tilde{c} > 0$ for some dimensionless positive constant $\tilde{c}$ (independent of $l$ and $L$). This is a sufficient condition for the energy-momentum tensor of the leading order 2-derivative theory to satisfy the Null Energy Condition (NEC). For Einstein-Maxwell theory without a scalar field, $c_1=1$, so this bounded below condition is also motivated on the grounds that we do not expect the scalar field to change the sign of coupling between $g_{\mu\nu}$ and $F_{\mu\nu}$ or make it arbitrarily small. If we were additionally to impose $V(\phi)\geq 0$ then the 2-derivative energy-momentum tensor would also satisfy the Dominant Energy Condition (DEC). However as discussed previously we only need the 2-derivative energy-momentum tensor minus any parts proportional to the metric to satisfy the DEC for the zeroth law to hold for the 2-derivative theory. Since $V(\phi)$ only appears in $T_{\mu\nu}$ multiplying $g_{\mu \nu}$, this is satisfied by our 2-derivative theory regardless of the sign of $V(\phi)$. Indeed, in the following proofs we will require no condition on the sign of $V(\phi)$.

The equations of motion for this action are
\begin{equation} \label{EoMs}
\begin{split}
    E_{\alpha\beta} \equiv E^{(0)}_{\alpha \beta} + l H_{\alpha \beta}[g_{\mu\nu}, F_{\gamma\delta},\phi] = O(l^N),  \quad & \quad  E_\alpha \equiv E^{(0)}_{\alpha} + l H_{\alpha}[g_{\mu\nu}, F_{\gamma\delta},\phi] = O(l^N)\\
    E \equiv E^{(0)} + l H&[g_{\mu\nu}, F_{\gamma\delta},\phi] = O(l^N)
\end{split}
\end{equation}
where $E^{(0)}_{\alpha \beta}, E^{(0)}_{\alpha}, E^{(0)}$ are the result of varying the 2-derivative terms from $\LL_{2}$, i.e.,
\begin{equation} \label{0th order EoM 1}
    E^{(0)}_{\alpha\beta}= R_{\alpha \beta} - \frac{1}{2} \nabla_{\alpha} \phi \nabla_{\beta} \phi -\frac{1}{2} c_1(\phi) F_{\alpha \delta} F_{\beta}^{\,\,\,\,\delta} - \frac{1}{2} g_{\alpha \beta} \left( R - V(\phi) -\frac{1}{2} \nabla_{\gamma}{\phi} \nabla^{\gamma}{\phi} - \frac{1}{4} c_1(\phi) F_{\gamma \delta} F^{\gamma\delta} \right)
\end{equation}
\begin{equation}
    E^{(0)}_\alpha = \nabla^{\beta}{\Big[c_1(\phi)F_{\alpha\beta} - 4 c_2(\phi) F^{\gamma \delta} \epsilon_{\alpha \beta \gamma \delta}\Big]}
\end{equation}
\begin{equation} \label{0th order EoM 2}
    E^{(0)} = \nabla^{\alpha}{\nabla_{\alpha}{\phi}}-V'(\phi)-\frac{1}{4} c'_1(\phi) F_{\alpha \beta} F^{\alpha \beta} + c'_2(\phi) F_{\alpha \beta} F_{\gamma \delta} \epsilon^{\alpha \beta \gamma \delta}
\end{equation}
The important equations of motion will be $E_{\tau A}$ and $E_\tau$ evaluated on $C$ in Killing vector GNCs, which give respectively
\begin{equation}\label{EMEtauA}
    \partial_{A}\kappa = - \frac{1}{2} c_{1}(\phi) \left( F_{A B} h^{B C} - F_{\tau \rho} \delta_{A}^{C} \right) F_{\tau C}\big|_{C} + l H_{\tau A}[g_{\mu\nu}, F_{\gamma\delta},\phi]\big|_{C} + O_{\infty}(l^N)
\end{equation}
\begin{equation} \label{EMEtau}
    h^{A B} \DD_{A}\left[c_1(\phi)F_{\tau B} - 8 c_2(\phi) \epsilon_{B}^{\,\,\,\,C}F_{\tau C}\right]\big|_{C}  = -l H_{\tau}[g_{\mu\nu}, F_{\gamma\delta},\phi]\big|_{C} + O_{\infty}(l^N)
\end{equation}

where $\DD_A$ is the covariant derivative with respect to $h_{AB}$.

\subsection{Assumptions}

We assume we have a stationary black hole solution $(g_{\mu\nu},F_{\mu\nu},\phi)$ with horizon Killing vector, $\xi= \frac{\partial}{\partial\tau}$, that Lie derives the Maxwell field, $\LL_\xi F = 0$, and the scalar field, $\LL_\xi \phi =0$, i.e., they are independent of $\tau$ in Killing vector GNCs. Furthermore, we assume the fields lie in the regime of validity of the EFT as given in \ref{assumptionslbehaviour}, and that $V(\phi),c_1(\phi),c_2(\phi) = O_{\infty}(1)$ on $C$. 

We will also make additional assumptions on the topology of the horizon cross-section $C$ to ensure our arguments below hold. The first is that $C$ is closed (compact without boundary), which will be used to neglect boundary terms when integrating by parts. The second is that $C$ is simply connected, which implies that every closed 1-form on $C$ is exact. This will be used to ensure that the electric potential, $\Phi = -\xi^{\mu}A_{\mu}$, can be defined\footnote{It seems probable that we can drop the simply connectedness of $C$ whilst still ensuring the electric potential is defined on the whole of the horizon. In GR with matter satisfying appropriate energy conditions, the topological censorship theorem \cite{Friedman:1993ty} implies that the domain of outer communication is simply connected. This is sufficient to ensure the electric potential is defined throughout the domain of outer communication, including on its boundary, the horizon. A small EFT correction should not change the topology, so this theorem likely applies to our setting as well. I thank Stefan Hollands for pointing this out.} globally on $C$. These assumptions hold e.g. if $C$ has spherical $S^{d-2}$ topology with $d\geq 4$ but not e.g. if $C$ has the topology of a black ring $S^1\times S^{d-3}$. 

Key steps in the proof below will use the Sobolev embedding theorem, the Poincar\'e inequality, and elliptic estimates on the compact Riemannian manifold $(C,h_{AB})$. The constants relating the various norms in these inequalities implicitly depend on $l$ (or $L$ if we treat it as variable) because they depend on the Riemannian metric $h_{A B}$. For small $l/L$, we expect both $h_{A B}$ and $h^{AB}$ should be bounded above in some sense, but since they are inverses of each other they should also be bounded below. Thus in the regime of validity of the EFT we should expect $h_{A B}$ to be $O(1)$, not larger or smaller. It seems reasonable to assume the constants in the aforementioned inequalities are also $O(1)$ in this case, and so do not affect the order of either side of the inequalities. To prove this rigorously however, would require delving into analysis that is rather tangential to the main purpose of this thesis, and hence we shall sweep it under the rug and take it as another assumption.

\subsection{Zeroth Law Formulation}

When electromagnetic fields are included in a black hole theory, the zeroth law is usually generalized to include a statement about their behaviour on the horizon. We can view $F_{\tau A}|_{C}$ as a 1-form on the submanifold $C$, which we shall call $V_A=F_{\tau A}|_{C}$. Our zeroth law formulation in this theory is

\begin{equation} \label{0thLawCond}
    \partial_{A} \kappa = O_{\infty}(l^{N-1/2}), \quad \text{and} \quad V_A = O_{\infty}(l^{N-1/2})
\end{equation}

For technical reasons, the proof does not quite get us to the $O(l^N)$ we had in the vacuum case, but $O(l^{N-1/2})$ is still beyond the accuracy to which we know the EFT Lagrangian. The interpretation of the second condition can be seen as follows. By Cartan's formula, $\LL_\xi F = \text{d} (\iota_\xi F ) + \iota_\xi \text{d}F$. $\text{d}F=0$ because $F$ is a Maxwell field. We have also assumed $\LL_\xi F = 0$ above. Hence $\text{d} (\iota_\xi F ) = 0$, and so at least locally, $\iota_\xi F = \text{d} \Phi$ for some scalar $\Phi$. This scalar is $-1$ times the \textit{electric potential} from the definition of the first law. The condition $V_A = O_{\infty}(l^{N-1/2})$ is then equivalent to $\partial_A \Phi|_C=O_{\infty}(l^{N-1/2})$, which says that the electric potential is constant on the horizon up to $O(l^{N-1/2})$.

From the equations of motion (\ref{EMEtauA}) and (\ref{EMEtau}) we can see the two conditions in (\ref{0thLawCond}) are not independent. In fact, we can see that proving $V_A = O_{\infty}(l^{N-1/2})$ is key to proving $\partial_{A} \kappa = O_{\infty}(l^{N-1/2})$.

\subsection{Positive Boost Weight Quantities on the Horizon}\label{PBWQHF}

As in the vacuum gravity case, we will make use of the coordinate transformation $\rho = r( \kappa v+1)$, $\tau = \frac{1}{\kappa} \log\left(\kappa v+1\right)$. We saw in \ref{PBWQH} that we can split the metric in these coordinates into $\Tilde{g}_{\mu\nu}=g^{aff}_{\mu\nu} + \partial_{A} \kappa \sigma^{A}_{\mu\nu}$, where $\sigma^{A}_{\mu\nu}$ and all its $v$ and $r$-derivatives are $O_{\infty}(1)$ on $C$ and all positive boost weight quantities made from $g^{aff}_{\mu\nu}$ vanish on the horizon. We also saw in \ref{IncMatter} that all positive boost weight quantities made from $\phi$ vanish on the horizon, without needing to do a split. Let us now get a result regarding quantities made from $F$. 

In Killing vector GNCs, $F_{\mu\nu}$ has no dependence on $\tau$ because of the assumption $\LL_{\xi}F=0$. By smoothness, 
\begin{equation}
    F_{\tau A}(\rho,x^C) = V_A(x^C) + \rho f_A(\rho,x^C) 
\end{equation}
for some $f_A(\rho,x^C)$ regular on the horizon. Then, suppressing $x^A$ dependence of all quantities,
\begin{equation}
    F = (V_A + \rho f_{A}(\rho)) \text{d}\tau \wedge \text{d}x^A + F_{\tau \rho}(\rho) \text{d}\tau \wedge \text{d}\rho + F_{\rho A}(\rho) \text{d}\rho \wedge \text{d}x^A +  F_{A B}(\rho) \text{d}x^A \wedge \text{d}x^B
\end{equation}
Now make the coordinate transformation $\rho = r( \kappa v+1)$, $\tau = \frac{1}{\kappa} \log\left(\kappa v+1\right)$ once again. The components in the new coordinate system, $\tilde{F}_{\mu\nu}$, can be split into the form
\begin{equation}
    \tilde{F}_{\mu\nu} = F^{aff}_{\mu\nu} + V_A \theta^A_{\mu\nu} + \partial_A \kappa \zeta^A_{\mu\nu}
\end{equation}
where $\theta^A_{\mu\nu}$, $\zeta^A_{\mu\nu}$ and all their $v$ and $r$-derivatives are $O_{\infty}(1)$ on $C$, and 
\begin{align}
    F^{aff}_{v r}(r,v) = F_{\tau \rho}\big(r(\kappa v+1)\big), \quad & \quad F^{aff}_{A B}(r,v) = F_{A B}\big(r(\kappa v+1)\big),\nonumber\\ 
    F^{aff}_{r A}\big(r, v\big) = (\kappa v+1)&F_{\rho A}\big(r(\kappa v+1)\big),\nonumber\\
    F^{aff}_{v A}\big(r, v\big) = r\Big[ \kappa F_{\rho A}&\big(r(\kappa v+1)\big) +f_{A}\big(r(\kappa v+1)\big)\Big],
\end{align}
The auspicious naming of $F^{aff}_{\mu\nu}$ is because it is precisely of the correct form such that all positive boost weight quantities made from it vanish on the horizon. To see this, note that positive boost weight building blocks made from $F^{aff}_{\mu\nu}$ are of the form 
\begin{equation} \label{PBBF}
    \partial_{A_1}... \partial_{A_n} \partial_{v}^q (\partial_r\partial_{v})^{p}  \varphi \,\, \text{with}\,\, \varphi \in \{F^{aff}_{v A}, \partial_{v}F^{aff}_{v r}, \partial_{v}F^{aff}_{A B},  \partial_{v}^2 F^{aff}_{r A} \} \,\, \text{and} \,\, q\geq p
\end{equation}
One can show that all $\varphi \in \{F^{aff}_{v A}, \partial_{v}F^{aff}_{v r}, \partial_{v}F^{aff}_{A B},  \partial_{v}^2 F^{aff}_{r A} \}$ are of the form 
\begin{equation}
    \varphi = r f_{\varphi}\big(r(\kappa v+1)\big)
\end{equation}
and so using the same arguments as in \ref{PBWQH}, one can show all quantities in (\ref{PBBF}) come with an overall factor of $r$ and hence vanish on the horizon.

We make a sidenote here (as we did in the vacuum case) to say that if the zeroth law formulation holds exactly, i.e., $\partial_A \kappa \equiv 0$, $V_A \equiv 0$, then $\tilde{F}_{\mu\nu} = F_{\mu\nu}^{aff}$ and hence all positive boost weight quantities made from $\tilde{F}_{\mu\nu}$ would vanish on the horizon. We also saw that all positive boost weight quantities made from $\phi$ vanish on the horizon in the same choice of affinely parameterized GNCs. Lemma 2.1 of \cite{Hollands:2022} can trivially be extended to include $F_{\mu\nu}$ and $\phi$ in its proof that boost weight is independent of choice of affinely parameterized GNCs, and hence all positive boost weight quantities made from $g_{\mu\nu}$, $F_{\mu\nu}$ and $\phi$ vanish on the horizon in \textit{any} affinely parameterized GNCs if the zeroth law formulation holds exactly. This is a new proof of this result, and avoids resorting to regularity on the bifurcation surface as done in e.g. \cite{Wall:2024lbd}.

\subsection{The Inductive Loop}\label{InductiveLoop}

Returning to our main argument, we shall take as our inductive hypothesis
\begin{equation}
    \partial_A \kappa = O_{\infty}(l^{k/2}), \quad V_A = O_{\infty}(l^{k/2})
\end{equation}
on $C$ for $k/2 \leq N-1$. The base case $k=0$ follows from our regime of validity assumptions. 

Under the usual coordinate transformation to $(v,r,x^A)$ coordinates, 
\begin{align}
    H_{\tau A}[g_{\mu\nu}, F_{\gamma\delta},\phi]|_C & = H_{vA}[\tilde{g}_{\mu\nu} , \tilde{F}_{\gamma\delta}, \phi]|_C\\
    & = H_{vA}[g^{aff}_{\mu\nu}+ \partial_B \kappa \sigma^{B}_{\mu\nu}, F^{aff}_{\gamma\delta} + V_C \theta^C_{\gamma\delta} + \partial_C \kappa \zeta^C_{\gamma\delta}, \phi]|_C\\
    & = H_{vA}[g^{aff}_{\mu\nu}, F^{aff}_{\gamma\delta}, \phi]|_C + O_{\infty}(l^{k/2})
\end{align}
where in the last step we used that $\partial_B \kappa \sigma^{B}_{\mu\nu}$ and $V_C \theta^C_{\gamma\delta} + \partial_C \kappa \zeta^C_{\gamma\delta}$ and all their $v$ and $r$-derivatives are $O_{\infty}(l^{k/2})$ on $C$ by our inductive hypothesis. As we have shown above and in the previous section, all positive boost weight quantities built from $g^{aff}_{\mu\nu}$, $F^{aff}_{\gamma\delta}$ and $\phi$ vanish on the horizon, and so $H_{vA}[g^{aff}_{\mu\nu}, F^{aff}_{\gamma\delta},\phi]|_C = 0$. Therefore 
\begin{equation}
    H_{\tau A}[g_{\mu\nu}, F_{\gamma\delta},\phi]|_C = O_{\infty}(l^{k/2})
\end{equation}
In precisely the same fashion, one shows that
\begin{equation}
    H_{\tau}[g_{\mu\nu}, F_{\gamma\delta},\phi]|_C = O_{\infty}(l^{k/2})
\end{equation}
Therefore the equations of motion (\ref{EMEtauA}) and (\ref{EMEtau}) are respectively
\begin{equation}\label{EMEtauAinduct}
    \partial_{A}\kappa = - \frac{1}{2} c_1(\phi) \left( F_{A B} h^{B C} - F_{\tau \rho} \delta_{A}^{C} \right) V_C + O_{\infty}(l^{1+k/2})
\end{equation}
\begin{equation} \label{EMEtauinduct}
    h^{A B} \DD_A \left[c_1(\phi)V_B - 8 c_2(\phi) \epsilon_{B}^{\,\,\,\,C} V_C \right]  = O_{\infty}(l^{1+k/2})
\end{equation}
on $C$. In the rest of this section, we will show that the second equation implies $V_A = O_{\infty}(l^{(k+1)/2})$. Plugging this into the first equation, we get $\partial_A \kappa = O_{\infty}(l^{(k+1)/2})$ and hence the induction proceeds. 

We now turn to the analysis of equation (\ref{EMEtauinduct}). Note that since $F_{\alpha \beta}$ is a Maxwell field, it satisfies $\partial_{[\alpha} F_{\beta \gamma]} = 0$. We can combine this with $\partial_{\tau} F_{\alpha \beta} = 0$ to get
\be
\partial_{[\tau} F_{A B]} = 0 \implies \partial_A F_{\tau B} - \partial_B F_{\tau A} = 0 \implies \text{d}V=0
\ee
where $\text{d}$ is the exterior derivative on $C$. Our assumptions on the topology of $C$ imply that every closed 1-form on $C$ is exact, hence there exists some function $f(x^A)$ defined on the whole of $C$ such that 
\be
V = \text{d}f
\ee
$f$ is just the electric potential up to some unknown constant. Our inductive hypothesis says $\partial_A f = O_{\infty}(l^{k/2})$. Note we can also add an arbitrary constant to $f$, which we can use to bound the size of $f$ as follows. WLOG we can let $f(x_0^A)=0$ at some fixed ($l$,$L$-independent) point $x_0^A$ on $C$. For each point $x^A\in C$, take some fixed curve $x^A(\lambda)$ with $x^A(0)=x_0^A$ and $x^A(1) = x^A$. Call its tangent vector $T^{A}$. Then 
\begin{equation}\label{curveargument}
    f(x^A) = \int_0^1 \frac{\diff}{\diff \lambda} f(x^A(\lambda)) \diff \lambda = \int_0^1 T^A \partial_A f \diff \lambda
\end{equation}
and hence by our inductive hypothesis and the compactness of $C$ we have $f = O_{\infty}(l^{k/2})$.

Now define $\braket{f}$ to be the mean value of $f$ over $C$:
\begin{equation}
    \braket{f} = \frac{1}{A(C)} \int_C f \diff A
\end{equation}
where $A(C) = \int_C \diff A$ and we take the shorthand $\diff A = \sqrt{h} \diff^{d-2}x$. Let $\tilde{f}$ be the difference of $f$ from its mean:
\begin{equation}
    \tilde{f}(x^A) = f(x^A) - \braket{f} 
\end{equation}
Since we can add an arbitrary constant to $f$, we now drop the tilde and use this as our choice of $f$, i.e., we have 
\begin{equation}
    V_A = \DD_A f, \quad \quad f = O_{\infty}(l^{k/2}), \quad \quad \int_C f \diff A = 0
\end{equation}
For non-negative integers $m$, define the $H_m$ norm of $f$ by
\begin{equation}
    ||f||^2_{H_m} = \sum_{n=0}^m \int_C |\DD^n f|_h^2 \, \diff A
\end{equation}
We will use equation (\ref{EMEtauinduct}) to show that $||f||_{H_m} = O(l^{(k+1)/2})$ for all $m=0,1,2,..$. Firstly, $m=1$ and $m=0$. Integrate (\ref{EMEtauinduct}) against the $O(l^{k/2})$ quantity $f \sqrt{h}$ over $C$:
\be
\int_{C} f \DD^A \left[ c_1(\phi) \DD_A f - 8 c_2(\phi) \epsilon_{A}^{\,\,\,\,B} \DD_B f \right] \diff A  = O(l^{1+k})
\ee
Apply the divergence theorem to get
\be
-\int_{C} \left[ c_1(\phi) \DD^A f \DD_A f - 8 c_2(\phi) \epsilon^{A B} \DD_A f \DD_B f \right] \diff A = O(l^{1+k})
\ee
There is no boundary term by the closedness of $C$. The second term in the brackets is 0 by the antisymmetry of $\epsilon^{A B}$. Hence we can use this to bound 
\be
    \int_C |\DD f|_h^2 \, \diff A \leq \frac{1}{\tilde{c}} \int_{C} c_1(\phi) \DD^A f \DD_A f \diff A  = O(l^{1+k}) 
\ee
using our assumption $c_1(\phi)\geq \tilde{c} > 0$ for some positive constant $\tilde{c}$ independent of $l$. We also have the Poincar\'e inequality for compact $C^{\infty}$ manifolds (Corollary 4.3 in \cite{Aubin:1982}) which says that
\begin{equation}
    \int_C f \diff A = 0 \implies \int_C f^2 \diff A \leq \frac{1}{\lambda_1} \int_C |\DD f|^2 \, \diff A
\end{equation}
where $\lambda_1$ is the smallest nonzero eigenvalue of the Laplacian $-h^{A B}\DD_A \DD_B$ (and which we assume to be $O(1)$ in $l$). Combining these results gives
\begin{equation}
    ||f||_{H_0} = O(l^{(k+1)/2}), \quad \quad ||f||_{H_1} = O(l^{(k+1)/2})
\end{equation}
Let us now look at equation (\ref{EMEtauinduct}) again and expand the derivatives by the product rule:
\begin{equation}\label{Etauexpand}
    c_1(\phi) \DD^A \DD_A f + \DD^A c_1 \, \DD_A f - 8 \epsilon^{A B} \DD_A c_2 \, \DD_B f = O_{\infty}(l^{1+k/2}) 
\end{equation}
The left hand side is a second order elliptic operator $P = c_1(\phi) \DD^A \DD_A + \DD^A c_1 \, \DD_A - 8 \epsilon^{A B} \DD_A c_2 \, \DD_B$ acting on $f$, and hence we can use standard elliptic estimates on compact manifolds given in e.g. Chapter III, Theorem 5.2 of \cite{Lawson:1990}, which state that for each non-negative integer $s$, there exists a constant $C_{s}$ such that
\begin{equation}
    ||f||_{H_{s+2}} \leq C_s ( ||f||_{H_{s}} + ||Pf||_{H_{s}})
\end{equation}
The constant $C_s$ is independent of $f$, but does depend on $h_{AB}$ and $P$. Again we assume that in the regime of validity, $C_s = O(1)$. We can repeatedly use this inequality on the $||f||_{H_{s}}$ on the right hand side, reducing $s$ by 2 each time until we are left with $||f||_{H_{1}}$ or $||f||_{H_{0}}\leq ||f||_{H_{1}}$. Therefore, noting $||Pf||_{H_{s-2}}\leq ||Pf||_{H_{s}}$ etc., there exists a constant $C'_s = O(1)$ such that
\begin{equation}
    ||f||_{H_{s+2}} \leq C'_s ( ||f||_{H_{1}} + ||Pf||_{H_{s}})
\end{equation}
But from above, $||f||_{H_{1}} = O(l^{(k+1)/2})$, and by (\ref{Etauexpand}), $||Pf||_{H_{s}} = O(l^{1+ k/2})$. Therefore
\begin{equation}
    ||f||_{H_m} = O(l^{(k+1)/2}) 
\end{equation}
for all $m=0,1,2,..$. To relate the $H_m$ norms to the maximum value of $f$ and its derivatives, we use the Sobolev embedding theorem for compact Riemannian manifolds (Theorem 2.7 of \cite{Hebey:2000}), which says that for all non-negative integers $m,r$ with $m>\frac{d-2}{2} + r$ (where $d-2$ is the dimension of $C$), there exists a constant $C_{m,r}$ depending only on $h_{AB}$ (which we again assume is $C_{m,r}=O(1)$) such that
\begin{equation}
    \max_{0\leq n \leq r} \sup_C |\DD^n f|_h \leq C_{m,r} ||f||_{H_m}
\end{equation}
Therefore, $f = O_{\infty}(l^{(k+1)/2})$ and the induction proceeds.

This induction gets us as far as $\partial_A \kappa = O_{\infty}(l^{N-1/2}), \quad V_A = O_{\infty}(l^{N-1/2})$. We cannot quite go an extra $l^{1/2}$ higher because at this point, $lH_{\tau A}, lH_{\tau} = O_{\infty}(l^{N+1/2})$, which is higher than the unknown $O_{\infty}(l^{N})$ terms in (\ref{EMEtauA}) and (\ref{EMEtau}). This means the right hand side of (\ref{EMEtauinduct}) is $O_{\infty}(l^{N})$, which is only $O(l^{1/2})$ higher than $f$ rather than the full $O(l)$ we had in the induction. This leads to $||f||_{H_0}, ||f||_{H_1} = O(l^{N-1/4})$ rather than $O(l^N)$, and so we can only progress to $V_A = O_{\infty}(l^{N-1/4})$, $\partial_A \kappa = O_{\infty}(l^{N-1/4})$. One can actually repeat this inductively to get
\begin{equation}
    \partial_A \kappa = O_{\infty}(l^{N-\frac{1}{2^k}}), \quad V_A = O_{\infty}(l^{N-\frac{1}{2^k}})
\end{equation}
for any $k\in \mathbb{N}$, but we shall not bother to write this out.

\section{The Zeroth Law in the EFT of Gravity, Electromagnetism and a Charged Scalar Field}\label{ChargedScalar}

With a few modifications and additional technical assumptions, this proof of the zeroth law can also be applied to the EFT of gravity, electromagnetism and a \textit{charged} scalar field. In this scenario, we assume we have a global gauge potential $A_\mu$ with $F=\text{d}A$. The dynamical fields are $\Phi_I = (g_{\mu \nu}, A_\mu, \phi)$, the scalar field $\phi$ is complex with some charge $\lambda$, and $A_\mu$ and $\phi$ transform under an electromagnetic gauge transformation as
\be \label{gaugetransform}
A_\mu \rightarrow \Tilde{A}_\mu = A_\mu + \partial_\mu \chi, \quad \quad \phi \rightarrow \Tilde{\phi} = e^{i \lambda \chi} \phi
\ee
with $\chi$ an arbitrary real-valued function. We generalize our leading order Lagrangian to 
\begin{equation} \label{0and2-derivcharged}
    \LL_{2} = R - V(|\phi|^2) - g^{\alpha \beta} \left(\mathfrak{D}_{\alpha}{\phi}\right)^{*} \mathfrak{D}_{\beta}{\phi} - \frac{1}{4} c_1(|\phi|^2) F_{\alpha\beta} F^{\alpha\beta} + c_2(|\phi|^2) F_{\alpha \beta} F_{\gamma \delta} \epsilon^{\alpha \beta \gamma \delta}
\end{equation}
where $\mathfrak{D}_\alpha$ is the gauge covariant derivative $\fD_\alpha = \nabla_\alpha-i \lambda A_\mu$. $\LL_{2}$ is invariant under the gauge transform (\ref{gaugetransform}). Since the charge $\lambda$ adds a new scale into the theory, the EFT series is now a joint series in derivatives and powers of $\lambda$. For the charge to be relevant from an EFT perspective, we assume $\lambda \leq 1/L$ where $L$ is a typical length scale of the solution, so that $\lambda$ is comparable to a 1-derivative term. This is reasonable if we want the classical approximation to be valid. The EFT Lagrangian is
\be
\LL = \LL_{2} + \sum_{n=1}^{N-1} l^n \LL_{n+2} + O(l^N)
\ee
where the $\LL_{n+2}$ contains all gauge-independent terms with $n+2$ derivatives or powers of $\lambda$. 

The equations of motion are 
\begin{equation}
\begin{split}
    E_{\alpha\beta} \equiv E^{(0)}_{\alpha \beta} + l H_{\alpha \beta}[g_{\mu\nu}, A_{\gamma},\phi] = O(l^N),  \quad & \quad  E_\alpha \equiv E^{(0)}_{\alpha} + l H_{\alpha}[g_{\mu\nu}, A_{\gamma},\phi] = O(l^N)\\
    E \equiv E^{(0)} + l H&[g_{\mu\nu}, A_{\gamma},\phi] = O(l^N)
\end{split}
\end{equation}
where the parts arising from the leading order theory are 
\begin{multline} \label{0th order EoM 1 charged}
    E^{(0)}_{\alpha\beta}= R_{\alpha \beta} - \left(\fD_{(\alpha}\phi\right)^{*} \fD_{\beta)} \phi -\frac{1}{2} c_1(|\phi|^2) F_{\alpha \delta} F_{\beta}^{\,\,\,\,\delta}\\
    - \frac{1}{2} g_{\alpha \beta} \left( R - V(|\phi|^2) - g^{\gamma \delta}\left(\fD_{\gamma}{\phi}\right)^{*} \fD_{\delta}{\phi} - \frac{1}{4} c_1(|\phi|^2) F_{\gamma \delta} F^{\gamma\delta} \right)
\end{multline}
\begin{equation}
    E^{(0)}_\alpha = i\lambda\left[\phi^* \fD_\alpha \phi - \phi \left(\fD_{\alpha}\phi\right)^{*} \right] + \nabla^{\beta}{\Big[c_1(|\phi|^2)F_{\alpha\beta} - 4 c_2(|\phi|^2) F^{\gamma \delta} \epsilon_{\alpha \beta \gamma \delta}\Big]}
\end{equation}
\begin{equation} \label{0th order EoM 2 charged}
    E^{(0)} = g^{\alpha\beta}\fD_{\alpha}{\fD_{\beta}{\phi}}-\phi V'(|\phi|^2)-\frac{1}{4} \phi c'_1(|\phi|^2) F_{\alpha \beta} F^{\alpha \beta} + \phi c'_2(|\phi|^2) F_{\alpha \beta} F_{\gamma \delta} \epsilon^{\alpha \beta \gamma \delta}
\end{equation}

\subsection{Assumptions - Electromagnetic Gauge}\label{assumegauge}

We again assume we have a stationary black hole solution to these equations, with a Killing horizon $\cH$ with closed and simply connected spatial cross-section $C$. What needs more subtlety is our assumption of the invariance of the matter fields on the symmetry corresponding to the Killing vector $\xi = \frac{\partial}{\partial \tau}$. The assumption that $\phi$ and $A_\mu$ are independent of $\tau$ can no longer be applied generally because the conditions $\partial_\tau \phi = 0$ and $\partial_\tau A_\mu=0$ are not invariant under an electromagnetic gauge transformation. Instead, we need to modify our notion of a symmetry of the system.

For the metric, a Killing vector symmetry corresponds to invariance under the diffeomorphism $\tau \rightarrow \tau+s$ for all $s$, i.e., $g_{\mu \nu}(\tau + s) = g_{\mu \nu}(\tau)$. This diffeomorphism can be viewed as a one-parameter coordinate gauge transformation of the metric, labelled by $s$. A complete gauge transformation of our matter fields $\phi$ and $A_\mu$ would be a \textit{combined} diffeomorphism and electromagnetic gauge transformation. Hence we assume the notion of symmetry for $\phi$ and $A_\mu$ is their invariance under a one-parameter combined diffeomorphism and electromagnetic gauge transformation. More precisely, given any fixed gauge of $A_\mu$ and $\phi$, we assume there exists some one-parameter family of functions $\theta_s$ with $\theta_0 = 0$ such that
\be
    A_{\mu}(\tau+s)+ \partial_\mu \theta_s = A_{\mu}(\tau), \quad \quad e^{i\lambda \theta_s}\phi(\tau+s) = \phi(\tau)
\ee
for all $s$. We can take the derivative of this with respect to $s$ and set $s=0$ to obtain the conditions
\be \label{symmetryconditions}
    \partial_\tau A_\mu = -\partial_\mu \Theta, \quad \quad \partial_\tau \phi = -i \lambda \Theta \phi
\ee
where $\Theta = \frac{\text{d} \theta_s}{\text{d} s }\Big|_{s=0}$.
These are the conditions\footnote{These conditions can be proved to be equivalent to the conditions assumed in (2.9) of \cite{An:2023} (a recent paper discussing stationary black hole solutions with charged scalar hair). The formulation above avoids the need to define the phase of $\phi$ however.} which constrain the $\tau$-dependence of $A_\mu$ and $\phi$. Note the first condition implies $\partial_\tau F_{\mu \nu} = 0$, which was the condition we assumed in the real scalar field case.

Let us now make an electromagnetic gauge transformation of the form (\ref{gaugetransform}). Using (\ref{symmetryconditions}), the $\tau$-dependence of $\Tilde{A}_\mu$ and $\Tilde{\phi}$ can be found to be
\be
    \partial_\tau \Tilde{A}_\mu = - \partial_\mu ( \Theta - \partial_\tau \chi), \quad \quad \partial_\tau \Tilde{\phi} = - i \lambda ( \Theta - \partial_\tau \chi ) \Tilde{\phi}
\ee
From this we see that conditions (\ref{symmetryconditions}) are preserved under an electromagnetic gauge transformation, so long as we relabel $\Tilde{\Theta} = \Theta - \partial_\tau \chi$. In particular, we can take $\chi = \int^\tau \Theta(\tau') \text{d}\tau'$ to find a gauge in which 
\be
    \partial_\tau \Tilde{A}_\mu = 0, \quad \quad \partial_\tau \Tilde{\phi} = 0
\ee
We will drop the tildes and work in this gauge to prove the zeroth law. We assume the regime of validity of the EFT assumptions as detailed previously hold for $\phi$ and $A_\mu$ in this gauge. We also assume $c_1(|\phi|^2)\geq \tilde{c}>0$ for some constant $\tilde{c}$ independent of $l$, as before.

For technical reasons, we will also take an additional assumption on the charged scalar field, which is that 
\begin{equation}\label{assumephi}
    \int_C |\phi|^2 \diff A \geq C_\phi >0
\end{equation}
for some constant $C_\phi$ independent of $l$. This loosely says that $\phi$ is an $O(1)$ quantity and not higher order in $l$ on the horizon, so it is on the same scale as the metric. Therefore we are studying some so-called "hairy" black hole in the limit $l\rightarrow 0$ (or equivalently large $L$), rather than one that tends to a black hole with a vanishing scalar field e.g. Kerr-Newman. For a complete proof one should also allow $\int_C |\phi|^2 \diff A$ to get arbitrarily small with $l$. Appendix \ref{appendixzeroth} deals with the case where $\phi$ is proportional to $l^m$ for some $m>0$, $m\in \mathbb{R}$, however the proof in full generality remains elusive.

\subsection{Zeroth Law Formulation}

We can view $A_{\tau}|_C$ as a scalar on $C$, which we shall call $-f(x^A) = A_{\tau}|_C$. The zeroth law formulation we shall prove in this gauge is 
\begin{equation}
    \partial_A \kappa = O_{\infty}(l^{N-1/2}), \quad \text{and} \quad f = O_{\infty}(l^{N-1/2})
\end{equation}
Note that $F_{\tau A} = - \partial_A A_\tau$ in this gauge, so in the notation of the previous section, $V_A = \DD_A f$ and this is the same $f$ as before up to some unknown constant. $V_A$ and $\partial_A \kappa$ are electromagnetic gauge invariant, so proving the formulation above in this specific gauge proves the formulation (\ref{0thLawCond}) in all gauges.

\subsection{Positive Boost Weight Quantities on the Horizon} \label{PBWQHCharged}

The important difference between the charged scalar field theory and the uncharged scalar field theory of the previous section is that there is now explicit appearance of $A_\mu$ rather than just $F_{\mu\nu}$. There are positive boost weight quantities made from $A_\mu$ that could never appear through $F_{\mu\nu}$, such as $A_v$ and its $v$-derivatives. Therefore we must redo our analysis in \ref{PBWQHF}, but study positive boost weight quantities made from $A_\mu$ rather than $F_{\mu\nu}$.

In our choice of electromagnetic gauge, $A_\mu$ has no dependence on $\tau$ in Killing vector GNCs. By smoothness,
\begin{equation}
    A_{\tau}(\rho,x^A) = f(x^A) + \rho B(\rho,x^A)
\end{equation}
for some $B(\rho,x^A)$ regular on the horizon. Then, suppressing $x^A$ dependence of all quantities,
\be \label{AKilling}
    A= (f + \rho B(\rho)) \text{d}\tau + A_\rho(\rho) \text{d}\rho + A_A(\rho) \text{d}x^A
\ee
Make the coordinate transformation $\rho = r( \kappa v+1)$, $\tau = \frac{1}{\kappa} \log\left(\kappa v+1\right)$ once again. The components in the new coordinate system, $\tilde{A}_{\mu}$, can be split into the form
\begin{equation}\label{Asplit}
    \tilde{A}_{\mu} = A^{aff}_{\mu} + f \theta_\mu + \partial_A \kappa \zeta^A_{\mu}
\end{equation}
where $\theta_\mu, \zeta^A_{\mu}$ and all their $v$ and $r$-derivatives are $O_{\infty}(1)$ on $C$, and 
\begin{equation} \label{Atransform}
    \begin{split}
        A^{aff}_v(r,v)= r \Big[ B\big(r\left(\kappa v+ 1\right)&\big)+ \kappa A_\rho\big(r\left(\kappa v+1\right)\big) \Big],\\
        A^{aff}_r(r,v)=(\kappa v+1) A_\rho\big(r\left(\kappa v+1\right)\big), \quad & \quad A^{aff}_A(r,v)=A_A\big(r\left(\kappa v+1\right)\big)
    \end{split}
\end{equation}
Once again we have auspiciously named $A^{aff}_{\mu}$ because all positive boost weight quantities made from it vanish on the horizon. This follows from identical arguments to those used in \ref{PBWQHF}.

(\ref{Asplit}) is the result we will use for $A_{\mu}$. In this gauge $\partial_\tau \phi = 0$, so we still have that all positive boost weight quantities made from $\phi$ vanish on the horizon.

We again make a brief sidenote to say the above steps prove that if the zeroth law formulation holds exactly, i.e., $\partial_A \kappa \equiv 0$, $f \equiv 0$, then all positive boost weight quantities made from $g_{\mu\nu}$, $A_{\mu}$ and $\phi$ in this electromagnetic gauge vanish on the horizon in any choice of affinely parameterized GNCs by a trivial extension of Lemma 2.1 in \cite{Hollands:2022}. However, what happens if we change the electromagnetic gauge? We have for example,
\begin{equation}
    \partial_v \tilde{\phi} = e^{i\lambda \chi} \partial_v\phi + i\lambda \partial_v \chi e^{i\lambda \chi} \phi
\end{equation}
and hence we see that $\partial_v \tilde{\phi}$ does not vanish if we take a $v$-dependent $\chi$. Therefore, we do not have the same simple result in all electromagnetic gauges. However, so long as the theory is electromagnetic gauge-invariant overall, then we can always express dependence on $\phi$ and $A_{\mu}$ in terms of $\fD_{\mu}...\fD_{\nu} \phi$ and derivatives of $F_{\mu\nu}$, which vanish in all gauges if they vanish in one. Thus the useful result we have is that all positive boost weight quantities made from $g_{\mu\nu}$, $F_{\mu\nu}$ and $\fD_\mu$ derivatives of $\phi$ vanish on the horizon in any affinely parameterized GNCs and any electromagnetic gauge if the zeroth law formulation holds exactly.

\subsection{The Inductive Loop}

Again, the important equations of motion are $E_{\tau A}$ and $E_\tau$ evaluated on $C$ in Killing vector GNCs in our electromagnetic gauge, which give respectively
\begin{multline}\label{EtauAcharge}
    \partial_A \kappa = - \frac{1}{2} i \lambda f \left(\phi^{*} \partial_A \phi - \phi \partial_A \phi^* - 2i\lambda A_A |\phi|^2\right) - \frac{1}{2} c_1(|\phi|^2)\left( F_{A B} h^{B C} - F_{\tau \rho} \delta_{A}^{C} \right) \DD_C f\\
    + l H_{\tau A}[g_{\mu\nu}, A_{\gamma},\phi]\big|_C + O_{\infty}(l^N)
\end{multline}
\begin{equation}\label{Etaucharge}
    -2\lambda^2 |\phi|^2 f + h^{A B} \DD_{A}\Big[c_1(|\phi|^2)\DD_B f - 8 c_2(|\phi|^2) \epsilon_{B}^{\,\,\,\,C}\DD_C f \Big] = lH_{\tau}[g_{\mu\nu}, A_{\gamma},\phi]\big|_C + O_{\infty}(l^N) 
\end{equation}

Our inductive hypothesis is 
\begin{equation}\label{inductivehypo}
    \partial_A \kappa = O_{\infty}(l^{k/2}), \quad \text{and} \quad f = O_{\infty}(l^{k/2})
\end{equation}
for $k/2\leq N-1$. The base case $k=0$ follows from our regime of validity assumptions.

In the same fashion as before, now using our result (\ref{Asplit}), we can show
\begin{equation}
    H_{\tau A}[g_{\mu\nu}, A_{\gamma},\phi]\big|_C = O(l^{k/2}), \quad \quad H_{\tau}[g_{\mu\nu}, A_{\gamma},\phi]\big|_C = O(l^{k/2})
\end{equation}
by transforming to $(v,r,x^A)$ coordinates, considering our results regarding positive boost weight quantities on the horizon, and using the inductive hypothesis for $\partial_A \kappa$ and $f$. 

Integrate (\ref{Etaucharge}) against $f \sqrt{h}$ and integrate by parts to get
\begin{equation}
    \int_C \left[ 2\lambda^2 |\phi|^2 f^2 + c_1(|\phi|^2) \DD^A f \DD_A f \right ] \diff A = O(l^{1+k})
\end{equation}
Therefore,
\begin{equation}
    \int_C c_1(|\phi|^2) \DD^A f \DD_A f \diff A = O(l^{1+k})
\end{equation}
from which we can again get
\begin{equation}\label{H1norm}
    \int_C |\DD f|_h^2 \diff A = O(l^{1+k})
\end{equation}
We also have
\begin{equation}\label{fphisquared}
    \int_C |\phi|^2 f^2 \diff A = O(l^{1+k})
\end{equation}
from which we would like to infer $\int_C f^2 \diff A = O(l^{1+k})$, but we cannot since $|\phi|$ may be arbitrarily small at any point. Now, define $\braket{f}$ and $\tilde{f}$ as before:
\begin{equation}
    \braket{f} = \frac{1}{A(C)} \int_C f \diff A, \quad \tilde{f}(x^A) = f(x^A) - \braket{f} \implies \int_C \tilde{f} \diff A = 0
\end{equation}
We cannot simply relabel $\tilde{f}$ to be $f$ as we did before. This is because $f = -A_\tau|_C$ now appears without derivatives in our theory, and so we cannot add an arbitrary constant to it. We can try to make the gauge transformation $A_\mu \rightarrow A_\mu + \partial_\mu ( K \tau )$ for constant $K$, which does shift $f$ by $-K$, but $\phi$ transforms as $\phi \rightarrow e^{i\lambda K\tau} \phi$ and so is no longer independent of $\tau$. Therefore, this would take us out of the gauge defined in \ref{assumegauge}, which would change the form of the above equations and add extra positive boost weight quantities to be considered - something we do not want to do. Instead, we will work with $\tilde{f}$ directly. Using the Poincar\'e inequality again with (\ref{H1norm}), we have
\begin{equation}\label{ftildenorms}
    ||\tilde{f}||_{H_0} = O(l^{(k+1)/2}), \quad \quad ||\tilde{f}||_{H_1} = O(l^{(k+1)/2})
\end{equation}
Now substitute $f = \tilde{f}+ \braket{f}$ into (\ref{Etaucharge}) and expand the derivatives to get
\begin{equation}\label{Etauexpandcharge}
    c_1(\phi) \DD^A \DD_A \tilde{f} + \DD^A c_1 \, \DD_A \tilde{f} - 8 \epsilon^{A B} \DD_A c_2 \, \DD_B \tilde{f} - 2\lambda^2 |\phi|^2 \tilde{f} = 2\lambda^2 |\phi|^2 \braket{f} + O_{\infty}(l^{1+k/2}) 
\end{equation}
The left hand side is an elliptic operator $Q = c_1(\phi) \DD^A \DD_A + \DD^A c_1 \, \DD_A - 8 \epsilon^{A B} \DD_A c_2 \, \DD_B  - 2\lambda^2 |\phi|^2$ acting on $\tilde{f}$, and thus we can again estimate the $H_{s+2}$ norms of $\tilde{f}$ by
\begin{equation}\label{tildefellipticest}
    ||\tilde{f}||_{H_{s+2}} \leq C'_s ( ||\tilde{f}||_{H_{1}} + ||Q\tilde{f}||_{H_{s}}) \leq 2C'_s  \lambda^2  \big|\big||\phi|^2\big|\big|_{H_{s}} |\braket{f}|  + O(l^{(k+1)/2}) 
\end{equation}
Again, the constant $C_s'$ depends implicitly on $l$ because it depends on $Q$, but we assume that it is $O(1)$ in the regime of validity of the EFT.
We can bound the size of $|\braket{f}|$ by integrating (\ref{Etaucharge}) against $\sqrt{h}$. The total derivative term vanishes and we are left with
\begin{equation}
    \int_C f |\phi|^2 \diff A = O(l^{1+k/2}) \implies \braket{f} \int_C |\phi|^2 \diff A = - \int_C \tilde{f} |\phi|^2 \diff A + O(l^{1+k/2})
\end{equation}
But by Cauchy-Schwarz, 
\begin{equation}
    \Big| \int_C \tilde{f} |\phi|^2 \diff A \Big| \leq \left(\int_C \tilde{f}^2 \diff A\right)^{1/2} \left(\int_C |\phi|^4 \diff A\right)^{1/2} \leq C_h' ||\tilde{f}||_{H_0} \left(\int_C |\phi|^4 \diff A\right)^{1/2} = O(l^{(k+1)/2})
\end{equation}
by (\ref{ftildenorms}). Therefore
\begin{equation}\label{fhatsize}
    \braket{f} \int_C |\phi|^2 \diff A = O(l^{(k+1)/2})
\end{equation}
This is where we need the technical assumption (\ref{assumephi}) on $\phi$ to deduce 
\begin{equation}
    \braket{f} = O(l^{(k+1)/2})
\end{equation}
Plug this into (\ref{tildefellipticest}) to get
\begin{equation}
    ||\tilde{f}||_{H_{m}} = O(l^{(k+1)/2})
\end{equation}
for all $m=0,1,2,...$. We again use the Sobolev embedding theorem on compact Riemannian manifolds to get
\begin{equation}
    \tilde{f} = O_{\infty}(l^{(k+1)/2})
\end{equation}
Combining this with $\braket{f} = O(l^{(k+1)/2})$, we get $f= \tilde{f}+ \braket{f} =O_{\infty}(l^{(k+1)/2})$, then plugging this into (\ref{EtauAcharge}) gives $\partial_A \kappa = O_{\infty}(l^{(k+1)/2})$ which completes the inductive loop.

\section{Appendix}\label{appendixzeroth}

\subsection{Removing the Assumption on the Charged Scalar Field}

The technical assumption we needed during the induction in the charged scalar field case was that 
\begin{equation}\label{appendixphiassume}
    \int_C |\phi|^2 \diff A \geq C_\phi >0
\end{equation}
for some constant $C_\phi$ independent of $l$. This was so we could deduce that $\braket{f}$, the mean value of $f$ over $C$, was sufficiently small to advance the induction. If instead $\int_C |\phi|^2 \diff A$ gets arbitrarily small as $l$ goes to 0, then we cannot deduce anything about the size of $\braket{f}$ from (\ref{fhatsize}). Indeed, one can see that in the extreme case $\phi \equiv 0$ then we have no control over the size of $\braket{f}$ whatsoever because $f$ only ever appears differentiated in the equations.

The reason we needed $f$ to be small was so that we could split $A_\mu$ into $A_{\mu}^{aff}$ plus terms that were sufficiently small: $f \theta_\mu + \partial_A \kappa \zeta^A_{\mu}$. That allows us to get $H_{\tau A}|_C, H_\tau|_C = O(l^{k/2})$ at each order in the induction. However, the only appearance of $A_{\mu}$ outside of $F_{\mu\nu}$ in these equations is through the gauge covariant derivative $\fD_\mu= \nabla_\mu - i \lambda A_\mu$ acting on $\phi$. Indeed, since $H_{\mu\nu}$ and $H_{\mu}$ are electromagnetic gauge invariant, any appearance of $A_\mu$ outside of $F_{\mu\nu}$ must come with at least one factor of $\phi$ and one factor of $\phi^{*}$ (or their derivatives) in the schematic combination
\begin{equation}\label{schematicphiappearance}
    \partial^a \phi^* \partial^b A_\mu \partial^c \phi
\end{equation}
Thus, morally, if $\phi$ can get arbitrarily small so that (\ref{appendixphiassume}) doesn't hold, then it may not matter that $f$ is not small by itself because appearances of $A_\mu$ are suppressed by $|\phi|^2$ anyway. 

We shall use this fact to get around the problem in specific cases of $\phi$ that are excluded by the technical assumption (\ref{appendixphiassume}).  Namely, we shall consider
\begin{equation}
    \phi  = l^{m/2} \hat{\phi}
\end{equation}
for some $m>0$, $m\in \mathbb{R}$ and where $\hat{\phi}(\rho,x^A)$ and all its $\rho$-derivatives are $O_{\infty}(1)$ on $C$, and 
\begin{equation}\label{appendixtildephiassume}
    \int_C |\hat{\phi}|^2 \diff A \geq C_{\hat{\phi}} >0
\end{equation}
for some constant $C_{\hat{\phi}}$ independent of $l$. Loosely speaking, this is the case where $\phi$ is an $O(l^{m/2})$ quantity and not higher. This does not cover all possible behaviour in $l$ that is excluded by (\ref{appendixphiassume}) but where $\phi$ and its derivatives are still uniformly bounded by any means - for example $\phi$ could go to 0 faster than any power of $l$, or $\hat{\phi}$ could be $O_{\infty}(1)$ on $C$ but its $\rho$-derivatives could be $O(l^{-m/2})$. However, it does include the case where $\phi$ is an asymptotic series in $l$ with the first $m-1$ terms exactly zero. Such $\phi$ have been constructed, for example, for the dilatonic scalar field of the string theory low energy effective action \cite{Kanti:1995vq}. 

We now take as our inductive hypothesis
\begin{equation}
    \partial_A \kappa = O_{\infty}(l^{k/2}), \quad V_A = O_{\infty}(l^{k/2}) \quad \text{and} \quad f = O(l^{k/2 - m})
\end{equation}
for $k/2\leq N-1$. Separate out the dependence on $A_\mu$ and $F_{\mu\nu}$ in $H_{\mu\nu}[g_{\alpha\beta}, F_{\gamma\delta}, A_{\sigma},\phi]$, $H_{\mu}[g_{\alpha\beta}, F_{\gamma\delta}, A_{\sigma},\phi]$ so that $A_\mu$ only refers to those that appear through gauge covariant derivatives. Then, in the transformation to $(v,r,x^A)$ coordinates, we can write
\begin{align}
    H_{\tau}[g_{\alpha\beta}, F_{\gamma\delta}, A_{\sigma}&, \phi]|_C = H_{v}[\tilde{g}_{\alpha\beta} , \tilde{F}_{\gamma\delta}, \tilde{A}_\sigma, \phi]|_C\nonumber\\
    & = H_{v}[g^{aff}_{\mu\nu}+ \partial_B \kappa \sigma^{B}_{\mu\nu}, F^{aff}_{\gamma\delta} + V_C \theta^C_{\gamma\delta} + \partial_C \kappa \zeta^C_{\gamma\delta}, A^{aff}_{\mu} + f \theta_\mu + \partial_A \kappa \zeta^A_{\mu}, \phi]|_C\nonumber\\
    & = H_{v}[g^{aff}_{\mu\nu}, F^{aff}_{\gamma\delta}, A^{aff}_{\mu} + f \theta_\mu, \phi]|_C + O_{\infty}(l^{k/2})
\end{align}
where we used our inductive hypothesis on $\partial_A \kappa$ and $V_A$ to get the $O_{\infty}(l^{k/2})$ terms. Now, $H_v$ is positive boost weight, and we know that all positive boost weight quantities made from $g^{aff}_{\mu\nu}$, $F^{aff}_{\gamma\delta}$, $A^{aff}_{\mu}$ and $\phi$ vanish on the horizon. Therefore our only obstruction to simplifying the last line is the $f \theta_\mu$ dependence. However, as mentioned above, all appearances of $A_\mu$ will come in the combination (\ref{schematicphiappearance}) and hence any positive boost weight quantity made from $f \theta_\mu$ will have factors
\begin{equation}
    \partial^a \phi^* \partial^b (f \theta_\mu) \partial^c \phi = l^m \partial^a \hat{\phi}^* \partial^b (f \theta_\mu) \partial^c \hat{\phi} = O_{\infty}(l^{k/2})
\end{equation}
on $C$ using our inductive hypothesis on $f$. Thus 
\begin{equation}
    H_{\tau}[g_{\alpha\beta}, F_{\gamma\delta}, A_{\sigma}, \phi]|_C = O_{\infty}(l^{k/2})
\end{equation}
and similarly
\begin{equation}
    H_{\tau A}[g_{\alpha\beta}, F_{\gamma\delta}, A_{\sigma}, \phi]|_C = O_{\infty}(l^{k/2})
\end{equation}
which is what we desired. Now we perform a similar analysis on (\ref{Etaucharge}) and (\ref{EtauAcharge}) to advance the induction. Define $\braket{f}$ and $\tilde{f} = f-\braket{f}$ as before. Take any fixed point $x_0^A$ on $C$ and define $f_0 = f(x^A_0)$. We can use the same argument involving a curve from $x_0^A$ to $x^A$ (see around equation (\ref{curveargument})) to prove that our inductive hypothesis $V_A = O_{\infty}(l^{k/2})$ implies $f-f_0 = O(l^{k/2})$. Then note $\tilde{f} = f-f_0 - \braket{f-f_0}$, hence $\tilde{f} = O_{\infty}(l^{k/2})$. Integrate (\ref{Etaucharge}) against $\sqrt{h}$ again to get
\begin{align} \label{fhatpower}
    \braket{f} \int_C |\phi|^2 \diff A =& - \int_C \tilde{f} |\phi|^2 \diff A + O(l^{1+k/2})\nonumber\\
    \implies l^m \braket{f} \int_C |\hat{\phi}|^2 \diff A =& O(l^{k/2+m}) + O(l^{1+k/2})\nonumber\\
    \implies \braket{f} =& O(l^{k/2}) + O(l^{k/2 + 1 - m})
\end{align}
where we used our assumptions on $\hat{\phi}$ in the last line. If $0<m\leq 1$ then this implies $\braket{f}= O(l^{k/2})$ and hence $f=\tilde{f} + \braket{f} = O_{\infty}(l^{k/2})$. But this gets us exactly to the inductive hypothesis (\ref{inductivehypo}) from the previous section, and so precisely the same steps can be performed to advance the induction. Therefore, suppose $m>1$, in which case (\ref{fhatpower}) implies $\braket{f} = O(l^{k/2 -m + 1})$. Hence $f=\tilde{f} + \braket{f} = O(l^{k/2 - m + 1})$ which advances the inductive hypothesis on $f$ (actually we have raised the power of $f$ by a whole $l$ rather than $l^{1/2}$ but this is fine). Now integrate (\ref{Etaucharge}) against $\tilde{f} \sqrt{h}$ and swap $f=\tilde{f} + \braket{f}$ to get
\begin{equation}
    \int_C \left[ 2\lambda^2 |\phi|^2 \tilde{f}^2 + 2\lambda^2 |\phi|^2 \braket{f}\tilde{f} + c_1(|\phi|^2) \DD^A \tilde{f} \DD_A \tilde{f} \right ] \diff A = O(l^{1+k})
\end{equation}
The only term that is not manifestly non-negative is the middle one, however this is $|\phi|^2 \braket{f}\tilde{f} = O(l^{m})O(l^{k/2 - m + 1})O(l^{k/2}) = O(l^{1+k})$ and hence we can absorb it into the right hand side. Therefore we still have 
\be
    \int_C |\DD f|_h^2 \, \diff A \leq \frac{1}{\tilde{c}} \int_{C} c_1(\phi) \DD^A f \DD_A f \diff A  = O(l^{1+k}) 
\ee
and using the Poincar\'e inequality on $\tilde{f}$ again gives us 
\begin{equation}
    ||\tilde{f}||_{H_0} = O(l^{(k+1)/2}), \quad \quad ||\tilde{f}||_{H_1} = O(l^{(k+1)/2})
\end{equation}
We can again write (\ref{Etaucharge}) in terms of the elliptic operator $Q$:
\begin{equation}
    Q \tilde{f} = 2\lambda^2 |\phi|^2 \braket{f} + O_{\infty}(l^{1+k/2}) = O_{\infty}(l^{1+k/2})
\end{equation}
and hence can estimate the $H_{s+2}$ norms of $\tilde{f}$ as before (again assuming $C_s'=O(1)$)
\begin{equation}
    ||\tilde{f}||_{H_{s+2}} \leq C'_s ( ||\tilde{f}||_{H_{1}} + ||Q\tilde{f}||_{H_{s}}) = O(l^{(k+1)/2}) 
\end{equation}
and using the Sobolev embedding theorem on compact manifolds we get
\begin{equation}
    \tilde{f} = O_{\infty}(l^{(k+1)/2})
\end{equation}
Since $V_A = \DD_A \tilde{f}$, this advances the induction on $V_A$. Finally we consider equation (\ref{EtauAcharge})
\begin{multline}
    \partial_A \kappa = - \frac{1}{2} i \lambda l^m f \left(\hat{\phi}^{*} \partial_A \hat{\phi} - \hat{\phi} \partial_A \hat{\phi}^* - 2i\lambda A_A |\hat{\phi}|^2\right) - \frac{1}{2} c_1(|\phi|^2)\left( F_{A B} h^{B C} - F_{\tau \rho} \delta_{A}^{C} \right) \DD_C \tilde{f}\\
    + O_{\infty}(l^{1+k/2})
\end{multline}
to see $\partial_A \kappa = O_{\infty}(l^{(k+1)/2})$ which advances the induction on $\partial_A \kappa$.

This induction gets us as far as
\begin{equation}
    \partial_A \kappa = O_{\infty}(l^{N-1/2}), \quad V_A = O_{\infty}(l^{N-1/2}) \quad \text{and} \quad f = O(l^{N-1/2 - m})
\end{equation}
Thus we cannot control the size of $f$ (the electric potential in this gauge) as much as before, but we can still prove the zeroth law in the formulation that we wrote down for the uncharged scalar field case, which is the physically important one.

It seems likely that the zeroth law can still be proved even in the case where $\phi$ has more general behaviour in $l$. This could possibly be achieved by a more thorough examination of the structure of positive boost weight quantities made from $A_{\mu}$ and $\phi$, or by considering the scalar field equation of motion $E = O(l^N)$.
\chapter{The Second Law of Black Hole Mechanics in Effective Field Theory}\label{ChapterSecondLaw}

The contents of this chapter are the results of original research conducted by the author of this thesis in collaboration with Harvey Reall. It is based on work published in \cite{Davies:2024fut} and \cite{Davies:2023qaa}.

\section{The Second Law}

The second law of black hole mechanics is the statement that the entropy of dynamical (i.e., non-stationary and therefore out of equilibrium) black hole solutions is non-decreasing in time. This is assumed to be the classical limit of the second law of thermodynamics, which states that the total thermodynamic entropy of the whole system (i.e., the black hole and its surroundings) is non-decreasing. 

In Chapter \ref{ChapterLaws}, we saw that in standard 2-derivative GR coupled to matter satisfying the Null Energy Condition, it can be proved that the area $A(v)$ of a spacelike cross section of the horizon is always non-decreasing in $v$. This supports a natural interpretation of the entropy of a black hole as proportional to its area. However, when we include higher derivative terms in the Lagrangian, $A(v)$ is no longer necessarily non-decreasing. Therefore we need a generalization of the definition of entropy in order to satisfy a second law. Whilst there has been no answer that applies to all situations, a fruitful avenue has been to study dynamical black holes that settle down to equilibrium at late times and that are in the regime of validity of EFT. In particular, we saw that Hollands, Kov\'acs and Reall (HKR) \cite{Hollands:2022} have made a proposal for dynamical black hole entropy
\be
    S_{HKR}(v) = 4\pi \int_{C(v)}\text{d}^{d-2}x \sqrt{\mu} s^{v}_{HKR}
\ee
constructed from an entropy density $s_{HKR}^v$ in affinely parameterized GNCs. This satisfies the second law to quadratic order in perturbations around a stationary black hole, up to $O(l^N)$ terms for any EFT of gravity and a scalar field known up to order $l^{N-1}$:
\begin{equation}
    \delta^2 \dot{S}_{HKR} \geq -O(l^N)
\end{equation}

In this Chapter we make a new proposal for dynamical black hole entropy
\be
    S(v) = 4\pi \int_{C(v)}\text{d}^{d-2}x \sqrt{\mu} s^{v}
\ee
which adds particular terms that are quadratic or higher in positive boost weight quantities to $s^{v}_{HKR}$ to construct a new entropy density $s^v$ in affinely parameterized GNCs. This definition improves upon $S_{HKR}(v)$ in two ways. The first is that $S(v)$ satisfies the second law \textit{non-perturbatively} up to $O(l^N)$ terms in the regime of validity of the EFT:
\begin{equation}
    \dot{S} \geq -O(l^N)
\end{equation}
This entirely avoids the need for perturbation theory around a stationary black hole and thus can apply to highly dynamical situations, so long as the black hole horizon remains smooth for all future time (which is needed for us to define GNCs). Therefore, situations like the period after merger or gravitational collapse, or a black hole interacting with weak gravitational waves are applicable.

The second avenue of improvement is that we widen the class of gravitational EFTs for which this entropy is defined, by including EFTs describing gravity coupled to electromagnetism and a real scalar field as defined in the previous Chapter. As discussed previously, the electromagnetic field is the obvious matter field to which we should generalize our definition to, as it is the only other field for which we have observed the classical approximation to be valid.

\subsection{The Second Law in EFT and Regime of Validity}

In EFT, the effective action is made up of potentially infinitely many higher derivative terms. However, as discussed in the previous Chapter, in practice we will only ever know finitely many of the coefficients of these terms, and so there will be some maximal $N$ for which we know all the terms with $N+1$ or fewer derivatives. In this case we only fully know part of the equations of motion, which in the regime of validity of the EFT satisfies
\be \label{DynEFTEOM}
E_I\equiv E_I^{(0)} + \sum_{n=1}^{N-1}l^{n}E^{(n)}_I = O(l^N)
\ee
Since we only know our theory up to some accuracy of order $l^N$, it is reasonable to expect our second law to only be provable up to order $l^N$ terms. This is indeed what we will show, i.e.,
\be \label{2ndLawSense}
    \dot{S}(v) \geq -O(l^N)
\ee
where the RHS of the inequality signifies that $\dot{S}(v)$ might be negative but only by an $O(l^N)$ amount. This means that the better we know our EFT, the closer we can construct an entropy satisfying a complete second law. The entropy $S(v)$ will contain terms of up to $N-2$ derivatives. 

We shall not be interested in arbitrary black hole solutions of our Einstein-Maxwell-Scalar EFT. In general, there will be pathological solutions that blow up in time or exhibit rapid oscillations and are considered unphysical. See Section IV of  \cite{Flanagan:1996gw} for a discussion around the existence of such solutions, which should not be expected to satisfy the second law.

Instead, we shall consider only black hole solutions that lie within the regime of validity of the EFT, as we did in the zeroth law. Since we are now in a dynamical situation, we no longer have the Killing vector GNCs with which we used to precisely formulate this condition in \ref{assumptionslbehaviour}. Therefore, we reformulate it here in a manner that can be applied to this scenario.

We again assume we have a 1-parameter family  $g_{\mu\nu}(x^\mu,L)$, $F_{\rho\sigma}(x^\mu,L)$, $\phi(x^\mu,L)$ of dynamical black hole solutions to the EFT equations of motion, with $L$ representing the smallest length/timescale on which the solution varies. Let $C$ be some fixed cross-section of the horizon $\cH$, and let $\cH_+$ denote the part of the horizon to the future of $C$. We work in some choice of affinely parameterized GNCs $x^\mu = (v,r,x^A)$ with $C$ at $r=v=0$, and $C(v)$ a foliation of $\cH_+$. We assume that in this choice of affinely parameterized GNCs, any $n$-derivative quantity made from the GNC metric quantities $\alpha$, $\beta_A$, $\mu_{AB}$, $\mu^{A B}$, the matter fields $F_{\mu\nu}$, $\phi$, or their derivatives (w.r.t. $x^\mu$) is uniformly bounded on $\cH_+$ by $c_n/L^n$ for some dimensionless $L$-independent constant $c_n$. We again formulate this in a way that is covariant in $A, B,... $ indices by enforcing the uniform boundedness in the sense of the norm $|\cdot|_{\mu}$ (defined similarly to (\ref{hABnorm})). Then the solution is said to be in the regime of validity of the EFT if $L\gg l$. 

Again, we henceforth suppress any explicit dependence on $L$, and write a quantity's order with respect to $l/L$ purely in terms of $l$. Factors of $L$ can be reinstated through dimensional analysis. 

Furthermore, we shall only consider black holes that settle down to equilibrium at late times. As discussed in the previous Chapter, if one assumes the zeroth law holds exactly then all positive boost weight quantities vanish on the horizon of a stationary black hole. Therefore, we assume that positive boost weight quantities vanish on the horizon of our dynamical black hole at late times. In addition, we will need the following assumptions. For $v>v_0>0$, define 
\be
    a(v_0,v,x^A)= \sqrt{\frac{\mu(v, r=0, x^A)}{\mu(v_0,r=0,x^A)}}
\ee
which measures the change in the area element from $v_0$ to $v$. Then we assume there is some dimensionless constant $C_a\geq 1$ such that
\begin{equation}
    \frac{1}{C_a} \leq a(v_0,v,x^A) \leq C_a
\end{equation}
for all $v,v_0,x^A$, and that any $n$-th derivative of $a(v_0,v,x^A)$ is bounded uniformly by $c_n/L^n$ in the sense described above in the regime of validity of EFT. These assumptions roughly state that the black hole does not grow, shrink or oscillate an arbitrarily large amount over the course of the dynamics. This should be expected when we combine the notion of the regime of validity of the EFT with the notion that the black hole settles down to equilibrium at late times.

Additionally, we assume $c_1(\phi)\geq \tilde{c} > 0$ for some positive constant $\tilde{c}$ independent of $l,L$, as before. Finally, we assume every horizon cross-section $C(v)$ is compact and without boundary.

\section{The Main Results}

We will work exclusively in affinely parameterized GNCs. Rewriting the metric here for convenience,
\begin{equation}
    g = 2 \text{d}v \text{d}r - r^2 \alpha(r,v,x^C) \text{d}v^2 -2 r\beta_{A}(r,v,x^C) \text{d}v \text{d}x^A + \mu_{A B}(r,v,x^C) \text{d}x^A \text{d}x^B.
\end{equation}
We raise and lower $A,B,C,...$ indices with $\mu_{A B}$ and denote the covariant derivative with respect to $\mu_{A B}$ by $D_{A}$. As well as $K_{A B} \equiv \frac{1}{2} \partial_{v}{\mu_{A B}}, \bar{K}_{A B} \equiv \frac{1}{2} \partial_{r}{ \mu_{A B} }$ defined previously, it will be useful to define
\begin{equation} \label{Maxwellquants}
    K_{A} \equiv F_{v A}, \quad \bar{K}_{A} \equiv F_{r A}, \quad \psi \equiv F_{v r}
\end{equation}
$K_A$ has boost weight $+1$, $\bar{K_A}$ has boost weight $-1$ and $\psi$ has boost weight $0$.

The construction of the entropy is performed in two parts. The first part consists of generalizing the HKR entropy to allow for the inclusion of the electromagnetic field in our EFT. We do this by studying the $E_{v v}$ component of the equations of motion in affinely parameterized GNCs on $\cH_+$. Our generalization of equation (\ref{HKRForm}) is as follows. We show that on-shell (i.e., by using other components of the equations of motion $E_I = O(l^N)$) we can bring $E_{v v}$ on $\cH_+$ into the form
\begin{multline} \label{EvvForm}
    - E_{v v}\Big|_{\cH_+} = \partial_{v}\left[\frac{1}{\sqrt{\mu}} \partial_{v}\left(\sqrt{\mu} s^{v}_{HKR}\right) + D_{A}{ s^A }\right] + \left(K_{A B} + X_{A B}\right) \left(K^{A B} + X^{A B}\right)\\
    + \frac{1}{2} c_1(\phi) \left( K_{A} + X_{A} \right)\left(K^{A} + X^{A} \right) + \frac{1}{2} \left( \partial_v \phi + X\right)^2 + D_A Y^A + O(l^N)
\end{multline}
where $X = \sum_{n=1}^{N-1} l^n X^{(n)}, X_A = \sum_{n=1}^{N-1} l^n X_A^{(n)}, X_{A B} = \sum_{n=1}^{N-1} l^n X^{(n)}_{A B}$ (boost weights +1) are linear or higher in positive boost weight quantities, and $Y^A = \sum_{n=1}^{N-1} l^n Y^{(n) A}$ (boost weight +2) are quadratic or higher. $s^v_{HKR} = \sum_{n=0}^{N-1} l^n s^{(n) v}_{HKR}$ has boost weight 0 and $s^A=\sum_{n=1}^{N-1} l^n s^{(n) A}$ has boost weight +1. They will be invariant upon change of electromagnetic gauge.

The generalization of the HKR entropy of the spacelike cross section $C(v)$ is then defined to be
\be
    S_{HKR}(v) = 4\pi \int_{C(v)}\text{d}^{d-2}x \sqrt{\mu} s^{v}_{HKR}
\ee
which satisfies
\begin{equation}
        \dot{S}_{HKR}(v)= 4 \pi \int_{C(v)} \text{d}^{d-2} x \sqrt{\mu} \int_{v}^{\infty} \text{d}v' \left[W^2 + D_{A}{Y^{A}} + O(l^{N}) \right](v',x)
\end{equation}
with $W^2= \left(K_{A B} + X_{A B}\right) \left(K^{A B} + X^{A B}\right) + \frac{1}{2} \left( \partial_v \phi + X\right)^2 + \frac{1}{2} c_1(\phi) \left( K_{A} + X_{A} \right)\left(K^{A} + X^{A} \right)$. Note that $\delta^2 \dot{S}_{HKR} \geq -O(l^N)$ follows in the same way as detailed in \ref{HKRExplanation}, because the additional term in $W^2$ is still a positive definite form. The obstruction to $\dot{S}_{HKR}$ being non-negative up to $O(l^N)$ \textit{non-perturbatively} is $D_{A}{Y^{A}}(v',x)$. Despite being a divergence term, it does not integrate to zero because it is evaluated at the integration variable $v'$, whereas the area element $\sqrt{\mu(v,x)}$ is evaluated at $v$.

The second part of the construction consists of getting around this problem. Specifically, we shall perform further manipulations on $D_{A}{Y^{A}}$ in order to write $\dot{S}_{HKR}(v)$ in the form
\begin{multline} \label{HKRGeneralization}
        \dot{S}_{HKR}(v)= -\frac{d}{dv} \left(4\pi \int_{C(v)} \text{d}^{d-2}x \sqrt{\mu(v)} \sigma^{v}(v) \right)\\
        +4 \pi \int_{C(v)} \text{d}^{d-2} x \sqrt{\mu} \int_{v}^{\infty} \text{d}v' \Big[(K_{A B} + Z_{A B})(K^{A B} + Z^{A B}) + \frac{1}{2} c_1(\phi) \left( K_{A} + Z_{A} \right)\left(K^{A} + Z^{A} \right)\\
        + \frac{1}{2} \left( \partial_v \phi + Z\right)^2 + O(l^{N}) \Big](v,v',x)
\end{multline}
for $Z_{A B}, Z_{A}$ and $Z$ that are "bilocal", meaning they depend on both $v$ and the integration variable $v'$. The first term on the right hand side is the $v$-derivative of something that looks like an entropy, hence we can move it over to the left hand side. The remaining right hand side is now manifestly non-negative up to $O(l^N)$ terms because there is no term like $D_{A}{Y^{A}}$. Hence, defining
\be
    S(v) = 4\pi \int_{C(v)}\text{d}^{d-2}x \sqrt{\mu} s^v
\ee
with $s^v=s^{v}_{HKR}+\sigma^v$ implies $S(v)$ satisfies $\dot{S}(v)\geq -O(l^N)$ as desired.

\section{Entropy Construction Part I: Generalizing the HKR Entropy}

\subsection{Leading Order Einstein-Maxwell-Scalar Theory}

Let us start with the $2$-derivative Einstein-Maxwell-Scalar terms arising from $\LL_{2}$. Their contribution to $E_{\alpha \beta}$ is
\begin{equation}
    E^{(0)}_{\alpha\beta}= R_{\alpha \beta} - \frac{1}{2} \nabla_{\alpha} \phi \nabla_{\beta} \phi -\frac{1}{2} c_1(\phi) F_{\alpha \delta} F_{\beta}^{\,\,\,\,\delta} - \frac{1}{2} g_{\alpha \beta} \left( R - V(\phi) -\frac{1}{2} \nabla_{\gamma}{\phi} \nabla^{\gamma}{\phi} - \frac{1}{4} c_1(\phi) F_{\gamma \delta} F^{\gamma\delta} \right)
\end{equation}
In affinely parameterized GNCs on $\cH_+$, we have
\begin{equation}
    E^{(0)}_{v v}\big|_{\cH_+}= R_{v v} - \frac{1}{2} (\partial_{v} \phi)^2 -\frac{1}{2} c_1(\phi) K_A K^A
\end{equation}
Using $R_{v v}|_{\cH_+} = -\mu^{A B} \partial_{v} K_{A B} + K_{A B} K^{A B}$ and $\partial_{v} \sqrt{\mu} = \sqrt{\mu}\, \mu^{A B} K_{A B}$, we can write this as
\begin{equation} \label{Orderl^0}
    -E^{(0)}_{v v}\big|_{\cH_+}= \partial_{v}\left[\frac{1}{\sqrt{\mu}} \partial_{v}\left(\sqrt{\mu}\right) \right] + K_{A B} K^{A B} + \frac{1}{2} c_1(\phi) K_A K^A + \frac{1}{2} (\partial_{v} \phi)^2 
\end{equation}
This is of the form (\ref{EvvForm}) with $N=1$, $s^{(0)v}_{HKR} = 1$ and $s^{(0) A} = X^{(0)} = X^{(0)}_A = X^{(0)}_{A B} = Y_A^{(0)} = 0$. Because there is no total derivative term $D_A Y^{(0)A}$, there are no further manipulations needed to get to (\ref{HKRGeneralization}) with $\sigma^v=0$, $Z^{(0)}=X^{(0)}$, $Z_{A}^{(0)}=X_{A}^{(0)}$ and $Z_{A B}^{(0)}=X_{A B}^{(0)}$. Thus we have proved our theory satisfies the second law non-perturbatively at leading order $l^0$ (which of course can be proved by the usual proof of the second law on the 2-derivative theory).

We will ultimately work through the higher order terms order-by-order to mould them into the correct form. To get to that point however, we must start with the Iyer-Wald-Wall entropy for this theory.

\subsection{The Iyer-Wald-Wall Entropy}\label{BDKEntropy}

In \ref{IWWExplanation}, we discussed the Iyer-Wald-Wall (IWW) \cite{Wall:2015} proposal for dynamical black hole entropy and its generalization by \cite{Biswas:2022} to arbitrary theories of gravity, electromagnetism and a real (uncharged) scalar field with diffeomorphism-invariant and electromagnetic gauge-independent Lagrangian. In the construction of this proposal, it is proved that the $E_{v v}$ component of the equations of motion of such a theory can be brought into the following form on the horizon\footnote{In \cite{Biswas:2022}, they have an additional $T_{v v}$ in this defining equation, which is the part of the energy momentum tensor arising from the minimal coupling part of the matter sector Lagrangian. However they also show that $T_{v v}$ is quadratic in positive boost weight quantities if $T_{\mu \nu}$ satisfies the NEC, which is the case for our 2-derivative Einstein-Maxwell-Scalar theory and hence we can absorb $T_{v v}$ into the ellipsis.}:
\begin{equation} \label{IWW}
    -E_{v v}\Big|_{\cH_+} = \partial_{v}\left[\frac{1}{\sqrt{\mu}} \partial_{v}\left(\sqrt{\mu} s^{v}_{IWW}\right) + D_{A}{ s^A }\right] + ...
\end{equation}
where the ellipsis denote terms at least quadratic in positive boost weight quantities. We will call the quantity $(s^{v}_{IWW}, s^A)$ the IWW entropy current. In \cite{Biswas:2022} it is proved to be electromagnetic gauge-invariant. $s^A$ is a vector in $A, B,..$ indices on $C(v)$ whilst $s^v_{IWW}$ is a scalar. They are only defined uniquely up to linear order in positive boost weight quantities, as any higher order terms can be absorbed into the ellipsis. 

For our Einstein-Maxwell-Scalar EFT, we can calculate $s^v_{IWW}$ up to $O(l^N)$ and take (\ref{IWW}) as our starting point. We group all the remaining terms in the ellipsis and define
\begin{equation} \label{Qdef}
    \Delta\equiv - E_{v v}\Big|_{\cH_+} - \partial_{v}\left[\frac{1}{\sqrt{\mu}} \partial_{v}\left(\sqrt{\mu} s^{v}_{IWW}\right) + D_{A}{ s^A }\right]
\end{equation}

We will use the fact that $\Delta$ is at least quadratic in positive boost weight terms to show we can manipulate it so that (\ref{IWW}) becomes (\ref{EvvForm}). The resulting generalization of the HKR entropy density $s^v_{HKR}$ will be 
\be
    s^v_{HKR} = s^v_{IWW} + \sum_{n=0}^{N-1}l^n \varsigma^{(n) v}
\ee
where the $\varsigma^{(n) v}$ are at least quadratic in positive boost weight quantities. We will not need to add any terms to $s^A$.

From (\ref{Orderl^0}), we can see that for the leading order theory $\LL_{2}$, the IWW entropy density is $s^{(0) v}_{IWW}=1$ and we need no correction, $\varsigma^{(0) v} = 0$. 

\subsection{Reducing to Allowed Terms}

To generalize the HKR entropy we study the possible quantities $\Delta$ is made out of. $\Delta$ is electromagnetic gauge-invariant and a scalar with respect to $A,B,...$ indices. Therefore it is made from electromagnetic gauge-invariant affinely parameterized GNC quantities of the metric and matter fields that are covariant in $A, B,...$ indices, namely
\be \label{covariant list}
    D^k{ \partial_{v}^{p} \partial_{r}^q \varphi } \,\,\, \text{for} \,\,\, \varphi \in \{\alpha, \beta_A, \mu_{A B}, R_{A B C D}[\mu], \epsilon_{A_1 ... A_{d-2}}, \phi, F_{A B}, K_A, \bar{K}_A, \psi\}
\ee
where $k,p,q\geq 0$ and we have suppressed the indices $D^k = D_{A_1}...D_{A_k}$. $K_A, \bar{K}_A$ and $\psi$ are defined in (\ref{Maxwellquants}), and $\epsilon_{A_1 ... A_{d-2}}$ and $R_{A B C D}[\mu]$ are the induced volume form and induced Riemann tensor on $C(v)$. WLOG we have kept all $D_A$ derivatives on the left using the commutation rules
\begin{equation} \label{CommutationRules}
\begin{split}
    [\partial_{v}, D_{A}]t_{B_1 ... B_n} =& \sum_{i=1}^{n} \mu^{C D} ( D_{D}{ K_{A B_i} } -D_{A}{ K_{D B_i} } - D_{B_i}{ K_{A D} } ) t_{B_1... B_{i-1} C B_{i+1} ... B_n},\\
    [\partial_{r}, D_{A}]t_{B_1 ... B_n} =& \sum_{i=1}^{n} \mu^{C D} ( D_{D}{ \Bar{K}_{A B_i} } -D_{A}{ \Bar{K}_{D B_i} } - D_{B_i}{ \Bar{K}_{A D} } ) t_{B_1... B_{i-1} C B_{i+1} ... B_n}
\end{split}
\end{equation}

We can immediately eliminate $\partial_{v}$ and $\partial_{r}$ derivatives of $\epsilon_{A_1 ... A_{d-2}}$ and $R_{A B C D}[\mu]$ via the formulae
\begin{equation}\label{EliminationOfREpsilon}
    \begin{split}
        \partial_{v}{\epsilon_{A_1 ... A_{d-2}}} =& \epsilon_{A_1 ... A_{d-2}} K,\\
        \partial_{r}{\epsilon_{A_1 ... A_{d-2}}} =& \epsilon_{A_1 ... A_{d-2}} \Bar{K},\\
        \partial_{v}{ R_{A B C D}[\mu]} =& K^{E}\,_{B} R_{A E C D}[\mu] - K^{E}\,_{A} R_{B E C D}[\mu] +D_{C}{D_{B}{K_{A D}}}-\\
        &D_{C}{D_{A}{K_{B D}}}-D_{D}{D_{B}{K_{A C}}}+D_{D}{D_{A}{K_{B C}}},\\
        \partial_{r}{ R_{A B C D}[\mu]} =& \Bar{K}^{E}\,_{B} R_{A E C D}[\mu] - \Bar{K}^{E}\,_{A} R_{B E C D}[\mu] +D_{C}{D_{B}{\Bar{K}_{A D}}}-\\
        &D_{C}{D_{A}{\Bar{K}_{B D}}}-D_{D}{D_{B}{\Bar{K}_{A C}}}+D_{D}{D_{A}{\Bar{K}_{B C}}}
    \end{split}
\end{equation}

We will now show that we can further reduce the set of objects that can appear on the horizon by using the equations of motion. It is worth emphasising, this reduction holds in an EFT sense, meaning \textit{it is only done up to $O(l^N)$ terms}. In the original HKR procedure of \cite{Hollands:2022} for Einstein-Scalar EFT, they show how to reduce the metric and scalar field terms to the set $\mu_{A B}$, $\epsilon_{A_1 ... A_{d-2}}$, $D^k R_{A B C D}[\mu]$, $D^k \beta_{A}$, $D^k\partial_{v}^p K_{A B}$, $D^k\partial_{r}^p \bar{K}_{A B}$, $D^k\partial_{v}^p \phi$, $D^k\partial_{r}^p \phi$ with $p\geq 0$. This procedure still holds in our Einstein-Maxwell-Scalar EFT. To focus on where we need to generalize the HKR procedure, we only detail how to reduce the Maxwell terms.

We aim to reduce the set of Maxwell terms on the horizon to
\be
    D^k \psi,\,\, D^k F_{A B},\,\, D^k\partial_{v}^p K_{A},\,\, D^k\partial_{r}^q \bar{K}_{A}
\ee
To do this we must eliminate any $\partial_v$ and $\partial_r$ derivative of both $\psi$ and $F_{A B}$. We must also eliminate any $\partial_r$ derivative of $K_A$, and any $\partial_v$ derivative of $\bar{K}_A$. 

To begin we use the fact that
\be \label{Fantisym}
\partial_{\alpha}F_{\beta \gamma} + \partial_{\beta}F_{\gamma\alpha} + \partial_{\gamma}F_{\alpha \beta} = 0
\ee
which follows from $F=\text{d}A$. Taking $\alpha = v, \beta = A, \gamma = B$, we can rearrange this to\footnote{If we had explicitly picked a gauge $A_\mu$, then this relation would be trivially true and we would have fewer terms to eliminate. However, we would like to keep the entropy current manifestly gauge invariant, and hence we do not pick a gauge.}
\be
    \partial_v F_{A B} = D_A K_B - D_B K_A
\ee
Similarly, taking $\alpha = v, \beta = A, \gamma = B$ gives
\be
    \partial_r F_{A B} = D_A \bar{K}_B - D_B \bar{K}_A
\ee
These two relations allow us to eliminate all $\partial_v$ and $\partial_r$ derivatives of $F_{A B}$ in favour of other Maxwell and metric terms.

Furthermore, taking $\alpha = v, \beta = r, \gamma = A$, we get
\be \label{dvKA}
    \partial_v \bar{K}_{A} = \partial_r K_A - D_A \psi
\ee
which allows us to eliminate any $\partial_v$ derivative or mixed $\partial_v$ and $\partial_r$ derivative of $\bar{K}_A$.

To go further, we will have to use the equations of motion for the Maxwell field:
\be
    E_\alpha \equiv \nabla^{\beta}{\Big[c_1(\phi)F_{\alpha\beta} - 4 c_2(\phi) F^{\gamma \delta} \epsilon_{\alpha \beta \gamma \delta}\Big]} + lH_{\alpha} = O(l^N)
\ee
We can use $\epsilon_{\alpha \beta \gamma \delta} \nabla^{\beta} F^{\gamma \delta} = 0$  to rewrite this as
\begin{equation} \label{Fswap}
    \nabla^{\beta}F_{\alpha\beta} = \frac{1}{c_1(\phi)}\left[ 4 c'_2(\phi) \nabla^{\beta}\phi F^{\gamma \delta} \epsilon_{\alpha \beta \gamma \delta} - c'_1(\phi) \nabla^{\beta}\phi F_{\alpha\beta} \right] + O(l)
\end{equation}
The order $l^0$ terms on the right hand only involve Maxwell terms that we are not trying to eliminate. Let us now evaluate the $v$ component of $\nabla^{\beta}F_{\alpha\beta}$ in affinely parameterized GNCs:
\be \label{Ev0}
    \nabla^{\beta}F_{v \beta} = \partial_{v} \psi + D^{A}{K_{A}} +\psi K + ...
\ee
where the ellipsis denotes terms that vanish on $\cH_+$. We can substitute this into (\ref{Fswap}) to get an expression for $\partial_v \psi$ on the horizon up to terms higher order in $l$:
\begin{multline} \label{dvpsi}
    \partial_v \psi\big|_{\cH_+} = - D^{A}{K_{A}} - \psi K + \\
    \frac{1}{c_1(\phi)}\left[ 4c'_2(\phi)\epsilon^{A B}\left(2 D_{A} \phi K_B - \partial_v \phi F_{A B} \right) - c'_1(\phi)\left( \psi \partial_v \phi + K_A D^A \phi \right) \right] + O(l)
\end{multline} 
Therefore, wherever we find a $\partial_v \psi$ in $\Delta$, we can swap it out order-by-order in $l$, pushing it to higher order with each step. Eventually it will only appear at $O(l^N)$, at which point it is not relevant to our analysis since we do not know the equations of motion at that order.

Similarly we can evaluate $\nabla^\beta F_{r \beta}$ in affinely parameterized GNCs:
\be \label{Er0}
    \nabla^{\beta}F_{r \beta} = -\partial_{r}{\psi} + D^{A}{\bar{K}_{A}} + \bar{K}^{A} \beta_{A}- \psi \bar{K} + ...
\ee
where, again, the terms in the ellipsis vanish on the horizon. We can substitute this into (\ref{Fswap}) to get an expression for $\partial_r \psi$ on the horizon:
\begin{multline} \label{drpsi}
    \partial_r \psi\big|_{\cH_+} = D^{A}{\bar{K}_{A}} + \bar{K}^{A} \beta_{A}- \psi \bar{K} +\\
    \frac{1}{c_1(\phi)}\left[ c'_1(\phi)\left( \bar{K}_A D^A \phi - \psi \partial_r \phi  \right) - 4c'_2(\phi)\epsilon^{A B}\left( \partial_r \phi F_{A B} - 2 D_{A} \phi \bar{K}_B  \right)  \right] + O(l) 
\end{multline}
This allows us to eliminate $\partial_r \psi$ up to $O(l^N)$ in a similar fashion.

We can take $\partial_v$ derivatives of  (\ref{dvpsi}) and (\ref{drpsi}) in order to eliminate $\partial^p_v \partial_r^q \psi$ for $p\geq 1$ and $q = 0,1$. However, we cannot naively take $\partial_r$ derivatives because these expressions are evaluated on the horizon $r=0$. Instead, we must take successive $\partial_r$ derivatives of (\ref{Fswap}) and (\ref{Er0}), and then evaluate them on $r=0$, possibly using substitution rules already calculated for lower order derivatives. This will involve taking care of the terms in the ellipsis in (\ref{drpsi}), which are given in full in Appendix \ref{dFAppendix}. However, these only ever involve lower order derivatives, for which we already have substitution rules and hence do not cause an issue. Therefore, we can eliminate all $\partial_v$ and $\partial_r$ derivatives of $\psi$ up to order $O(l^N)$.

This just leaves $\partial_r$ derivatives of $K_A$ to be eliminated, for which we look at $\nabla^\beta F_{A \beta}$:
\begin{multline} \label{EA0}
    \nabla^{\beta}F_{A \beta} = -2\partial_{r}{K_{A}}+D_{A}{\psi} + D^{B}{F_{A B}}+2\bar{K}^{B} K_{A B} + 2K^{B} \bar{K}_{A B} -\\
    \psi \beta_{A} - \bar{K}_{A} K - K_{A} \bar{K} + F_{A B} \beta^{B} + ...
\end{multline}
Substituting this into (\ref{Fswap}) gives us an expression which we can use to eliminate $\partial_r K_A$ on the horizon. Taking $\partial_r$ derivatives of (\ref{EA0}) again allows us to eliminate higher $\partial_r$ derivatives of $K_A$ because the terms in the ellipsis only involve lower order derivatives. This completes the reduction of Maxwell terms.

Combining the Maxwell terms with the metric and scalar field terms already reduced through the HKR procedure, we are left with a small set of "allowed terms":
\begin{empheq}[box=\fbox]{align}\label{AllowedTermsIncMaxwell}
    \text{Allowed terms:} \,\, &\mu_{A B}, \,\, \mu^{A B}, \,\, \epsilon_{A_1 ... A_{d-2}}, \,\, D^k R_{A B C D}[\mu], \,\, D^k \beta_{A}, \,\, 
    D^k\partial_{v}^p K_{A B}, \,\, D^k\partial_{r}^q \bar{K}_{A B},\nonumber\\
    & D^k \psi, \,\, D^k F_{A B}, \,\, 
    D^k\partial_{v}^p K_{A}, \,\, D^k\partial_{r}^q \bar{K}_{A}, \,\,
    D^k\partial_{v}^p \phi, \,\, D^k\partial_{r}^q \phi 
\end{empheq}
In particular, the only allowed positive boost weight terms are of the form $D^k\partial_{v}^p K_{A B}$ and $ D^k\partial_{v}^p K_{A}$ with $p\geq 0$, and $D^k\partial_{v}^p \phi$ with $p\geq 1$. This will be the crucial fact that allows us to manipulate the terms in $\Delta$.

\subsection{Manipulating Terms Order-by-Order} \label{ManipTermsObO}

Let us return to $\Delta$. We use the above procedures to eliminate any non-allowed terms up to $O(l^N)$. Once this is done, we group together terms with the same overall power of $l$:  
\begin{equation}\label{Qseries}
    \Delta = \Delta^{(0)} + \sum_{n=1}^{N-1} l^n \Delta^{(n)} + O(l^N)
\end{equation}
By construction, the $\Delta^{(n)}$ are quadratic or higher in positive boost weight terms. Furthermore, $\Delta^{(0)}$ are the terms calculated from the leading order part of the equation of motion in (\ref{Orderl^0}):
\be \label{Q0}
    \Delta^{(0)} = K_{A B} K^{A B} + \frac{1}{2} c_1(\phi) K_A K^A + \frac{1}{2} (\partial_{v} \phi)^2
\ee

We now proceed by induction order-by-order in $l$. Our inductive hypothesis is that we have manipulated the terms in $\Delta$ up to $O(l^m)$ into the form
\begin{multline}\label{Qinduction}
    \Delta= \partial_{v}\left[\frac{1}{\sqrt{\mu}} \partial_{v}\left(\sqrt{\mu} \sum_{n=0}^{m-1} l^n \varsigma^{(n) v} \right)\right] + \left(K_{A B}+\sum_{n=0}^{m-1} l^n X^{(n)}_{A B}\right) \left(K^{A B}+\sum_{n=0}^{m-1} l^n X^{(n) A B}\right) +\\
     \frac{1}{2} c_1(\phi) \left(K_{A}+\sum_{n=0}^{m-1} l^n X^{(n)}_{A}\right) \left(K^{A}+\sum_{n=0}^{m-1} l^n X^{(n) A}\right) + \frac{1}{2}\left( \partial_v \phi + \sum_{n=0}^{m-1} l^n X^{(n)} \right)^2 +\\
     D_{A}{\sum_{n=0}^{m-1} l^n Y^{(n) A}} + \sum_{n=m}^{N-1} l^n \Delta^{(n)} + O(l^{N})
\end{multline}
where the $\Delta^{(n)}$ may have gained extra terms compared to (\ref{Qseries}) but are still quadratic or higher in positive boost weight terms.

By (\ref{Q0}), this is true for $m=1$ with $\varsigma^{(0) v} = X^{(0)}_{A B} = X^{(0)}_{A} = X^{(0)} = Y^{(0) A} = 0$. So assume it is true for some $1\leq m\leq N-1$. 

We now consider $\Delta^{(m)}$. It is quadratic or higher in positive boost weight quantities. However, we have reduced the set of allowed positive boost weight quantities. Therefore we can write it as a sum 
\be
    \Delta^{(m)} = \sum_{k_1, k_2, p_1, p_2, P_1, P_2} (D^{k_1}{\partial_{v}^{p_1} P_1}) \, (D^{k_2}{\partial_{v}^{p_2} P_2}) \, Q_{k_1, k_2, p_1, p_2, P_1, P_2} 
\ee
where $P_1, P_2 \in \{ K_{A B}, K_{A}, \partial_v \phi \}$ and $Q_{k_1, k_2, p_1, p_2, P_1, P_2}$ is some linear combination of allowed terms. Note that we have dropped $A, B,...$ indices here for notational ease, and the terms can be contracted in any way.

We aim to manipulate this sum so that everything in it is proportional to some $P_1 \in \{ K_{A B}, K_{A}, \partial_v \phi \}$. First we move over the $D^{k_1}$ derivatives in the first factor of each term using the product rule $D(f) g = - f D(g) + D(f g)$. This will produce some total derivative $D_{A} Y^{(m)A}$ with $Y^{(m) A}$ at least quadratic in positive boost weight quantities:
\begin{equation}
    \Delta^{(m)} = \sum_{k, p_1, p_2, P_1, P_2} (\partial_{v}^{p_1} P_1) \, (D^{k}{\partial_{v}^{p_2} P_2}) \, Q_{k, p_1, p_2, P_1, P_2}  + D_{A} Y^{(m)A } 
\end{equation}

Secondly we move over the $\partial_{v}^{p_1}$ derivatives in each term. We do this in such a way as to produce terms of the form $\partial_{v}\left(\frac{1}{\sqrt{\mu}} \partial_{v}\left(\sqrt{\mu} \varsigma^{v}\right)\right)$, where $\varsigma^v$ will contribute to the HKR entropy density. It is proved in \cite{Hollands:2022} that for any $k\geq0$ and $p_1, p_2\geq 1$, there exist unique numbers $a_j$ such that
\be \label{vintegrationbyparts}
    (\partial_{v}^{p_1} P_1) \, (D^{k}{\partial_{v}^{p_2}P_2}) Q = \partial_{v}\left\{\frac{1}{\sqrt{\mu}} \partial_{v}\left[\sqrt{\mu} \sum_{j=1}^{\mathclap{p_1+p_2-1}} a_j (\partial_{v}^{p_1+p_2-1-j} P_1) \, (D^{k}{\partial_{v}^{j-1}P_2 }) Q \right]\right\} + ...
\ee
where the ellipsis denotes terms of the form
$(\partial_{v}^{\Bar{p}_1} P_1)$ $(D^{\bar{k}}{\partial_{v}^{\Bar{p}_2}P_2}) \tilde{Q}$ with $\Bar{p}_1+\Bar{p}_2 < p_1+p_2$ or $\Bar{p}_1=0$ or $\Bar{p}_2=0$. We provide a proof of this statement in Appendix \ref{app:aj}. The new $\tilde{Q}$ include terms like $\partial_{v}{ Q }$ which will involve non-allowed terms. These must be swapped out using the elimination rules and equations of motion, and so will generate more $O(l)$ terms that can be absorbed into $\sum_{n=m+1}^{N-1} l^n \Delta^{(n)}$.

We repeat this procedure on the terms in the ellipsis with $\Bar{p}_1+\Bar{p}_2 < p_1+p_2$ until eventually $\Bar{p}_1=0$ or $\Bar{p}_2=0$ for all terms. This must eventually happen because $\Bar{p}_1+\Bar{p}_2$ must decrease by at least 1 if the new $\bar{p}_1\neq 0$ and $\bar{p}_2\neq 0$, and hence $\Bar{p}_1+\Bar{p}_2$ eventually falls below 2, meaning one of $\bar{p}_1$ and $\bar{p}_2$ must be 0. For the terms with $\bar{p}_2 = 0$ we move over the $D^{\bar{k}}$ using the product rule as above, and relabel $P_1 \leftrightarrow P_2$.

Thus we can repeat this procedure until we have manipulated $\Delta^{(m)}$ into the form
\be
    \Delta^{(m)} = \sum_{k, p, P_1, P_2} P_1 \, (D^{k}{\partial_{v}^{p} P_2}) \, Q_{k, p,P_1,P_2} + \partial_{v}\left[\frac{1}{\sqrt{\mu}} \partial_{v}\left(\sqrt{\mu} \varsigma^{(m) v} \right)\right] + D_{A}Y^{(m) A}
\ee
with $\varsigma^{(m) v}$ and $Y^{(m) A}$ at least quadratic in positive boost weight quantities. 

We now split the sum over $P_1 \in \{ K_{A B}, K_{A}, \partial_v \phi \}$, write the remaining sums as $2X^{(m) A B}$, $c_1(\phi) X^{(m) A}$ and $X^{(m)}$, and substitute this into (\ref{Qinduction}):
\begin{multline}\label{Qinductionfinal}
    \Delta= \partial_{v}\left[\frac{1}{\sqrt{\mu}} \partial_{v}\left(\sqrt{\mu} \sum_{n=0}^{m} l^n \varsigma^{(n) v} \right)\right] +\\
    \left(K_{A B}+\sum_{n=0}^{m-1} l^n X^{(n)}_{A B}\right) \left(K^{A B}+\sum_{n=0}^{m-1} l^n X^{(n) A B}\right) +2 l^m K_{A B} X^{(m) A B} +\\
     \frac{1}{2} c_1(\phi) \left(K_{A}+\sum_{n=0}^{m-1} l^n X^{(n)}_{A}\right) \left(K^{A}+\sum_{n=0}^{m-1} l^n X^{(n) A}\right) + l^m c_1(\phi) K_{A} X^{(m) A} + \\
     \frac{1}{2}\left( \partial_v \phi + \sum_{n=0}^{m-1} l^n X^{(n)} \right)^2 + l^m \partial_v \phi X^{(m)} +   D_{A}{\sum_{n=0}^{m} l^n Y^{(n) A}} + \sum_{n=m+1}^{N-1} l^n \Delta^{(n)} + O(l^{N})
\end{multline}
We can complete the three squares to bring $l^m X^{(m) A B}$, $l^m X^{(m) A}$ and $l^m X^{(m)}$ into the sums. The extra terms produced are $O(l^{m+1})$ because $X^{(0)}_{A B} = X^{(0)}_A = X^{(0)} = 0$, and are at least quadratic in positive boost weight quantities so can be absorbed into $\sum_{n=m+1}^{N-1} l^n \Delta^{(n)}$. This completes the inductive step. 

This can be repeated until all terms up to $O(l^N)$ are of the correct form. Substituting this back into the definition of $\Delta$ in (\ref{Qdef}), we can now write $E_{v v}\big|_{\cH_+}$ in the desired form (\ref{EvvForm}) with 
\be
    s^v_{HKR} = s^v_{IWW} + \sum_{n=0}^{N-1}l^n \varsigma^{(n) v}
\ee
This completes the generalization of the HKR entropy
\be
    S_{HKR}(v) = 4\pi \int_{C(v)}\text{d}^{d-2}x \sqrt{\mu} s^{v}_{HKR}
\ee
which satisfies $\delta^2 \dot{S}_{HKR} \geq -O(l^N)$.

\section{Entropy Construction Part II: To a Non-Perturbative Second Law}\label{FurtherModifications}

We now further modify this to construct an entropy that satisfies a non-perturbative second law. Performing on $S_{HKR}$ the same steps used to get to (\ref{dotSHKR}) we have

\begin{equation} \label{HKREMS}
        \dot{S}_{HKR}(v_0)= 4 \pi \int_{C(v_0)} \text{d}^{d-2} x \sqrt{\mu(v_0)} \int_{v_0}^{\infty} \text{d}v \left[W^2 + D_{A}{Y^{A}} + O(l^{N}) \right](v)
\end{equation}
where $W^2= \left(K_{A B} + X_{A B}\right) \left(K^{A B} + X^{A B}\right) + \frac{1}{2} c_1(\phi) \left(K_{A}+ X_{A}\right) \left(K^{A}+X^{A}\right) + \frac{1}{2} \left( \partial_v \phi + X\right)^2$. We have suppressed all $x$-dependence, and switched notation to $v_0$ and $v$ to match \cite{Davies:2023qaa}. The obstruction to this integral being non-negative up to $O(l^N)$ is $D_{A}{Y^{A}}(v)$. Define, as before,
\be
    a(v_0,v)= \sqrt{\frac{\mu(v)}{\mu(v_0)}}
\ee
which measures the change in the area element from $v_0$ to $v$. Then, if we try to integrate $D_{A}{Y^{A}}(v)$ by parts we get\footnote{All cross sections $C(v)$ are diffeomorphic to each other, and thus we write them all as $C$ in the limits of the integrals for this section.}
\be \label{DYbyparts}
\begin{split}
    \int_{C} \text{d}^{d-2} x \sqrt{\mu(v_0)} \int_{v_0}^{\infty} \text{d}v D_{A}{Y^{A}}(v) =& \int_{v_0}^{\infty} \text{d}v \int_{C}  \text{d}^{d-2} x \sqrt{\mu(v)} a^{-1}(v_0,v) D_{A}{Y^{A}}(v)\\
    =& -\int_{v_0}^{\infty} \text{d}v \int_{C}  \text{d}^{d-2} x \sqrt{\mu(v)} Y^{A}(v) D_A a^{-1}(v_0,v)\\
    =& \int_{C} \text{d}^{d-2} x \sqrt{\mu(v_0)} \int_{v_0}^{\infty} \text{d}v Y^{A}(v) D_A \log a(v_0,v)
\end{split}
\ee
$Y^A(v)$ is quadratic or higher in positive boost weight quantities and so is a sum of terms of the form $(D^{k_1}{\partial_{v}^{p_1} P_1}) $ $ (D^{k_2}{\partial_{v}^{p_2} P_2}) $ $ Q(v)$ where, as before, $P_1, P_2 \in \{ K_{A B}, K_{A}, \partial_v \phi \}$ and $Q(v)$ is some linear combination of allowed terms. Therefore this integrand closely resembles the terms we manipulated in the previous section, except with factors of $D_A \log a(v_0,v)$. We will show these terms can still be absorbed into the positive definite terms in (\ref{HKREMS}). We will do this via a similar induction over powers of $l$.

Our inductive hypothesis is that we have manipulated $\dot{S}_{HKR}(v_0)$ up to $O(l^m)$ into the form
\begin{multline} \label{Hypothesis}
    \dot{S}_{HKR}(v_0) = -\frac{d}{dv} \left(4\pi \int_{C(v)} \text{d}^{d-2}x \sqrt{\mu(v)} \sigma_m^{v}(v) \right)+\\
    4\pi \int_{C} \text{d}^{d-2} x \sqrt{\mu(v_0)} \int_{v_0}^{\infty} \text{d}v  \Big[ \left(K^{A B} + Z_{m}^{A B}\right) \left( K_{A B} + Z_{m A B} \right)+ \frac{1}{2} c_1(\phi) \left(K_{A}+ Z_{m A}\right) \left(K^{A}+Z^{A}_m\right)\\
    + \frac{1}{2}\left( \partial_v \phi + Z_m \right)^2 + R_m  + O(l^N) \Big](v_0,v)
\end{multline}
where $Z^{A B}_{m}(v_0,v)$, $Z^{A}_{m}(v_0,v)$ and $Z_{m}(v_0,v)$ are $O(l)$ and at least linear in positive boost weight quantities, and $R_m(v_0,v)$ is of the form
\be
    R_m(v_0,v) = \sum_{n=m}^{N-1} l^n \sum_{k_1, k_2, p_1, p_2, P_1, P_2} (D^{k_1}{\partial_{v}^{p_1} P_1}) \, (D^{k_2}{\partial_{v}^{p_2} P_2}) \, Q_{k_1, k_2, p_1, p_2,P_1,P_2,m, n} (v_0,v)
\ee
and, in particular, $Z_{m}^{A B}(v_0,v)$, $Z_{m}^{A}(v_0,v)$, $Z_{m}(v_0,v)$ and $Q_{k_1, k_2, p_1, p_2,P_1,P_2, m, n}(v_0,v)$ is each a linear combination of terms, where each term is a product of factors of two possible types: (i) allowed terms evaluated at $v$ and (ii) $D^q \log a(v_0,v)$ with $q\ge 1$ ($D_A$ evaluated at time $v$). If a factor of type (ii) is present then the term is bilocal, otherwise it is local. All covariant derivatives $D$ are constructed from $\mu_{A B}(v)$, and all $P_1, P_2$ terms are evaluated at $v$. 

By (\ref{HKREMS}) and (\ref{DYbyparts}), the base case $m=0$ is satisfied with $\sigma^v_0 = 0$, $Z^{A B}_{0} = X^{A B}$, $Z^{A}_{0} = X^{A}$, $Z_{0} = X$ and $R_0= Y^A D_A \log a$. Assuming true for $m$, the obstruction to proceeding is the order $l^{m}$ terms in the sum in $R_m$, which are of the form $l^m (D^{k_1}{\partial_{v}^{p_1} P_1}) $ $ (D^{k_2}{\partial_{v}^{p_2} P_2}) $ $ Q(v_0,v)$. We aim to remove the $D^{k_1}\partial_{v}^{p_1}$ from each term and then complete the square.

We first reduce $k_1$ by 1 in each term via a spatial integration by parts:
\begin{equation} \label{spatialibp}
    \begin{split}
        &\int_{C} \text{d}^{d-2} x \sqrt{\mu(v_0)} \int_{v_0}^{\infty} \text{d}v (D^{k_1}{\partial_{v}^{p_1} P_1}) \, (D^{k_2}{\partial_{v}^{p_2} P_2}) \, Q\\
        = & - \int_{v_0}^{\infty} \text{d}v \int_{C}  \text{d}^{d-2} x \sqrt{\mu(v)} \, (D^{k_1-1}{\partial_{v}^{p_1} P_1}) \, D \left[ a^{-1} (D^{k_2}{\partial_{v}^{p_2} P_2}) Q \right]\\
        = & - \int_{C} \text{d}^{d-2} x \sqrt{\mu(v_0)} \int_{v_0}^{\infty} \text{d}v \, (D^{k_1-1}{\partial_{v}^{p_1} P_1}) \, \Big[ (D^{k_2+1}{\partial_{v}^{p_2} P_2}) \, Q \\
        & + (D^{k_2}{\partial_{v}^{p_2} P_2}) \, DQ - (D^{k_2}{\partial_{v}^{p_2} P_2}) \, Q\, D \log{a} \Big]
    \end{split}
\end{equation}
where in the last step we used $a D(a^{-1}) = -D \log a$. We repeat to bring $k_1$ to 0 in all terms, leaving us with terms of the form $l^m (\partial_{v}^{p_1} P_1) \, (D^{k}{\partial_{v}^{p_2} P_2}) \, Q$, with $Q$ still made exclusively from local allowed terms and factors of $D^q \log a(v_0,v)$.

We now aim to reduce $p_1$ to 0 by $v$-integration by parts. However, to avoid surface terms we must treat local and bilocal terms separately. 

\subsection{Bilocal Terms}

Bilocal terms have at least one factor of $D^q \log a(v_0,v)$. Their $v$-integration by parts follows simply:
\begin{equation} \label{vint}
\begin{split}
    \int_{v_0}^{\infty} \text{d}v (\partial_{v}^{p_1} P_1) \, (D^{k}{\partial_{v}^{p_2} P_2}) \, Q D^{q} \log a(v_0,v) =& \left[(\partial_{v}^{p_1-1} P_1) (D^{k}{\partial_{v}^{p_2} P_2}) Q D^{q} \log a(v_0,v) \right]^{\infty}_{v_0}\\
    &- \int_{v_0}^{\infty} \text{d}v (\partial_{v}^{p_1-1} P_1) \, \partial_v \left[(D^{k}{\partial_{v}^{p_2} P_2}) \, Q D^{q} \log a(v_0,v) \right]
\end{split}
\end{equation}
The boundary term vanishes at $v=\infty$ because we assume the black hole settles down to stationarity, and so positive boost weight quantities vanish. The boundary term also vanishes at $v=v_0$ because $a(v_0,v_0) \equiv 1$ and hence $D^q \log a = 0$. 
In the remaining $v$-integral, we can commute the $\partial_v$ past any $D$ derivatives using the formula 
\be
    [\partial_{v}, D_{A}]t_{B_1 ... B_n} = \sum_{i=1}^{n} \mu^{C D} ( D_{D}{ K_{A B_i} } - D_{A}{ K_{D B_i} } - D_{B_i}{ K_{A D} } ) t_{B_1... B_{i-1} C B_{i+1} ... B_n}
\ee
which will produce additional terms proportional to some $D^{k'} K$. Commuting $\partial_v$ past $D^q$ will leave $D^q \partial_v \log a$, which initially looks like a new type of bilocal term, however one can calculate that
\begin{equation} \label{dvloga}
    \partial_v \log a = \mu^{A B} K_{A B}
\end{equation}
and so this term is actually proportional to $D^q K$. Similarly in $\partial_v Q$, any $v$ derivative of $D^{q'} \log a$ can be dealt with by commuting and then using $(\ref{dvloga})$, and any non-allowed terms such as $\partial_v \beta$ or $\partial_{v r} \phi$ can be swapped out to $O(l^N)$ using the equations of motion, which will generate additional terms in $R_m$ of $O(l^{m+1})$. 

Therefore we are left with two types of terms at order $l^m$: i) terms that retain their factor of $D^q \log a$, which will be of the form $(\partial_{v}^{p_1-1} P_1) (D^{k}{\partial_{v}^{p_2} P_2}) Q D^{q} \log a(v_0,v)$ (with possibly changed $k$,$p_2$ and $Q$), and ii) terms that had $D^q \log a$ hit by $\partial_v$, which will be of the form $(D^{k'}{ K }) \,(\partial_{v}^{p_1-1} P_1) Q$ for some $k'$ and $Q$. This second type of term can potentially be local.

The $v$-integration by parts can be repeated on terms of type (i) until $p_1$ is reduced to 0, producing more terms of type (ii) along the way (which will have varying $p_1$'s). To terms of type (ii) we move the $D^{k'}$ derivatives off of $K$ via the same spatial integration by parts as in (\ref{spatialibp}). This brings them proportional to $K$, and hence, after relabelling this $K$ as $P_1$ and the old $P_1$ as $P_2$, they also effectively have $p_1$ reduced to 0. 

\subsection{Local Terms}

Local terms are of the form $(\partial_{v}^{p_1} P_1) \, (D^{k}{\partial_{v}^{p_2} P_2}) \, Q(v)$ with $Q(v)$ made exclusively from allowed terms evaluated at $v$. We can no longer simply do a $v$-integration by parts on this because there is no $D^q \log a$ to make the boundary term vanish at $v=v_0$. However, we can manipulate these terms in the same fashion as when we were trying to lift off the $\partial_{v}^{p_1}$ in the HKR procedure, namely by using (\ref{vintegrationbyparts}) repeatedly to write them as a sum of terms of the form i) $\partial_{v}\left\{\frac{1}{\sqrt{\mu}} \partial_{v}\left[\sqrt{\mu} \sigma^{v}(v) \right] \right\}$ with $\sigma^v$ local and quadratic or higher in positive boost weight quantities, ii) $P_1 \, (D^{\bar{k}}{\partial_{v}^{\bar{p}_2} P_2}) \, \tilde{Q}(v)$, and iii) $(\partial_{v}^{\bar{p}_1} P_1) \, (D^{\bar{k}} P_2 ) \, \tilde{Q}(v)$.

We have successfully reduced $k_1=p_1=0$ to 0 in terms of type (ii). For terms of type (iii), we can relabel $P_1 \leftrightarrow P_2$ and then remove the $D^{\bar{k}}$ derivatives from $P_1$ by using spatial integration by parts, as in (\ref{spatialibp}). This will introduce bilocal factors of $D^q \log a$, but this is fine: the resulting terms will all be of the desired form $P_1 \, (D^{k}{\partial_{v}^{\tilde{p}} P_2}) \, Q(v_0,v)$, i.e., they also have $k_1=p_1=0$.

Let us look at what happens to terms of type (i) when they are placed in the integral:
\be
\begin{split}
    \int_{C} \text{d}^{d-2} x \sqrt{\mu(v_0)} \int_{v_0}^{\infty} \text{d}v \,  \partial_{v}\left\{\frac{1}{\sqrt{\mu}}\partial_{v}\left[\sqrt{\mu} \sigma^{v}(v) \right] \right\} =& - \int_{C} \text{d}^{d-2}x \partial_{v}\left(\sqrt{\mu(v)} \sigma^{v}(v) \right)\Big|_{v=v_0}\\
    =& - \frac{\diff}{\diff v} \left(\int_{C} \text{d}^{d-2}x \sqrt{\mu(v)} \sigma^{v}(v) \right)\Big|_{v=v_0}
\end{split}
\ee
where in the first line the boundary term at $v=\infty$ vanishes because we assume the black hole settles down to equilibrium. These are the terms which will modify our definition of the entropy. 

\subsection{Completion of the Induction}

To summarise, we have rewritten all order $l^{m}$ terms in $R_m$ as
\begin{multline}
    - \frac{d}{dv} \left(l^{m} \int_{C} \text{d}^{d-2}x \sqrt{\mu(v)} \sigma^{v}(v) \right)\Big|_{v=v_0}
    + \\
    l^{m} \int_{C} \text{d}^{d-2}x \sqrt{\mu(v_0)} \int_{v_0}^{\infty} \text{d}v \sum_{k, p,P_1,P_2} \, P_1 \, (D^{k}{\partial_{v}^{p} P_2}) \, Q_{k, p,P_1,P_2}(v_0,v)
\end{multline}
where $Q_{k, p,P_1,P_2}(v_0,v)$ is a linear combination of allowed terms evaluated at $v$ and factors of $D^q \log a(v_0,v)$ with $q\geq 1$. 

Similarly to Section \ref{ManipTermsObO}, the final step in the induction is to split the sum over $P_1\in\{K_{A B}, K_A, \partial_v \phi\}$ and write the remaining sums as $2 l^{m} \tilde{Z}^{A B}$, $l^m c_1(\phi) \tilde{Z}^{A}$ and $l^m \tilde{Z}$. We then absorb them into the positive definite terms in (\ref{Hypothesis}) by completing the squares and setting $Z^{A B}_{m+1} = Z^{A B}_m + l^m \tilde{Z}^{A B}$, $Z^{A}_{m+1} = Z^{A}_m + l^m \tilde{Z}^{A}$ and $Z_{m+1} = Z_m + l^m \tilde{Z}$. The remainder terms will be $O(l^{m+1})$ (because $Z^{A B}_m$ etc. are $O(l)$), at least quadratic in positive boost weight (because $Z^{A B}_m$ etc. are at least linear), and linear combinations of local allowed terms and factors of $D^q \log a$ (because $Z^{A B}_m$ etc. are such linear combinations). Furthermore, $Z^{A B}_{m+1}$, $Z^{A}_{m+1}$ and $Z_{m+1}$ retain these properties. Finally we label $\sigma_{m+1}^v=\sigma_m^v+l^m\sigma^v$. Thus the induction proceeds.

We continue the induction until $m=N$, at which point $R_N$ is $O(l^N)$. Therefore, the entropy defined by
\be
    S(v) := \int_{C} \text{d}^{d-2}x \sqrt{\mu(v)} s^v(v)
\ee
with $s^v = s^{v}_{HKR} + \sigma^{v}_{N}$ satisfies a non-perturbative second law up to $O(l^N)$. 

\subsection{Difference from HKR}\label{DiffFromHKR}

Our entropy differs from the HKR entropy because of the terms $\sigma_{m+1}^v$, arising from a $v$-integration by parts of a \textit{local} term in $R_m$. For what $m$ is this first required? 

Initially, for $m=0$, all terms in $R_0=YD\log a$ are bilocal. Local terms are generated by integration by parts w.r.t. $v$ of a bilocal term. This is first required when a term in $R_m$ contains the $5$-derivative factor $(\partial_v P_1) \, (\partial_v P_2) D \log a$. However, every term in $R_m$ has overall boost weight $+2$ so there must an additional factor with boost weight $-2$, such as $\bar{K}\bar{K}$ or $\partial_r \phi \bar{K}_A$, which adds $2$ more derivatives to make 7 in total. $R_m$ is a sum of terms with at least $m+2$ derivatives, so  $m \ge 5$, e.g. consider a term in $R_{5}$ of the form $(\partial_v P_1) \, (\partial_v P_2) \, \bar{K} \, \bar{K} \, D \log a$ (other terms behave similarly). When we perform the $v$ integration by parts, we get a local term when $\partial_v$ hits $D \log a$, producing $P_1 \, (\partial_v P_2) \, \bar{K} \, \bar{K} \, D K$. This is still fine, because it is proportional to $P_1$ and so can be absorbed into the square, generating a $8$-derivative ``error term'' in $R_{6}$. This error term is proportional to $DK$, and so integration by parts w.r.t. $x^A$ renders it proportional to $K$. Therefore, it too can be absorbed by completing the square at $m=6$. It is only in $R_{7}$ (9 derivatives, or equivalently $O(l^7)$) that we might first need to perform a $v$ integration by parts on a local term. Thus $\sigma_{m+1}^v=0$ for $m \leq 6$, i.e., our entropy agrees with the HKR entropy for theories with up to 8 derivatives ($N \le 7$, or equivalently to order $l^6$ and below). Thus the HKR entropy satisfies a non-perturbative second law up to this order.

Note that in even spacetime dimension and in EFTs of just gravity and a scalar field (i.e., no electromagnetic field) then only even numbers of derivatives appear in the effective action. In this case, every term in $R_m$ must have an even total number of derivatives, which slightly changes the above analysis, e.g. the 7-derivative term studied above cannot appear. The final result is that $S(v)$ agrees with the HKR entropy for theories with up to 10 derivatives ($N \le 10$, to order $l^8$) in this case.

\section{Discussion of The Second Law for a Charged Scalar Field}

We can ask, can we generalize our proof of the second law to the EFT of gravity, electromagnetism and a \textit{charged} scalar field as defined in Section \ref{ChargedScalar}? Our starting point in the above was the IWW entropy density defined in Section \ref{BDKEntropy}. Such an entropy density has only been defined for a real uncharged scalar, and its generalization to a charged scalar does not exist in the literature. Proving such a generalization exists is beyond the scope of this paper as it would involve delving into covariant phase space formalism, and therefore this section is merely a discussion. However, it seems reasonable that such a generalization would exist, in which case the following discussion would complete the construction of an entropy that satisfies a non-perturbative second law as above.

In the analysis of the real scalar field EFT, we could use positive boost weight quantities as a proxy for order of perturbation around a stationary black hole because in \ref{PBWQH} and \ref{PBWQHF} we proved all such quantities vanish on the horizon for a stationary black hole (so long as we assume the zeroth law). However, in the charged scalar case we have the gauged fields $A_\mu$ and $\phi$. Here things are more subtle because whilst e.g. $\partial_{v}{\phi}$ may vanish in one electromagnetic gauge, it does not in another. As discussed at the end of \ref{PBWQH}, for a generic gauge we can only state that positive boost weight quantities made from $F_{\mu\nu}$ and the gauged derivatives of $\phi$,
\be \label{gaugephiderivs}
    (\partial_{\mu_1}-i\lambda A_{\mu_1})...(\partial_{\mu_n}-i\lambda A_{\mu_n})\phi,
\ee
vanish on the horizon of a stationary black hole.

This is fine, however, because we can apply this to a choice of gauge particularly suited to our affinely parameterized GNCs. By a suitable gauge transformation, we can always achieve \cite{Hollands:2022}
\be
    A= r \eta \text{d}v + A_A \text{d}x^A
\ee
for some function $\eta(r,v,x^A)$ regular on the horizon. $\eta$ and $A_A$ have boost weight 0. In this gauge on $\cH$,
\be
    \partial_{r}^p \eta = \partial_r^p F_{r v}, \quad \partial_r^q A_A = \partial_{r}^{q-1}F_{r A}, \quad \partial_v A_A = F_{v A}
\ee
for $p\geq 0, q\geq 1$, and hence all positive boost weight derivatives of $\eta$ and $A_A$ can be written as positive boost weight derivatives of $F_{\mu \nu}$ on the horizon. Similarly
\be
    \partial^p_v \partial^q_r \phi\big|_{\cH_+} = (\partial_{v}-i\lambda A_{v})^p (\partial_{r}-i\lambda A_{r})^q\phi\big|_{\cH_+}
\ee
and hence all positive boost weight derivatives of $\phi$ can be written as positive boost weight components of (\ref{gaugephiderivs}) (or their $\partial_A$ derivatives) on the horizon. Therefore, in this gauge all positive boost weight quantities still vanish on the horizon in equilibrium and hence can still be used as a proxy for perturbations around a stationary black hole. 

In this gauge, the leading order 2-derivative part of $E_{v v}|_{\cH_+}$ can be written as
\begin{equation}
    -E^{(0)}_{v v}\big|_{\cH_+}= \partial_{v}\left[\frac{1}{\sqrt{\mu}} \partial_{v}\left(\sqrt{\mu}\right) \right] + K_{A B} K^{A B} + \frac{1}{2} c_1(|\phi|^2) h^{A B} \partial_v A_A \partial_v A_B + |\partial_{v} \phi|^2 
\end{equation}

For the higher derivative terms, let us now assume that we can generalize the IWW entropy to the charged scalar case. I.e., we assume we can write 
\begin{equation} \label{BDKCharged}
    -E_{v v}\Big|_{\cH_+} = \partial_{v}\left[\frac{1}{\sqrt{\mu}} \partial_{v}\left(\sqrt{\mu} s^{v}_{IWW}\right) + D_{A}{ s^A }\right] + ...
\end{equation}
for some real entropy current $(s^v_{IWW}, s^A)$ and where the ellipsis denotes terms that are quadratic or higher in positive boost weight quantities. 

We can now generalize the HKR procedure as follows. We first reduce, up to $O(l^N)$, to a set of "allowed terms" given by
\begin{empheq}[box=\fbox]{align}\label{AllowedTermsCharged}
    \text{Allowed terms:} \,\, &\mu_{A B}, \,\, \mu^{A B}, \,\, \epsilon_{A_1 ... A_{d-2}}, \,\, D^k R_{A B C D}[\mu], \,\, D^k \beta_{A}, \,\, 
    D^k\partial_{v}^p K_{A B}, \,\, D^k\partial_{r}^q \bar{K}_{A B},\nonumber\\
    & D^k \eta, \,\, D^k\partial_{v}^p A_{A}, \,\, D^k\partial_{r}^q A_{A}, \,\,
    D^k\partial_{v}^p \phi, \,\, D^k\partial_{r}^q \phi, \,\, D^k\partial_{v}^p \phi^*, \,\, D^k\partial_{r}^q \phi^* 
\end{empheq}
The reduction of the metric terms follows straightforwardly in the same way as vacuum gravity by using the $E^{(0)}_{\mu\nu}=O(l)$ equations of motion. We can eliminate mixed $v$ and $r$ derivatives of $\phi$ by using $E^{(0)}=O(l)$ and evaluating $E^{(0)}$ in affinely parameterized GNCs in this gauge:
\begin{multline}
    E^{(0)}=2\partial_{r}\partial_v{\phi}+ D^{A}{D_{A}{\phi}} +K^{A}_{A} \partial_{r}{\phi}+\beta^A D_{A}{\phi} +\bar{K}^{A}_{A} \partial_{v}{\phi}\\
    -2i \lambda A^{A} D_{A}{\phi} - i \lambda \phi  D^{A}{A_{A}} - i \lambda \eta \phi - i \lambda \beta^{A} A_{A}  \phi - \lambda^2 A_{A} A^{A} \phi+...
\end{multline}
where the ellipsis denotes terms that vanish on the horizon.
The reduction of the Maxwell terms is achieved by using the equation of motion
\begin{multline}
    E^{(0)}_\mu = c_1(|\phi|^2) \nabla^{\nu} F_{\mu \nu} + F_{\mu \nu} \nabla^{\nu}[c_1(|\phi|^2)] - 4 \epsilon_{\mu\nu\alpha\beta}F^{\alpha\beta} \nabla^{\nu}[c_2(|\phi|^2)] +\\
    i\lambda \left[ \phi^* \fD_\mu \phi-\phi(\fD_\mu \phi)^* \right] = O(l) 
\end{multline}
    
and by substituting our choice of gauge $\psi=\eta+r\partial_r \eta$, $K_A=\partial_v A_A-r D_A \eta$, $\bar{K}_A = \partial_r A_A$ into our affinely parameterized GNC expressions for $\nabla^{\nu} F_{\mu\nu}$ given in Appendix \ref{dFAppendix}. These allow us to eliminate $v$ and $r$ derivatives of $\eta$, and mixed $v$ and $r$ derivatives of $A_A$.

In particular, the only positive boost weight allowed terms are $D^k\partial_{v}^p K_{A B}$ with $p\geq 0$, and $D^k\partial_{v}^p A_{A}$, $D^k\partial_{v}^p \phi$ and $D^k\partial_{v}^p \phi^*$ with $p\geq 1$. Therefore we can rewrite all the terms in the ellipsis in (\ref{BDKCharged}), which we label $\Delta$, up to $O(l^N)$ as a sum of terms of the form 
\be
    (D^{k_1}{\partial_{v}^{p_1} P_1}) \, (D^{k_2}{\partial_{v}^{p_2} P_2}) \, Q
\ee
with $P_1,P_2\in \{K_{A B}, \partial_{v} A_{A}, \partial_{v} \phi, \partial_{v} \phi^*  \}$, and where $Q$ is made from allowed terms. 

Now, since $(s^v_{IWW}, s^A)$ is real, the overall sum of these terms, $\Delta$, is real. Hence we can pair each of these terms up with its complex conjugate (or itself if it is real) and write $\Delta$ as
\begin{multline}
    \Delta = \sum_{k_1, k_2, p_1, p_2, P_1, P_2}\Big[ (D^{k_1}{\partial_{v}^{p_1} P_1}) \, (D^{k_2}{\partial_{v}^{p_2} P_2}) \, Q_{k_1, k_2, p_1, p_2, P_1, P_2} +\\
    (D^{k_1}{\partial_{v}^{p_1} P_1^*}) \, (D^{k_2}{\partial_{v}^{p_2} P_2^*}) \, Q^*_{k_1, k_2, p_1, p_2, P_1, P_2} \Big]
\end{multline}
with $P_1,P_2\in \{K_{A B}, \partial_{v} A_{A}, \partial_{v} \phi\}$.

We now generalise our inductive hypothesis (\ref{Qinduction}) to 
\begin{multline}\label{Qinductioncharged}
    \Delta= \partial_{v}\left[\frac{1}{\sqrt{\mu}} \partial_{v}\left(\sqrt{\mu} \sum_{n=0}^{m-1} l^n \varsigma^{(n) v} \right)\right] + \left(K_{A B}+\sum_{n=0}^{m-1} l^n X^{(n)}_{A B}\right) \left(K^{A B}+\sum_{n=0}^{m-1} l^n X^{(n) A B}\right) +\\
     \frac{1}{2} c_1(|\phi|^2) \left(\partial_v A_{A}+\sum_{n=0}^{m-1} l^n X^{(n)}_{A}\right) \left(\partial_v A^{A}+\sum_{n=0}^{m-1} l^n X^{(n) A}\right) +\\
     \left( \partial_v \phi + \sum_{n=0}^{m-1} l^n X^{(n)} \right)\left( \partial_v \phi + \sum_{n=0}^{m-1} l^n X^{(n)} \right)^* +
     D_{A}{\sum_{n=0}^{m-1} l^n Y^{(n) A}} + \sum_{n=m}^{N-1} l^n \Delta^{(n)} + O(l^{N})
\end{multline}
where the $\Delta^{(n)}$, $X^{(n)}_{A B}$ etc., are real. To proceed the induction we manipulate the terms in $\Delta^{(m)}$ exactly as in Section \ref{ManipTermsObO}, except we always keep complex conjugates paired up and perform identical operations on them. This will ensure that when we get to the equivalent of (\ref{Qinductionfinal}) we can split the sum over $P_1\in \{K_{A B}, \partial_{v} A_{A}, \partial_{v} \phi\}$ and get
\begin{multline}\label{Qinductionchargedfinal}
    \Delta= \partial_{v}\left[\frac{1}{\sqrt{\mu}} \partial_{v}\left(\sqrt{\mu} \sum_{n=0}^{m-1} l^n \varsigma^{(n) v} \right)\right] +\\
    \left(K_{A B}+\sum_{n=0}^{m-1} l^n X^{(n)}_{A B}\right) \left(K^{A B}+\sum_{n=0}^{m-1} l^n X^{(n) A B}\right) + 2l^m K_{A B} X^{(m) A B}+\\
     \frac{1}{2} c_1(|\phi|^2) \left(\partial_v A_{A}+\sum_{n=0}^{m-1} l^n X^{(n)}_{A}\right) \left(\partial_v A^{A}+\sum_{n=0}^{m-1} l^n X^{(n) A}\right) +l^m c_1(|\phi|^2) \partial_v A_A X^{(m) A}+\\
     \left( \partial_v \phi + \sum_{n=0}^{m-1} l^n X^{(n)} \right)\left( \partial_v \phi + \sum_{n=0}^{m-1} l^n X^{(n)} \right)^* +l^m \partial_v\phi X^{(m)*}+l^m\partial_v \phi^*X^{(m)}+\\
     D_{A}{\sum_{n=0}^{m} l^n Y^{(n) A}} + \sum_{n=m+1}^{N-1} l^n \Delta^{(n)} + O(l^{N})
\end{multline}
and thus we can still absorb the order $l^m$ terms into the positive definite terms by completing the squares. The remainder terms are real, and thus the induction can proceed. This would complete the generalization of the HKR entropy. Generalizing the further modifications of \ref{FurtherModifications} would follow similarly.

Thus we can get a non-perturbative second law for a charged scalar field if we assume an IWW entropy density exists in such a scenario. The procedure outlined here does not produce an entropy that is manifestly electromagnetic gauge-independent like in the real scalar field case. However, it seems reasonable this could be achieved if the hypothesized IWW entropy density was gauge invariant. One could take a more careful approach to the gauge field, for example by keeping derivatives of $\phi$ in terms of gauged derivatives $(D_A -i\lambda A_A)$, $(\partial_v-i\lambda A_v)$ etc in the hope of achieving this.

\section{Appendix}

\subsection{Full expressions for \texorpdfstring{$\nabla^{\beta} F_{\alpha \beta}$}{derivatives of F} in affinely parameterized GNCs} \label{dFAppendix}

The components of $\nabla^{\beta} F_{\alpha \beta}$ in affinely parameterized GNCs are as follows:

\be
\begin{split}
    \nabla^{\beta}F_{v \beta} = &\partial_{v}{\psi} + D^{A}{K_{A}}+K \psi+ r \big(-D^{A}{\psi} \beta_{A}+\beta^{A} \partial_{r}{K_{A}}-D^{A}{\beta^{B}} F_{A B}-\beta^{A} \beta_{A} \psi-\bar{K}^{A} \partial_{v}{\beta_{A}}+\\
    &K^{A} \bar{K} \beta_{A}-D^{A}{\beta_{A}} \psi-2K^{A} \bar{K}_{A}\,^{B} \beta_{B}\big) +{r}^{2} \big(-\alpha \partial_{r}{\psi}-\beta^{A} \beta_{A} \partial_{r}{\psi}-F^{A B} \beta_{A} \partial_{r}{\beta_{B}}+\\
    &D^{A}{\beta^{B}} \bar{K}_{A} \beta_{B}+D^{A}{\alpha} \bar{K}_{A}-D^{A}{\beta^{B}} \bar{K}_{B} \beta_{A}+\bar{K}^{A} \alpha \beta_{A}-\beta^{A} \partial_{r}{\beta_{A}} \psi-\bar{K} \alpha \psi-\bar{K} \beta^{A} \beta_{A} \psi+\\
    &2\bar{K}^{A B} \beta_{A} \beta_{B} \psi\big) +{r}^{3} \left(-\bar{K}^{A} \alpha \partial_{r}{\beta_{A}}-\bar{K}^{A} \beta^{B} \beta_{B} \partial_{r}{\beta_{A}}+\bar{K}^{A} \beta_{A} \beta^{B} \partial_{r}{\beta_{B}}+\bar{K}^{A} \beta_{A} \partial_{r}{\alpha}\right)
\end{split}
\ee

\be 
    \nabla^{\beta}F_{r \beta} = -\partial_{r}{\psi}+D^{A}{\bar{K}_{A}}+\bar{K}^{A} \beta_{A}-\bar{K} \psi+ r \left(\beta^{A} \partial_{r}{\bar{K}_{A}}-2\bar{K}^{A} \bar{K}_{A}\,^{B} \beta_{B}+\bar{K}^{A} \partial_{r}{\beta_{A}}+\bar{K} \bar{K}^{A} \beta_{A}\right)
\ee

\be \label{FullFA}
\begin{split}
    \nabla^{\beta}F_{A \beta} = &-\partial_{r}{K_{A}}-\partial_{v}{\bar{K}_{A}} + D^{B}{F_{A B}}+2K_{A B} \bar{K}^{B}+2K^{B} \bar{K}_{A B}-\beta_{A} \psi-K \bar{K}_{A}-K_{A} \bar{K}+\\
    &F_{A}\,^{B} \beta_{B}+ r \big(-F^{B C} \bar{K}_{A B} \beta_{C}+D^{B}{\beta_{A}} \bar{K}_{B}+\bar{K}_{A}\,^{B} \beta_{B} \psi-\partial_{r}{\beta_{A}} \psi+\frac{1}{2}\bar{K}^{B} \beta_{A} \beta_{B}+\\
    &F_{A}\,^{B} \bar{K} \beta_{B}- D^{B}{\beta_{B}} \bar{K}_{A}+F_{A}\,^{B} \partial_{r}{\beta_{B}}-\bar{K}_{A} \beta^{B} \beta_{B}-2\bar{K}_{A} \alpha\big) +{r}^{2} \big(\bar{K}^{B} \bar{K}_{A B} \alpha+\\
    &\bar{K}^{B} \bar{K}_{A B} \beta^{C} \beta_{C}+\frac{1}{2}\bar{K}^{B} \beta_{B} \partial_{r}{\beta_{A}}-\bar{K} \bar{K}_{A} \alpha-\bar{K} \bar{K}_{A} \beta^{B} \beta_{B}-\bar{K}_{A} \beta^{B} \partial_{r}{\beta_{B}}-\bar{K}_{A} \partial_{r}{\alpha}\big)
\end{split}
\ee

These were calculated using the symbolic computer algebra system Cadabra \cite{Cadabra1}\cite{Cadabra2} \cite{Cadabra3}. To get equation (\ref{EA0}), we use (\ref{dvKA}) to eliminate $\partial_{v}{\bar{K}_{A}}$ in (\ref{FullFA}).

\subsection{Calculation of \texorpdfstring{$\varsigma^{(m)v}$}{sigma n} Terms} \label{app:aj}

The HKR algorithm involves finding numbers $a_j$ such that

\begin{equation} \label{dKDdK}
    (\partial_{v}^{p_1} P_1) \, (D^{k}{\partial_{v}^{p_2}P_2}) Q = \partial_{v}\left[\frac{1}{\sqrt{\mu}} \partial_{v}\left(\sqrt{\mu} \sum_{j=1}^{p_1+p_2-1} a_j (\partial_{v}^{p_1+p_2-1-j} P_1) \, (D^{k}{\partial_{v}^{j-1}P_2 }) Q \right)\right] + ...
\end{equation}
where the ellipsis denotes terms of the form $(\partial_{v}^{\Bar{p}_1} P_1) \, (D^{\bar{k}}{\partial_{v}^{\Bar{p}_2}P_2}) \tilde{Q}$ with $\Bar{p}_1+\Bar{p}_2 < p_1+p_2$ or $\Bar{p}_1=0$ or $\Bar{p}_2=0$. 

How to calculate the $a_j$ for any $k, p_1, p_2$ is as follows. When the derivative term on the RHS is expanded out we get a set of $p_1 + p_2 -1$ linear equations on the $a_j$ in order to satisfy the required conditions. The linear equations can be written in matrix form as 
\begin{equation}
    M_{p_1 + p_2 -1} \textbf{a} = \textbf{v}_{p_2}
\end{equation}
where
\begin{equation}
    M_{p_1 + p_2 -1} = \begin{pmatrix}
        2 & 1 & 0 & 0 & ... & 0 \\
        1 & 2 & 1 & 0 & ... & 0 \\
        0 & 1 & 2 & 1 & ... & 0 \\
        ... & ... & ... & ... & ... & ... \\
        0 & ... & ... & 0 & 1 & 2 
    \end{pmatrix}, \quad
    \textbf{a}= \begin{pmatrix}
        a_1\\
        a_2\\
        ...\\
        a_{p_1 + p_2 -1}
    \end{pmatrix}, \quad
    \textbf{v}_{p_2} = \begin{pmatrix}
        0\\
        ...\\
        0\\
        1\\
        0\\
        ...\\
        0
    \end{pmatrix}
\end{equation}
where $(\textbf{v}_{p_2})_{p_2} = 1$ and $(\textbf{v}_{p_2})_{j} = 0$ for $j\neq p_2$. $M_{p_1+p_2-1}$ can be shown to have non-vanishing determinant, and so the system of equations has a unique solution. 
\chapter{Dynamical Black Hole Entropy in Effective Field Theory}\label{ChapterBHEntropy}

The contents of this chapter are the results of original research conducted by the author of this thesis in collaboration with Harvey Reall. It is based on work published in \cite{Davies:2023}.

\section{Introduction}

As we saw in \ref{WaldEntropy}, the work of Wald in the early 1990s provided a definition of entropy, the Wald entropy, for any stationary black hole solution in any theory of gravity with diffeomorphism-invariant Lagrangian. This definition depends only on the geometry of a cross-section of the event horizon and satisfies the first law of black hole mechanics. Thus it provides a completely satisfactory definition of the entropy of a stationary black hole. When making a proposal for dynamical black hole entropy therefore, it is natural to ask whether it has the following three properties: (i) it depends only on the geometry of a cross-section of the event horizon, (ii) it agrees with the Wald entropy in equilibrium, and (iii) it satisfies the second law of black hole mechanics.

In the previous Chapter, we made a proposal for the definition of dynamical black hole entropy in EFTs of gravity coupled to electromagnetism and a scalar field:
\be \label{EntropyDef}
    S(v) = 4\pi \int_{C(v)}\text{d}^{d-2}x \sqrt{\mu} s^{v}
\ee
We demonstrated that it satisfies property (iii) non-perturbatively up to $O(l^N)$ terms. It also satisfies property (ii) by construction. This is because the entropy density $s^v$ was constructed in three parts:
\begin{equation}
    s^v = s_{IWW}^v + \varsigma_{HKR}^v + \sigma^v
\end{equation}
where $s_{IWW}^v$ is the Iyer-Wald-Wall (IWW) entropy density of the theory, $\varsigma_{HKR}^v$ are the corrections from the Hollands-Kov\'acs-Reall (HKR) procedure, and $\sigma^v$ are from the further modifications needed to make it satisfy the second law non-perturbatively. $\varsigma_{HKR}^v$ and $\sigma^v$ are at least quadratic in positive boost weight quantities by construction. Thus they vanish for a stationary black hole, as we saw in the proof of the zeroth law. Furthermore, $s_{IWW}^v$ can be split into \cite{Bhattacharyya:2021jhr}\cite{Biswas:2022}
\begin{equation}
    s_{IWW}^v = s^v_{IW} + \sigma^v_{IWW}
\end{equation}
where $\sigma^v_{IWW}$ is at least linear in positive boost weight quantities (and hence also vanishes in equilibrium) and $s^v_{IW}$ is the Iyer-Wald entropy density. Thus, for a stationary black hole
\be
    S(v) = 4\pi \int_{C(v)}\text{d}^{d-2}x \sqrt{\mu} s^{v}_{IW}
\ee
which is precisely defined to be the Wald entropy (see \ref{IyerWaldEntropy}).

The aim of this Chapter is to investigate whether $S(v)$ satisfies property (i). As with with the method of Wall used to construct $s_{IWW}^v$, both the HKR procedure and further modifications performed in the previous Chapter make use of affinely parameterized GNCs defined near the black hole horizon. One of these coordinates is an affine parameter along the horizon generators. There is the freedom to rescale this affine parameter by a different amount along each generator. None of the Wall, HKR or further modification approaches maintain covariance w.r.t. such a rescaling. Nevertheless, HKR proved in \cite{Hollands:2022} that $s_{IWW}^v$ can be made suitably gauge invariant under the rescaling\footnote{For completeness, \cite{Hollands:2022} only proved this in EFTs of gravity and a real scalar field. We shall assume and not prove this remains true if we include the electromagnetic field in our EFT.}. However, they did not demonstrate that the new terms generated by their approach must also be gauge invariant. By classifying the possible form of terms that can appear in $\varsigma_{HKR}^v$ we shall show that non-gauge-invariant terms cannot arise for theories with $6$ or fewer derivatives, and hence $S(v)$ is gauge invariant for EFTs known up to $6$-derivative terms (the further modification terms $\sigma^v$ only start appearing at very high derivative order, as we saw in \ref{DiffFromHKR}, so do not affect this). However, our classification shows that non-gauge-invariant terms might appear for theories with $7$ or more derivatives. By considering a particular $8$-derivative theory we confirm that such terms do appear. Hence our proposal for dynamical black hole entropy is not gauge invariant for this theory.

Undeterred, a further aim of this Chapter is to study examples for which the HKR terms $\varsigma_{HKR}^v$ are non-trivial. The leading EFT corrections to Einstein-Maxwell-scalar theory have $4$ derivatives (in $d\geq 4)$. However, HKR terms are only generated for theories with $6$ or more derivatives \cite{Hollands:2022}, i.e., $S(v)$ coincides with the IWW entropy for a $5$-derivative theory. To find examples for which $S(v)$ differs from the IWW entropy one needs to consider theories with at least $6$ derivatives. For simplicity, we shall consider the EFT of vacuum gravity in 4d, in which case only even numbers of derivatives can appear. In vacuum, one can use a field redefinition to eliminate (non-topological) $4$-derivative terms from the action and so the leading EFT corrections to the Einstein-Hilbert Lagrangian have $6$ derivatives. We shall determine the HKR terms in the entropy for this EFT. 

Our final objective is to consider a particularly interesting theory, namely Einstein-Gauss-Bonnet (EGB) theory. In $d>4$ dimensional vacuum gravity, field redefinitions can be used to bring terms with up to $4$-derivatives to the EGB form. Thus EGB can be regarded as an EFT. However, 
since this theory has second order equations of motion, it is often regarded as a self-contained classical theory (e.g. it admits a well-posed initial value problem if the curvature is small enough \cite{Kovacs:2020ywu}). It is interesting to ask whether the HKR approach can be used to determine the entropy for this theory. To do this, we can regard it as a very special EFT for which the coefficients of all terms with more than $4$ derivatives are exactly zero. We can then apply the HKR method to this EFT to determine the entropy to any desired order in the Gauss-Bonnet coupling constant. We shall use this method to calculate the entropy to quadratic order in this coupling constant. 

This Chapter is organized as follows. Section \ref{Classify} presents a classification of terms that can arise in the HKR entropy, explaining why this is necessarily gauge invariant for theories with up to $6$ derivatives. In section \ref{Implementation} we explain how we calculate the $S(v)$ in practice using computer algebra. Section \ref{Examples} presents the results of our calculations for EGB theory, $6$-derivative theories, and a particular $8$-derivative theory. In section \ref{GaugeInvariance} we demonstrate gauge non-invariance of the HKR entropy for the $8$-derivative theory. Section \ref{sec:discuss} contains a brief discussion. 

We use units such that $16 \pi G=1$.

\section{Classification of Terms}\label{Classify}

\subsection{Gauge Transformations}\label{GaugeInvarianceLaws}

There are two types of gauge in the EFT of gravity coupled to electromagnetism and a scalar field: the choice of electromagnetic gauge, and our choice of coordinates.

By construction, at least in the real uncharged scalar field case, $s^v$ only depends on Maxwell quantities through $F_{\mu \nu}$, which is invariant under a change of electromagnetic gauge. Therefore the entropy $S(v)$ is independent of electromagnetic gauge.

As for coordinate independence, our procedure was performed in affinely parameterized GNCs with $r=v=0$ on a given spacelike horizon cross section\footnote{The GNCs can be defined starting from any horizon cross-section, so the restriction $r=v=0$ is not restricting the choice of cross-section considered} $C$ which determines a foliation $C(v)$ of $\cH_+$. However, as discussed in Section \ref{APGNCs}, such affinely parameterized GNCs are not unique.

If we fix $C$ then there are two freedoms in our choice of GNCs: (a) picking a different set of co-ordinates $x'^A$ on $C$, and (b) rescaling the affine parameter on each generator $v'=v/a(x^A)$ with $a(x^A)>0$. The procedure for calculating $s^v$ is manifestly covariant in $A, B, C,...$ indices and so $S(v)$ is invariant under (a). If instead we make the rescaling (b) with non-constant $a(x^A)$ then as well as changing to a new set of affinely parameterized GNCs $(v',r',x'^A$), our foliation will change $C'(v')\neq C(v)$. The exception is the surface $C$, i.e., $v=v'=0$, which belongs to both foliations. 

This freedom in rescaling the affine parameter is essential if we wish to compare the entropy of two arbitrary horizon cuts $C$, $C'$ with $C'$ strictly to the future of $C$ \cite{Hollands:2022}. Introducing GNCs based on $C$ we can use the rescaling of affine parameter to ensure that $C'$ is a surface of constant $v$, and then apply the second law as proved in the previous Chapter. However, 
we do not want our definition of the entropy of $C$ to depend on the choice of $C'$ and so 
we want the entropy of $C$ to be gauge invariant under the rescaling of affine parameter. This is what we mean by the desired property (i) for a definition of dynamical black hole entropy: that the entropy of $C$ depend only on the geometry of $C$. Investigating whether or not this is true for $S(v)$ is one of the aims of this Chapter.

In general, affinely parameterized GNC quantities have complicated transformation laws under this rescaling. However, we will only be concerned with how they transform on the horizon cut $C$, i.e., on $r=v=0$. On $C$, some quantities satisfy simple transformation laws \cite{Hollands:2022}:
\begin{align}\label{GNCTransform}
    x'^{A} = x^A, \quad v'=&\, 0 \quad r' = 0,\nonumber\\
    \mu'_{A B} = \mu_{A B}, \quad \epsilon'_{A_1 ... A_{d-2}}= \epsilon_{A_1 ... A_{d-2}}, \quad &R_{A B C D}[\mu'] = R_{A B C D}[\mu], \quad D'_A = D_A,\nonumber\\
    \partial_{v'}^p K'_{A B} = a^{p+1} \partial_{v}^p K_{A B}, \,\,\,\,& \partial_{r'}^p \Bar{K}'_{A B} = a^{-p-1} \partial_{r}^p \Bar{K}_{A B},\\
    \partial_{v'}^p K'_{A} = a^{p+1} \partial_{v}^p K_{A}, \,\,\,\,& \partial_{r'}^p \Bar{K}'_{A} = a^{-p-1} \partial_{r}^p \Bar{K}_{A},\nonumber\\
    \partial_{v'}^p \phi' = a^{p} \partial_{v}^p \phi, \,\,\,\,& \partial_{r'}^p \phi' = a^{-p} \partial_{r}^p \phi,\nonumber\\
    \beta'_{A} = \beta_A +& 2 D_{A} \log a\nonumber\\
    \text{on} \,\, v=&r=0\nonumber
\end{align}
Note in particular that $\beta_A$ transforms inhomogeneously, in the same way as a gauge field (this is because $\beta_A$ is a connection on the normal bundle of $C$ \cite{Hollands:2022}). Additionally, a tensor component $T^{\mu_1 ... \mu_n}_{\beta_1 ... \beta_m}$ transforms homogeneously $T'^{\mu_1 ... \mu_n}_{\nu_1 ... \nu_m} = a^b T^{\mu_1 ... \mu_n}_{\nu_1 ... \nu_m}$ on $C$, where $b$ is the component's boost weight.

By construction, $s^v$ can be split into two parts: the Iyer-Wald-Wall part $s^v_{IWW}$ (which is only defined uniquely up to linear order in positive boost weight quantities), and the modification terms $\varsigma_{HKR}^{v}+\sigma^v$ (which are quadratic or higher order in positive boost weight terms). It is proved in \cite{Hollands:2022} that for Einstein-Scalar EFTs, $s^v_{IWW}$ is gauge invariant on $C$ to linear order, and can be made gauge invariant non-perturbatively by adjusting the non-unique higher order terms. We expect the proof can be extended to the Einstein-Maxwell-Scalar EFT case. However, to delve into the covariant phase space formalism of the proof would divert somewhat from the material here, so we assume it without proof.

\subsection{Terms Produced by the HKR Procedure}

We now discuss what possible terms can appear in the terms produced by the HKR procedure, $\varsigma_{HKR}^{v}$, and their gauge invariance, at each derivative order for the EFT of gravity coupled to electromagnetism and a real scalar field.

We will be interested in the number of derivatives associated with a quantity on $\cH$. We can write GNC quantities on $\cH$ as derivatives of the metric: $\alpha = -\frac{1}{2} \partial^2_{r}{ g_{v v}}\big|_{\cH}$, $\beta_{A} = -\partial_{r}{g_{v A}}\big|_{\cH}$ and $\mu_{A B} = g_{A B}$. Hence $\alpha, \beta_{A}, \mu_{A B}$ are associated with 2, 1 and 0 derivatives respectively. Similarly, the Maxwell terms $K_{A} \equiv F_{v A}$, $\bar{K}_{A} \equiv F_{r A}$, $\psi \equiv F_{v r}$ and $F_{AB}$ are 1-derivative quantities because they correspond to first derivatives of the Maxwell potential. This motivates a definition \cite{Hollands:2022} of ``dimension'' with which we can count derivatives, corresponding to the standard notion of mass dimension in quantum field theory:\\

\textbf{Definition}: \textit{The “dimension” of $\alpha$, $\beta_A$ and $\mu_{A B}$ are 2, 1, 0 respectively. The "dimension" of $K_{A}$, $\bar{K}_{A}$, $\psi$ and $F_{AB}$ are all 1. The "dimension" of $\phi$ is 0. Taking a derivative w.r.t. $v$, $r$ or $x_A$ increases the dimension by 1. Dimension is additive under products.}\\

Note $K_{A B}$ and $\Bar{K}_{A B}$ both have dimension 1. $\epsilon_{A_1 ... A_{d-2}}$ has dimension 0. We also define the dimension of $l$ to be $-1$, and the dimension of $\Lambda$ and the scalar potential $V(\phi)$ to be $+2$. $\diff/\diff \phi$ derivatives do not change the dimension, and we assume any other general function of $\phi$ in the effective action has dimension $0$. These definitions mean that the terms in the equations of motion on $\cH$ have consistent dimension, with the overall dimension of $E_{\mu \nu}$, $E_{\mu}$ and $E$ being 2. These also mean the elimination rules for non-allowed terms are dimensionally consistent. A quantity with boost weight $b$ must involve at least $|b|$ derivatives, and so has dimension at least $|b|$.

$\varsigma^v_{HKR}$ is constructed from $E_{vv}$ in a form that appears with two extra derivatives
\begin{equation}
    -E_{vv}=\partial_{v}\left[\frac{1}{\sqrt{\mu}} \partial_{v}\left(\sqrt{\mu} \varsigma_{HKR}^{v}\right)\right] + ...
\end{equation}
and hence $\varsigma^v_{HKR}$ has dimension 0. It is also boost weight 0 since $E_{vv}$ is boost weight $+2$.

By construction, we can write $\varsigma^v_{HKR}$ as a series
\begin{equation}
    \varsigma^v_{HKR} = \sum_{n=0}^{N-1} l^n \varsigma^v_{n}
\end{equation}
and hence each $\varsigma^v_{n}$ has dimension $n$. Also, $\varsigma_n^v$ is a sum of terms of the form $(\partial_{v}^{p} P_1) \, (D^{k}{\partial_{v}^{p'}P_2 }) Q_{n, k, p, p'}$ with $P_1,P_2 \in \{ K_{A B}, K_A, \partial_v \phi \}$ (all dimension 1, boost weight $+1$ quantities), $k,p,p'\geq 0$ and where $Q_{n, k, p, p'}$ is made exclusively out of allowed terms. Suppose $Q_{n, k, p, p'}$ has dimension $d_{n, k, p, p'}$ and boost weight $b_{n, k, p, p'}$. To match the dimension and boost weight of $\varsigma_n^v$, we have two conditions:
\begin{equation}
    \begin{split}
        d_{n, k, p, p'} &= n-2-p-p'-k \\
        b_{n, k, p, p'} &= -2-p-p'
    \end{split}
\end{equation}
But $d_{n, k, p, p'}\geq|b_{n, k, p, p'}|$ which we can rearrange to give
\begin{equation}
    2p+2p'+k\leq n-4 
\end{equation}

If $n<4$ then the RHS is negative, which is impossible for $k,p,p'\geq0$. Thus $\varsigma_{n}^{v}=0$, and so there are no HKR terms at order $l^3$ or below, i.e., for theories with up to 5 derivatives, as mentioned above.

If $n=4$ then the RHS is 0, and so we must have $k=p=p'=0$. Hence $\varsigma_{4}^v$ must be a sum of terms of the form $P_1 P_2 Q_{4,0,0,0}$ with $P_1,P_2 \in \{ K_{A B}, K_A, \partial_v \phi \}$. We have $d_{4, 0, 0, 0}=-b_{4, 0, 0, 0}=2$ so $Q_{4, 0, 0, 0}$ must contain two $r$ derivatives and no other derivatives. The only combinations of allowed terms that have this are $N_1 N_2 T$ and $\partial_{r}N_1 T$, with $N_1,N_2 \in \{ \bar{K}_{A B}, \bar{K}_A, \partial_r \phi \}$ and $T$ any combination of the 0-dimension quantities $\mu_{A B}$, $\epsilon_{A_1 ... A_{d-2}}$ and $\phi$. In other words, $\varsigma_{4}^v$ is a sum of terms of the form $P_1 P_2 N_1 N_2$ or $P_1 P_2 \partial_r N_1$ with their indices completely contracted in some way with $\mu^{A B}$ or $\epsilon^{A_1 ... A_{d-2}}$ and possibly multiplied by a function of $\phi$. The transformation laws (\ref{GNCTransform}) imply that such terms are gauge invariant on $C$. Precisely the same analysis can be applied to the further modification terms $\sigma^v$ (although recall that these do not start appearing until order $l^7$ anyway), and so this implies $S(v)$ is gauge invariant up to and including order $l^4$ terms, i.e., it is gauge invariant for EFTs known up to and including $6$ derivatives. 

If $n\geq 5$ then the above conditions do not rule out gauge non-invariant terms appearing in $\varsigma_n^v$. For example the term $K K \partial_r \phi \bar{K}_A \beta^{A}$ has dimension 5, boost weight 0 and is quadratic in positive boost weight quantities so might appear in $\varsigma_{5}^v$. But due to the appearance of $\beta^{A}$, which transforms inhomogeneously, this term is not gauge invariant on $C$. To investigate whether such non-gauge invariant terms do appear in reality, we will construct $S(v)$ for specific EFTs and inspect the results.

\section{Implementation of Algorithm}\label{Implementation}

We shall now discuss how one calculates $S(v)$ for a particular EFT in practice. Since the equations involved get extremely lengthy, a symbolic computer algebra program is needed to do this. The program of choice of the author is Cadabra \cite{Cadabra1}\cite{Cadabra2}\cite{Cadabra3} due to its ability to split $\mu, \nu, ...$ indices into $r$, $v$ and $A, B, C, ...$ indices, and the ease with which expressions can be canonicalised using symmetry or anti-symmetry of indices.  

For simplicity, let us consider a vacuum gravity EFT (i.e., no Maxwell or scalar field), in even spacetime dimension, in which case only even numbers of derivatives appear. We assume that we know the terms with up to $N$ derivatives in the EFT Lagrangian. The Lagrangian can be written $\mathcal{L}+O(l^N)$ where $O(l^N)$ is the contribution from the unknown terms with $N+2$ or more derivatives and  
\begin{equation}\label{HDLag}
    \mathcal{L} = -2\Lambda + R + \mathcal{L}_{higher} 
\end{equation}
where $\mathcal{L}_{higher} = \sum_{n=2}^{N-2} l^{n} \mathcal{L}_{n+2}$ are the known higher derivative terms.

\subsubsection{Calculate Equations of Motion}
First we calculate the equations of motion for this Lagrangian. This gives
\begin{equation}
    E_{\mu \nu}\equiv \frac{1}{\sqrt{-g}} \frac{\delta(\sqrt{-g}\mathcal{L})}{\delta g^{\mu \nu}} = \Lambda g_{\mu \nu} + R_{\mu \nu} - \frac{1}{2}R g_{\mu \nu} + H_{\mu \nu}
\end{equation}
where $H_{\mu \nu}$ is the contribution from $\mathcal{L}_{higher}$. 

\subsubsection{Calculate IWW Entropy Current}
Second we must find the IWW entropy current $(s^{v}_{IWW},s^A)$ (of equation \eqref{IWW}) for this theory. Since they are not the main subject of this thesis, we will only touch briefly on how to calculate them. Full procedures are given in \cite{Bhat:2020}\cite{Bhattacharyya:2021jhr}, but since the algorithms have many steps there are no simple formulas for general $\mathcal{L}$. However, if the Lagrangian depends only on the Riemann tensor and not its derivatives, then there is a formula for $s^{v}_{IWW}$ calculated by Wall in \cite{Wall:2015}, which reproduces a formula for holographic entanglement entropy derived previously by Dong \cite{Dong:2013qoa}. This formula involves taking partial derivatives of the Lagrangian with respect to Riemann components\footnote{We define $\partial/(\partial R_{\mu \nu \rho \sigma})$ to have the same symmetries as $R_{\mu \nu \rho \sigma}$ and to be normalised such that a first variation of some quantity $X(R_{\mu \nu \rho \sigma})$ will give $\delta X = \delta R_{\mu \nu \rho \sigma} \frac{\partial X}{\partial R_{\mu \nu \rho \sigma}}$.}, and then discarding terms that are quadratic (or higher order) in positive boost weight quantities:
\begin{equation}\label{WallIWW}
    s^{v}_{IWW} = -2\left( \frac{\partial \mathcal{L}}{\partial R_{r v r v }}\bigg|_{\cH} + 4 \frac{\partial^2 \mathcal{L}}{\partial R_{v A v B } \partial R_{r C r D}}\bigg|_{\cH} K_{A B} \bar{K}_{C D} \right) - s^{v}_{quadratic}
\end{equation}
Here $s^{v}_{quadratic}$ is purely there to cancel any terms that are quadratic or higher order in positive boost weight quantities when the Riemann components are expanded out in GNCs on $\cH$, so that overall $s^{v}_{IWW}$ has only linear or zero order terms. Technically, Wall does not discard these terms, and one can include them without affecting the validity of the linearized second law for the IWW entropy, or affecting the end result for $S(v)$. However, we take our prescription of $s^v_{IWW}$ to require that such terms are discarded. The gauge invariance of this particular prescription for $s^v_{IWW}$ is discussed in Section \ref{IWWGaugeInvariance}.

We can apply this to our EFT Lagrangian (\ref{HDLag}) if $\mathcal{L}_{higher}$ is a polynomial in $R_{\mu \nu \rho \sigma}$. The contribution from the Einstein-Hilbert Lagrangian is $1$, and so
\begin{equation}\label{WallIWWEFT}
    s^{v}_{IWW} = 1 -2\left( \frac{\partial \mathcal{L}_{higher}}{\partial R_{r v r v }}\bigg|_{\cH} + 4 \frac{\partial^2 \mathcal{L}_{higher}}{\partial R_{v A v B } \partial R_{r C r D}}\bigg|_{\cH} K_{A B} \bar{K}_{C D} \right) - s^{v}_{quadratic} 
\end{equation}
None of the Lagrangians we consider in Section \ref{Examples} contain derivatives of Riemann components, and hence (\ref{WallIWWEFT}) is the formula we use to calculate $s^{v}_{IWW}$. 

There is no simple formula for $s^{A}$ in general, and so we must follow the procedure in \cite{Bhattacharyya:2021jhr} to calculate it. The method requires finding the total derivative term $\Theta^{\mu}$ given by $\delta (\sqrt{-g} \mathcal{L}) = \sqrt{-g} (E^{\mu \nu} \delta g_{\mu \nu} + D_{\mu} \Theta^{\mu}[\delta g])$, and the Noether charge for diffeomorphisms $Q^{\mu \nu}$ given by $\nabla_{\nu}{ Q^{\mu \nu}} = 2 E^{\mu \nu} \zeta_{\nu} + \Theta^{\mu}[\partial \zeta] - \zeta^{\mu} \mathcal{L} $ where we have set $\delta g_{\mu \nu} = \nabla_{\mu}{ \zeta_{\nu} } + \nabla_{\nu}{ \zeta_{\mu}}$ in the argument of $\Theta^{\mu}$. For any given theory, both of these quantities can be calculated via theorems given in \cite{Bhattacharyya:2021jhr}.

\subsubsection{Calculate Remaining Terms in \texorpdfstring{$E_{v v}$}{Evv} and Reduce to Allowed Terms}
Now that we have $E_{v v}$, $s^{v}_{IWW}$ and $s^A$, we proceed by expanding them out in GNCs\footnote{Expanding out $E_{v v}$ in GNCs requires expanding out Riemann components and possibly their covariant derivatives in GNCs. The appendix of \cite{Hollands:2022} lists the expansions of all Riemann components, which is sufficient for the EGB Lagrangian calculation in Section \ref{Examples}. However, the cubic and quartic Lagrangians below have equations of motion that depend on $\nabla_{\mu}{ R_{\nu \rho \sigma \kappa} }$ and $\nabla_{\mu}{ \nabla_{\nu}{ R_{\rho \sigma \kappa \lambda} } }$, and so these must also be calculated in terms of GNCs. Beware, even for Cadabra this is an extensive computation.} on $\cH$ and calculating the terms denoted by the ellipsis in equation (\ref{IWW}), which we denote by $\Delta$:
\begin{equation}
    \Delta \equiv -E_{v v}\Big|_{\cH} - \partial_{v}\left[\frac{1}{\sqrt{\mu}} \partial_{v}\left(\sqrt{\mu} s^{v}_{IWW}\right) + D_{A}{ s^A }\right]
\end{equation}
By construction, these terms will be of quadratic or higher order in positive boost weight quantities. For example, in standard GR (i.e., $N=2$) we have $s^{v}_{IWW}=1$, $s^A=0$, and $-E_{v v}|_{\cH} = \mu^{A B} \partial_{v}{K_{A B}}-K_{A B} K^{A B} $. We can use $\partial_{v}{\sqrt{\mu}} = \sqrt{\mu} K$ to obtain $\Delta = K_{A B} K^{A B}$ for standard GR.

In our vacuum gravity EFT, $\Delta$ will be some polynomial in affinely parameterized GNC metric quantities that are covariant in $A, B,...$ indices: $\alpha, \beta_{A}, \mu_{A B}, \mu^{A B}, \Lambda, \epsilon_{A_1 ... A_{d-2}}, R_{A B C D}[\mu]$ and their $\partial_{v}, \partial_{r}$ and $D_{A}$ derivatives. Note that when considering $d=4$ dimensions, $C(v)$ is 2-dimensional which implies that $R_{A B C D}[\mu]= \frac{R[\mu]}{2} (\mu_{A C} \mu_{B D} - \mu_{A D} \mu_{B C})$.

We now reduce the above set of terms to a much smaller set of ``allowed'' terms:
\begin{empheq}[box=\fbox]{align}\label{AllowedTerms}
    \text{Allowed terms:} \,\, \mu_{A B}, \,\, &\mu^{A B}, \,\, \epsilon_{A_1 ... A_{d-2}}, \,\, D_{A_1}{... D_{A_n}{ R_{A B C D}[\mu] } },\nonumber\\
    D_{A_1}{... D_{A_n}{ \beta_{A} }}, \,\, D_{A_1}&... D_{A_n}{\partial_{v}^p K_{A B} }, \,\, D_{A_1}{... D_{A_n}{\partial_{r}^q \bar{K}} }, \,\, \Lambda
\end{empheq}
In particular, the only positive boost weight allowed terms are of the form $D^k\partial_{v}^p K_{A B}$.

To do this, first we commute any $D$ derivatives to the left in a term of the form $\partial_{v}^{p} \partial_{r}^q D_{A_1}{... D_{A_n}{ \varphi } }$ using the commutation rules (\ref{CommutationRules}), and eliminate any $\partial_{v}$ and $\partial_{r}$ derivatives of $\epsilon_{A_1 ... A_{d-2}}$ and $R_{A B C D}[\mu]$ using (\ref{EliminationOfREpsilon}). The remaining reduction is achieved through careful inspection of the affinely parameterized GNC expressions for Ricci components and application of the equations of motion. We saw how this worked when reducing the Maxwell quantities in the previous Chapter. We now briefly explain how to do this for metric quantities.

Consider the GNC expression for $R_{v A}$ on $\cH$:
\begin{equation}
    R_{v A}\Big|_{\cH} = \frac{1}{2}\partial_{v}{\beta_{A}} + D_{B}{K_{A C}} \mu^{B C}-D_{A}{K_{B C}} \mu^{B C}+\frac{1}{2}K_{B C} \beta_{A} \mu^{B C}
\end{equation}
We can rearrange this to get $\partial_{v}{ \beta_A }$ in terms of allowed terms and the Ricci component $R_{v A}$:
\begin{equation} \label{dvbeta}
    \partial_{v}{\beta_{A}} = - 2 D_{B}{K_{A C}} \mu^{B C}+2 D_{A}{K_{B C}} \mu^{B C}-K_{B C} \beta_{A} \mu^{B C} + 2 R_{v A}\Big|_{\cH}
\end{equation}
We can rewrite the equation of motion as $R_{\mu\nu} = \frac{2}{d-2} \Lambda g_{\mu \nu} + \frac{1}{d-2} g^{\rho \sigma} H_{\rho \sigma} g_{\mu \nu} - H_{\mu \nu} + O(l^N)$. Since $H_{\mu \nu}$ is at least $O(l^2)$, we can therefore swap $R_{\mu \nu}$ for $\frac{2}{d-2} \Lambda g_{\mu \nu}$ plus higher order terms in $l$. Thus we can use (\ref{dvbeta}) to eliminate $\partial_v \beta_A$ in favour of allowed terms plus terms of higher order in $l$. Working order by order in $l$ we can therefore eliminate occurrences of $\partial_{v}{ \beta_A }$, pushing them to higher order at each step. Eventually we reach $O(l^N)$, at which point we stop since we do not know the terms in the equation of motion at this order. 

We can find a similar expression for $\partial_{r}{ \beta_A }$ by considering $R_{r A}\Big|_{\cH}$:
\begin{equation}
    \partial_{r}{ \beta_{A} } = D_{B}{\bar{K}_{A C}} \mu^{B C}-D_{A}{\bar{K}_{B C}} \mu^{B C}+\bar{K}_{A B} \beta_{C} \mu^{B C} - \frac{1}{2}\bar{K}_{B C} \beta_{A} \mu^{B C} - R_{r A}\Big|_{\cH}
\end{equation}
Again we can use the equations of motion to swap out $R_{r A}$ and push any occurrence of $\partial_{r}{ \beta_A }$ to higher order in $l$, eventually reaching $O(l^N)$. 

We can similarly eliminate $\partial_{v}{ \Bar{K}_{A B} }$ using $R_{A B}$, and eliminate $\alpha$ using $R_{v r}$. We can take $D$ and $\partial_{v}$ derivatives of these expressions to eliminate further terms. We can't take $\partial_{r}$ derivatives since the expressions are evaluated on $r=0$, so instead we look at the $\nabla_{r}$ derivatives of the Ricci tensor. For example, to eliminate $\partial_{r}{\alpha}$, we look at $\nabla_{r}{ R_{v r} }$.

When looking at a specific theory, one will only need to calculate a finite number of these elimination rules, as there will be only so many derivatives involved. The set we need for the EGB, cubic and quartic Lagrangians in Section \ref{Examples} are given in Appendix \ref{app:elimination}.

\subsection{Manipulate Terms Order-by-Order}

By construction, $\Delta$ is quadratic (or higher) in positive boost weight terms. The only positive boost weight \text{allowed} terms are of the form $D^k \partial_{v}^p K_{A B}$. Therefore, once we have eliminated all non-allowed terms up to $O(l^N)$, $\Delta$ will be of the schematic form (with indices suppressed)
\begin{equation}
    \Delta = \sum_{k_1, k_2, p_1, p_2} (D^{k_1}{\partial_{v}^{p_1} K}) \, (D^{k_2}{\partial_{v}^{p_2}K}) \, Q_{k_1, k_2, p_1, p_2} + O(l^N) 
\end{equation}
where the $Q_{k_1, k_2, p_1, p_2}$ (boost weight $-p_1-p_2$) are made up of allowed terms. The $Q_{k_1, k_2, p_1, p_2}$ can in principle include more $D^{k}{\partial_{v}^{p} K }$ terms, so for simplicity when performing the algorithm we always order the terms in the following priority: $p_1$ and $p_2$ as small as possible, $p_1\leq p_2$, and then $k_1$ and $k_2$ as small as possible.

We then proceed through the algorithm described in the previous Chapter order-by-order in $l$. We first lift off the $D^{k_1}$ derivatives, producing a total derivative term $D_A Y^A$. Second, we remove the $\partial_{v}^{p_1}$ derivatives in each term, producing terms of the form $\partial_{v}\left(\frac{1}{\sqrt{\mu}} \partial_{v}\left(\sqrt{\mu} \varsigma^{v}\right)\right)$ by repeatedly using equation (\ref{vintegrationbyparts}). When performing this second step for specific Lagrangians in practice, one should investigate for which values of $k, p_1, p_2$ do terms of the form on the LHS of (\ref{vintegrationbyparts}) appear in $\Delta$, and pre-calculate the corresponding RHS of (\ref{vintegrationbyparts}), in particular the numbers $a_j$. For the EGB, cubic and quartic Lagrangians we shall study below, the only relevant terms that appear are $\partial_{v} K \partial_{v} K$, $\partial_{v} K D \partial_{v} K$, $\partial_{v} K D D\partial_{v} K$ which have $(p_1, p_2)=(1,1)$ and $a_1=1$, and $\partial_{v} K \partial_{v v} K$ which has $(p_1, p_2)=(1,2)$ and $(a_1, a_2)= (-\frac{1}{3}, \frac{2}{3})$.

Thus at each order in $l$, we can manipulate $\Delta^{(m)}$ into the form
\begin{equation}
    \Delta^{(m)} = \partial_{v}\left[\frac{1}{\sqrt{\mu}} \partial_{v}\left(\sqrt{\mu} \varsigma^{(m)v}\right)\right] + 2 K_{A B} X^{(m)A B} + D_{A} Y^{(m)A}
\end{equation}
We take $X^{(m)A B}$ to be symmetric. It will be at least linear in positive boost weight quantities, and $\varsigma^{(m)v}$ and $Y^{(m)A}$ will be at least quadratic in positive boost weight quantities. 
We add these to the terms lower order in $l$ and complete the square as described in the previous Chapter. We perform this procedure through each order of $l$ until all terms have been dealt with up to $O(l^N)$:
\begin{multline}
    \Delta= \partial_{v}\left[\frac{1}{\sqrt{\mu}} \partial_{v}\left(\sqrt{\mu} \sum_{n=2}^{N-2} l^n \varsigma^{(n)v} \right)\right] + \left(K_{A B}+\sum_{n=2}^{N-2} l^n X^{(n)}_{A B}\right) \left(K^{A B}+\sum_{n=2}^{N-2} l^n X^{(n)A B}\right) +\\
     D_{A}{\sum_{n=2}^{N-2} l^n Y^{(n)A}} + O(l^{N})
\end{multline}
The HKR entropy density is then defined as
\begin{equation}
\label{HKRdef}
    s^{v}_{HKR} = s_{IWW}^{v} + \sum_{n=2}^{N-2} l^n \varsigma^{(n)v}
\end{equation}

\section{Examples of HKR Entropy Density}\label{Examples}

\subsection{Einstein-Gauss-Bonnet}\label{EGBCalc}

For our first example, we shall consider Einstein-Gauss-Bonnet (EGB) theory. Recall that this describes the leading $4$-derivative EFT corrections to Einstein gravity in $d>4$ dimensions. Specifically, in \ref{EFTCorstoGR} we argued that using field redefinitions and dropping total derivatives, the Lagrangian for the EFT for vacuum gravity can be brought to the form $\mathcal{L}_{EGB}+O(l^4)$ where
\begin{equation}
    \mathcal{L}_{EGB} = -2 \Lambda + R + \frac{1}{16} k l^2 \delta^{\rho_1 \rho_2 \rho_3 \rho_4}_{\sigma_1 \sigma_2 \sigma_3 \sigma_4} R_{\rho_1 \rho_2}\,^{\sigma_1 \sigma_2} R_{\rho_3 \rho_4}\, ^{\sigma_3 \sigma_4}
\end{equation}
The equation of motion is $E_{\mu \nu} = O(l^4)$ with 
\begin{equation}
    E_{\mu \nu} = \Lambda g_{\mu \nu} + G_{\mu \nu} - \frac{1}{32} k l^2 g_{\mu \tau} \delta^{\tau \rho_1 \rho_2 \rho_3 \rho_4}_{\nu \sigma_1 \sigma_2 \sigma_3 \sigma_4} R_{\rho_1 \rho_2}\,^{\sigma_1 \sigma_2} R_{\rho_3 \rho_4}\, ^{\sigma_3 \sigma_4} 
\end{equation}
If we view this theory as an EFT with $N=4$ then, as explained above, the HKR entropy coincides with the IWW entropy (as for any $N=4$ theory). However, EGB has the special property of possessing second order equations of motion, despite arising from a 4-derivative Lagrangian. This property means EGB gravity can be considered as a self-contained classical theory rather than just the $N=4$ truncation of an EFT. The HKR procedure can be used to define an entropy for this classical theory as an expansion in $l^2$. The idea is to treat this EGB theory as a very special EFT for which the coefficients of the terms with more than $4$ derivatives are exactly zero (i.e., the Lagrangian is exactly $\mathcal{L}_{EGB}$) and use the HKR procedure to define an entropy order by order in $l^2$. We shall show explicitly how this works to order $l^4$. 

Let us proceed with the HKR procedure for this theory. We have calculated the equations of motion above. The next step is to find the IWW entropy current. It is given in \cite{Bhattacharyya:2021jhr} with a different normalisation. In our units it is\footnote{For a stationary black hole, the IWW (or HKR) entropy reduces to the EGB entropy defined in Ref. \cite{Jacobson:1993xs} (which is reproduced by the method of \cite{Wald:1993nt}). For EGB, the IWW entropy density involves only terms of vanishing boost weight and so, for a non-stationary EGB black hole, the IWW entropy is the same as the Iyer-Wald entropy.
} 
\begin{equation}
    s^{v}_{IWW} = 1 + \frac{1}{2} k l^2 R[\mu], \quad \quad \quad \quad s^{A} = k {l}^{2} \left(D^{A}{K}-D^{B}{K^{A}\,_{B}}\right),
\end{equation}
We then calculate the remaining terms in $E_{v v}$ that are quadratic (or higher) in positive boost weight quantities: 
\begin{equation}
\begin{split}
    \Delta \equiv& -E_{v v} - \partial_{v}\left[\frac{1}{\sqrt{\mu}} \partial_{v}\left(\sqrt{\mu} s^{v}_{IWW}\right) + s^A\right]\\
    =& K_{A B} K^{A B} + k {l}^{2} \Big[D^{A}{K} D_{A}{K}-2D_{A}{K} D^{B}{K_{B}\,^{A}}+D^{A}{K_{A B}} D^{C}{K_{C}\,^{B}}-D^{A}{K^{B C}} D_{A}{K_{B C}}+\\
    &D^{A}{K^{B C}} D_{B}{K_{A C}} + K_{A B}\Big( \partial_{v}{K_{C E}}\bar{K}^{A B} \mu^{C E}- \partial_{v}{K_{C E}} \bar{K}\mu^{A B} \mu^{C E}-2 \partial_{v}{K_{C E}}\bar{K}^{B C} \mu^{A E}+\\
    & \partial_{v}{K_{C E}} \bar{K}\mu^{A C} \mu^{B E}+ \partial_{v}{K_{C E}} \bar{K}^{C E}\mu^{A B}+D^{A}{K} \beta^{B}-D^{C}{K} \beta_{C} \mu^{A B}-D^{C}{K^{A}\,_{C}} \beta^{B}+\\
    &D^{C}{K_{C}\,^{E}} \beta_{E} \mu^{A B}+D^{C}{K^{A B}} \beta_{C}-D^{A}{K^{B C}} \beta_{C}+D^{A}{D^{C}{K^{B}\,_{C}}}-D^{A}{D^{B}{K}}-D^{C}{D_{C}{K^{A B}}}+\\
    &D^{C}{D^{A}{K^{B}\,_{C}}}+D^{C}{D_{C}{K}} \mu^{A B}-D^{C}{D^{E}{K_{C E}}} \mu^{A B}+\frac{1}{2}K^{A B} R[\mu]-K^{A B} K^{C E} \bar{K}_{C E}-\\
    &2K^{A C} R^{B}\,_{C}[\mu] - \frac{1}{2}K^{C E} \beta_{C} \beta_{E} \mu^{A B}+\frac{1}{4}K \beta^{C} \beta_{C} \mu^{A B}+\frac{1}{2}K^{A C} \beta^{B} \beta_{C} - \frac{1}{4}K^{A B} \beta^{C} \beta_{C}-\\
    &K^{C E} R^{A}\,_{C}\,^{B}\,_{E}[\mu]+K^{C E} R_{C E}[\mu] \mu^{A B}+K^{C E} K_{C E} \bar{K} \mu^{A B}+2K^{A C} K^{B E} \bar{K}_{C E}-\\
    &K^{A C} K^{B}\,_{C} \bar{K}-K^{C E} K_{C}\,^{F} \bar{K}_{E F} \mu^{A B} \Big) \Big]
\end{split}
\end{equation}
It happens that no non-allowed terms appear in $\Delta$ for this Lagrangian, so we do not yet need to use equations of motion to eliminate such terms. We proceed to move one of the $D$ derivatives over in the terms of the form $(DK) (DK)$, producing a total derivative $D_A(l^2 Y^{(2)A})$. We find that all remaining terms are proportional to $K_{A B}$, and we can express $\Delta$ in the desired form
\begin{equation} \label{EGBl2}
    \Delta = \left(K_{A B}+l^2 X_{A B}^{(2)}\right) \left(K^{A B}+l^2 X^{(2)A B}\right) + D_{A}(l^2 Y^{(2)A}) - l^4 X_{ AB}^{(2)} X^{(2)A B}
\end{equation}
where
\begin{equation}
    \begin{split}
        X^{(2)A B} = &\frac{1}{8} k \Big(4 \partial_{v}{K_{C E}}\bar{K}^{A B} \mu^{C E}-4 \partial_{v}{K_{C E}} \bar{K}\mu^{A B} \mu^{C E}-4 \partial_{v}{K_{C E}} \bar{K}^{B C}\mu^{A E}+\\
        & 4 \partial_{v}{K_{C E}} \bar{K}\mu^{A C} \mu^{B E}+4 \partial_{v}{K_{C E}} \bar{K}^{C E} \mu^{A B}+2D^{A}{K} \beta^{B}-4D^{C}{K} \beta_{C} \mu^{A B}-\\
        &2D^{C}{K^{A}\,_{C}} \beta^{B}+4D^{C}{K_{C}\,^{E}} \beta_{E} \mu^{A B}+4D^{C}{K^{A B}} \beta_{C}-2D^{A}{K^{B C}} \beta_{C}+2D^{A}{D^{B}{K}}-\\
        &4D^{C}{D^{E}{K_{C E}}} \mu^{A B}+2K^{A B} R[\mu]-4K^{A B} K^{C E} \bar{K}_{C E}-4K^{A C} R[\mu]^{B}\,_{C}-\\
        &2K^{C E} \beta_{C} \beta_{E} \mu^{A B}+K \beta^{C} \beta_{C} \mu^{A B}+K^{A C} \beta^{B} \beta_{C} -K^{A B} \beta^{C} \beta_{C}-4K^{C E} R[\mu]^{A}\,_{C}\,^{B}\,_{E}+\\
        &4K^{C E} R[\mu]_{C E} \mu^{A B}+4K^{C E} K_{C E} \bar{K} \mu^{A B}+8K^{A C} K^{B E} \bar{K}_{C E}-4K^{A C} K^{B}\,_{C} \bar{K}-\\
        &4K^{C E} K_{C}\,^{F} \bar{K}_{E F} \mu^{A B}-4\bar{K}^{A C} \partial_{v}{K_{C E}} \mu^{B E}+2D^{B}{K} \beta^{A}-2D^{C}{K^{B}\,_{C}} \beta^{A}-\\
        &2D^{B}{K^{A C}} \beta_{C}+2D^{B}{D^{A}{K}}-4K^{B C} R[\mu]^{A}\,_{C}+K^{B C} \beta^{A} \beta_{C}\Big),
    \end{split}
\end{equation}
\begin{equation}
    Y^{(2)A} =k\left( D^{A}{K} K-2D_{B}{K} K^{A B}+D^{B}{K_{B C}} K^{A C}-D^{A}{K_{B C}} K^{B C}+D_{B}{K^{A}\,_{C}} K^{B C} \right)
\end{equation}
As expected from Section \ref{Classify}, we find $\varsigma^{(2)v}=0$, i.e., the HKR entropy coincides with the IWW entropy to $O(l^2)$ \cite{Hollands:2022}.

In completing the square in (\ref{EGBl2}) we have produced $O(l^4)$ terms, namely $
- l^4 X_{A B}^{(2)} X^{(2)A B}$. We denote these as $l^4 \Delta^{(4)}$. We can expand these out and proceed with the HKR algorithm at order $l^4$. We find
\begin{equation}\label{F4EGB}
    \Delta^{(4)}=\partial_{v}\left[\frac{1}{\sqrt{\mu}} \partial_{v}\left(\sqrt{\mu} \varsigma^{(4)v}_{4}\right)\right] + 2 K_{A B} X^{(4)A B} + D_{A}{Y^{(4)A}} + O(l^2)
\end{equation}
where\footnote{
For $d=4$ the GB term is topological. In this case
 one can show that $\zeta^{(4)v}$ vanishes, using special identities satisfied by $2\times 2$ matrices.
} 
\begin{equation}
    \begin{split}
        \varsigma^{(4)v} =& \frac{1}{8} k^{2} \Big[\left(6-d\right) K K \bar{K} \bar{K}-K K \bar{K}^{A B} \bar{K}_{A B}+4K K^{A B} \bar{K}_{A}\,^{C} \bar{K}_{B C}+\\
        &\left(-14+2d\right) K K^{A B} \bar{K} \bar{K}_{A B}-2K^{A B} K_{A}\,^{C} \bar{K}_{B}\,^{E} \bar{K}_{C E}-2K^{A B} K^{C E} \bar{K}_{A C} \bar{K}_{B E}+\\
        &\left(6-d\right) K^{A B} K^{C E} \bar{K}_{A B} \bar{K}_{C E}+4K^{A B} K_{A}\,^{C} \bar{K} \bar{K}_{B C}-K^{A B} K_{A B} \bar{K} \bar{K}\Big]
    \end{split}
\end{equation}
and $X^{(4)A B}$ and $Y^{(4)A}$ are very lengthy expressions of boost weight $+1$ and $+2$ respectively. Non-allowed terms do appear at this order after extracting $\varsigma^{(4)v}$, and swapping them out using the equations of motion produces the $O(l^2)$ terms in (\ref{F4EGB}).

We then group like terms together and complete the square to write $\Delta$ in the desired form:
\begin{multline}
    \Delta = \partial_{v}\left[\frac{1}{\sqrt{\mu}} \partial_{v}\left(\sqrt{\mu}l^4 \varsigma^{(4)v}\right)\right] +\left(K_{A B}+l^2 X_{A B}^{(2)} + l^4 X_{A B}^{(4)}\right) \left(K^{A B}+l^2 X^{(2)A B} + l^4 X^{(4)A B}\right) +\\
     D_{A}{ (l^2 Y^{(2)A} + l^4 Y^{(4)A}) } + O(l^6)
\end{multline}
The HKR entropy density is then
\begin{equation}
    s^{v}_{HKR}= s^{v}_{IWW} + l^4 \varsigma^{(4)v} + O(l^6)
\end{equation}
To this order, $s^{v}_{HKR}$ is gauge invariant as expected (section \ref{Classify}), as it only contains $\mu^{A B}, R[\mu], K_{A B}$ and $\bar{K}_{A B}$ which transform homogeneously under a change of GNCs on $C$.  

We can continue the algorithm to the next order by expanding out all the $O(l^6)$ terms. We find that the $O(l^6)$ part of $s^v_{HKR}$ is extremely lengthy. It is also not gauge invariant, however since it is so unwieldy we leave the discussion of gauge non-invariance to the considerably shorter expression arising from the quartic Lagrangian below.

\subsection{Cubic Order Riemann Lagrangians}\label{CubicCalc}

Let us now specialise to $d=4$. As mentioned above, the EGB term is purely topological in this dimension so we shall ignore it, and we can eliminate all other 4-derivative corrections through field redefinitions and total derivatives. At 6-derivative order, we can similarly reduce the number of corrections to just two \cite{Endlich:2017, Cano:2019}. The Lagrangian is $\mathcal{L}+O(l^6)$ where
\begin{equation}\label{4DEFT}
    \mathcal{L} = -2 \Lambda + R + l^4( k_{1} \mathcal{L}_{even} + k_{2} \mathcal{L}_{odd} ) 
\end{equation}
where $\mathcal{L}_{even}$ and $\mathcal{L}_{odd}$ are even and odd parity terms respectively, given by 
\begin{equation}
    \begin{split}
        \mathcal{L}_{even} =& R_{\mu \nu \kappa \lambda} R^{\kappa \lambda \chi \eta} R_{\chi \eta}\,^{\mu \nu} \\
        \mathcal{L}_{odd} =& R_{\mu \nu \kappa \lambda} R^{\kappa \lambda \chi \eta} R_{\chi \eta \rho \sigma} \epsilon^{\mu \nu \rho \sigma}
    \end{split}
\end{equation}
Let us follow our implementation of the HKR algorithm to find the entropy density at order $l^4$. The equation of motion for this Lagrangian is $E_{\mu \nu} = O(l^6)$ with 
\begin{equation}
    \begin{split}
        E_{\mu \nu} = & \Lambda g_{\mu \nu} + G_{\mu \nu} -\\
        & l^4 \bigg[ k_{1} \Big( \frac{1}{2}R^{\kappa \lambda \alpha \beta} R_{\kappa \lambda}\,^{\rho \sigma} R_{\alpha \beta \rho \sigma} g_{\mu \nu}-3R_{\mu}\,^{\kappa \alpha \beta} R_{\nu \kappa}\,^{\rho \sigma} R_{\alpha \beta \rho \sigma} - 3R_{\mu}\,^{\alpha \beta \rho} \nabla_{\alpha}\nabla^{\sigma}{R_{\nu \sigma \beta \rho}}-\\
        & 6\nabla^{\alpha}{R_{\mu \alpha}\,^{\beta \rho}} \nabla^{\sigma}{R_{\nu \sigma \beta \rho}}-6\nabla^{\alpha}{R_{\mu}\,^{\beta \rho \sigma}} \nabla_{\beta}{R_{\nu \alpha \rho \sigma}} - 3R_{\mu}\,^{\alpha \beta \rho} \nabla^{\sigma}\nabla_{\alpha}{R_{\nu \sigma \beta \rho}}-\\
        & 3R_{\nu}\,^{\alpha \beta \rho} \nabla_{\alpha}\nabla^{\sigma}{R_{\mu \sigma \beta \rho}}-3R_{\nu}\,^{\alpha \beta \rho} \nabla^{\sigma}\nabla_{\alpha}{R_{\mu \sigma \beta \rho}} \Big) + \\
        & k_{2} \Big( - R_{\mu}\,^{\eta}\,_{\kappa \lambda} R_{\nu \eta}\,^{\alpha \beta} R_{\alpha \beta \rho \sigma} \epsilon^{\kappa \lambda \rho \sigma} - \frac{1}{2}R_{\mu \eta}\,^{\kappa \lambda} R_{\alpha \beta}\,^{\rho \sigma} R_{\kappa \lambda \rho \sigma} \epsilon^{\eta \alpha \beta}\,_{\nu} - R_{\mu}\,^{\kappa}\,_{\lambda \alpha} \epsilon^{\lambda \alpha \beta \rho} \nabla_{\kappa}\nabla^{\sigma}{R_{\nu \sigma \beta \rho}}-\\
        & 2\epsilon^{\kappa \lambda \alpha \beta} \nabla^{\rho}{R_{\mu \rho \kappa \lambda}} \nabla^{\sigma}{R_{\nu \sigma \alpha \beta}} - 2\epsilon^{\kappa \lambda \alpha \beta} \nabla^{\rho}{R_{\mu}\,^{\sigma}\,_{\kappa \lambda}} \nabla_{\sigma}{R_{\nu \rho \alpha \beta}} - R_{\mu}\,^{\kappa}\,_{\lambda \alpha} \epsilon^{\lambda \alpha \beta \rho} \nabla^{\sigma}\nabla_{\kappa}{R_{\nu \sigma \beta \rho}} + \\
        & R_{\kappa \lambda}\,^{\alpha \beta} \epsilon^{\kappa \lambda \rho}\,_{\nu} \nabla^{\sigma}\nabla_{\rho}{R_{\mu \sigma \alpha \beta}} + 2\epsilon^{\kappa \lambda \alpha}\,_{\nu} \nabla_{\kappa}{R_{\mu}\,^{\beta \rho \sigma}} \nabla_{\beta}{R_{\lambda \alpha \rho \sigma}} + 2\epsilon^{\kappa \lambda \alpha}\,_{\nu} \nabla^{\beta}{R_{\mu \beta}\,^{\rho \sigma}} \nabla_{\kappa}{R_{\lambda \alpha \rho \sigma}} + \\
        & R_{\mu}\,^{\kappa \lambda \alpha} \epsilon^{\beta \rho \sigma}\,_{\nu} \nabla_{\kappa}\nabla_{\beta}{R_{\lambda \alpha \rho \sigma}} + R_{\kappa \lambda}\,^{\alpha \beta} \epsilon^{\kappa \lambda \rho}\,_{\nu} \nabla_{\rho}\nabla^{\sigma}{R_{\mu \sigma \alpha \beta}} + R_{\mu}\,^{\kappa \lambda \alpha} \epsilon^{\beta \rho \sigma}\,_{\nu} \nabla_{\beta}\nabla_{\kappa}{R_{\lambda \alpha \rho \sigma}}-\\
        & R_{\mu}\,^{\eta \kappa \lambda} R_{\nu \eta \alpha \beta} R_{\kappa \lambda \rho \sigma} \epsilon^{\alpha \beta \rho \sigma} - \frac{1}{2}R_{\nu \eta}\,^{\kappa \lambda} R_{\alpha \beta}\,^{\rho \sigma} R_{\kappa \lambda \rho \sigma} \epsilon^{\eta \alpha \beta}\,_{\mu} - R_{\nu}\,^{\kappa}\,_{\lambda \alpha} \epsilon^{\lambda \alpha \beta \rho} \nabla_{\kappa}\nabla^{\sigma}{R_{\mu \sigma \beta \rho}} - \\
        & R_{\nu}\,^{\kappa}\,_{\lambda \alpha} \epsilon^{\lambda \alpha \beta \rho} \nabla^{\sigma}\nabla_{\kappa}{R_{\mu \sigma \beta \rho}} + R_{\kappa \lambda}\,^{\alpha \beta} \epsilon^{\kappa \lambda \rho}\,_{\mu} \nabla^{\sigma}\nabla_{\rho}{R_{\nu \sigma \alpha \beta}} + 2\epsilon^{\kappa \lambda \alpha}\,_{\mu} \nabla_{\kappa}{R_{\nu}\,^{\beta \rho \sigma}} \nabla_{\beta}{R_{\lambda \alpha \rho \sigma}} + \\
        & 2\epsilon^{\kappa \lambda \alpha}\,_{\mu} \nabla^{\beta}{R_{\nu \beta}\,^{\rho \sigma}} \nabla_{\kappa}{R_{\lambda \alpha \rho \sigma}} + R_{\nu}\,^{\kappa \lambda \alpha} \epsilon^{\beta \rho \sigma}\,_{\mu} \nabla_{\kappa}\nabla_{\beta}{R_{\lambda \alpha \rho \sigma}} + R_{\kappa \lambda}\,^{\alpha \beta} \epsilon^{\kappa \lambda \rho}\,_{\mu} \nabla_{\rho}\nabla^{\sigma}{R_{\nu \sigma \alpha \beta}} + \\
        & R_{\nu}\,^{\kappa \lambda \alpha} \epsilon^{\beta \rho \sigma}\,_{\mu} \nabla_{\beta}\nabla_{\kappa}{R_{\lambda \alpha \rho \sigma}} \Big) \bigg]
    \end{split}
\end{equation}
We now must find the IWW entropy current for this theory. $\mathcal{L}$ depends only on the Riemann tensor and not its derivatives up to and including order $l^4$, and hence we can calculate the IWW entropy density $s^{v}_{IWW}$ using the formula (\ref{WallIWWEFT}). Splitting this into the individual contributions from $\mathcal{L}_{even}$ and $\mathcal{L}_{odd}$ we get
\begin{equation}
    s^{v}_{IWW} = 1 + l^4( k_{1} s^{v}_{even} + k_{2} s^{v}_{odd} ) - s^v_{quadratic} 
\end{equation}
where
\begin{equation}
\label{evenodd}
    \begin{split}
    s^{v}_{even} =&-6R_{r v A B} R_{r v C D} \mu^{A C} \mu^{B D}-24R_{r v r A} R_{r v v B} \mu^{A B}+12R_{r v r v} R_{r v r v}-24 R_{r A v B} K^{A C} \bar{K}^{B}\,_{C} \\
    s^{v}_{odd} =& -4R_{A B C E} R_{r v F G} \epsilon^{A B} \mu^{C F} \mu^{E G}-8R_{r A B C} R_{r v v D} \epsilon^{B C} \mu^{A D}-8R_{r v r A} R_{v B C D} \epsilon^{C D} \mu^{A B}-\\
    &16R_{r v r A} R_{r v v B} \epsilon^{A B}+16R_{r v A B} R_{r v r v} \epsilon^{A B}+16 R_{r A v B} K^{A C} \bar{K}_{C D} \epsilon^{B D} +\\
    & 16 R_{r A v B} K^{A}\,_{C} \bar{K}^{B}\,_{D} \epsilon^{C D}-16 R_{r A v B} K_{C D} \bar{K}^{B D} \epsilon^{A C}
    \end{split}
\end{equation}
and where $s^{v}_{quadratic}$ is there to cancel any terms that are quadratic or higher in positive boost weight quantities when the Riemann components above are expanded out in GNCs, so that $s^v_{IWW}$ is only linear or zero order in positive boost weight. After expanding, we find
\begin{equation}
\begin{split}
    s^{v}_{quadratic} =& l^4\Big[ k_1 (12K^{A B} K^{C D} \bar{K}_{A C} \bar{K}_{B D}-36K^{A B} K_{A}\,^{C} \bar{K}_{B}\,^{D} \bar{K}_{C D}) + \\
    &k_2( 32K_{A}\,^{B} K_{B}\,^{C} \bar{K}_{E}\,^{F} \bar{K}_{C F} \epsilon^{A E}-16K_{A}\,^{B} K^{C E} \bar{K}_{B E} \bar{K}_{C F} \epsilon^{A F} ) \Big]
\end{split}
\end{equation}
The quantities $s^v_{even}$, $s^v_{odd}$ and $s^v_{quadratic}$ are separately gauge invariant. This is because each of them is a zero boost weight quantity depending only on Riemann components, $\mu^{A B}$, $\epsilon^{A B}$, $K_{A B}$ and $\bar{K}_{A B}$, all of which transform homogeneously on $C$. Hence $s^v_{IWW}$ is also gauge invariant without the need to add terms of quadratic or higher order in positive boost weight quantities (which one is free to do without affecting the linearized second law that the IWW entropy satisfies). The gauge invariance of $s^v_{even}$ and $s^v_{odd}$ follows from the gauge invariance of the first part of the formula \eqref{WallIWW}. What is perhaps surprising here is that $s^v_{quadratic}$ is also gauge invariant. We shall discuss this further in section \ref{GaugeInvariance}.

We can expand out the Riemann components in GNCs on $\cH$ using the expressions in the Appendix of \cite{Hollands:2022} to get
\begin{equation}
    \begin{split}
        s^{v}_{IWW} =& 1 + \\
        &l^4 \bigg[ k_{1} \Big( 12{\alpha}^{2} -3D^{A}{\beta^{B}} D_{A}{\beta_{B}}+3D^{A}{\beta^{B}} D_{B}{\beta_{A}}+ 6\alpha \beta^{A} \beta_{A}+ \frac{3}{4}\beta^{A} \beta_{A} \beta^{B} \beta_{B} + \\
        &24D^{A}{\beta^{B}} K_{B}\,^{C} \bar{K}_{A C} - 12D^{A}{\beta^{C}} K_{A}\,^{B} \bar{K}_{C B}-12\mu^{A B} \partial_{r}{\beta_{A}} \partial_{v}{\beta_{B}}-\\
        &12K^{A B} \beta_{A} \partial_{r}{\beta_{B}}+ 6\bar{K}^{A B} \beta_{A} \partial_{v}{\beta_{B}}+ 12K^{A B} \bar{K}_{A}\,^{C} \beta_{B} \beta_{C} +24K^{A B} \bar{K}_{A}\,^{C} \partial_{v}{\bar{K}_{B C}}\Big)\\
        & k_{2} \Big( -4D_{A}{\beta_{B}} R[\mu] \epsilon^{A B} + 16D_{A}{\beta_{B}} \alpha \epsilon^{A B}+4D_{A}{\beta_{B}} \beta^{C} \beta_{C} \epsilon^{A B} +\\
        &8D^{A}{\beta^{B}} K_{A C} \bar{K}_{B E} \epsilon^{C E}+8D_{A}{\bar{K}_{B}\,^{C}} \epsilon^{A B} \partial_{v}{\beta_{C}}+8D_{A}{\bar{K}_{B}\,^{C}} K_{C}\,^{E} \beta_{E} \epsilon^{A B}-\\
        &4\bar{K}_{A}\,^{B} \beta_{C} \epsilon^{A C} \partial_{v}{\beta_{B}}+16D_{A}{K_{B}\,^{C}} \epsilon^{A B} \partial_{r}{\beta_{C}}+8K_{A}\,^{B} \beta_{C} \epsilon^{A C} \partial_{r}{\beta_{B}}-\\
        &8D_{A}{K_{B}\,^{C}} \bar{K}_{C}\,^{E} \beta_{E} \epsilon^{A B}-8K_{A}\,^{B} \bar{K}_{B}\,^{C} \beta_{C} \beta_{E} \epsilon^{A E}-8\epsilon^{A B} \partial_{r}{\beta_{A}} \partial_{v}{\beta_{B}}+\\
        &8K_{A}\,^{B} \beta_{B} \epsilon^{A C} \partial_{r}{\beta_{C}}+4\bar{K}_{A}\,^{B} \beta_{B} \epsilon^{A C} \partial_{v}{\beta_{C}}-8K_{A}\,^{B} \bar{K}_{C}\,^{E} \beta_{B} \beta_{E} \epsilon^{A C}-\\
        &8K_{A}\,^{B} \bar{K}_{B C} R[\mu] \epsilon^{A C}-16D^{A}{\beta^{B}} K_{B C} \bar{K}_{A E} \epsilon^{C E} +32K_{A}\,^{B} \bar{K}_{B C} \alpha \epsilon^{A C}+\\
        &8K_{A}\,^{B} \bar{K}_{B C} \beta^{E} \beta_{E} \epsilon^{A C}-8D_{A}{\beta^{B}} K_{B}\,^{C} \bar{K}_{C E} \epsilon^{A E}+16K^{A B} \bar{K}_{A C} \epsilon^{C E} \partial_{v}{\bar{K}_{B E}}-\\
        &16K_{A}\,^{B} \bar{K}_{C}\,^{E} \epsilon^{A C} \partial_{v}{\bar{K}_{B E}}+8D^{A}{\beta_{B}} K_{C}\,^{E} \bar{K}_{A E} \epsilon^{B C}-16K_{A}\,^{B} \bar{K}_{B}\,^{C} \epsilon^{A E} \partial_{v}{\bar{K}_{C E}} \Big) \bigg] 
    \end{split}
\end{equation}
In this expression, the terms involving only boost-weight $0$ quantities give the Iyer-Wald entropy. The other terms, which are linear in positive boost weight quantities, are the Wall terms. We can calculate $s^A$ using the method in \cite{Bhattacharyya:2021jhr}. The expression is very lengthy and is given in Appendix \ref{app:sAcubic}.

Proceeding with the HKR algorithm, we calculate $\Delta \equiv -E_{v v} - \partial_{v}\left[\frac{1}{\sqrt{\mu}} \partial_{v}\left(\sqrt{\mu} s^{v}_{IWW}\right) + D_{A}{ s^A }\right] $ in GNCs on $\cH$, swap out any non-allowed terms using the equations of motion and then manipulate the order $l^4$ terms into the required form:
\begin{equation} \label{finalR3}
    \Delta=\partial_{v}\left[\frac{1}{\sqrt{\mu}} \partial_{v}\left(\sqrt{\mu} l^4 \varsigma^{(4)v}\right)\right] +\left(K_{A B}+l^4 X_{A B}^{(4)}\right) \left(K^{A B}+l^4 X^{(4)A B}\right) + D_{A}(l^4 Y^{(4)A}) + O(l^6)
\end{equation}
where we find
\begin{equation}
    \begin{split}
   \varsigma^{(4)v} =& k_{1} \Big(-36K^{A B} K_{A}\,^{C} \bar{K}_{B}\,^{E} \bar{K}_{C E}+6K K^{A B} \bar{K}_{A}\,^{C} \bar{K}_{B C}+6K^{A B} K_{A}\,^{C} \bar{K} \bar{K}_{B C} \Big) + \\
   & k_{2} \Big( 32K_{A}\,^{B} K_{B}\,^{C} \bar{K}_{E}\,^{F} \bar{K}_{C F} \epsilon^{A E}-8K K_{A}\,^{B} \bar{K}_{C}\,^{E} \bar{K}_{B E} \epsilon^{A C}+8K_{A}\,^{B} K^{C E} \bar{K}_{B F} \bar{K}_{C E} \epsilon^{A F}-\\
   &8K_{A}\,^{B} K_{B}\,^{C} \bar{K} \bar{K}_{C E} \epsilon^{A E} \Big)
    \end{split}
\end{equation}
and $X^{(4)A B}$ and $Y^{(4)A}$ are very lengthy expressions. The HKR entropy density is then
\begin{equation}
    s^{v}_{HKR} = s^{v}_{IWW} + l^4 \varsigma^{(4)v} 
\end{equation}
As expected (section \ref{Classify}), $\varsigma^{v}_4$ is gauge invariant since it involves only $\mu^{A B}$, $\epsilon^{A B}$, $K_{A B}$ and $\bar{K}_{A B}$. Thus, the leading order ($6$-derivative) EFT corrections to vacuum gravity in four dimensions produce a $S(v)$ that is gauge invariant on $C$. $s^v_{HKR}$ can be written in a manifestly gauge invariant way by re-instating the Riemann components in $s^v_{IWW}$: 
\begin{equation}
    \begin{split}
        s^v_{HKR} =& 1 + l^4 \Big[k_1 \Big(-6R_{r v A B} R_{r v C D} \mu^{A C} \mu^{B D}-24R_{r v r A} R_{r v v B} \mu^{A B}+12R_{r v r v} R_{r v r v}-\\
        &24 R_{r A v B} K^{A C} \bar{K}^{B}\,_{C}- 12K^{A B} K^{C D} \bar{K}_{A C} \bar{K}_{B D}+6K K^{A B} \bar{K}_{A}\,^{C} \bar{K}_{B C}+6K^{A B} K_{A}\,^{C} \bar{K} \bar{K}_{B C} \Big)+\\
        &k_2 \Big( -4R_{A B C E} R_{r v F G} \epsilon^{A B} \mu^{C F} \mu^{E G}-8R_{r A B C} R_{r v v D} \epsilon^{B C} \mu^{A D}-8R_{r v r A} R_{v B C D} \epsilon^{C D} \mu^{A B}-\\
        &16R_{r v r A} R_{r v v B} \epsilon^{A B}+16R_{r v A B} R_{r v r v} \epsilon^{A B}+16 R_{r A v B} K^{A C} \bar{K}_{C D} \epsilon^{B D} +\\
        & 16 R_{r A v B} K^{A}\,_{C} \bar{K}^{B}\,_{D} \epsilon^{C D}-16 R_{r A v B} K_{C D} \bar{K}^{B D} \epsilon^{A C} -8 K K_{A}\,^{B} \bar{K}_{C}\,^{E} \bar{K}_{B E} \epsilon^{A C}+\\
        &8K_{A}\,^{B} K^{C E} \bar{K}_{B F} \bar{K}_{C E} \epsilon^{A F}-8K_{A}\,^{B} K_{B}\,^{C} \bar{K}_{C E} \bar{K} \epsilon^{A E}+16K_{A}\,^{B} K^{C E} \bar{K}_{B E} \bar{K}_{C F} \epsilon^{A F} \Big) \Big] 
    \end{split}
\end{equation}
This satisfies the second law non-perturbatively, modulo terms of order $l^6$.

We have followed a ``strict'' definition of the IWW entropy density, in which it contains terms only of up to linear order in positive boost weight quantities. An alternative definition might allow terms of quadratic or higher order in such quantities (which do not affect the linearized second law). For example, Wall defines the entropy for the above theory to be $s^v_{IWW} + s^v_{quadratic}$, corresponding to the just the first part of \eqref{WallIWW}. Adopting this alternative definition for the IWW entropy would not affect our final answer for $s^v_{HKR}$ because adding a quantity to $s^v_{IWW}$ and then running the HKR algorithm simply results in subtracting the same quantity from $\sum_n l^n \varsigma^{(n)v}$ and so this quantity cancels out in the definition of $s^v_{HKR}$, equation \eqref{HKRdef}.

\subsection{An Example Quartic Riemann Theory}\label{QuarticCalc}

We can go beyond 6-derivative, i.e., $O(l^4)$, terms in our EFT Lagrangian in $d=4$ and consider the next order, 8-derivative terms, i.e., $O(l^6)$. To do this in the most general case, we should include all possible 8-derivative terms in the Lagrangian, up to field redefinitions and topological terms. We should also be aware that the algorithm used to construct $\varsigma^{(4)v}$ above produces higher order terms via two sources: (a) swapping non-allowed terms using the higher order equations of motion and (b) the remainder, $-l^8 X_{A B}^{(4)} X^{(4)A B}$, from completing the square in (\ref{finalR3}). However since there are no $O(l^2)$ terms in our theory, both of these sources produce terms of order at least $(l^4)^2 = l^8$. Therefore, the $O(l^6)$ terms in the entropy arise only from the 8-derivative terms in the Lagrangian. 

The minimal set of 8-derivative terms in EFT after field redefinitions and neglecting total derivatives is discussed in \cite{Endlich:2017}. Here we will restrict our attention to just one term since this is enough to demonstrate that the HKR entropy can be gauge non-invariant. Therefore let us consider adding to our $d=4$ EFT Lagrangian (\ref{4DEFT}) the following 8-derivative term
\begin{equation}
    l^6 \mathcal{L}_8 = k l^6 R_{\mu \nu \rho \sigma} R^{\mu \nu \rho \sigma} R_{\kappa \lambda \chi \eta} R^{\kappa \lambda \chi \eta} 
\end{equation}
This contributes the following to the equation of motion:
\begin{equation}
    \begin{split}
    l^6 E^{(6)}_{\mu \nu} =&- k l^6 \Big[ \frac{1}{2}R^{\chi \eta \kappa \lambda} R_{\chi \eta \kappa \lambda} R^{\alpha \beta \rho \sigma} R_{\alpha \beta \rho \sigma} g_{\mu \nu}-4R_{\mu}\,^{\eta \kappa \lambda} R_{\nu \eta \kappa \lambda} R^{\alpha \beta \rho \sigma} R_{\alpha \beta \rho \sigma}-\\
    & 4R^{\kappa \lambda \alpha \beta} R_{\kappa \lambda \alpha \beta} \nabla^{\rho}\left(\nabla^{\sigma}{R_{\mu \rho \nu \sigma}}\right)-16R^{\kappa \lambda \alpha \beta} \nabla^{\rho}{R_{\mu \rho \nu}\,^{\sigma}} \nabla_{\sigma}{R_{\kappa \lambda \alpha \beta}}-\\
    &8R_{\mu}\,^{\kappa}\,_{\nu}\,^{\lambda} R^{\alpha \beta \rho \sigma} \nabla_{\kappa}\left(\nabla_{\lambda}{R_{\alpha \beta \rho \sigma}}\right)-16R_{\mu}\,^{\kappa}\,_{\nu}\,^{\lambda} \nabla_{\kappa}{R^{\alpha \beta \rho \sigma}} \nabla_{\lambda}{R_{\alpha \beta \rho \sigma}}-\\
    &4R^{\kappa \lambda \alpha \beta} R_{\kappa \lambda \alpha \beta} \nabla^{\rho}\left(\nabla^{\sigma}{R_{\mu \sigma \nu \rho}}\right)-16R^{\kappa \lambda \alpha \beta} \nabla^{\rho}{R_{\mu}\,^{\sigma}\,_{\nu \rho}} \nabla_{\sigma}{R_{\kappa \lambda \alpha \beta}}-\\
    &8R_{\mu}\,^{\kappa}\,_{\nu}\,^{\lambda} R^{\alpha \beta \rho \sigma} \nabla_{\lambda}\left(\nabla_{\kappa}{R_{\alpha \beta \rho \sigma}}\right) \Big]
    \end{split}
\end{equation}
We can calculate the contribution to the IWW entropy current:
\begin{equation} \label{IWWR4}
    -l^6 E^{(6)}_{v v} = \partial_{v}\left[\frac{1}{\sqrt{\mu}} \partial_{v}\left(\sqrt{\mu} l^6 s^{(6)v}_{IWW}\right) + D_{A}{ \left(l^6 s^{(6)A}\right) }\right] + ...
\end{equation}
where the ellipsis denotes terms at least quadratic in positive boost weight quantities. The lengthy expression for $s^{(6)A}$ is given in Appendix \ref{app:s6A}. For $s^{(6)v}_{IWW}$ we find
\begin{equation}\label{sIWW6}
    \begin{split}
        s^{(6)v}_{IWW}=&k \Big(128R_{r v r A} R_{r v r v} R_{r v v B} \mu^{A B} -8R_{A B C E} R_{F G H I} R_{r v r v} \mu^{A F} \mu^{B G} \mu^{C H} \mu^{E I}-\\
        &64R_{r A B C} R_{r v r v} R_{v E F G} \mu^{A E} \mu^{B F} \mu^{C G}-64R_{r A v B} R_{r C v E} R_{r v r v} \mu^{A E} \mu^{B C}-\\
        &64R_{r A r B} R_{r v r v} R_{v C v E} \mu^{A C} \mu^{B E}+32R_{r v A B} R_{r v C E} R_{r v r v} \mu^{A C} \mu^{B E}-\\
        &32R_{r v r v} R_{r v r v} R_{r v r v}-8R_{A B C E} R_{F G H I} K^{J P} \bar{K}_{J P} \mu^{A F} \mu^{B G} \mu^{C H} \mu^{E I}-\\
        &64R_{r A B C} R_{v E F G} K^{H I} \bar{K}_{H I} \mu^{A E} \mu^{B F} \mu^{C G}-64R_{r A v B} R_{r C v E} K^{F G} \bar{K}_{F G} \mu^{A E} \mu^{B C}-\\
        &64R_{r A r B} R_{v C v E} K^{F G} \bar{K}_{F G} \mu^{A C} \mu^{B E}-64R_{r A r B} R_{v C v E} K^{A B} \bar{K}^{C E}+\\
        &32R_{r v A B} R_{r v C E} K^{F G} \bar{K}_{F G} \mu^{A C} \mu^{B E}+128R_{r v r A} R_{r v v B} K^{C E} \bar{K}_{C E} \mu^{A B}-\\
        &32R_{r v r v} R_{r v r v} K^{A B} \bar{K}_{A B} \Big) - s^{(6)v}_{quadratic}
\end{split}
\end{equation}
where we cancel the terms that are of quadratic or higher order in positive boost weight quantities with

\begin{equation}
\label{s6quad}
\begin{split}
        s^{(6)v}_{quadratic} =& k  \Big(-32K^{A B} K_{A B} \bar{K}^{C E} \bar{K}_{C E} \alpha-8K^{A B} K_{A B} \bar{K}^{C E} \bar{K}_{C E} \beta^{F} \beta_{F}-\\
        &32K^{A B} K^{C E} \bar{K}_{A B} \bar{K}_{C E} \alpha+64K^{A B} K_{A}\,^{C} \bar{K}_{B}\,^{E} \bar{K}_{C E} \alpha+\\
        &16K^{A B} K_{A}\,^{C} \bar{K}_{B}\,^{E} \bar{K}_{C E} \beta^{F} \beta_{F}-128K^{A B} K^{C E} \bar{K}_{A C} \bar{K}_{B E} \alpha+\\
        &64K^{A B} K_{A}\,^{C} \alpha \partial_{r}{\bar{K}_{B C}}+16K^{A B} K_{A}\,^{C} \beta^{E} \beta_{E} \partial_{r}{\bar{K}_{B C}}+32K K^{B C} \bar{K}_{B C} \bar{K} R[\mu]-\\
        &32K^{A B} K^{C E} \bar{K}_{A B} \bar{K}_{C E} R[\mu]-32K^{A B} K_{A B} K^{C E} \bar{K}_{C E} \bar{K}^{F G} \bar{K}_{F G}-\\
        &32K^{A B} K^{C E} K^{F G} \bar{K}_{A B} \bar{K}_{C E} \bar{K}_{F G}+64K^{A B} K_{A}\,^{C} K^{E F} \bar{K}_{B}\,^{G} \bar{K}_{C G} \bar{K}_{E F}-\\
        &128D^{A}{K^{B C}} D_{A}{\bar{K}_{B C}} K^{E F} \bar{K}_{E F}+128D^{A}{K^{B C}} D_{B}{\bar{K}_{A C}} K^{E F} \bar{K}_{E F}+\\
        &64D^{A}{\bar{K}^{B C}} K_{B C} K^{E F} \bar{K}_{E F} \beta_{A}-64D^{A}{\bar{K}^{B C}} K_{A B} K^{E F} \bar{K}_{E F} \beta_{C}-\\
        &64D^{A}{K^{B C}} K^{E F} \bar{K}_{B C} \bar{K}_{E F} \beta_{A}+64D^{A}{K^{B C}} K^{E F} \bar{K}_{A B} \bar{K}_{E F} \beta_{C}-\\
        &32K^{A B} K^{C E} \bar{K}_{A B} \bar{K}_{C}\,^{F} \beta_{E} \beta_{F}+128D^{A}{\beta^{B}} K_{A}\,^{C} K^{E F} \bar{K}_{B C} \bar{K}_{E F}+\\
        &128K^{A B} K^{C E} \bar{K}_{A B} \bar{K}_{C}\,^{F} \partial_{v}{\bar{K}_{E F}}-128K^{A B} K^{C E} K^{F G} \bar{K}_{A B} \bar{K}_{C F} \bar{K}_{E G}-\\
        &64K^{A B} \bar{K}_{A B} \mu^{C E} \mu^{F G} \partial_{r}{\bar{K}_{C F}} \partial_{v}{K_{E G}}+64K^{A B} K_{A}\,^{C} K^{E F} \bar{K}_{E F} \partial_{r}{\bar{K}_{B C}}+\\
        &64K^{A B} \bar{K}_{A B} \bar{K}^{C E} \bar{K}_{C}\,^{F} \partial_{v}{K_{E F}}-64K^{A B} \bar{K}^{C E} \partial_{r}{\bar{K}_{A B}} \partial_{v}{K_{C E}}+\\
        &64K^{A B} K_{A}\,^{C} K^{E F} \bar{K}_{B C} \partial_{r}{\bar{K}_{E F}}+64K^{A B} \bar{K}_{A}\,^{C} \bar{K}_{B C} \bar{K}^{E F} \partial_{v}{K_{E F}}-\\
        &64K^{A B} K_{A}\,^{C} K^{E F} \bar{K}_{B C} \bar{K}_{E}\,^{G} \bar{K}_{F G}-64D^{A}{\beta^{B}} K_{B}\,^{C} K^{E F} \bar{K}_{A C} \bar{K}_{E F}+\\
        &64K^{A B} \bar{K}_{A B} \mu^{C E} \partial_{r}{\beta_{C}} \partial_{v}{\beta_{E}}+64K^{A B} K^{C E} \bar{K}_{A B} \beta_{C} \partial_{r}{\beta_{E}}-\\
        &32K^{A B} \bar{K}_{A B} \bar{K}^{C E} \beta_{C} \partial_{v}{\beta_{E}} -32K^{A B} K^{C E} \bar{K}_{A C} \bar{K}_{B E} \beta^{F} \beta_{F}\\
        &+24K^{A B} K^{C E} \bar{K}_{A B} \bar{K}_{C E} \beta^{F} \beta_{F}\Big)
    \end{split}
\end{equation}
We shall prove that $s^{(6)v}_{IWW}$ is gauge invariant in the next section. Proceeding with the HKR algorithm once again with $l^6 \Delta^{(6)} = -l^6 E^{(6)}_{v v} - \partial_{v}\left[\frac{1}{\sqrt{\mu}} \partial_{v}\left(\sqrt{\mu} l^6 s^{(6)v}_{IWW}\right) + D_{A}{\left(l^6 s^{(6)A}\right) }\right]$, we find
\begin{equation} \label{finalR4}
    \Delta^{(6)} = \partial_{v}\left[\frac{1}{\sqrt{\mu}} \partial_{v}\left(\sqrt{\mu} \varsigma^{(6)v}\right)\right] + 2 K_{A B} X^{(6)A B} + D_{A} Y^{(6)A} + O(l^2)
\end{equation}
where

\begin{equation}
\label{varsigma6}
    \begin{split}
   \varsigma^{(6)v} =& \frac{k}{3} \Big[ - 64 K^{A B} \bar{K}^{C E} \partial_{r}{\bar{K}_{A B}} \partial_{v}{K_{C E}}+64K^{A B} \bar{K}_{A}\,^{C} \bar{K}_{B C} \bar{K}^{E F} \partial_{v}{K_{E F}} -\\
   & 64K^{A B} \bar{K}_{A B} \mu^{C E} \mu^{F G} \partial_{r}{\bar{K}_{C F}} \partial_{v}{K_{E G}}+64K^{A B} \bar{K}_{A B} \bar{K}^{C E} \bar{K}^{F}\,_{E} \partial_{v}{K_{C F}}+\\
   &96 D^{A}{D_{A}{K}} K^{B C} \partial_{r}{\bar{K}_{B C}}-96 D^{A}{D_{A}{K}} K^{B C} \bar{K}_{B}\,^{E} \bar{K}_{C E}-96 D^{A}{D_{A}{K^{B C}}} K \partial_{r}{\bar{K}_{B C}}+\\
   &96 D^{A}{D_{A}{K^{B C}}} K \bar{K}_{B}\,^{E} \bar{K}_{C E}-96 D^{A}{K^{B C}} K^{E F} \bar{K}_{B C} \bar{K}_{E F} \beta_{A}+48 D^{A}{K} K^{B C} \bar{K}_{B}\,^{E} \bar{K}_{C E} \beta_{A}-\\
   &192 D^{A}{K^{B C}} D_{A}{\bar{K}_{B C}} K^{E F} \bar{K}_{E F}+192 D^{A}{K^{B C}} D_{B}{\bar{K}_{A C}} K^{E F} \bar{K}_{E F}-48 D^{A}{K} K^{B C} \beta_{A} \partial_{r}{\bar{K}_{B C}}+\\
   &96 D^{A}{K^{B C}} K^{E F} \bar{K}_{A C} \bar{K}_{E F} \beta_{B}-96 D^{A}{K_{A}\,^{B}} K^{C E} \bar{K}_{B F} \bar{K}_{C E} \beta^{F}+96 D^{A}{K} K^{B C} \bar{K}_{A E} \bar{K}_{B C} \beta^{E}-\\
   &48 D^{A}{K^{B C}} K \beta_{A} \partial_{r}{\bar{K}_{B C}}+48 D^{A}{K^{B C}} K \bar{K}_{B}\,^{E} \bar{K}_{C E} \beta_{A}-192 D^{A}{K_{A}\,^{B}} D^{C}{\bar{K}_{B C}} K^{E F} \bar{K}_{E F}+\\
   &192 D^{A}{K_{A}\,^{B}} D_{B}{\bar{K}} K^{C E} \bar{K}_{C E}+96 D^{A}{K_{A}\,^{B}} K^{C E} \bar{K} \bar{K}_{C E} \beta_{B}+192 D^{A}{K} D^{B}{\bar{K}_{A B}} K^{C E} \bar{K}_{C E}-\\
   &192 D^{A}{K} D_{A}{\bar{K}} K^{B C} \bar{K}_{B C}-96 D^{A}{K} K^{B C} \bar{K} \bar{K}_{B C} \beta_{A}+96 D^{A}{K_{A}\,^{B}} D_{B}\left(\partial_{r}{\bar{K}_{C E}}\right) K^{C E}-\\
   &192 D^{A}{K_{A}\,^{B}} D_{B}{\bar{K}_{C E}} K^{F C} \bar{K}_{F}\,^{E}+96 D^{A}{K} D_{A}\left(\partial_{r}{\bar{K}_{B C}}\right) K^{B C}-192 D^{A}{K} D_{A}{\bar{K}_{B C}} K^{E B} \bar{K}_{E}\,^{C}+\\
   &96 D_{A}{K^{B C}} D_{E}\left(\partial_{r}{\bar{K}_{B C}}\right) K^{A E}-192 D_{A}{K^{B C}} D_{E}{\bar{K}_{C F}} K^{A E} \bar{K}_{B}\,^{F}-48 D_{A}{K^{B C}} K^{A E} \beta_{E} \partial_{r}{\bar{K}_{B C}}-\\
   &48 D^{A}{K_{A}\,^{B}} K^{C E} \beta_{B} \partial_{r}{\bar{K}_{C E}}+48 D_{A}{K^{B C}} K^{A E} \bar{K}_{B}\,^{F} \bar{K}_{C F} \beta_{E}+48 D^{A}{K_{A}\,^{B}} K^{C E} \bar{K}_{C}\,^{F} \bar{K}_{E F} \beta_{B}-\\
   &96 D^{A}{K^{B C}} D_{A}\left(\partial_{r}{\bar{K}_{B C}}\right) K+192 D^{A}{K^{B C}} D_{A}{\bar{K}_{C E}} K \bar{K}_{B}\,^{E} - 16 K K^{A B} R[\mu] \partial_{r}{\bar{K}_{A B}} -\\
   &16 D_{A}{\beta_{B}} K^{A B} K^{C E} \partial_{r}{\bar{K}_{C E}}+16 K^{A B} K^{C E} \beta_{A} \beta_{B} \partial_{r}{\bar{K}_{C E}}+32 K K^{A B} \Lambda \partial_{r}{\bar{K}_{A B}}+\\
   &16 K K^{A B} \bar{K}_{A}\,^{C} \bar{K}_{B C} R[\mu]+16 D_{A}{\beta_{B}} K^{A B} K^{C E} \bar{K}_{C}\,^{F} \bar{K}_{E F} - 16 K^{A B} K^{C E} \bar{K}_{A}\,^{F} \bar{K}_{B F} \beta_{C} \beta_{E} -\\
   &32 K K^{A B} \bar{K}_{A}\,^{C} \bar{K}_{B C} \Lambda-48 D^{A}{\beta_{A}} K K^{B C} \partial_{r}{\bar{K}_{B C}}+48 D^{A}{\beta_{A}} K K^{B C} \bar{K}_{B}\,^{E} \bar{K}_{C E}+\\
   &216 K^{A B} K^{C E} \bar{K}_{A B} \bar{K}_{C E} R[\mu]-60 K K^{A B} \bar{K} \bar{K}_{A B} R[\mu]+48 D_{A}{\bar{K}_{B C}} K^{B C} K^{E F} \bar{K}_{E F} \beta^{A} -\\
   &64 D_{A}{\bar{K}_{B C}} K^{A B} K^{E F} \bar{K}_{E F} \beta^{C}+28 K^{A B} K^{C E} \bar{K}_{A B} \bar{K}_{C E} \beta^{F} \beta_{F}-48 K K^{A B} \bar{K}_{A B} \bar{K}_{C E} \beta^{C} \beta^{E} -\\
   &112 D^{A}{\beta_{B}} K^{B C} K^{E F} \bar{K}_{A C} \bar{K}_{E F}+112 D_{A}{\beta^{B}} K^{A C} K^{E F} \bar{K}_{B C} \bar{K}_{E F}+8 D_{A}{\beta_{B}} K^{A B} K^{C E} \bar{K} \bar{K}_{C E} -\\
   &304 K^{A B} K^{C E} \bar{K}_{A B} \bar{K}_{C E} \Lambda+104 K K^{A B} \bar{K} \bar{K}_{A B} \Lambda-96 D^{A}{\bar{K}_{A B}} K K^{C E} \bar{K}_{C E} \beta^{B}+\\
   &96 D_{A}{\bar{K}} K K^{B C} \bar{K}_{B C} \beta^{A}+48 K K^{A B} \bar{K} \bar{K}_{A B} \beta^{C} \beta_{C}+64 D^{A}{\bar{K}_{A B}} K^{C B} K^{E F} \bar{K}_{E F} \beta_{C}-\\
   &48 D_{A}{\bar{K}} K^{B A} K^{C E} \bar{K}_{C E} \beta_{B} - 28 K^{A B} K^{C E} \bar{K} \bar{K}_{A B} \beta_{C} \beta_{E} - 16 D^{A}{D_{A}{\bar{K}_{B C}}} K^{B C} K^{E F} \bar{K}_{E F}+\\
   &16 D^{A}{D_{B}{\bar{K}_{A C}}} K^{B C} K^{E F} \bar{K}_{E F} - 8 D^{A}{\beta_{A}} K^{B C} K^{E F} \bar{K}_{B C} \bar{K}_{E F} - 16 D_{A}{D^{B}{\bar{K}_{B C}}} K^{A C} K^{E F} \bar{K}_{E F}+\\
   &16 D_{A}{D_{B}{\bar{K}}} K^{A B} K^{C E} \bar{K}_{C E}-48 D_{A}\left(\partial_{r}{\bar{K}_{B C}}\right) K K^{B C} \beta^{A}+96 D_{A}{\bar{K}_{B C}} K K^{E B} \bar{K}_{E}\,^{C} \beta^{A}-\\
   &96 K^{A B} K_{A}\,^{C} R[\mu] \partial_{r}{\bar{K}_{B C}}+96 D_{A}{D_{B}\left(\partial_{r}{\bar{K}_{C E}}\right)} K^{A B} K^{C E}-192 D_{A}{D_{B}{\bar{K}_{C E}}} K^{A B} K^{F C} \bar{K}_{F}\,^{E}-\\
   &192 D_{A}{\bar{K}_{B}\,^{C}} D_{E}{\bar{K}_{C F}} K^{A E} K^{B F}+192 K^{A B} K^{C E} \bar{K}_{A C} \bar{K}_{B E} R[\mu]+48 K^{A B} K_{A B} \bar{K}^{C E} \bar{K}_{C E} R[\mu]-\\
   &96 K^{A B} K_{A}\,^{C} \bar{K}_{B}\,^{E} \bar{K}_{C E} R[\mu]-48 D_{A}\left(\partial_{r}{\bar{K}_{B C}}\right) K^{B C} K^{E A} \beta_{E}+96 D_{A}{\bar{K}_{B C}} K^{E A} K^{F B} \bar{K}_{F}\,^{C} \beta_{E} \Big]
    \end{split}
\end{equation}
and $X^{(6)A B}$ and $Y^{(6)A}$ are (also) very lengthy expressions. $S(v)$ for this 8-derivative theory is given by adding $s^{(6)v}_{IWW}+l^6 \varsigma^{(6)v}$ to the result for the general 6-derivative theory in the previous section. The main reason for performing this calculation is to investigate whether or not the result is gauge invariant. This will be done in the next section.

\section{Gauge (Non-)Invariance} \label{GaugeInvariance}

We now, at last, concern ourselves with how $S(v)$ changes at $v=0$ under the rescaling of the affine parameter on each generator $v'=v/a(x^A)$ with $a(x^A)>0$.
The transformation laws of tensorial components and all allowed GNC quantities on $C$ are given in Section \ref{GaugeInvarianceLaws}. Most terms transform homogeneously: $T^{\mu_1 ... \mu_n}_{\nu_1 ... \nu_m}, \mu_{A B}, \epsilon_{A B}, R_{A B C D}[\mu], D_A, \partial_{v}^p K_{A B}$ and $\partial_{r}^p \Bar{K}_{A B}$ just gain a factor of $a^b$, where $b$ is their boost weight. However, $\beta_A$ transforms as
\begin{equation}
    \beta'_{A}= \beta_A+2D_{A} \log a
\end{equation}
and so the presence of $\beta_A$ is a warning sign of gauge non-invariance. Also note that a quantity such as $D_{A}{ K_{B C}}$ transforms as
\begin{equation}
\begin{split}
    D'_{A}{ K'_{B C} } =& D_{A}(a K_{B C})\\
    =& a (D_A K_{B C} + D_{A}( \log a ) K_{B C})
\end{split}
\end{equation}
which is also inhomogeneous. This will be the case for all $D$ derivatives of $\partial_{v}^p K_{A B}$ or $\partial_{r}^p \Bar{K}_{A B}$. We can get round this by swapping $D_A$ derivatives for gauge covariant derivatives $\mathfrak{D}_A$ which we define by \cite{Hollands:2022}
\begin{equation}
    \mathfrak{D}_A T = D_A T -(b/2) \beta_A T
\end{equation}
for $T=\mathfrak{D}_{A_1}...\mathfrak{D}_{A_n} \partial_{v}^p K_{A B}$ or $\mathfrak{D}_{A_1}...\mathfrak{D}_{A_n} \partial_{r}^p \Bar{K}_{A B}$ with boost weight $b$. This can be shown to transform homogeneously as $\mathfrak{D}'_A T'= a^b \mathfrak{D}_A T$ on $C$. 

\subsection{IWW entropy} \label{IWWGaugeInvariance}
We shall start by discussing how the IWW entropy behaves under a gauge transformation. For EGB theory, the IWW entropy is the same as the Iyer-Wald entropy, which is manifestly gauge invariant. For the cubic and quartic Riemann Lagrangians, the IWW entropy is determined by equation \eqref{WallIWW}. The first part of this equation is manifestly gauge invariant. Hence $s^v_{IWW} + s^v_{quadratic}$ must be gauge invariant, as confirmed by equations (\ref{evenodd}) and (\ref{sIWW6}). For the cubic theories we found that $s^v_{quadratic}$ is also gauge invariant, and hence so is $s^v_{IWW}$. The same is true for the quartic theory considered above. This can be shown as follows. In $s^{(6)v}_{quadratic}$ (equation \eqref{s6quad}) we swap out non-allowed terms for allowed terms plus Ricci components using the elimination rules in Appendix A. Here we do not eliminate the Ricci components using equations of motion, we are simply working with Ricci components since, as tensor components, they transform homogeneously under gauge transformations. We also swap $D$ derivatives for $\mathfrak{D}$ derivatives on $K_{A B}$ and $\Bar{K}_{A B}$ terms. Of the terms that are left, the only ones that can transform inhomogeneously under a gauge transformation are those involving $D_{A_1}...D_{A_2} \beta_A$. We find that
\begin{equation}
    s^{(6)v}_{quadratic} = 192 k K^{A B} K^{C E} \bar{K}_{A B} \Bar{K}_{C}\,^{F} D_{[E}{\beta_{F]}}+ \text{terms independent of $\beta_A$}
\end{equation}
But $D_{[A} \beta_{B]}$ is clearly gauge invariant (it is the ``field strength'' of the connection $\beta_A$). Hence $s^{(6)v}_{quadratic}$ is gauge invariant. In summary, for both the cubic and quartic Lagrangians, we have found that $s^v_{quadratic}$, and hence also $s^v_{IWW}$, is gauge invariant. 

This is puzzling. HKR proved that, without modification, the IWW entropy is gauge invariant to linear order in perturbation theory. Since the IWW entropy is of at most linear order in positive boost weight quantities, one might think that this automatically implies that it is gauge invariant to all orders. That this is not true can be seen as follows. The easiest way to perform a gauge transformation of an expression involving (derivatives of) $\alpha,\beta_A$ etc is to rewrite it in terms of a new set of quantities (e.g. Ricci components) that transform nicely under gauge transformations. In general, this rewriting does not preserve the property of being ``linear in positive boost weight quantities.'' (For example the linear term $D_A \partial_v \beta_B$ can be eliminated in favour of $\nabla_A R_{vB}$ but this introduces nonlinear terms of the schematic form $K^2 \bar{K}$.) Hence after applying a gauge transformation, when we transform back to the original set of quantities, the difference $s^{v'}_{IWW}-s^v_{IWW}$ will involve not just terms of linear order in positive boost weight quantities, but also possibly terms of higher order. The former must vanish by the linear argument of \cite{Hollands:2022} but the latter may not. This problem is what the ``improvement'' terms of \cite{Hollands:2022} are designed to fix. But surprisingly, in the cases we have studied, such terms are not required, and $s^v_{IWW}$ is gauge invariant without improvement. 

To solve this puzzle, we shall now show that this result holds for any Lagrangian that depends only on the Riemann tensor (and not its derivatives). For such a Lagrangian, $s^v_{IWW}$ is given by \eqref{WallIWW}, which involves the Riemann tensor but not its derivatives. Since the Riemann tensor has dimension $2$, this implies that $s^v_{IWW}$ depends only on quantities with dimension of $2$ or less. Any such quantity is built from  ``primitive factors'' (in the terminology of \cite{Hollands:2022}) belonging to one of the following sets, where the subscript refers to the boost weight: $S_2=\{\partial_v K_{AB} \}$, $S_{-2} = \{ \partial_r \bar{K}_{AB} \}$, $S_1 = \{D_A K_{BC}, K_{AB}, \partial_v \beta_A\}$, $S_{-1}=\{ D_A \bar{K}_{BC}, \bar{K}_{AB}, \partial_r \beta_A \}$ and 
\begin{equation}
S_0=\{ \mu_{AB},\mu^{AB},\epsilon_{AB},\alpha,\beta_A, D_A \beta_B,R[\mu]_{ABCD}, \partial_v \partial_r \mu_{AB} \}.
\end{equation}
Furthermore, $s^v_{IWW}$ depends at most linearly on elements of $S_2$ and $S_1$. We can write $s^v_{IWW} =s^{v}_0 + s^{v}_1$ where $s^{v}_0$ is built only from terms of zero boost weight and $s^{v}_1$ is a sum of terms, each containing exactly one primitive factor of positive boost weight. Now, $s^{v}_{0}$ is simply the Iyer-Wald entropy density, which is gauge-invariant by definition. So we just need to understand how $s^{v}_{1}$ transforms. 

Using a formula from Appendix \ref{app:elimination}, $\partial_v \beta_A$ can be eliminated in favour of $R_{vA}$ and other elements of the above sets. Importantly, $\partial_v \beta_A$ depends linearly on $R_{vA}$ and other quantities with positive boost weight so when we eliminate it, we do not introduce any nonlinear dependence on positive boost weight quantities (unlike what happens for a term like $D_A \partial_v \beta_B$, mentioned above, which might arise from a more general Lagrangian involving derivatives of the Riemann tensor). Similarly we can eliminate $\partial_r \beta_A$ in favour of $R_{rA}$ and other quantities listed above. We can also eliminate $D_A K_{BC}$ and $D_A \bar{K}_{BC}$ in favour of $\mathfrak{D}_A K_{BC}$ and $\mathfrak{D}_A \bar{K}_{BC}$ respectively. The result is to replace $S_1$ and $S_{-1}$ with $S'_1 = \{\mathfrak{D}_A K_{BC}, K_{AB}, R_{vA}\}$ and $S_{-1}'=\{ \mathfrak{D}_A \bar{K}_{BC}, \bar{K}_{AB}, R_{rA} \}$ respectively. So now $s^v_1$ is a sum of terms built from quantities belonging to $S_{\pm 2}$, $S_{\pm 1}'$ and $S_0$, and each term in $s^v_1$ contains exactly one element of $S_2$ or $S_1'$. We can write 
 \begin{equation}
 s^{v}_{1}=\sum_i P_i N_i Z_i
 \end{equation}
 where $P_i$ is an element of $S_2$ or $S_1'$, with boost weight $b_i \in \{2,1\}$, $N_i$ has boost weight $-b_i$ and is either an element of $S_{-2}$ or $S_{-1}'$ or a product of two elements in $S_{-1}'$. $Z_i$ has boost weight $0$ and is a product of elements of $S_0$. Now $P_i$ and $N_i$ are quantities that transform homogeneously under a gauge transformation and $P_i N_i$ has boost weight zero so it is gauge invariant. Hence, under a gauge transformation the change in $s^v_1$ is
 \begin{equation}
 \label{Deltas}
 \Delta s^{v}_{1}=\sum_i P_i N_i \Delta Z_i
 \end{equation}
 where $\Delta Z_i$ is the change in $Z_i$, which arises from the dependence of $Z_i$ on the quantities $\alpha,\beta_A,D_A \beta_B$ and $\partial_v \partial_r \mu_{AB}$ which do not transform homogeneously under a gauge transformation. Importantly, the transformation laws for these quantities involve only other quantities of boost weight $0$ and not, say, quantities like $K \bar{K}$. This is obvious for $\beta_A$ and $D_A \beta_B$. We can write $\partial_v \partial_r \mu_{AB}$ in terms of $R_{AB}$ (Appendix \ref{app:elimination}) to deduce how it transforms. The result is that $\Delta (\partial_v \partial_r \mu_{AB})$ depends only on the first and second derivatives of $\log a$, and on $\beta_A$. By writing $\alpha$ in terms of $R_{vr}$ and $R_{AB}$ one sees that the same is true for\footnote{
An alternative way of obtaining these results is to observe that $\Delta (\partial_v \partial_r \mu_{AB})$ and $\Delta \alpha$ are of boost weight $0$ and dimension $2$ and are a sum of terms, each of which involves at least one factor of a (first or second) derivative of $\log a$. A derivative of $\log a$ has boost weight $0$ and dimension at least $1$, and so must multiply a boost weight $0$ term of dimension at most $1$. A term with boost weight $0$ that contains primitive factors with non-zero boost weight must have dimension at least $2$. So primitive factors of non-zero boost weight cannot appear in these quantities.
} $\Delta \alpha$. Thus $\Delta Z_i$ depends only on elements of $S_0$ and on the first and second derivatives of $\log a$. 

We can now rewrite $P_i$ and $N_i$ of \eqref{Deltas} in terms of our original basis, i.e., in terms of $S_{\pm 1}$ instead of $S_{\pm 1}'$. Recall that this does not spoil the property of having exactly one primitive factor of positive boost weight. This rewriting may generate extra factors (e.g. $\beta_A$) belonging to $S_0$. The result is that we have shown
\begin{equation}
 \label{Deltas2}
 \Delta s^{v}_{1}=\sum_i \hat{P}_i \hat{N}_i \hat{Z}_i
 \end{equation}
where $\hat{P}_i$ is an element of $S_2$ or $S_1$, with boost weight $b_i$,
$\hat{N}_i$ has boost weight $-b_i$ and is either an element of $S_{-2}$ or $S_{-1}$ or a product of two elements of $S_{-1}$, and $\hat{Z}_i$ depends only on elements of $S_0$ and on the first and second derivatives of $\log a$. 

Given an arbitrary dynamical black hole with metric $g_{\mu\nu}$, Ref. \cite{Hollands:2022} explains (section 3.3) how to construct a ``background'' black hole metric $\tilde{g}_{\mu\nu}$ (not necessarily satisfying any equations of motion) such that, on $C$, all background quantities of positive boost weight vanish whereas background quantities of non-positive boost weight agree with those of $g_{\mu\nu}$. Let $\delta g_{\mu\nu} = g_{\mu\nu}- \tilde{g}_{\mu\nu}$, for which all quantities of non-positive boost weight vanish on $C$. Consider the $1$-parameter family of metrics $g_{\mu\nu}(\lambda) = \tilde{g}_{\mu\nu} + \lambda \delta g_{\mu\nu}$, for which all quantities of non-positive boost weight agree with the corresponding quantities of $\tilde{g}_{\mu\nu}$ and $g_{\mu\nu}$ on $C$ and hence $\hat{N}_i[g(\lambda)]=\hat{N}_i[\tilde{g}]=N_i[g]$ and $\hat{Z}_i[g(\lambda)]=\hat{Z}_i[\tilde{g}]=Z_i[g]$. Since $\hat{P}_i$ is an element of $S_2$ or $S_1$ we have $\hat{P}_i[g(\lambda)] \propto \lambda$, i.e., the linear approximation to $\hat{P}_i[g(\lambda)]$ is exact: there are no terms of order $\lambda^2$ or higher. Hence we have $\hat{P}_i[g(\lambda)] \hat{N}_i[g(\lambda)]\hat{Z}_i[g(\lambda)] \propto \lambda$ so $\Delta s^v_1[g(\lambda)] \propto \lambda$. However, Ref. \cite{Hollands:2022} proved that the IWW entropy is gauge invariant to linear order in perturbations around any background solution defined as above, i.e., $\Delta s^v_1[g(\lambda)] =O(\lambda^2)$. Combining these results we have $\Delta s^v_1[g(\lambda)]=0$. Setting $\lambda=1$ we obtain $\Delta s^v_1[g]=0$ and we have proved that the IWW entropy is non-perturbatively gauge invariant for this class of theories.

\subsection{HKR entropy}

Now we shall discuss gauge invariance of the HKR entropy. In Section \ref{Classify}, we explained why the HKR entropy density $s^v_{HKR}$ must be gauge invariant up to and including order $l^4$ terms, simply because there are no gauge-non-invariant terms that can appear at this order. This is confirmed by our calculations for the EGB and cubic Lagrangians. We shall now discuss the quantities calculated at order $l^6$ for the quartic Lagrangian above. As just discussed, the IWW part of the entropy is gauge invariant so we just need to discuss the transformation of the quantity $\varsigma^{(6)v}$ given in equation \eqref{varsigma6}. This quantity is made out of  allowed terms, so we already know how all the terms transform. We again swap all $D$ derivatives for $\mathfrak{D}$ derivatives on $K_{A B}$ and $\Bar{K}_{A B}$ terms. We then find 
\begin{equation}
    \begin{split}
        \varsigma^{(6)v} =& \frac{k}{3} \bigg[ D_{(A}{\beta_{B)}} \Big(-48K K^{H I} \partial_{r}{\bar{K}_{H I}} \mu^{A B}+48K K^{H J} \bar{K}_{H}\,^{P} \bar{K}_{J P} \mu^{A B}-112K^{A B} K^{H I} \partial_{r}{\bar{K}_{H I}}+\\
        &112K^{A B} K^{H J} \bar{K}_{H}\,^{P} \bar{K}_{J P}\Big)+240D_{[A}{\beta_{B]}}K^{A I} K^{J P} \bar{K}^{B}\,_{I} \bar{K}_{J P}+\\
        &\beta_{A} \Big(-48K^{H I} \mathfrak{D}^{A}{K} \partial_{r}{\bar{K}_{H I}}+48K^{H J} \mathfrak{D}^{A}{K} \bar{K}_{H}\,^{P} \bar{K}_{J P}-48K \mathfrak{D}^{A}{K^{H I}} \partial_{r}{\bar{K}_{H I}}+\\
        &48K \mathfrak{D}^{A}{K^{H J}} \bar{K}_{H}\,^{P} \bar{K}_{J P}-32K^{G H} K^{J P} \bar{K}_{G H} \mathfrak{D}^{A}{\bar{K}_{J P}}+32K^{G H} K^{I P} \bar{K}_{G H} \mathfrak{D}_{I}{\bar{K}^{A}\,_{P}}-\\
        &32K^{A J} K^{G H} \bar{K}_{G H} \mathfrak{D}^{P}{\bar{K}_{J P}}+32K^{A I} K^{G H} \bar{K}_{G H} \mathfrak{D}_{I}{\bar{K}}-144K^{H I} \mathfrak{D}^{G}{K^{A}\,_{G}} \partial_{r}{\bar{K}_{H I}}-\\
        &144K^{A G} K^{H I} \mathfrak{D}_{G}\left(\partial_{r}{\bar{K}_{H I}}\right)+144K^{H J} \mathfrak{D}^{G}{K^{A}\,_{G}} \bar{K}_{H}\,^{P} \bar{K}_{J P}+288K^{A I} K^{G J} \bar{K}_{G}\,^{P} \mathfrak{D}_{I}{\bar{K}_{J P}}-\\
        &48K K^{H I} \mathfrak{D}^{A}\left(\partial_{r}{\bar{K}_{H I}}\right)+96K K^{G J} \bar{K}_{G}\,^{P} \mathfrak{D}^{A}{\bar{K}_{J P}}-144K^{A E} \mathfrak{D}_{E}{K^{H I}} \partial_{r}{\bar{K}_{H I}}+\\
        &144K^{A E} \mathfrak{D}_{E}{K^{H J}} \bar{K}_{H}\,^{P} \bar{K}_{J P}\Big)+\\
        &\beta_{A} \beta_{B} \left(16K^{A B} K^{H I} \partial_{r}{\bar{K}_{H I}}-16K^{A B} K^{H J} \bar{K}_{H}\,^{P} \bar{K}_{J P}\right) \bigg] + \text{homogeneous terms}
    \end{split}
\end{equation}
The $\beta_A$-dependence strongly suggests that this is not gauge invariant. To confirm this, we apply a gauge transformation and find
\begin{equation}
    \begin{split}
        \varsigma'^{(6)v'} =& \varsigma^{(6)v} + \frac{k}{3} \bigg[ 2D_{A}D_{B} \log a \Big(-48K K^{H I} \partial_{r}{\bar{K}_{H I}} \mu^{A B}+48K K^{H J} \bar{K}_{H}\,^{P} \bar{K}_{J P} \mu^{A B}-\\
        &112K^{A B} K^{H I} \partial_{r}{\bar{K}_{H I}}+112K^{A B} K^{H J} \bar{K}_{H}\,^{P} \bar{K}_{J P}\Big)+\\
        &2D_{A} \log a \Big(-48K^{H I} \mathfrak{D}^{A}{K} \partial_{r}{\bar{K}_{H I}}+48K^{H J} \mathfrak{D}^{A}{K} \bar{K}_{H}\,^{P} \bar{K}_{J P}-48K \mathfrak{D}^{A}{K^{H I}} \partial_{r}{\bar{K}_{H I}}+\\
        &48K \mathfrak{D}^{A}{K^{H J}} \bar{K}_{H}\,^{P} \bar{K}_{J P}-32K^{G H} K^{J P} \bar{K}_{G H} \mathfrak{D}^{A}{\bar{K}_{J P}}+32K^{G H} K^{I P} \bar{K}_{G H} \mathfrak{D}_{I}{\bar{K}^{A}\,_{P}}-\\
        &32K^{A J} K^{G H} \bar{K}_{G H} \mathfrak{D}^{P}{\bar{K}_{J P}}+32K^{A I} K^{G H} \bar{K}_{G H} \mathfrak{D}_{I}{\bar{K}}-144K^{H I} \mathfrak{D}^{G}{K^{A}\,_{G}} \partial_{r}{\bar{K}_{H I}}-\\
        &144K^{A G} K^{H I} \mathfrak{D}_{G}\left(\partial_{r}{\bar{K}_{H I}}\right)+144K^{H J} \mathfrak{D}^{G}{K^{A}\,_{G}} \bar{K}_{H}\,^{P} \bar{K}_{J P}+288K^{A I} K^{G J} \bar{K}_{G}\,^{P} \mathfrak{D}_{I}{\bar{K}_{J P}}-\\
        &48K K^{H I} \mathfrak{D}^{A}\left(\partial_{r}{\bar{K}_{H I}}\right)+96K K^{G J} \bar{K}_{G}\,^{P} \mathfrak{D}^{A}{\bar{K}_{J P}}-144K^{A E} \mathfrak{D}_{E}{K^{H I}} \partial_{r}{\bar{K}_{H I}}+\\
        &144K^{A E} \mathfrak{D}_{E}{K^{H J}} \bar{K}_{H}\,^{P} \bar{K}_{J P}\Big)+\\
        &4\left( \beta_{A} D_{B} \log a + D_A \log a D_B \log a \right) \left(16K^{A B} K^{H I} \partial_{r}{\bar{K}_{H I}}-16K^{A B} K^{H J} \bar{K}_{H}\,^{P} \bar{K}_{J P}\right) \bigg]
    \end{split}
\end{equation}
The entropy of $C$ involves the above expression integrated over the horizon cross-section $C$. Integration by parts can be used to simplify the dependence on $a(x^A)$ in this integral:
\begin{equation}
    \begin{split}
        S'=&S + 4 \pi\int_C d^2 x \frac{kl^6}{3} \sqrt{\mu} \bigg[2 D_{A}{\log a} \Big(-32K^{C E} D^{B}{K^{A}\,_{B}} \partial_{r}{\bar{K}_{C E}}-\\
        &32K^{A B} D_{B}{K^{C E}} \partial_{r}{\bar{K}_{C E}}-32K^{A B} K^{C E} D_{B}\left(\partial_{r}{\bar{K}_{C E}}\right)+\\
        &32K^{C E} D^{B}{K^{A}\,_{B}} \bar{K}_{C}\,^{F} \bar{K}_{E F}+32K^{A B} D_{B}{K^{C E}} \bar{K}_{C}\,^{F} \bar{K}_{E F}+\\
        &64K^{A B} K^{C E} \bar{K}_{C}\,^{F} D_{B}{\bar{K}_{E F}}-32K^{B C} K^{E F} \bar{K}_{B C} D^{A}{\bar{K}_{E F}}-\\
        &16\beta^{A} K^{B C} K^{E F} \bar{K}_{B C} \bar{K}_{E F}+32K^{C E} K^{B F} \bar{K}_{C E} D_{B}{\bar{K}^{A}\,_{F}}+\\
        &16\beta^{B} K_{B}\,^{C} K^{E F} \bar{K}^{A}\,_{C} \bar{K}_{E F}-32K^{A C} K^{E F} \bar{K}_{E F} D^{B}{\bar{K}_{C B}}-\\
        &16\beta^{F} K^{A B} K^{C E} \bar{K}_{F B} \bar{K}_{C E}+32K^{A B} K^{C E} \bar{K}_{C E} D_{B}{\bar{K}}+\\
        &16\beta^{B} K^{A}\,_{B} K^{C E} \bar{K}_{C E} \bar{K}+32\beta^{B} K^{A}\,_{B} K^{C E} \partial_{r}{\bar{K}_{C E}}-\\
        &32\beta^{B} K^{A}\,_{B} K^{C E} \bar{K}_{C}\,^{F} \bar{K}_{E F}\Big)\\
        &+4D_{A}{\log a}D_{B}{\log a} \left(16K^{A B} K^{C E} \partial_{r}{\bar{K}_{C E}}-16 K^{A B} K^{C E} \bar{K}_{C}\,^{F} \bar{K}_{E F}\right) \bigg]
    \end{split}
\end{equation}
For gauge invariance to hold the coefficients of the terms linear and quadratic in $D_A \log a$ must vanish independently. However, these coefficients depend in a complicated way on expressions of quadratic order in positive boost weight quantities. There is no reason why they will vanish for a generic perturbation. Therefore the HKR entropy (and by extension $S(v)$) of this $8$-derivative theory is not gauge invariant at order $l^6$. 

This statement concerns non-perturbative gauge invariance. However, one could ask merely for gauge invariance up to quadratic order in perturbation theory around a stationary black hole, modulo terms of order $l^8$, in the same sense as the second law was originally formulated for the HKR entropy. Positive boost weight quantities are of at least linear order in perturbation theory so, to quadratic order, we can evaluate the negative boost weight quantities above in the unperturbed stationary black hole geometry. This might lead to extra cancellations. To investigate this, we shall focus on the $D_A \log a D_B \log a$ term above. It has coefficient proportional to $l^6 K^{AB} K^{CD} R_{rCrD}$ (using the expression for $R_{rCrD}$ in \cite{Hollands:2022}). To quadratic order, we can evaluate $R_{rCrD}$ in the stationary black hole geometry. Gauge invariance in the sense just discussed would require that $R_{rCrD}=O(l^2)$ on $C$. Using the equation of motion $R_{rr} = O(l^2)$, this gives $C_{rCrD} = O(l^2)$ on $C$, where the LHS is a component of the Weyl tensor. This is the statement that $n=\partial/\partial r$ is a principal null direction of the unperturbed black hole, modulo terms of order $l^2$. Recall that, by definition, $n$ is orthogonal to $C$. For a generic choice of $C$ there is no reason why $n$ should be close to being a principal null direction (although it can be in special cases e.g. a spherically symmetric cross-section of a spherically symmetric black hole). 
In particular, for a rotating black hole, one would expect an ``ingoing'' principal null direction to have non-zero rotation at the horizon (e.g. this is true for a Kerr black hole), whereas $n$ has vanishing rotation by definition. We conclude that, in general, the $D_A \log a D_B \log a$ term above is generically non-vanishing, of order $l^6$, to quadratic order in perturbation theory. Thus, for this $8$-derivative theory, the HKR entropy (and by extension $S(v)$) is not gauge invariant in the above sense.

\section{Discussion}\label{sec:discuss}

We have explained why our proposal for dynamical black hole entropy is gauge invariant to order $l^4$, i.e., up to and including $6$-derivative terms in the Lagrangian. But we have seen that it is not gauge invariant at order $l^6$ for a specific $8$-derivative term in the Lagrangian. It is conceivable that if we allowed all possible $8$-derivative terms then demanding gauge invariance might lead to non-trivial relations between their coefficients, i.e., it might function as a selection rule for such theories. However we think this is unlikely, and that the lack of gauge invariance is a flaw of the HKR prescription. Nevertheless, we repeat that our proposal {\it is} gauge invariant for terms with up to $6$ derivatives in the Lagrangian, i.e., for the leading order EFT corrections to 4d vacuum gravity (and next to leading order for higher dimensional gravity). It is only for the next to leading order corrections that the problem arises so this is probably not a serious issue for practical applications. 

It is also worth noting that, as discussed in Section \ref{EFTValidityDef}, the concept of regime of validity of EFT is not covariant to start with. We assumed the solution lies in the regime of validity in some particular choice of affinely parameterized GNCs, but one can always make a rescaling $v'=v/a(x^A)$ on each generator that takes the solution out of the regime of validity. In these choices, we should not expect a second law to hold, and thus it is perhaps too much to hope that a definition of dynamical black hole entropy should be invariant under all possible rescalings. However, it seems reasonable to hope it is invariant under rescalings that preserve the regime of validity, which seems unlikely from the above.


\section{Appendix}
\subsection{Elimination Rules for Non-allowed Terms} \label{app:elimination}

In our affinely parameterized GNC expansion of $\Delta\equiv -E_{v v} - \partial_{v}\left[\frac{1}{\sqrt{\mu}} \partial_{v}\left(\sqrt{\mu} s^{v}_{IWW}\right) + D_{A}{ s^A }\right]$ on $\cH$, we want to reduce the set of terms that appear up to $O(l^N)$ to the set of "allowed terms" given in (\ref{AllowedTerms}). To eliminate non-allowed terms we study Ricci components and their covariant derivatives evaluated on $\cH$. For the EGB, cubic and quartic Lagrangians above, we need the following:

\begin{equation}
    \begin{split}
        \partial_{v}{\beta_{A}} =& -2D^{B}{K_{A B}}+2D_{A}{K}-K \beta_{A}+2R_{v A}\\
        \partial_{r}{\beta_{A}} =& D^{B}{\bar{K}_{A B}}-D_{A}{\bar{K}}+\bar{K}_{A}\,^{B} \beta_{B} - \frac{1}{2}\bar{K} \beta_{A}-R_{r A}\\
        \partial_{v}{\bar{K}_{A B}} =& \frac{1}{2}R_{A B}[\mu]+K_{B}\,^{C} \bar{K}_{A C}+K_{A}\,^{C} \bar{K}_{B C} - \frac{1}{2}K \bar{K}_{A B} - \frac{1}{2}K_{A B} \bar{K} -\\
        &\frac{1}{4}D_{B}{\beta_{A}} - \frac{1}{4}\beta_{A} \beta_{B} - \frac{1}{4}D_{A}{\beta_{B}} - \frac{1}{2}R_{A B}\\
        \alpha =&  - \frac{1}{2}D^{A}{\beta_{A}}-\mu^{A B} \partial_{v}{\bar{K}_{A B}}+K^{A B} \bar{K}_{A B} - \frac{1}{2}\beta^{A} \beta_{A}-R_{r v}\\
        \end{split}
\end{equation}
\begin{equation}
    \begin{split}
        \partial_{r r}{\beta_{A}} =& \frac{2}{3}D^{B}\left(\partial_{r}{\bar{K}_{A B}}\right) - \frac{4}{3}D^{B}{\bar{K}_{A}\,^{C}} \bar{K}_{B C}+2D_{A}{\bar{K}^{B C}} \bar{K}_{B C}-2D^{B}{\bar{K}_{B}\,^{C}} \bar{K}_{A C}+\frac{4}{3}D^{B}{\bar{K}} \bar{K}_{A B} - \\
        &\frac{2}{3}D_{A}\left(\partial_{r}{\bar{K}_{B C}}\right) \mu^{B C} - \frac{2}{3}\bar{K} \partial_{r}{\beta_{A}}+\bar{K} \bar{K}_{A}\,^{B} \beta_{B} - \frac{2}{3}\beta_{A} \mu^{B C} \partial_{r}{\bar{K}_{B C}} - \frac{10}{3}\bar{K}_{A}\,^{B} \bar{K}_{B}\,^{C} \beta_{C}+\\
        &\frac{4}{3}\beta^{B} \partial_{r}{\bar{K}_{A B}}+2\bar{K}_{A}\,^{B} \partial_{r}{\beta_{B}}+\bar{K}^{B C} \bar{K}_{B C} \beta_{A}-\nabla_{r}{R_{A r}}
    \end{split}
\end{equation}
\begin{equation}
    \begin{split}
        \partial_{r v}{\bar{K}_{A B}} =& \frac{1}{2}D_{B}{D^{C}{\bar{K}_{A C}}} - \frac{1}{2}D_{B}{D_{A}{\bar{K}}} - \frac{1}{2}D^{C}{D_{C}{\bar{K}_{A B}}}+\frac{1}{2}D^{C}{D_{A}{\bar{K}_{B C}}}+\frac{3}{4}D^{C}{\beta_{A}} \bar{K}_{B C} -\\
        &\frac{1}{4}D_{B}{\beta_{A}} \bar{K}+\frac{1}{4}D_{B}{\beta^{C}} \bar{K}_{A C}+\frac{3}{4}D^{C}{\beta_{B}} \bar{K}_{A C}+2\bar{K}_{A}\,^{C} \partial_{v}{\bar{K}_{B C}}+K_{B}\,^{C} \partial_{r}{\bar{K}_{A C}}+\\
        &2\bar{K}_{B}\,^{C} \partial_{v}{\bar{K}_{A C}}+K_{A}\,^{C} \partial_{r}{\bar{K}_{B C}}+\frac{1}{4}D_{A}{\beta^{C}} \bar{K}_{B C} - \frac{1}{2}D^{C}{\beta_{C}} \bar{K}_{A B} - \frac{1}{2}\bar{K}_{A B} \mu^{C E} \partial_{v}{\bar{K}_{C E}} -\\
        &\frac{1}{2}K \partial_{r}{\bar{K}_{A B}} - \frac{1}{2}\bar{K} \partial_{v}{\bar{K}_{A B}} - \frac{1}{2}K_{A B} \mu^{C E} \partial_{r}{\bar{K}_{C E}} - \frac{1}{4}D_{A}{\beta_{B}} \bar{K}+\frac{1}{2}K_{B}\,^{C} \bar{K} \bar{K}_{A C}+\\
        &K \bar{K}_{A}\,^{C} \bar{K}_{B C}-2K^{C E} \bar{K}_{A C} \bar{K}_{B E}-3K_{B}\,^{C} \bar{K}_{A}\,^{E} \bar{K}_{C E}+\frac{1}{4}D^{C}{\bar{K}_{B C}} \beta_{A} - \frac{1}{4}D_{B}{\bar{K}} \beta_{A}+\\
        &\bar{K}_{B}\,^{C} \beta_{A} \beta_{C} - \frac{1}{4}\bar{K} \beta_{A} \beta_{B}-3K_{A}\,^{C} \bar{K}_{B}\,^{E} \bar{K}_{C E}+K^{C E} \bar{K}_{A B} \bar{K}_{C E}+K_{A B} \bar{K}^{C E} \bar{K}_{C E} -\\
        &\frac{3}{2}D^{C}{\bar{K}_{A B}} \beta_{C}+D_{B}{\bar{K}_{A}\,^{C}} \beta_{C}-\bar{K}_{A B} \beta^{C} \beta_{C}+\bar{K}_{A}\,^{C} \beta_{B} \beta_{C} - \frac{1}{2}\bar{K}_{B}\,^{C} R_{A C}[\mu]+\\
        &\frac{1}{2}K_{A}\,^{C} \bar{K} \bar{K}_{B C}+\frac{1}{4}D^{C}{\bar{K}_{A C}} \beta_{B} - \frac{1}{4}D_{A}{\bar{K}} \beta_{B} - \frac{1}{2}\bar{K}^{C E} R_{A C B E}[\mu]+D_{A}{\bar{K}_{B}\,^{C}} \beta_{C} -\\
        &\frac{1}{2}D_{B}\left(\partial_{r}{\beta_{A}}\right)-\bar{K}_{A B} \alpha - \frac{3}{4}\beta_{A} \partial_{r}{\beta_{B}} - \frac{3}{4}\beta_{B} \partial_{r}{\beta_{A}} - \frac{1}{2}D_{A}\left(\partial_{r}{\beta_{B}}\right) - \frac{1}{2}\nabla_{r}{R_{A B}}\\
    \end{split}
\end{equation}
\begin{equation}
    \begin{split}
        \partial_{r}{\alpha} =& \frac{1}{3}D^{A}{\beta^{B}} \bar{K}_{A B} - \frac{1}{4}\bar{K} \beta^{A} \beta_{A} - \frac{1}{3}D^{A}\left(\partial_{r}{\beta_{A}}\right)+\frac{1}{2}D^{A}{\bar{K}_{A}\,^{B}} \beta_{B} - \frac{1}{3}D^{A}{\bar{K}} \beta_{A} - \frac{1}{3}\bar{K} \alpha -\\
        &\frac{1}{3}\mu^{A B} \partial_{r v}{\bar{K}_{A B}}+\bar{K}^{A B} \partial_{v}{\bar{K}_{A B}}+\frac{1}{3}K^{A B} \partial_{r}{\bar{K}_{A B}} - \frac{4}{3}K^{A B} \bar{K}_{A}\,^{C} \bar{K}_{B C}+\frac{5}{6}\bar{K}^{A B} \beta_{A} \beta_{B} -\\
        &\frac{7}{6}\beta^{A} \partial_{r}{\beta_{A}}+\frac{1}{3}\nabla_{r}{R_{v r}}
    \end{split}
\end{equation}
We also need elimination rules for $\partial_{v v}{\beta_{A}}, \partial_{v r}{\beta_{A}}, \partial_{v v v}{\beta_{A}}, \partial_{v v r}{\beta_{A}}, \partial_{v v}{\Bar{K}_{A B}}$ and $\partial_{v v}{\alpha}$, but these can be found by taking $\partial_{v}$ derivatives of the above. We can then use the equation of motion $R_{\mu\nu} = \frac{2}{d-2} \Lambda g_{\mu \nu} - \frac{1}{d-2} g^{\rho \sigma} H_{\rho \sigma} g_{\mu \nu} + H_{\mu \nu} + O(l^N)$ to exchange $R_{\mu \nu}$ for $\frac{2}{d-2} \Lambda g_{\mu \nu}$ plus terms that are at least $O(l^2)$. We can repeat this process until all non-allowed terms are of order $l^N$, at which point we can neglect them for our calculation of the HKR entropy.

\subsection{\texorpdfstring{$s^{A}$}{sA} for Cubic Riemann Lagrangians} \label{app:sAcubic}

\begin{equation}
    s^{A} = l^4( k_{1} s^{A}_{even} + k_{2} s^{A}_{odd} )
\end{equation}
where
\begin{equation}
    \begin{split}
   s^{A}_{even} =& -18D^{B}{K^{A C}} D_{B}{\beta_{C}}+12D^{B}{K^{A C}} D_{C}{\beta_{B}}+9D_{B}{\beta_{C}} K^{A C} \beta^{B}-6D_{B}{\beta_{C}} K^{A B} \beta^{C}+\\
   &12D^{B}{K^{A}\,_{C}} K^{C E} \bar{K}_{B E}-12D_{B}{K^{A C}} K^{B E} \bar{K}_{C E}-12K^{A B} K_{B}\,^{C} \bar{K}_{C E} \beta^{E}+\\
   &6K^{A B} K^{C E} \bar{K}_{B E} \beta_{C}-24\mu^{A B} \mu^{C E} \partial_{r}{\beta_{C}} \partial_{v}{K_{B E}}+12K^{A B} K_{B}\,^{C} \partial_{r}{\beta_{C}}+\\
   &12\bar{K}^{B}\,_{C} \beta^{C} \mu^{A E} \partial_{v}{K_{B E}}-6D^{A}{\beta^{B}} \partial_{v}{\beta_{B}}-12\mu^{A B} \mu^{C E} \partial_{v}{\bar{K}_{B C}} \partial_{v}{\beta_{E}}+\\
   &6K^{B C} \bar{K}^{A}\,_{C} \partial_{v}{\beta_{B}}+6K^{B C} K_{B}\,^{E} \bar{K}^{A}\,_{E} \beta_{C}-3\beta^{A} \beta^{B} \partial_{v}{\beta_{B}}-12\alpha \mu^{A B} \partial_{v}{\beta_{B}}-\\
   &12K^{A B} \alpha \beta_{B}-3\beta^{B} \beta_{B} \mu^{A C} \partial_{v}{\beta_{C}}-6D^{B}{K_{B}\,^{C}} D^{A}{\beta_{C}}-12D^{B}{K_{B}\,^{C}} \mu^{A E} \partial_{v}{\bar{K}_{C E}}-\\
   &3D^{B}{K_{B}\,^{C}} \beta^{A} \beta_{C}-6D^{B}{D_{B}{\beta_{C}}} K^{A C}-12D^{B}\left(\partial_{v}{\bar{K}_{B C}}\right) K^{A C}-3D^{B}{\beta_{B}} K^{A C} \beta_{C}-\\
   &12D^{B}{K^{A C}} \partial_{v}{\bar{K}_{B C}}-3D^{B}{K^{A C}} \beta_{B} \beta_{C}+12K^{A B} \beta^{C} \partial_{v}{\bar{K}_{B C}}-6D_{B}{D^{A}{\beta_{C}}} K^{B C}-\\
   &12D_{B}\left(\partial_{v}{\bar{K}_{C E}}\right) K^{B C} \mu^{A E}-3D_{B}{\beta_{C}} K^{B C} \beta^{A}-3D_{B}{\beta^{A}} K^{C B} \beta_{C}
   \end{split}
\end{equation}
\begin{equation}
    \begin{split}
    s^{A}_{odd} =& -8D_{B}{\beta_{C}} \epsilon^{B C} \mu^{A E} \partial_{v}{\beta_{E}}-12D_{B}{\beta_{C}} K^{A E} \beta_{E} \epsilon^{B C}-8K_{B}\,^{C} \bar{K}_{C E} \epsilon^{B E} \mu^{A F} \partial_{v}{\beta_{F}}-\\
   &8K_{B}\,^{C} K^{A E} \bar{K}_{C F} \beta_{E} \epsilon^{B F}-16\epsilon^{B C} \mu^{A E} \partial_{r}{\beta_{B}} \partial_{v}{K_{C E}}-8K_{B}\,^{C} K^{A}\,_{C} \epsilon^{B E} \partial_{r}{\beta_{E}}+\\
   &8\bar{K}_{B C} \beta^{B} \epsilon^{C E} \mu^{A F} \partial_{v}{K_{E F}}+ 8K_{B}\,^{C} K^{A}\,_{C} \bar{K}_{E F} \beta^{E} \epsilon^{B F}-4D^{A}{\beta_{B}} \epsilon^{B C} \partial_{v}{\beta_{C}}-\\
   &8\epsilon^{B C} \mu^{A E} \partial_{v}{\bar{K}_{B E}} \partial_{v}{\beta_{C}} + 4K_{B}\,^{C} \bar{K}^{A}\,_{C} \epsilon^{B E} \partial_{v}{\beta_{E}}+4K_{B}\,^{C} K_{E}\,^{F} \bar{K}^{A}\,_{C} \beta_{F} \epsilon^{B E}-\\
   &2\beta_{B} \beta^{A} \epsilon^{B C} \partial_{v}{\beta_{C}} + 16D_{B}{K_{C}\,^{A}} \alpha \epsilon^{B C}+8K_{B}\,^{A} \alpha \beta_{C} \epsilon^{B C}+ 4D_{B}{K_{C}\,^{A}} \beta^{E} \beta_{E} \epsilon^{B C} -\\
   &12D^{B}{K_{C}\,^{E}} D_{B}{\beta_{E}} \epsilon^{A C}+8D^{B}{K_{C}\,^{E}} D_{E}{\beta_{B}} \epsilon^{A C}+6D_{B}{\beta_{C}} K_{E}\,^{C} \beta^{B} \epsilon^{A E}-\\
   &4D_{B}{\beta_{C}} K_{E}\,^{B} \beta^{C} \epsilon^{A E}+8D^{B}{K_{C E}} K^{C F} \bar{K}_{B F} \epsilon^{A E}-8D_{B}{K_{C}\,^{E}} K^{B F} \bar{K}_{E F} \epsilon^{A C}-\\
   &8K_{B}\,^{C} K_{C}\,^{E} \bar{K}_{E F} \beta^{F} \epsilon^{A B}+4K_{B}\,^{C} K^{E F} \bar{K}_{C F} \beta_{E} \epsilon^{A B}-16\epsilon^{A B} \mu^{C E} \partial_{r}{\beta_{C}} \partial_{v}{K_{B E}}+\\
   &8K_{B}\,^{C} K_{C}\,^{E} \epsilon^{A B} \partial_{r}{\beta_{E}}+8\bar{K}^{B}\,_{C} \beta^{C} \epsilon^{A E} \partial_{v}{K_{B E}}-4D_{B}{\beta^{C}} \epsilon^{A B} \partial_{v}{\beta_{C}}-\\
   &8\epsilon^{A B} \mu^{C E} \partial_{v}{\bar{K}_{B C}} \partial_{v}{\beta_{E}}+4K^{B C} \bar{K}_{C E} \epsilon^{A E} \partial_{v}{\beta_{B}}+4K^{B C} K_{B}\,^{E} \bar{K}_{E F} \beta_{C} \epsilon^{A F}-\\
   &2\beta_{B} \beta^{C} \epsilon^{A B} \partial_{v}{\beta_{C}}-8\alpha \epsilon^{A B} \partial_{v}{\beta_{B}}-8K_{B}\,^{C} \alpha \beta_{C} \epsilon^{A B}-2\beta^{B} \beta_{B} \epsilon^{A C} \partial_{v}{\beta_{C}}-\\
   &8D_{B}{K_{C}\,^{A}} R[\mu] \epsilon^{B C}-4K_{B}\,^{A} R[\mu] \beta_{C} \epsilon^{B C}-8D^{B}{K^{A}\,_{C}} K_{E}\,^{C} \bar{K}_{B F} \epsilon^{E F}+\\
   &8D_{B}{K^{A C}} K_{E}\,^{B} \bar{K}_{C F} \epsilon^{E F}-4K_{B}\,^{C} K^{A E} \bar{K}_{E F} \beta_{C} \epsilon^{B F}+16D_{B}{\bar{K}_{C}\,^{E}} \epsilon^{B C} \mu^{A F} \partial_{v}{K_{E F}}-\\
   &8D_{B}{\bar{K}_{C E}} K^{A F} K_{F}\,^{C} \epsilon^{B E}-8\bar{K}_{B}\,^{C} \beta_{E} \epsilon^{B E} \mu^{A F} \partial_{v}{K_{C F}}+4K^{A B} K_{B}\,^{C} \bar{K}_{C E} \beta_{F} \epsilon^{E F}+\\
   &4D_{B}{K_{C}\,^{E}} D^{A}{\beta_{E}} \epsilon^{B C}+8D_{B}{K_{C}\,^{E}} \epsilon^{B C} \mu^{A F} \partial_{v}{\bar{K}_{E F}}-8D_{B}{K_{C E}} K^{C F} \bar{K}^{A}\,_{F} \epsilon^{B E}-\\
   &4K_{B}\,^{C} K_{C}\,^{E} \bar{K}^{A}\,_{E} \beta_{F} \epsilon^{B F}+2D_{B}{K_{C}\,^{E}} \beta^{A} \beta_{E} \epsilon^{B C}+4D^{B}{K_{B C}} D^{A}{\beta_{E}} \epsilon^{C E}+\\
   &8D^{B}{K_{B C}} \epsilon^{C E} \mu^{A F} \partial_{v}{\bar{K}_{E F}}+2D^{B}{K_{B C}} \beta_{E} \beta^{A} \epsilon^{C E}-4D^{B}{D_{B}{\beta_{C}}} K_{E}\,^{A} \epsilon^{C E}-\\
   &8D^{B}\left(\partial_{v}{\bar{K}_{B C}}\right) K_{E}\,^{A} \epsilon^{C E}+2D^{B}{\beta_{B}} K_{C}\,^{A} \beta_{E} \epsilon^{C E}+2D_{B}{\beta_{C}} K_{E}\,^{A} \beta^{C} \epsilon^{B E}+\\
   &4D^{B}{K_{C}\,^{A}} D_{B}{\beta_{E}} \epsilon^{C E}+8D^{B}{K_{C}\,^{A}} \epsilon^{C E} \partial_{v}{\bar{K}_{B E}}+2D^{B}{K_{C}\,^{A}} \beta_{B} \beta_{E} \epsilon^{C E}-\\
   &8K_{B}\,^{A} \beta^{C} \epsilon^{B E} \partial_{v}{\bar{K}_{C E}}-4D_{B}{D^{A}{\beta_{C}}} K_{E}\,^{B} \epsilon^{C E}-8D_{B}\left(\partial_{v}{\bar{K}_{C E}}\right) K_{F}\,^{B} \epsilon^{C F} \mu^{A E}-\\
   &2D_{B}{\beta_{C}} K_{E}\,^{B} \beta^{A} \epsilon^{C E}+2D_{B}{\beta^{A}} K_{C}\,^{B} \beta_{E} \epsilon^{C E}-4D_{B}{D_{C}{\beta_{E}}} K^{A E} \epsilon^{B C}-\\
   &6D_{B}{\beta_{C}} K^{A C} \beta_{E} \epsilon^{B E}-4D_{B}{K^{A C}} D_{E}{\beta_{C}} \epsilon^{B E}-8D_{B}{K^{A C}} \epsilon^{B E} \partial_{v}{\bar{K}_{C E}}-\\
   &2D_{B}{K^{A C}} \beta_{C} \beta_{E} \epsilon^{B E}+8K^{A B} \beta_{C} \epsilon^{C E} \partial_{v}{\bar{K}_{B E}}-4D_{B}{D^{A}{\beta_{C}}} K_{E}\,^{C} \epsilon^{B E}-\\
   &8D_{B}\left(\partial_{v}{\bar{K}_{C E}}\right) K_{F}\,^{C} \epsilon^{B F} \mu^{A E}-2D_{B}{\beta_{C}} K_{E}\,^{C} \beta^{A} \epsilon^{B E}-2D_{B}{\beta^{A}} K_{C}\,^{E} \beta_{E} \epsilon^{B C}-\\
   &8D^{B}{K_{C}\,^{E}} \epsilon^{A C} \partial_{v}{\bar{K}_{B E}}-2D^{B}{K_{C}\,^{E}} \beta_{B} \beta_{E} \epsilon^{A C}+8K_{B}\,^{C} \beta^{E} \epsilon^{A B} \partial_{v}{\bar{K}_{C E}}-\\
   &4D_{B}{D_{C}{\beta_{E}}} K^{B E} \epsilon^{A C}-8D_{B}\left(\partial_{v}{\bar{K}_{C E}}\right) K^{B C} \epsilon^{A E}-2D_{B}{\beta_{C}} K^{B C} \beta_{E} \epsilon^{A E}-\\
   &2D_{B}{\beta_{C}} K^{E B} \beta_{E} \epsilon^{A C}-4D^{B}{K_{B}\,^{C}} D_{E}{\beta_{C}} \epsilon^{A E}-8D^{B}{K_{B}\,^{C}} \epsilon^{A E} \partial_{v}{\bar{K}_{C E}}-\\
   &2D^{B}{K_{B}\,^{C}} \beta_{C} \beta_{E} \epsilon^{A E}-4D^{B}{D_{B}{\beta_{C}}} K_{E}\,^{C} \epsilon^{A E}-8D^{B}\left(\partial_{v}{\bar{K}_{B C}}\right) K_{E}\,^{C} \epsilon^{A E}-\\
   &2D^{B}{\beta_{B}} K_{C}\,^{E} \beta_{E} \epsilon^{A C}-8D_{B}\left(\partial_{v}{\bar{K}_{C E}}\right) K^{A C} \epsilon^{B E}
    \end{split}
\end{equation}

\subsection{\texorpdfstring{$s^{(6)A}$}{sA6} for Quartic Riemann Lagrangian} \label{app:s6A}

\begin{equation}
    \begin{split}
        s^{(6)A} =& k \Big[ 8\mu^{A B} \partial_{v}{\beta_{B}} {R[\mu]}^{2}-16K \bar{K} R[\mu] \mu^{A E} \partial_{v}{\beta_{E}}+16K^{B C} \bar{K}_{B C} R[\mu] \mu^{A E} \partial_{v}{\beta_{E}}-\\
        &16K^{A B} K \bar{K} R[\mu] \beta_{B}+16K^{A B} K^{C E} \bar{K}_{C E} R[\mu] \beta_{B}+16K^{B C} K_{B C} \bar{K}^{E F} \bar{K}_{E F} \mu^{A G} \partial_{v}{\beta_{G}}+\\
        &16K^{A B} K^{C E} K_{C E} \bar{K}^{F G} \bar{K}_{F G} \beta_{B}+16K^{B C} K^{E F} \bar{K}_{B C} \bar{K}_{E F} \mu^{A G} \partial_{v}{\beta_{G}}+\\
        &16K^{A B} K^{C E} K^{F G} \bar{K}_{C E} \bar{K}_{F G} \beta_{B}-32K^{B C} K_{B}\,^{E} \bar{K}_{C}\,^{F} \bar{K}_{E F} \mu^{A G} \partial_{v}{\beta_{G}}-\\
        &32K^{A B} K^{C E} K_{C}\,^{F} \bar{K}_{E}\,^{G} \bar{K}_{F G} \beta_{B}+64D^{B}{K^{C E}} D_{B}{\bar{K}_{C E}} \mu^{A F} \partial_{v}{\beta_{F}}-\\
        &64D^{B}{K^{C E}} D_{C}{\bar{K}_{B E}} \mu^{A F} \partial_{v}{\beta_{F}}-32D_{B}{\bar{K}_{C E}} K^{C E} \beta^{B} \mu^{A F} \partial_{v}{\beta_{F}}+\\
        &32D_{B}{\bar{K}_{C E}} K^{B C} \beta^{E} \mu^{A F} \partial_{v}{\beta_{F}}+64D^{B}{K^{C E}} D_{B}{\bar{K}_{C E}} K^{A F} \beta_{F}-64D^{B}{K^{C E}} D_{C}{\bar{K}_{B E}} K^{A F} \beta_{F}-\\
        &32D_{B}{\bar{K}_{C E}} K^{A F} K^{C E} \beta_{F} \beta^{B}+32D_{B}{\bar{K}_{C E}} K^{A F} K^{B C} \beta_{F} \beta^{E}+32D^{B}{K^{C E}} \bar{K}_{C E} \beta_{B} \mu^{A F} \partial_{v}{\beta_{F}}-\\
        &32D^{B}{K^{C E}} \bar{K}_{B E} \beta_{C} \mu^{A F} \partial_{v}{\beta_{F}}-16K^{B C} \bar{K}_{B C} \beta^{E} \beta_{E} \mu^{A F} \partial_{v}{\beta_{F}}+16K^{B C} \bar{K}_{C E} \beta_{B} \beta^{E} \mu^{A F} \partial_{v}{\beta_{F}}+\\
        &32D^{B}{K^{C E}} K^{A F} \bar{K}_{C E} \beta_{B} \beta_{F}-32D^{B}{K^{C E}} K^{A F} \bar{K}_{B E} \beta_{C} \beta_{F}-16K^{A B} K^{C E} \bar{K}_{C E} \beta_{B} \beta^{F} \beta_{F}+\\
        &16K^{A B} K^{C E} \bar{K}_{E F} \beta_{B} \beta_{C} \beta^{F}+32\mu^{A B} \mu^{C E} \mu^{F G} \partial_{r}{\bar{K}_{C F}} \partial_{v}{K_{E G}} \partial_{v}{\beta_{B}}-\\
        &32K^{B C} K_{B}\,^{E} \mu^{A F} \partial_{r}{\bar{K}_{C E}} \partial_{v}{\beta_{F}}+32K^{A B} \beta_{B} \mu^{C E} \mu^{F G} \partial_{r}{\bar{K}_{C F}} \partial_{v}{K_{E G}}-\\
        &32K^{A B} K^{C E} K_{C}\,^{F} \beta_{B} \partial_{r}{\bar{K}_{E F}}-32\bar{K}^{B C} \bar{K}^{E}\,_{C} \mu^{A F} \partial_{v}{K_{B E}} \partial_{v}{\beta_{F}}-32K^{A B} \bar{K}^{C E} \bar{K}^{F}\,_{E} \beta_{B} \partial_{v}{K_{C F}}-\\
        &16D^{B}{\beta^{C}} D_{B}{\beta_{C}} \mu^{A E} \partial_{v}{\beta_{E}}+32D^{B}{\beta^{C}} D_{C}{\beta_{B}} \mu^{A E} \partial_{v}{\beta_{E}}+32D^{B}{\beta_{C}} K^{C E} \bar{K}_{B E} \mu^{A F} \partial_{v}{\beta_{F}}+\\
        &32D^{B}{\beta_{C}} K^{A E} K^{C F} \bar{K}_{B F} \beta_{E}-64D_{B}{\beta^{C}} K^{B E} \bar{K}_{C E} \mu^{A F} \partial_{v}{\beta_{F}}-64D_{B}{\beta^{C}} K^{A E} K^{B F} \bar{K}_{C F} \beta_{E}+\\
        &64K^{B C} K^{E F} \bar{K}_{B E} \bar{K}_{C F} \mu^{A G} \partial_{v}{\beta_{G}}+64K^{A B} K^{C E} K^{F G} \bar{K}_{C F} \bar{K}_{E G} \beta_{B}+64D^{B}{\beta^{C}} \mu^{A E} \partial_{v}{\bar{K}_{B C}} \partial_{v}{\beta_{E}}+\\
        &16D_{B}{\beta_{C}} \beta^{B} \beta^{C} \mu^{A E} \partial_{v}{\beta_{E}}+64\mu^{A B} \mu^{C E} \mu^{F G} \partial_{v}{\bar{K}_{C F}} \partial_{v}{\bar{K}_{E G}} \partial_{v}{\beta_{B}}-64K^{B C} \bar{K}_{B}\,^{E} \mu^{A F} \partial_{v}{\bar{K}_{C E}} \partial_{v}{\beta_{F}}-\\
        &64K^{A B} K^{C E} \bar{K}_{C}\,^{F} \beta_{B} \partial_{v}{\bar{K}_{E F}}+32\beta^{B} \beta^{C} \mu^{A E} \partial_{v}{\bar{K}_{B C}} \partial_{v}{\beta_{E}}+6\beta^{B} \beta_{B} \beta^{C} \beta_{C} \mu^{A E} \partial_{v}{\beta_{E}}-\\
        &32\mu^{A B} \mu^{C E} \partial_{r}{\beta_{C}} \partial_{v}{\beta_{B}} \partial_{v}{\beta_{E}}-32K^{B C} \beta_{B} \mu^{A E} \partial_{r}{\beta_{C}} \partial_{v}{\beta_{E}}-32K^{A B} \beta_{B} \mu^{C E} \partial_{r}{\beta_{C}} \partial_{v}{\beta_{E}}-\\
        &32K^{A B} K^{C E} \beta_{B} \beta_{C} \partial_{r}{\beta_{E}}+16\bar{K}_{B}\,^{C} \beta^{B} \mu^{A E} \partial_{v}{\beta_{C}} \partial_{v}{\beta_{E}}+16K^{A B} \bar{K}_{C}\,^{E} \beta_{B} \beta^{C} \partial_{v}{\beta_{E}}+\\
        &32\mu^{A B} \partial_{v}{\beta_{B}} {\alpha}^{2}+16\alpha \beta^{B} \beta_{B} \mu^{A C} \partial_{v}{\beta_{C}}+8D^{B}{K^{A}\,_{B}} {R[\mu]}^{2}-16D^{B}{K^{A}\,_{B}} D^{C}{\beta^{E}} D_{C}{\beta_{E}}+\\
        &32D^{B}{K^{A}\,_{B}} D^{C}{\beta^{E}} D_{E}{\beta_{C}}+64D^{B}{K^{A}\,_{B}} D^{C}{\beta^{E}} \partial_{v}{\bar{K}_{C E}}+16D^{B}{K^{A}\,_{B}} D_{C}{\beta_{E}} \beta^{C} \beta^{E}+\\
        &64D^{B}{K^{A}\,_{B}} \mu^{C E} \mu^{F G} \partial_{v}{\bar{K}_{C F}} \partial_{v}{\bar{K}_{E G}}+32D^{B}{K^{A}\,_{B}} \beta^{C} \beta^{E} \partial_{v}{\bar{K}_{C E}}+6D^{B}{K^{A}\,_{B}} \beta^{C} \beta_{C} \beta^{E} \beta_{E}+\\
        &32D^{B}{K^{A}\,_{B}} {\alpha}^{2}+16D^{B}{K^{A}\,_{B}} \alpha \beta^{C} \beta_{C}+16D_{B}{R[\mu]} K^{A B} R[\mu]-32D_{B}{D_{C}{\beta_{E}}} D^{C}{\beta^{E}} K^{A B}+\\
        &64D_{B}{D_{C}{\beta_{E}}} D^{E}{\beta^{C}} K^{A B}+64D_{B}{D^{C}{\beta^{E}}} K^{A B} \partial_{v}{\bar{K}_{C E}}+16D_{B}{D_{C}{\beta_{E}}} K^{A B} \beta^{C} \beta^{E}+\\
        &64D^{B}{\beta^{C}} D_{E}\left(\partial_{v}{\bar{K}_{B C}}\right) K^{A E}+128D_{B}\left(\partial_{v}{\bar{K}_{C E}}\right) K^{A B} \mu^{C F} \mu^{E G} \partial_{v}{\bar{K}_{F G}}+\\
        &32D_{B}\left(\partial_{v}{\bar{K}_{C E}}\right) K^{A B} \beta^{C} \beta^{E}+16D_{B}{\beta^{C}} D_{C}{\beta_{E}} K^{A B} \beta^{E}+64D_{B}{\beta^{C}} K^{A B} \beta^{E} \partial_{v}{\bar{K}_{C E}}+\\
        &24D_{B}{\beta_{C}} K^{A B} \beta^{C} \beta_{E} \beta^{E}+16D_{B}{\beta^{C}} D_{E}{\beta_{C}} K^{A B} \beta^{E}+64D_{B}{\alpha} K^{A B} \alpha+\\
        &16D_{B}{\alpha} K^{A B} \beta^{C} \beta_{C}+32D_{B}{\beta_{C}} K^{A B} \alpha \beta^{C} \Big]
    \end{split}
\end{equation}
\chapter{Concluding Remarks}\label{ChapterConclusion}

This thesis has been concerned with proving the mathematical health and physical properties of gravitational EFTs. 

On the first goal, we have shown that a "modified harmonic" gauge that was originally formulated for use in scalar-tensor EFTs can be generalized to provide a well-posed initial value problem for the leading order Einstein-Maxwell EFT, so long as the initial data is weakly coupled, i.e., when the 4-derivative terms in the equations of motion are initially small compared to the 2-derivative terms. This provides more evidence that gravitational EFTs can be formulated in a mathematically healthy way, and that such a formulation is robust to the inclusion of physical matter fields. However, it is an open question as to how large the higher derivative terms can be before the formulation loses its strong hyperbolicity and is no longer well-posed. Moreover, we can only hope to apply this gauge to specific EFTs that produce second order equations of motion, for otherwise we cannot apply the same notion of strong hyperbolicity.

In very recent work \cite{Figueras:2024bba}, an alternative method for proving the well-posedness of EFTs in their regime of validity has been proposed. It applies to any vacuum gravitational EFT truncated at any order $l^N$. The idea is to perform field redefinitions on the truncated theory so as to produce specific "regularising" terms in the Lagrangian that recast the equations of motion in harmonic gauge into a system of non-linear second order wave equations that are manifestly well-posed, regardless of weak coupling. So long as the solution is in the regime of validity of the EFT, the regularising terms should not affect the dynamics at order $l^N$ or below, and thus the original truncated theory is well-posed to the same degree of accuracy to which we know the EFT. It would be interesting to apply these ideas to gravitational EFTs that include matter fields such as the electromagnetic field, to study whether the method is similarly robust.

On the second goal, we have made substantial progress on proving the laws of black hole mechanics. In particular, we have proved the zeroth and second laws for the EFT of gravity coupled to electromagnetism and a real uncharged scalar field, up to the order to which we know our EFT. These are proved merely under the assumption that the solution lies in the regime of validity of the EFT. Together with the work of Wald on the first law, they provide a complete set of proofs of the laws in these EFTs. The definition of dynamical black hole entropy which is constructed along the way is independent of electromagnetic gauge for theories with any number of derivatives, and is independent of the choice of coordinates for theories with up to 6 derivatives (order $l^4$). It reduces to the standard Bekenstein-Hawking entropy in 2-derivative GR, and reduces to the Wald entropy in equilibrium for any number of derivatives. In addition, we have shown the zeroth law continues to hold if the scalar is charged (under an additional technical assumption), and there is strong motivation to think the second law would hold. This suggests a more general result involving EFTs of gravity with any matter fields that satisfy the NEC at 2-derivative level may be provable. For example, it would be interesting to extend the result to Yang-Mills fields.

Our proofs of the zeroth and second laws are perhaps not as general as we would like them to be. For the zeroth law we excluded certain horizon topologies. For the second law we required our horizon to be smooth. Recent work \cite{Gadioux:2023} has considered the case of non-smooth horizons and suggested there may be additional contributions to black hole entropy motivated by quantum entanglement entropy. They also demonstrate that certain terms in the HKR entropy current can diverge when integrated over non-smooth features on the horizon. Furthermore, our definition of the entropy is dependent on our choice of GNCs above order $l^4$, as we saw for a specific 8-derivative EFT, which is an undesirable quality in a definition of entropy. 

Related to this is the criticism of \cite{Hollands:2024vbe} regarding the failure of our proposal for the entropy to be "cross-section continuous". This can be described as follows. Consider changing the GNC affine parameter $v'=a(x^A)v$ on $\cH$, as we did when discussing gauge invariance in the previous Chapter. If we take $a(x^a)$ to be very close to 1, then the foliations $C(v)$ and $C'(v')$ will describe surfaces that are very close. Our proposal for the entropy would then be said to be cross-section continuous if the corresponding $S(v)$ and $S'(v')$ were also necessarily very close. In Appendix C3 of \cite{Hollands:2024vbe}, they make the following argument as to why our proposal will not satisfy this. The GNC quantity $\beta_A$ transforms under the change of affine parameter as 
\begin{equation}
    \beta'_A = \beta_A + 2 D_A \log a - 2v K_{A}^{\,\,\,\,\,B}D_{B} \log a
\end{equation}
on $\cH$. Whilst still taking $a(x^A)$ very close to 1, we can take its $D_A$ derivatives to be extremely large, so $C'(v')$ is essentially a very wiggly perturbation of $C(v)$. Since $S'(v')$ will generally involve the integral of $\beta'_A$, we should then expect $S'(v')$ to be very different from $S(v)$. One can also make this argument with $\Bar{K}_{AB}$.

Whilst this is an interesting point of criticism, we dispute its relevance from an EFT perspective. This is because taking such a wiggly $C'(v')$ would take us out of the regime of validity of the EFT: $\beta_A$ and its derivatives (and other GNC quantities) would have to become extremely large, so the length/time scales $L'$ associated with the solution in these GNCs would be extremely small, eventually comparable to the UV-length scale $l$. Thus this choice of coordinates should be excluded from our analysis.

A final interesting open question regards the uniqueness of dynamical black hole entropy. The procedure we have outlined to produce our definition of black hole entropy is far from unique. At many of the steps there are choices to make, such as how to label $P_1$ and $P_2$, which $D_A$ or $\partial_v$ derivative to move over in which order, or whether to choose $X^{(m) A B}$ to be symmetric. Making different choices may produce different expressions for the entropy density $s^v$, even though the resulting entropy $S(v)$ will satisfy the second law in the same fashion. Furthermore, we mentioned that   EFT field redefinitions $g_{\mu \nu} \rightarrow g_{\mu \nu}  + l^2( a \Lambda g_{\mu \nu}+ b R g_{\mu \nu} + c R_{\mu \nu})+...$ could be used to make the Lagrangian simpler, for example to eliminate the 4-derivative terms in $d=4$ vacuum gravity. HKR \cite{Hollands:2022} noted that the HKR entropy is not invariant under these field redefinitions, even though the Lagrangians describe the same theory up to a given derivative order.

Is this non-uniqueness a problem? In general, thermodynamic entropy is known to suffer from ambiguities in non-equilibrium situations. For example, in relativistic viscous fluids, it is known there are multi-parameter families of entropy currents that satisfy the second law \cite{Bhattacharyya:2008xc}\cite{Romatschke:2009kr}\cite{Bhattacharyya:2013lha}. Maybe the non-uniqueness in our definition of dynamical black hole entropy just maps between different possible definitions of the entropy, all of which satisfy a second law. More work will need to be done in this area before a satisfactory conclusion is reached. For example, it would be interesting to study the Generalized Second Law for this definition of entropy.












\renewcommand\bibname{Bibliography}

\end{document}